\documentclass[11pt]{report}

\usepackage{natbib}
\usepackage{ifthen}
\newboolean{forlibrary}
\setboolean{forlibrary}{true}	
\usepackage{gmudissertation}
\usepackage{graphicx}
\usepackage{dcolumn}
\usepackage[usenames]{color}
\usepackage{colortbl}
\usepackage{psfrag}
\usepackage{alltt}
\usepackage{amsmath}
\usepackage{amsfonts}
\usepackage{amssymb}
\usepackage{amsthm}
\usepackage[compatible,notintoc]{nomencl}
\usepackage{ifthen}
\usepackage[normalem]{ulem}
\usepackage{setspace}
\usepackage{jbakosi_thesis}
\usepackage[breaklinks,colorlinks,letterpaper,dvips]{hyperref}
\hypersetup{citecolor=blue,
            linkcolor=blue,
	    urlcolor=blue,
	    pdftitle=PDF modeling of turbulent flows on unstructured grids,
	    pdfauthor=Jozsef Bakosi,
	    pdfkeywords=turbulent flow; scalar dispersion; probability density function method;
	                particle method; particle tracking; Langevin equation; Monte-Carlo method;
			unstructured grids; article-in-cell method; finite element method;
			parallel computing; OpenMP}
\usepackage{hypernat}

\makenomenclature

\beforedoc

\begin{document}

\title{PDF modeling of turbulent flows on unstructured grids}
\onelinetitle{PDF modeling of turbulent flows on unstructured grids}
\author{J\'ozsef Bakosi}
\degree{Doctor of Philosophy}
\doctype{Dissertation}
\dept{Computational and Data Sciences}
\field{Computational Sciences and Informatics}
\college{College of Science}

\firstdeg{Master of Science}
\firstdegschool{University of Miskolc}
\firstdegyear{1999}

\degreeyear{2008}
\degreesemester{Spring}

\advisor{Dr.\ Zafer Boybeyi}
\firstmember{Dr.\ Nash'at Ahmad}
\secondmember{Dr.\ Pasquale Franzese}
\thirdmember{Dr.\ Rainald L\"ohner}
\depthead{Dr.\ Dimitrios Papaconstantopoulos}
\programdirector{Dr.\ Peter Becker}
\deanCOS{Dr.\ Vikas Chandhoke}

\scannedsignaturepage	

\titlepage

\copyrightpage

\dedicationpage

\begin{center}
\emph{This work is dedicated to my family:\\my parents, my sister and my grandmother}
\end{center}

\acknowledgementspage

\noindent I would like to thank my advisor Dr.\ Zafer Boybeyi, who provided me the opportunity to embark upon this journey. Without his continuous moral and scientific support at every level, this work would not have been possible. Taking on the rather risky move of giving me 100\% freedom in research from the beginning certainly deserves my greatest acknowledgments. I am also thoroughly indebted to Dr.\ Pasquale Franzese for his guidance of this research, for the many lengthy -- sometimes philosophical -- discussions on turbulence and other topics, for being always available and for his painstaking drive with me through the dungeons of research, scientific publishing and aesthetics. My gratitude extends to Dr.\ Rainald L\"ohner from whom I had the opportunity to take classes in CFD and to learn how not to get lost in the details. I found his weekly seminars to be one of the best opportunities to learn critical and down-to-earth thinking. I am also indebted to Dr.\ Nash'at Ahmad for showing me the example that with a careful balance of school, hard work and family nothing is impossible.

I thank Dr.\ Thomas Dreeben for the many helpful and insightful discussions on the velocity model and elliptic relaxation. I am also grateful to the reviewers of our journal papers for their valuable comments and suggestions on our initial manuscripts.

I will always remember my professors at home at the Departments of Mathematics, Mechanics, Physics, and Fluid and Heat Engineering at the University of Miskolc, who taught me the foundations of engineering and science: Drs.\ Bert\'oti Edg\'ar, K\'alovics Ferenc, Koz\'ak Imre, M\'esz\'aros J\'ozsef, P\'aczelt Istv\'an, Ront\'o Mikl\'os, Szab\'o Szil\'ard, Szeidl Gy\"orgy, Tak\'acs Csaba and Vince Endre. I further extend my thanks to Prof.\ T\'oth L\'aszl\'o, who first gave me the opportunity to study abroad and introduced me to research. My admiration towards all of them turned into a passion for science.

I am also thankful to Dr.\ Sergei Ivanov, my mentor at Fraunhofer in Boston, who encouraged me to do a Ph.D.

I would also like to thank for the continous support of all my friends throughout the years. My friend from college, Feh\'er Zolt\'an, for keeping me sane with our long, challenging and fruitful discussions every weekend. Sunil Kumar Appanaboyina for the countless philosophical discussions on culture, life, religion, food and women. Jarek Pietrzykowski who has always been a partner in crime. Fernando Mut for the many discussions on coding, performance and for suggesting the algorithm for the particle search. My housemates Houmam Ali, Andrew Krapf and Chris Lafontaine who made me feel at home right back in college. The wonderful people here at the Laboratory for Atmospheric Hazard Modeling who have always been helpful and created an atmosphere so that I never felt alone: Dr.\ Guido Cervone, Laura Clemente, Yasemin Ezber, John Lindeman, Jacek Radzikowski and Priyanka Roy. Special thanks to my international student advisor, shuttle and jazz buddy, Amy Moffitt.

But my greatest thank you goes to my family and friends at home, who never ceased to believe in me, even at times when I myself almost did.

\tableofcontents
\listoftables
\listoffigures
\listnomenclature

\abstractpage

\noindent In probability density function (PDF) methods of turbulent flows, the joint PDF of several flow variables is computed by numerically integrating a system of stochastic differential equations for Lagrangian particles.  Because the technique solves a transport equation for the PDF of the velocity and scalars, a mathematically exact treatment of advection, viscous effects and arbitrarily complex chemical reactions is possible; these processes are treated without closure assumptions. A set of algorithms is proposed to provide an efficient solution of the PDF transport equation modeling the joint PDF of turbulent velocity, frequency and concentration of a passive scalar in geometrically complex configurations. An unstructured Eulerian grid is employed to extract Eulerian statistics, to solve for quantities represented at fixed locations of the domain and to track particles. All three aspects regarding the grid make use of the finite element method. Compared to \emph{hybrid methods}, the current methodology is \emph{stand-alone}, therefore it is consistent both numerically and at the level of turbulence closure without the use of consistency conditions. Since both the turbulent velocity and scalar concentration fields are represented in a stochastic way, the method allows for a direct and close interaction between these fields, which is beneficial in computing accurate scalar statistics.%
\nomenclature[A]{PDF}{probability density function}%
\nomenclature[A]{FEM}{finite element method}%

Boundary conditions implemented along solid bodies are of the free-slip and no-slip type\hfill

\abstractmultiplepage
\noindent without the need for ghost elements. Boundary layers at no-slip boundaries are either fully resolved down to the viscous sublayer, explicitly modeling the high anisotropy and inhomogeneity of the low-Reynolds-number wall region without damping or wall-functions or specified via logarithmic wall-functions. As in moment closures and large eddy simulation, these wall-treatments provide the usual trade-off between resolution and computational cost as required by the given application.

Particular attention is focused on modeling the dispersion of passive scalars in inhomogeneous turbulent flows. Two different micromixing models are investigated that incorporate the effect of small scale mixing on the transported scalar: the widely used interaction by exchange with the mean and the interaction by exchange with the conditional mean model. An adaptive algorithm to compute the velocity-conditioned scalar mean is proposed that homogenizes the statistical error over the sample space with no assumption on the shape of the underlying velocity PDF. The development also concentrates on a generally applicable micromixing timescale for complex flow domains.%
\nomenclature[A]{IEM}{interaction by exchange with the mean}%
\nomenclature[A]{IECM}{interaction by exchange with the conditional mean}%

Several newly developed algorithms are described in detail that facilitate a stable numerical solution in arbitrarily complex flow geometries, including a stabilized mean-pressure projection scheme, the estimation of conditional and unconditional Eulerian statistics and their derivatives from stochastic particle fields employing finite element shapefunctions, particle tracking through unstructured grids, an efficient particle redistribution procedure and techniques related to efficient random number generation.

The algorithm is validated and tested by computing three different turbulent flows: the fully developed turbulent channel flow, a street canyon (or cavity) flow and the turbulent wake behind a circular cylinder at a sub-critical Reynolds number.

The solver has been parallelized and optimized for shared memory and multi-core architectures using the OpenMP standard. Relevant aspects of performance and parallelism on cache-based shared memory machines are discussed and presented in detail. The methodology shows great promise in the simulation of high-Reynolds-number incompressible inert or reactive turbulent flows in realistic configurations.

\startofchapters

\chapter[Introduction]{\\Introduction}
\label{chap:introduction}

In engineering industry and atmospheric transport and dispersion modeling there is an increasing use of computational methods to calculate complex turbulent flow fields. Many of these computations depend on the \(k\)--\(\varepsilon\) turbulence model \citep{Jones_72,Bacon_00}, while some are based on second-moment closures \citep{Rotta_51,Launder_75,Hanjalic_72,Speziale_91}. The aim of these statistical methods is to predict the first and second moments of the turbulent velocity field, respectively. In large eddy simulation (LES) the large scale three-dimensional unsteady motions are represented exactly, while the small-scale motions are parameterized. As long as the transport-controlling processes of interest (eg.\ mass, momentum and heat transfer in shear flows) are resolved, LES predictions can be expected to be insensitive to the details of residual-scale modeling. In applications such as high-Reynolds-number turbulent combustion or near-wall flows, however, where the important rate-controlling processes occur below the resolved scales, the residual-scale models directly influence the model predictions. Since there is no universally `best' methodology that is applicable for every type of practical flow, it is valuable to develop improvements for the full range of turbulence modeling  approaches.%
\nomenclature[A]{LES}{large eddy simulation}%

The development of probability density function (PDF) methods is an effort to provide a higher-level statistical description of turbulent flows. The mean velocity and Reynolds stresses are statistics of (and can be obtained from) the PDF of velocity. In PDF methods, a transport equation is solved directly for the PDF of the turbulent velocity field, rather than for its moments as in Reynolds stress closures. Therefore, in principle, a more complete statistical description can be obtained. While for some flows (e.g.\ homogeneous turbulence) this higher-level description may provide little benefit over second moment closures, in general the fuller description is beneficial in allowing more processes to be treated exactly and in providing more information that can be used in the construction of closure models. Convection, for example, can be exactly represented mathematically in the PDF framework, eliminating the need for a closure assumption \citep{Pope_00}. Similarly, defining the joint PDF of velocity and species concentrations in a chemically reactive turbulent flow allows for the treatment of chemical reactions without the burden of closure assumptions for the highly nonlinear chemical source terms \citep{Fox_03}. This latter advantage has been one of the most important incentives for the development of PDF methods, since previous attempts to provide moment closures for this term resulted in errors of several orders of magnitude \citep{Pope_90}. 

The development of PDF methods has mostly been centered on chemically reactive turbulent flows on simple geometries, \citep[e.g.][]{Tang_00,Xu_00}, although applications to more complex configurations \citep{James_02,Subramaniam_00} as well as to atmospheric flows \citep{Heinz_98,Cassiani_et_al_05b} have also appeared. A large variety of compressible and incompressible laminar flows bounded by bodies of complex geometries have been successfully computed using unstructured grids \citep{Lohner_01}. The flexibility of these gridding techniques has also been exploited recently in mesoscale atmospheric modeling \citep{Bacon_00}. Significant advances in automatic unstructured grid generation \citep{Lohner_00}, sophisticated data structures and algorithms, automatic grid refinement and coarsening techniques \citep{Shostko_95} in recent years have made unstructured grids a common and convenient choice of spatial discretization in computational physics. The success of unstructured grids seems to warrant exploiting their advantages in conjunction with PDF modeling. For reasons to be elaborated on later, in PDF methods the usual choice of representation is the Lagrangian framework with a numerical method employing a large number of Lagrangian particles. A natural way to combine the advantages of existing traditional Eulerian computational fluid dynamics (CFD) codes with PDF methods, therefore, is to develop hybrid methods.%
\nomenclature[A]{CFD}{computational fluid dynamics}%

Using structured grids, a hybrid finite-volume (FV)/particle method has been developed by \citet{Muradoglu_99} and \citet{Jenny_01}, wherein the mean velocity and pressure fields are supplied by the FV code to the particle code, which in turn computes the Reynolds stress, scalar fluxes and reaction terms. Different types of hybrid algorithms are possible depending on which quantities are computed in the Eulerian and Lagrangian frameworks. For a list of approaches see \citet{Muradoglu_99}. Another line of research has been centered on the combination of LES with PDF methods \citep{Givi_89,Madnia_93}. This approach is based on the definition of the filtered density function (FDF) \citep{Pope_90} which is used to provide closure at the residual scale to the filtered LES equations. Depending on the flow variables included in the joint FDF, different variants of the method have been proposed providing a probabilistic treatment at the residual scale for species compositions \citep{Colucci_98}, velocity \citep{Gicquel_02} and velocity and scalars \citep{Sheikhi_03}. A common feature of these hybrid methods is that certain consistency conditions have to be met, since some fields are computed in both the Eulerian and Lagrangian frameworks. Additionally to the works cited above, further advances on consistency conditions and correction algorithms for hybrid FV/particle codes have been reported by \citet{Muradoglu_01} and \citet{Zhang_04}, whose authors also extend the hybrid formulation to unstructured grids. Following that line, a hybrid algorithm for unstructured multiblock grids has recently been proposed by \citet{Rembold_06}. Beside enforcing the consistency of redundantly computed fields, hybrid methods also have to be designed to ensure consistency at the level of the turbulence closure between the two frameworks. For example, the simplified Langevin model (SLM) \citep{Haworth_86} is equivalent to Rotta's model at the Reynolds stress level \citep{Pope_94}. Thus the use of a \(k\)--\(\varepsilon\) model in the Eulerian framework and of a SLM PDF model in the Lagrangian framework cannot be consistent \citep{Muradoglu_99}. In the current work, a different approach is taken by representing all turbulent fields by Lagrangian particles and employing the grid (a) to compute only inherently Eulerian quantities (that are only represented in the Eulerian sense), (b) to extract Eulerian statistics and (c) to locate particles throughout the domain. Because the resulting method is not a hybrid one, none of the fields are computed redundantly and the computation can remain fully consistent without the need for correction algorithms. We employ the finite element method (FEM) in all three aspects mentioned above in conjunction with Eulerian grids. The combined application of the FEM and the decoupling of the Eulerian and Lagrangian fields also have important positive consequences regarding particle boundary conditions as compared to the ``flux-view'' of FV methods.%
\nomenclature[A]{FV}{finite volume}%
\nomenclature[A]{FDF}{filtered density function}%
\nomenclature[A]{SLM}{simplified Langevin model}%

In the case of turbulent flows around complex geometries the presence of walls requires special treatment, since traditional turbulence models are developed for high Reynolds numbers and need to be modified in the vicinity of walls. This is necessary because the Reynolds number approaches zero at the wall, the highest shear rate occurs near the wall and the impermeability condition on the wall-normal velocity affects the flow up to an integral scale from the wall \citep{Hunt_78}. Possible modifications involve damping functions \citep{VanDriest_56,Lai_90,Craft_96,Rodi_93} or wall-functions \citep{Launder_74,Singhal_81,Rodi_80,Spalding_77}. In those turbulent flows where a higher level of statistical description is necessary close to walls, adequate representation of the near-wall anisotropy and inhomogeneity is crucial. \citet{Durbin_93} proposed a Reynolds stress closure to address these issues. In his model, the all-important process of pressure redistribution is modeled through an elliptic equation by analogy with the Poisson equation, which governs the pressure in incompressible flows. This represents the non-local effect of the wall on the Reynolds stresses through the fluctuating pressure terms. In an effort to extend PDF methods to wall-bounded turbulent flows, Durbin's elliptic relaxation method has been combined with the generalized Langevin model \citep{Haworth_86} by \citet{Dreeben_97,Dreeben_98}. Wall-function treatment has also been developed for the PDF framework by \citet{Dreeben_97b}, providing the option of the usual trade-off between computational expense and resolution at walls. With minor simplifications these wall-treatmets are closely followed throughout the present study. We compute fully resolved boundary layers with the elliptic relaxation technique and also apply wall-functions in order to investigate their effects on the results and the computational performance.

The dispersion of scalars (e.g.\ temperature, mass, etc.) in turbulent flows is relevant to a number of scientific phenomena including engineering combustion and atmospheric dispersion of pollutants. Reviews on the subject have been compiled by \citet{Shraiman_00} and \citet{Warhaft_00}. Several experimental studies have been carried out in order to better understand the behavior of transported scalars in homogeneous isotropic turbulence \citep{Warhaft_84,Stapountzis_86,Sawford_95}. A literature review of dispersion from a concentrated source in homogeneous but anisotropic turbulent shear flows is given by \citet{Karnik_89}. Inhomogeneous turbulence (e.g.\ the atmospheric boundary layer or any practical turbulent flow) adds a significant level of complexity to these cases. Extensive measurements of the mean, variance, intermittency, probability density functions and spectra of scalar have been made by \citet{Fackrell_82} in a turbulent boundary layer. One point statistics in turbulent channel flow have recently been reported by \citet{Lavertu_05}. In urban scale modeling of passive pollutants in the atmosphere, the simplest settings to study turbulent flow and dispersion patterns are street canyons. Different canyon configurations and release scenarios have been studied both experimentally \citep{Hoydysh_74,Wedding_77,Rafailids_95,Meroney_96,Pavageau_99} and numerically \citep{Lee_94,Johnson_95,Baik_99,Huang_00,Liu_02}. A widely studied case in both numerical and experimental fluid dynamics is the flow behind a circular cylinder. Despite its relative simplicity in domain geometry, a myriad of flow behaviors can be explored through a variety of physical circumstances in this flow \citep{Williamson_96}. Concentrating on the very near wake several aspects of the present PDF model will be explored by computing the turbulent velocity field behind a circular cylinder at a transitional Reynolds number. Direct numerical simulation (DNS) has served as an important counterpart to measurements of turbulence at moderate Reynolds numbers, shedding light on quantities that are difficult to measure (e.g.\ Lagrangian statistics) and at locations where it is nearly impossible to measure (e.g.\ close to walls). Turbulent velocity statistics extracted from DNS of channel flow have been reported by \citet{Moser_99} and \citet{Abe_04}, while \citet{Vrieling_03} performed a DNS study of dispersion of plumes from single and double line sources. We will draw several datasets from the above experimental and numerical studies to compare and validate our results pertaining to the channel flow, the street canyon and the wake behind a circular cylinder.%
\nomenclature[A]{DNS}{direct numerical simulation}%

A widely used model to incorporate the effects of small scale mixing on a scalar released in a turbulent flow in the PDF framework is the interaction by exchange with the mean (IEM) model of \citet{Villermaux_Devillon_72} and \citet{Dopazo_OBrien_74}. While this model has the virtue of being simple and efficient, it fails to comply with several physical constraints and desirable properties of an ideal mixing model \citep{Fox_03}. Although a variety of other mixing models have been proposed to satisfy these properties \citep{Dopazo_94}, the IEM model remains widely used in practice. Recently, increased attention has been devoted to the interaction by exchange with the conditional mean (IECM) model. \citet{Sawford_04} has done a comparative study of scalar mixing from line sources in homogeneous turbulence employing both the IEM and IECM models, wherein he demonstrated that the largest differences between the two models occur in the near-field. He also investigated the two models in a double scalar mixing layer \citep{Sawford_06} with an emphasis on those conditional statistics that frequently require closure assumptions. Based on the IECM model, PDF micromixing models have been developed for dispersion of passive pollutants in the atmosphere by \citet{Luhar_et_al_05} and \citet{Cassiani_et_al_05,Cassiani_et_al_05b,Cassiani_07,Cassiani_07b}. These authors compute scalar statistics in homogeneous turbulence and in neutral, convective and canopy boundary layer by assuming a joint PDF for the turbulent velocity field. However, no previous studies have been conducted on modeling the joint PDF of velocity and a passive scalar from a concentrated source in inhomogeneous flows.

The purpose of this research is to continue to widen the applicability of PDF methods in practical applications, especially to more realistic flow geometries by employing unstructured grids. The current work is a step in that direction, where we combine several models and develop a set of algorithms to compute the joint PDF of the turbulent velocity, characteristic frequency and scalar concentration in complex domains. Complementary to hybrid FV/particle and LES/FDF methods, we provide a different methodology to exploit the advantages of unstructured Eulerian meshes in conjunction with Lagrangian PDF methods. Three flows, a fully developed turbulent channel flow, a street canyon (or cavity) flow and the flow behind a circular cylinder are used to test several aspects of the algorithms.

A series of novel numerical algorithms are proposed to facilitate an efficient solution of the PDF transport equation. A modified pressure projection scheme that has traditionally been used to compute the pressure field in incompressible laminar flows is adapted to the Lagrangian Monte-Carlo solution to compute the mean pressure field in complex domains. Estimation of local Eulerian statistics and their derivatives employing finite element shapefunctions are presented. For the computation of the velocity-conditioned scalar mean required in the IECM model, we propose an adaptive algorithm that makes no assumption on the shape of the underlying velocity PDF and which, using a dynamic procedure, automatically homogenizes the statistical error over the sample space. An efficient particle-tracking procedure for two-dimensional triangles and three-dimensional tetrahedra is presented. Alternatively to particle splitting and merging algorithms, a particle redistribution algorithm is also proposed that ensures the stability of the numerical solution and reduces the need for high number of particles.

The solver has been optimized and parallelized for cache-based shared memory and multi-core machines using the OpenMP standard. Accordingly, the discussion on the numerical algorithms highlights several aspects of code design for these high-performance parallel architectures.

The remainder of the dissertation is organized as follows. In \autoref{chap:governing_equations} the exact and modeled governing equations are described. \autoref{chap:numerics} presents details of the solution algorithm with the underlying numerical methods. The method is tested and validated by computing scalar dispersion from concentrated sources in a fully developed turbulent channel flow in \autoref{chap:channel} and in a street canyon in \autoref{chap:canyon} and by computing the velocity field behind a circular cylinder in \autoref{chap:cylinder}. Finally, some conclusions are drawn and future directions are discussed in \autoref{chap:conclusions}. Several important aspects of the underlying algorithms are detailed in the Appendices.


\chapter[Governing equations]{\\Governing equations}
\label{chap:governing_equations}

The governing system of equations for a passive scalar released in a viscous, Newtonian, incompressible fluid can be derived from Newton's equations of motion \citep{Hirsch_88} and is written in the Eulerian framework as
\begin{eqnarray}
\frac{\partial U_i}{\partial x_i} &=& 0,\label{eq:continuity}\\
\frac{\partial U_i}{\partial t} + U_j\frac{\partial U_i}{\partial x_j} + \frac{1}{\rho}\frac{\partial P}{\partial x_i} &=&  \nu\nabla^2U_i \label{eq:NavierStokes},\\
\frac{\partial\phi}{\partial t} + U_i\frac{\partial\phi}{\partial x_i} &=& \Gamma\nabla^2\phi,\label{eq:scalar}
\end{eqnarray}%
\nomenclature[RU]{$U_i$}{Eulerian velocity ($U,V,W$)}%
\nomenclature[RP]{$P$}{pressure}%
\nomenclature[Gq]{$\rho$}{density}%
\nomenclature[Gl]{$\nu$}{kinematic viscosity}%
\nomenclature[Gu]{$\phi$}{scalar concentration}%
\nomenclature[Gc]{$\Gamma$}{scalar diffusivity}%
\nomenclature[Rx]{$x_i$}{position}%
\nomenclature[Rt]{$t$}{time}%
where \(U_i\), \(P\), \(\rho\), \(\nu\), \(\phi\) and \(\Gamma\) are the Eulerian velocity, pressure, constant density, kinematic viscosity, scalar concentration and scalar diffusivity, respectively. Based on these equations an exact transport equation can be derived for the one-point, one-time Eulerian joint PDF of velocity and concentration \(f(\bv{V},\psi;\bv{x},t)\) \citep{Pope_85,Pope_00},%
\nomenclature[Rf]{$f(\bv{V},\psi)$}{joint PDF of velocity and concentration}%
\begin{equation}
\frac{\partial f}{\partial t}+V_i\frac{\partial f}{\partial x_i}=-\frac{\partial}{\partial V_i}\bigg[\mean{\nu\nabla^2U_i-\frac{1}{\rho}\frac{\partial P}{\partial x_i}\bigg|\bv{V},\psi}f\bigg]-\frac{\partial}{\partial\psi}\Big[\mean{\Gamma\nabla^2\phi|\bv{V},\psi}f\Big],\label{eq:EulerianPDF}
\end{equation}%
\nomenclature[RV]{$\bv{V}$}{sample space variable corresponding to $\bv{U}$}%
\nomenclature[Gw]{$\psi$}{sample space variable corresponding to $\phi$}%
where \(\bv{V}\) and \(\psi\) denote the sample space variables of the stochastic velocity \(\bv{U}(\bv{x},t)\) and concentration \(\phi(\bv{x},t)\) fields, respectively. (\Eqre{eq:EulerianPDF} is derived in \autoref{app:derivation}.) A remarkable feature of \Eqre{eq:EulerianPDF} is that the effects of convection and viscous diffusion (processes of critical importance in wall-bounded turbulent flows) are in closed form, thus require no modeling assumptions. Other effects, however, require closure assumptions. They are the effects of dissipation of turbulent kinetic energy, pressure redistribution and the small-scale mixing of the transported scalar due to molecular diffusion. The joint PDF \(f(\bv{V},\psi;\bv{x},t)\) contains all one-point statistics of the velocity and scalar fields. The price to pay for the increased level of description (compared to traditional moment closures) is that in a general three-dimensional turbulent flow \(f(\bv{V},\psi;\bv{x},t)\) is a function of 8 independent variables. This effectively rules out the application of traditional techniques like the finite difference, finite volume or finite element methods for a numerical solution. While in principle this high-dimensional space could be discretized and (after appropriate modeling of the unclosed terms) \Eqre{eq:EulerianPDF} could be solved with the above methods, the preferred choice in the PDF framework is to use a Lagrangian Monte-Carlo formulation. As opposed to the other techniques mentioned, the computational requirements increase only linearly with increasing problem dimension with a Monte-Carlo method. Another advantage of employing a Lagrangian-particle based simulation is that the governing equations may take a significantly simpler form than \Eqre{eq:EulerianPDF}.

In a Lagrangian formulation, it is assumed that the motion of fluid particles along their trajectory is well represented by a diffusion process, namely a continuous-time Markov process with continuous sample paths \citep{vanKampen_04}. Such a process was originally proposed by \citet{Langevin_08} as a stochastic model of a microscopic particle undergoing Brownian motion. \citet{Pope_00} shows that Langevin's equation provides a good model for the velocity of a fluid particle in turbulence. It is important to appreciate that the instantaneous particle velocities, modeled by a Langevin equation, do not represent physical fluid particle velocities individually, rather their combined effect (i.e.\ their statistics) can model statistics of a turbulent flow. Therefore, the numerical particles can be thought of as an ensemble representation of turbulence, each particle embodying one realization of the flow at a given point in space and time. At a fundamental level, an interesting consequence of this view is that this definition does not require an external (spatial or temporal) filter explicitly, as the classical Reynolds averaging rules and large eddy simulation filtering do. For example, in unsteady homogeneous or steady inhomogeneous high-Reynolds-number flows, the natural Reynolds-average to define is the spatial and temporal average, respectively. In unsteady \emph{and} inhomogeneous flows however, one is restricted to employ temporal and/or spatial filters leading to the approaches of unsteady Reynolds-averaged Navier-Stokes (URANS) and LES methods, respectively \citep{Pope_04}. In the PDF framework statistics are defined based on a probability density function. In the current case, for example, the mean velocity and Reynolds stress tensor are obtained from the joint PDF \(f\) as%
\nomenclature[A]{URANS}{unsteady Reynolds-averaged Navier-Stokes}%
\begin{eqnarray}
\mean{U_i}(\bv{x},t)&\equiv&\int_{-\infty}^\infty\int_0^\infty V_i f(\bv{V},\psi;\bv{x},t)\mathrm{d}\psi\mathrm{d}\bv{V},\label{eq:firstmoment}\\
\mean{u_iu_j}(\bv{x},t)&\equiv&\int_{-\infty}^\infty\int_0^\infty (V_i-\mean{U_i})(V_j-\mean{U_j}) f(\bv{V},\psi;\bv{x},t)\mathrm{d}\psi\mathrm{d}\bv{V},\label{eq:secondmoment}
\end{eqnarray}%
\nomenclature[Ru]{$u_i$}{velocity fluctuation ($V_i-\mean{U_i}$)}%
\nomenclature[Rp]{$p$}{pressure fluctuation ($P-\mean{P}$)}%
where the velocity fluctuation is defined as \(u_i=V_i-\mean{U_i}\). These quantities are well-defined mathematically \citep{vanKampen_04,Pope_00}, independently of the underlying physics, the state of the flow (i.e.\ homogeneous or inhomogeneous, steady or unsteady), the numerical method and the spatial and temporal discretization. Therefore the promise of a probabilistic view of turbulence (as in PDF methods) at the fundamental level is a more rigorous statistical treatment.

An equivalent model to the Eulerian momentum equation \Eqr{eq:NavierStokes} in the Lagrangian framework is a system of governing equations for particle position \(\mathcal{X}_i\) and velocity \(\mathcal{U}_i\) increments \citep{Dreeben_97}%
\nomenclature[RX]{$\mathcal{X}_i$}{Lagrangian particle position}%
\nomenclature[RU]{$\mathcal{U}_i$}{Lagrangian particle velocity}%
\begin{eqnarray}
\mathrm{d}\mathcal{X}_i&=&\mathcal{U}_i\mathrm{d}t + \left(2\nu\right)^{1/2}\mathrm{d}W_i,\label{eq:Lagrangian-position-exact}\\
\mathrm{d}\mathcal{U}_i(t)&=&-\frac{1}{\rho}\frac{\partial P}{\partial x_i}\mathrm{d}t + 2\nu\frac{\partial^2U_i}{\partial x_j\partial x_j}\mathrm{d}t+\left(2\nu\right)^{1/2}\frac{\partial U_i}{\partial x_j}\mathrm{d}W_j,\label{eq:Lagrangian-velocity-exact}
\end{eqnarray}%
\nomenclature[RW]{$W_i$}{vector-valued Wiener process}%
where the isotropic Wiener process \(\mathrm{d}W_i\) \citep{Gardiner_04} is identical in both equations (numerically, the same exact series of Gaussian random numbers with zero mean and variance \(\mathrm{d}t\)) and it is understood that the Eulerian fields on the right hand side are evaluated at the particle locations \(\mathcal{X}_i\). Since \Eqre{eq:Lagrangian-velocity-exact} is a diffusion-type stochastic differential equation with a Gaussian white noise (i.e.\ a Wiener process), it is equivalent to the Fokker-Planck equation that governs the evolution of the probability distribution of the same process \citep{vanKampen_04,DreebenPhd_97}. \Eqres{eq:Lagrangian-position-exact} and \Eqr{eq:Lagrangian-velocity-exact} represent the viscous effects exactly in the Lagrangian framework. Particles governed by these equations are both advected and diffused in physical space. In other words, besides convection the particles diffuse in physical space with coefficient \(\nu\), thus they carry momentum as molecules do with identical statistics, as in Brownian motion \citep{Einstein_26}. After Reynolds decomposition is applied to the velocity and pressure, \Eqre{eq:Lagrangian-velocity-exact} results in
\begin{equation}
\begin{split}
\mathrm{d}\mathcal{U}_i(t)&=-\frac{1}{\rho}\frac{\partial\mean{P}}{\partial x_i}\mathrm{d}t + 2\nu\frac{\partial^2\mean{U_i}}{\partial x_j\partial x_j}\mathrm{d}t + \left(2\nu\right)^{1/2}\frac{\partial\mean{U_i}}{\partial x_j}\mathrm{d}W_j\\
&\quad- \frac{1}{\rho}\frac{\partial p}{\partial x_i}\mathrm{d}t + 2\nu\frac{\partial^2u_i}{\partial x_j\partial x_j}\mathrm{d}t + \left(2\nu\right)^{1/2}\frac{\partial u_i}{\partial x_j}\mathrm{d}W_j,
\end{split}
\label{eq:Lagrangian-velocity-decomposed}
\end{equation}
where the last three terms are unclosed. To model these terms, we adopt the generalized Langevin model (GLM) of \citet{Haworth_86}%
\nomenclature[A]{GLM}{generalized Langevin model}%
\begin{equation}
\begin{split}
\mathrm{d}\mathcal{U}_i(t)&=-\frac{1}{\rho}\frac{\partial\mean{P}}{\partial x_i}\mathrm{d}t + 2\nu\frac{\partial^2\mean{U_i}}{\partial x_j\partial x_j}\mathrm{d}t + \left(2\nu\right)^{1/2}\frac{\partial\mean{U_i}}{\partial x_j}\mathrm{d}W_j\\
&\quad+ G_{ij}\left(\mathcal{U}_j-\mean{U_j}\right)\mathrm{d}t + \left(C_0\varepsilon\right)^{1/2}\mathrm{d}W'_i,
\end{split}
\label{eq:Lagrangian-model}
\end{equation}%
\nomenclature[RG]{$G_{ij}$}{coefficient in the GLM, \Eqres{eq:Lagrangian-model} and \Eqrs{eq:Lagrangian-model-wf}}%
\nomenclature[RC]{$C_0$}{coefficient in the Langevin equations \Eqr{eq:Lagrangian-model} and \Eqr{eq:Lagrangian-model-wf}}%
\nomenclature[Ge]{$\varepsilon$}{dissipation rate of turbulent kinetic energy}%
\nomenclature[RW]{$W'_i$}{vector-valued Wiener process (independent of $W_i$)}%
where \(G_{ij}\) is a second-order tensor function of velocity statistics, \(C_0\) is a positive constant, \(\varepsilon\) denotes the rate of dissipation of turbulent kinetic energy and \(\mathrm{d}W'_i\) is another Wiener process. Because of the correspondence between stochastic Lagrangian models and Reynolds stress closures \citep{Pope_94}, different second order models can be realized with the Langevin equation \Eqr{eq:Lagrangian-model}, depending on how \(G_{ij}\) is specified. An advantage of the GLM family of models is that equation \Eqr{eq:Lagrangian-model} ensures realizability as a valid Reynolds stress closure, provided that \(C_0\) is non-negative and that \(C_0\) and \(G_{ij}\) are bounded \citep{Pope_00}. Compared to Reynolds stress closures, the terms in \(G_{ij}\) and \(C_0\) represent pressure redistribution and anisotropic dissipation of turbulent kinetic energy. Far from walls, these physical processes can be adequately modeled by appropriate local (algebraic) functions of the velocity statistics. However, such local representation is in contradiction with the large structures interacting with the wall and the viscous wall region \citep{Whizman_96}. The traditionally employed damping or wall-functions, therefore, are of limited validity in an approach aiming at a higher-level statistical description. To address these issues, \citet{Durbin_93} proposed a technique to incorporate the wall-effects on the Reynolds stress tensor in a more natural fashion. In his approach, an elliptic equation is employed to capture the non-locality of the pressure redistribution at the wall, based on the analogy with the Poisson equation which governs the pressure in incompressible flows. The methodology also provides more freedom on controlling the individual components of the Reynolds stress tensor at the wall, such as the suppression of only the wall-normal component representing wall-blocking. \citet{Dreeben_98} incorporated Durbin's elliptic relaxation technique into the PDF method using the constraint
\begin{equation}
\left(1+\tfrac{3}{2}C_0\right)\varepsilon+G_{ij}\mean{u_iu_j}=0,\label{eq:constraint-on-c0}
\end{equation}
which ensures that the kinetic energy evolves correctly in homogeneous turbulence \citep{Pope_00}. Introducing the tensor \(\wp_{ij}\) to characterize the non-local effects \(G_{ij}\) and \(C_0\) are defined as%
\nomenclature[Rp]{$\wp_{ij}$}{non-local tensor in the elliptic relaxation, \Eqre{eq:elliptic-relaxation-Lagrangian}}%
\begin{equation}
G_{ij} = \frac{\wp_{ij}-\frac{\varepsilon}{2}\delta_{ij}}{k} \quad\mathrm{and}\quad C_0 = \frac{-2\wp_{ij}\mean{u_iu_j}}{3k\varepsilon},\label{eq:GandC}
\end{equation}%
\nomenclature[Rk]{$k$}{turbulent kinetic energy}%
\nomenclature[Gd]{$\delta_{ij}$}{Kronecker symbol}%
where \(k=\frac{1}{2}\mean{u_iu_i}\) denotes the turbulent kinetic energy. The non-local quantity \(\wp_{ij}\) is specified with the following elliptic relaxation equation
\begin{equation}
\wp_{ij} - L^2\nabla^2\wp_{ij} = \frac{1-C_1}{2}k\mean{\omega}\delta_{ij} + kH_{ijkl}\frac{\partial\mean{U_k}}{\partial x_l},\label{eq:elliptic-relaxation-Lagrangian}
\end{equation}
where the fourth-order tensor \(H_{ijkl}\) is given by
\begin{equation}
H_{ijkl} = (C_2A_v + \frac{1}{3}\gamma_5)\delta_{ik}\delta_{jl} - \frac{1}{3}\gamma_5\delta_{il}\delta_{jk}+\gamma_5b_{ik}\delta_{jl} - \gamma_5b_{il}\delta_{jk},\label{eq:H}
\end{equation}
\begin{equation}
A_v = \min\left[1.0,C_v\frac{\det\mean{u_iu_j}}{\left(\frac{2}{3}k\right)^3}\right],
\end{equation}
and
\begin{equation}
b_{ij} = \frac{\mean{u_iu_j}}{\mean{u_ku_k}} - \frac{1}{3}\delta_{ij}\label{eq:bij}
\end{equation}
is the Reynolds stress anisotropy, \(\mean{\omega}\) denotes the mean characteristic turbulent frequency and \(C_1, C_2, \gamma_5, C_v\) are model constants. The characteristic lengthscale \(L\) is defined by the maximum of the turbulent and Kolmogorov lengthscales
\begin{equation}
L=C_L\max\left[C_\xi\frac{k^{3/2}}{\varepsilon},C_\eta\left(\frac{\nu^3}{\varepsilon}\right)^{1/4}\right],\label{eq:L}
\end{equation}
with
\begin{equation}
C_\xi=1.0+1.3n_in_i,\label{eq:Cxi}
\end{equation}%
\nomenclature[RH]{$H_{ijkl}$}{fourth-order tensor in the near-wall turbulence model, \Eqre{eq:elliptic-relaxation-Lagrangian}}%
\nomenclature[RA]{$A_v$}{scalar variable in the near-wall turbulence model, \Eqre{eq:elliptic-relaxation-Lagrangian}}%
\nomenclature[Rb]{$b_{ij}$}{normalized Reynolds stress anisotropy, \Eqre{eq:bij}}%
\nomenclature[Gz]{$\omega$}{sample space variable of turbulent frequency $\Omega$}%
\nomenclature[RC]{$C_1$}{constant in the turbulence model, \Eqre{eq:elliptic-relaxation-Lagrangian}}%
\nomenclature[RC]{$C_2$}{constant in the turbulence model, \Eqre{eq:H}}%
\nomenclature[Gc]{$\gamma_5$}{constant in the turbulence model, \Eqre{eq:H}}%
\nomenclature[RC]{$C_v$}{constant in the turbulence model, \Eqre{eq:H}}%
\nomenclature[RL]{$L$}{characteristic lengthscale in the near-wall turbulence model, \Eqre{eq:L}}%
\nomenclature[RC]{$C_\xi$}{wall-normal dependent variable in the near-wall turbulence model, \Eqre{eq:L}}%
\nomenclature[RC]{$C_\eta$}{constant in the near-wall turbulence model, \Eqre{eq:L}}%
\nomenclature[RC]{$C_L$}{constant in the near-wall turbulence model, \Eqre{eq:L}}%
\nomenclature[Rn]{$\bv{n}$}{outward pointing wall-normal}%
where \(n_i\) is the unit wall-normal of the closest wall-element pointing outward of the flow domain, while \(C_L\) and \(C_\eta\) are model constants. The definition of \(C_\xi\) in \Eqre{eq:Cxi} signifies a slight departure from the original model by attributing anisotropic and wall-dependent behavior to its value. In the case of a channel flow, for example, where the wall is aligned with \(x\), the wall-normal \(\bv{n}=(0,-1,0)\). This gives \(C_\xi=2.3\) in the computation of \(\wp_{22}\) in \Eqre{eq:elliptic-relaxation-Lagrangian} and \(C_\xi=1.0\) for all other components. The modification improves the channel-centerline behavior of the wall-normal Reynolds stress component \(\mean{v^{\scriptscriptstyle 2}}\) and in turn the cross-stream mixing of the passive scalar. Another departure from the original model is the application of the elliptic term \(L^2\nabla^2\wp_{ij}\) (as originally proposed by \citet{Durbin_93}) as opposed to \(L\nabla^2(L\wp_{ij})\). This simplification was adopted because no visible improvement has been found by employing the second, numerically more expensive term.

The right hand side of \Eqre{eq:elliptic-relaxation-Lagrangian} can be any local model for pressure redistribution; here we follow \citet{Dreeben_98} and use the stochastic Lagrangian equivalent of a modified isotropization of production (IP) model proposed by \citet{Pope_94}. It is apparent that \Eqre{eq:elliptic-relaxation-Lagrangian} acts like a blending function between the low-Reynolds-number near-wall region and the high-Reynolds-number free turbulence. Close to the wall, the elliptic term on the left hand side brings out the non-local, highly anisotropic behavior of the Reynolds stress tensor, whereas far from the wall the significance of the elliptic term vanishes and the local model on the right hand side is recovered.%
\nomenclature[A]{IP}{isotropization of production}%

The description of the computation of the mean-pressure gradient in \Eqre{eq:Lagrangian-model} is deferred to \autoref{chap:numerics}.

The above model needs to be augmented by an equation for a quantity that provides length-, or time-scale information for the turbulence. With traditional moment closures the most common approach is to solve a model equation for the turbulent kinetic energy dissipation rate \(\varepsilon\) itself as proposed by \citet{Hanjalic_72}. An alternative method is to solve an equation for the mean characteristic turbulent frequency \citep{Wilcox_93} \(\mean{\omega}\) and to define
\begin{equation}
\varepsilon = k\mean{\omega}.\label{eq:dissipation-from-frequency}
\end{equation}
In PDF methods, however, a fully Lagrangian description has been preferred. A Lagrangian stochastic model has been developed for the instantaneous particle frequency \(\omega\) by \citet{VanSlooten_98} of which different forms exist, but the simplest formulation can be cast into
\begin{equation}
\mathrm{d}\omega = -C_3\mean{\omega}\left(\omega-\mean{\omega}\right)\mathrm{d}t - S_\omega\mean{\omega}\omega\mathrm{d}t+\left(2C_3C_4\mean{\omega}^2\omega\right)^{1/2}\mathrm{d}W,\label{eq:frequency-model}
\end{equation}%
\nomenclature[RS]{$S_\omega$}{source/sink for the mean turbulent frequency, \Eqre{eq:frequency-model}}%
where \(S_\omega\) is a source/sink term for the mean turbulent frequency
\begin{equation}
S_\omega=C_{\omega2}-C_{\omega1}\frac{\mathcal{P}}{\varepsilon},\label{eq:frequency-source}
\end{equation}%
\nomenclature[RP]{$\mathcal{P}$}{production of turbulent kinetic energy $\left(-\mean{u_iu_j}\partial\mean{U_i}/\partial x_j\right)$}%
\nomenclature[RW]{$W$}{scalar-valued Wiener process}%
\nomenclature[RC]{$C_3$}{constant in the turbulent frequency model, \Eqre{eq:frequency-model}}%
\nomenclature[RC]{$C_4$}{constant in the turbulent frequency model, \Eqre{eq:frequency-model}}%
\nomenclature[RC]{$C_{\omega1}$}{constant in the turbulent frequency model, \Eqre{eq:frequency-model}}%
\nomenclature[RC]{$C_{\omega2}$}{constant in the turbulent frequency model, \Eqre{eq:frequency-model}}%
where \(\mathcal{P}=-\mean{u_iu_j}\partial\mean{U_i}/\partial x_j\) is the production of turbulent kinetic energy, \(\mathrm{d}W\) is a scalar-valued Wiener-process, while \(C_3,C_4,C_{\omega1}\) and \(C_{\omega2}\) are model constants. Since the no-slip condition would incorrectly force \(\varepsilon\) to zero at a no-slip wall, \Eqre{eq:dissipation-from-frequency} needs to be modified, thus the dissipation is defined as \citep{Dreeben_98}
\begin{equation}
\varepsilon = \mean{\omega}\left(k + \nu C_T^2\mean{\omega}\right),\label{eq:dissipation-from-frequency_wall}
\end{equation}
\nomenclature[RC]{$C_T$}{constant in the turbulent frequency model, \Eqre{eq:frequency-model}}%
where \(C_T\) is also a model constant. A simplification of the original model for the turbulent frequency employed by \citet{Dreeben_98} is the elimination of the ad-hoc source term involving an additional model constant, since in our case-studies we found no obvious improvements by including it. This completes the model for the joint PDF of velocity and the (now included) characteristic turbulent frequency \(\omega\). The specification of particle boundary conditions will be discussed in \autoref{chap:numerics}. The equations to model the joint PDF of velocity and turbulent frequency closely follow the work of \citet{Dreeben_98}. Slight modifications consist of
\begin{itemize}
\item the anisotropic definition of lengthscale \(L\) in \Eqres{eq:L} and \Eqr{eq:Cxi},
\item the application of the elliptic term \(L^2\nabla^2\wp_{ij}\) instead of \(L\nabla^2(L\wp_{ij})\) in \Eqre{eq:elliptic-relaxation-Lagrangian}, and
\item the elimination of an ad-hoc source term in \Eqre{eq:frequency-source}.
\end{itemize}

Since a passive scalar, by definition, has no effect on the turbulent velocity field, modeling the pressure redistribution and dissipation have been discussed independently from the scalar, i.e.\ it has been assumed that in \Eqre{eq:EulerianPDF} the following hold
\begin{equation}
\mean{\nu\nabla^2U_i-\frac{1}{\rho}\frac{\partial P}{\partial x_i}\bigg|\bv{V},\psi}=\mean{\nu\nabla^2U_i-\frac{1}{\rho}\frac{\partial P}{\partial x_i}\bigg|\bv{V}}.
\end{equation}
However, the opposite, that the micromixing of the scalar can be modeled independently of \(\bv{V}\), cannot be assumed in general \citep{Pope_98}. A simple mixing model is the interaction by exchange with the mean (IEM) model \citep{Villermaux_Devillon_72,Dopazo_OBrien_74}, which models the conditional scalar diffusion in \Eqre{eq:EulerianPDF} independent of the underlying velocity field, i.e.\ assuming
\begin{equation}
\mean{\Gamma\nabla^2\phi\big|\bv{V},\psi}\cong\mean{\Gamma\nabla^2\phi\big|\psi}.\label{eq:independent-scalar}
\end{equation}
In the Lagrangian framework, the IEM model is written as
\begin{equation}
\mathrm{d}\psi = -\frac{1}{t_\mathrm{m}}\left(\psi-\mean{\phi}\right)\mathrm{d}t,\label{eq:IEM}
\end{equation}%
\nomenclature[Rt]{$t_\mathrm{m}$}{micromixing timescale, \Eqres{eq:micromixing-timescale-channel}, \Eqrs{eq:micromixing-timescale-canyon}}%
where \(t_\mathrm{m}\) is a micromixing timescale. It has been pointed out, however, that the assumption that the scalar mixing is independent of the velocity, \Eqre{eq:independent-scalar}, bears no theoretical justification and is at odds with local isotropy of the scalar field \citep{Fox_96,Pope_98}. On the other hand, the interaction by exchange with the conditional mean (IECM) model does take the velocity field into consideration by employing the velocity-conditioned mean instead of the unconditional mean as
\begin{equation}
\mathrm{d}\psi=-\frac{1}{t_\mathrm{m}}\left(\psi-\mean{\phi|\bv{V}}\right)\mathrm{d}t.\label{eq:IECM}
\end{equation}
Both the IEM and IECM models represent the physical process of dissipation and reflect the concept of relaxation towards a scalar mean with the characteristic timescale \(t_\mathrm{m}\). The difference is that in the IEM model, all particles that have similar position interact with each other, while in the IECM model only those particles interact that also have similar velocities, e.g.\ fluid elements that belong to the same eddy.

It can be shown that in the case of homogeneous turbulent mixing with no mean scalar gradient the two models are equivalent since the velocity and scalar fields are uncorrelated \citep{Fox_96} and the micromixing timescale \(t_\mathrm{m}\) is proportional to the Kolmogorov timescale \(\tau=k/\varepsilon\). In an inhomogeneous case, e.g.\ a concentrated source, however, there are various stages of the spreading of the plume requiring different characterizations of \(t_\mathrm{m}\).  In this case, the formal simplicity of the IEM and IECM models is a drawback, since a single scalar parameter \(t_\mathrm{m}\) has to account for all the correct physics. The timescale should be inhomogeneous and should depend not only on the local turbulence characteristics but also on the source location, type, size, distribution and strength. Because of this complexity, a general flow-independent specification of \(t_\mathrm{m}\) has been elusive. We will define the micromixing timescale for a passive scalar in the following chapters corresponding to the flows modeled.%
\nomenclature[Gs]{$\tau$}{Kolmogorov timescale ($k/\varepsilon$)}%

This completes the model for the joint PDF of turbulent velocity, frequency and scalar. The model is `complete' in the sense, that the equations are free from flow-dependent specifications \citep{Pope_00}, thus, in principle, it is generally applicable to any transported passive scalar released into an incompressible, high-Reynolds-number flow.

Defining \(G_{ij}\) and \(C_0\) through (\ref{eq:GandC}) enables the model to adequately capture the near-wall effects in the higher-order statistics when the wall-region has sufficient resolution. In realistic simulations, however, full resolution of high-Reynolds-number boundary layers is not always possible (and may not be necessary), especially on the urban or meso-scale in atmospheric modeling. For such cases a second option is the use of wall-functions instead of the elliptic relaxation to model the near-wall turbulence. Employing wall-functions for no-slip walls provides a trade-off between the accuracy of fully resolved boundary layers and computational speed. The significantly more expensive full resolution is absolutely required in certain cases, such as computing the heat transfer at walls embedded in a flow or detaching boundary layers with high adverse pressure gradients. Conversely, a boundary layer representation by wall-functions is commonly used when the exact details close to walls are not important, and the analysis focuses on the boundary layer effects at farther distances. Wall-functions are widely applied in atmospheric simulations, where full wall-resolution is usually prohibitively expensive even at the micro- or urban-scale \citep{Bacon_00,Lien_04}. It is worth emphasizing that one of the main assumptions used in the development of wall-functions is that the boundary layer remains attached. This is not always the case in simulations of complex flows. However, since wall-functions are the only choice for realistic atmospheric simulations, they are still routinely employed with reasonable success.

To investigate the gain in performance and the effect on the results, we implemented the wall-treatment for complex flow geometries that have been developed for the PDF method by \citet{Dreeben_97b}. Since in this case the viscous effects are not explicitly modeled, the viscous terms do not appear in the particle equations for the position and velocity increments:
\begin{eqnarray}
\mathrm{d}\mathcal{X}_i&=&\mathcal{U}_i\mathrm{d}t,\label{eq:Lagrangian-position-wf}\\
\mathrm{d}\mathcal{U}_i(t)&=&-\frac{1}{\rho}\frac{\partial\mean{P}}{\partial x_i}\mathrm{d}t + G_{ij}\left(\mathcal{U}_j-\mean{U_j}\right)\mathrm{d}t + \left(C_0\varepsilon\right)^{1/2}\mathrm{d}W'_i.\label{eq:Lagrangian-model-wf}
\end{eqnarray}
Furthermore, in this case the tensor \(G_{ij}\) is defined by the simplified Langevin model (SLM) \citep{Haworth_86} and \(C_0\) is simply a constant:
\begin{equation}
G_{ij} = -\left(\tfrac{1}{2}+\tfrac{3}{4}C_0\right)\mean{\omega}\delta_{ij} \quad\text{with}\quad C_0=3.5.\label{eq:wall-GandC}
\end{equation}
In line with the purpose of wall-functions, boundary conditions have to be imposed on particles that hit the wall so that their combined effect on the statistics at the first gridpoint from the wall will be consistent with the universal logarithmic wall-function in equilibrium flows, i.e.\ in boundary layers with no significant adverse pressure gradients. The development of boundary conditions based on wall-functions rely on the self-similarity of attached boundary layers close to walls. These conditions are applied usually at the first gridpoint from the wall based on the assumption of constant or linear stress-distribution. This results in the well-known self-similar logarithmic profile for the mean velocity. For the sake of completeness the conditions on particles developed by \citet{Dreeben_97b} are reported here. The condition for the wall-normal component of the particle velocity reads
\begin{equation}
\mathcal{V}_{\scriptscriptstyle R} = -\mathcal{V}_{\scriptscriptstyle I},\label{eq:wall-normal-velocity}
\end{equation}%
\nomenclature[RV]{$\mathcal{V}_{\scriptscriptstyle I}$}{incident wall-normal particle velocity, \Eqre{eq:wall-normal-velocity}}%
\nomenclature[RV]{$\mathcal{V}_{\scriptscriptstyle R}$}{reflected wall-normal particle velocity, \Eqre{eq:wall-normal-velocity}}%
where the subscripts $R$ and $I$ denote reflected and incident particle properties, respectively. The reflected streamwise particle velocity is given by
\begin{equation}
\mathcal{U}_{\scriptscriptstyle R} = \mathcal{U}_{\scriptscriptstyle I} + \alpha\mathcal{V}_{\scriptscriptstyle I},\label{eq:particle-wallcondition-wf}
\end{equation}%
\nomenclature[RU]{$\mathcal{U}_{\scriptscriptstyle I}$}{incident streamwise particle velocity at a wall, \Eqre{eq:particle-wallcondition-wf}}%
\nomenclature[RU]{$\mathcal{U}_{\scriptscriptstyle R}$}{reflected streamwise particle velocity at a wall, \Eqre{eq:particle-wallcondition-wf}}%
\nomenclature[Ga]{$\alpha$}{coefficient in particle wall-velocity condition, \Eqre{eq:particle-wallcondition-wf}}%
where the coefficient $\alpha$ is determined by imposing consistency with the logarithmic law at the distance of the first gridpoint from the wall, $y_p$:%
\nomenclature[Ry]{$y_p$}{distance of the first gridpoint from the wall where wall-functions are applied}%
\begin{equation}
\alpha = \frac{2\Hat{u}_p^2\mean{U}_p|\mean{U}_p|}{\mean{v^2}_pU_e^2},\label{eq:alpha}
\end{equation}%
\nomenclature[Ru]{$\Hat{u}_p$}{characteristic velocity scale of the turbulence intensity in the vicinity of $y_p$, \Eqre{eq:hatu}}%
where $\Hat{u}_p$ is a characteristic velocity scale of the turbulence intensity in the vicinity of $y_p$, defined as
\begin{equation}
\Hat{u}_p=C_\mu^{1/4}k_p^{1/2},\label{eq:hatu}
\end{equation}%
\nomenclature[RC]{$C_\mu$}{constant in the $k$--$\varepsilon$ model}%
\nomenclature[RU]{$\mean{U}_p$}{mean streamwise velocity at $y_p$, \Eqre{eq:alpha}}%
\nomenclature[Rv]{$\mean{v^{\scriptscriptstyle 2}}_p$}{wall-normal component of the Reynolds stress tensor at $y_p$, \Eqre{eq:alpha}}%
\nomenclature[Rk]{$k_p$}{wall-normal component of the turbulent kinetic energy at $y_p$, \Eqre{eq:hatu}}%
\nomenclature[RU]{$U_e$}{magnitude of the equilibrium value of the mean velocity at $y_p$, \Eqre{eq:Ue}}%
with $C_\mu=0.09$. $\mean{U}_p$, $\mean{v^{\scriptscriptstyle 2}}_p$ and $k_p$ are, respectively, the mean streamwise velocity, the wall-normal component of the Reynolds stress tensor and the turbulent kinetic energy, all obtained from the particle fields at $y_p$. In \Eqre{eq:alpha} $U_e$ is the magnitude of the equilibrium value of the mean velocity at $y_p$ and is specified by the logarithmic law
\begin{equation}
U_e = \frac{u_\tau}{\kappa}\log\left(E\frac{y_pu_\tau}{\nu}\right),\label{eq:Ue}
\end{equation}
where $\kappa=0.41$ is the K\'arm\'an constant and the surface roughness parameter $E=8.5$ for a smooth wall. The friction velocity $u_\tau$ is computed from local statistics as%
\nomenclature[Gj]{$\kappa$}{K\'arm\'an constant in the law of the wall, \Eqre{eq:Ue}}%
\nomenclature[RE]{$E$}{surface roughness parameter in the wall-function, \Eqre{eq:Ue}}%
\nomenclature[Ru]{$u_\tau$}{friction velocity}%
\begin{equation}
u_\tau = \sqrt{\Hat{u}_p^2 + \gamma_\tau\left|\frac{y_p}{\rho}\frac{\partial\mean{P}}{\partial x}\right|} \quad \text{with} \quad \gamma_\tau=\max\left[0; \enskip \mathrm{sign}\left(\mean{uv}\frac{\partial\mean{P}}{\partial x}\right)\right].\label{eq:friction-velocity}
\end{equation}
In Equations (\ref{eq:wall-normal-velocity}-\ref{eq:friction-velocity}) the streamwise $x$ and wall-normal $y$ coordinate directions are defined according to the local tangential and normal coordinate directions of the particular wall-element in question. In other words, if the wall is not aligned with the flow coordinate system then the vectors $\mathcal{U}_i$ and $\partial\mean{P}/\partial x_i$, and the Reynolds stress tensor $\mean{u_iu_j}$, need to be appropriately transformed into the wall-element coordinate system before being employed in the above equations. The condition on the turbulent frequency is given by
\begin{equation}
\omega_{\scriptscriptstyle R} = \omega_{\scriptscriptstyle I}\exp\left[\beta\frac{\mathcal{V}_I}{y_p\mean{\omega}}\right]
\quad\text{with}\quad
\beta = -\frac{2}{\tfrac{1}{2}+\tfrac{3}{4}C_0+C_3+C_{\omega2}-C_{\omega1}}.\label{eq:wall-omega}
\end{equation}%
\nomenclature[Gz]{$\omega_{\scriptscriptstyle I}$}{incident turbulent frequency at a wall, \Eqre{eq:wall-omega}}%
\nomenclature[Gz]{$\omega_{\scriptscriptstyle R}$}{reflected turbulent frequency at a wall, \Eqre{eq:wall-omega}}%

In summary, the flow is modeled by a large number of Lagrangian particles representing a finite sample of all fluid particles in the domain which can be thought of as different realizations of the underlying stochastic fields. Numerically, each particle has position \(\mathcal{X}_i\) and with its velocity \(\mathcal{U}_i\) carries its turbulent frequency \(\omega\) and scalar concentration \(\psi\). For full wall-resolution the particle positions and velocities are advanced according to \Eqres{eq:Lagrangian-position-exact} and \Eqr{eq:Lagrangian-model} using Equations (\ref{eq:GandC}-\ref{eq:Cxi}). While in the wall-functions case the positions and velocities are advanced by \Eqres{eq:Lagrangian-position-wf} and \Eqr{eq:Lagrangian-model-wf} using Equations (\ref{eq:wall-GandC}-\ref{eq:wall-omega}). In both cases, the particle frequencies and scalar concentrations are governed by \Eqr{eq:frequency-model} and either \Eqr{eq:IEM} or \Eqr{eq:IECM}, respectively. The particle equations are discretized and advanced in time by the explicit forward Euler-Maruyama method \citep{Kloeden_99}. Even in the case of full wall-resolution using the elliptic relaxation technique, this method was preferred to the more involved exponential scheme that was originally suggested by \citet{Dreeben_98}, since the code is sufficiently stable with the simpler and computationally less expensive Euler-Maruyama method as well.


\chapter[Numerical implementation]{\\Numerical implementation}
\label{chap:numerics}

\section{Introduction}
The numerical solution algorithm is based on the time-dependent particle governing equations: \Eqr{eq:Lagrangian-position-exact} and \Eqr{eq:Lagrangian-model} in the full wall-resolution case and \Eqr{eq:Lagrangian-position-wf} and \Eqr{eq:Lagrangian-model-wf} in the wall-functions case, \Eqr{eq:frequency-model} and either \Eqr{eq:IEM} or \Eqr{eq:IECM}. An adaptive timestepping strategy to advance the system is described in \autoref{sec:timestepping}. All Eulerian statistics required in these equations need to be estimated at the particle locations at the given instant in time. This is performed by an unstructured Eulerian grid that discretizes the flow domain, which can be conveniently refined around regions where a higher resolution is necessary. The methods used to compute unconditional statistics, their derivatives and conditional statistics are described in Sections \ref{sec:Eulerian_statistics}, \ref{sec:derivatives} and \ref{sec:conditional_stats}, respectively. The grid is also used to solve the elliptic relaxation equation \Eqr{eq:elliptic-relaxation-Lagrangian} and to solve for the mean pressure required in \Eqre{eq:Lagrangian-model}. The main characteristics of the solution of these two Eulerian equations together with a projection method to obtain the mean pressure are described in \autoref{sec:Eulerian_equations}. In order to identify which particles contribute to local statistics, the particles need to be continuously followed as they travel throughout the domain. The particle tracking algorithm that is used for this purpose is described in \autoref{sec:particle_tracking}. In complex configurations, where the spatial resolution can differ significantly from one region to another, an algorithm is necessary to ensure that the number of particles in every computational element is above a certain threshold, so that meaningful statistics can be computed. We present and test an algorithm that accomplishes this task in \autoref{sec:particle-number-control} and \autoref{app:particle-redistribution} and further refine it in \autoref{app:particle-redistribution2}. The boundary conditions at no-slip walls applied to particles, to the elliptic relaxation equation \Eqr{eq:elliptic-relaxation-Lagrangian} and to the mean pressure are described in \autoref{sec:wall_conditions}. Some aspects of parallel random number generation are described in \autoref{sec:parallel_random_number_generation}. An overview of the solution procedure with the execution profile of a timestep is given in \autoref{sec:profile}.

\section{Timestepping procedure}
\label{sec:timestepping}
To discretize the governing equations in time we apply the explicit forward Euler-Maruyama scheme \citep{Kloeden_99}. The size of the timestep is estimated in every step based on the Courant-Friedrichs-Lewy (CFL) \citep{Courant_28} condition as%
\nomenclature[A]{$\mathrm{CFL}$}{Courant-Friedrichs-Lewy (\autoref{sec:timestepping})}%
\begin{equation}
\Delta t = C_\mathrm{CFL}\cdot\min\limits_n\frac{\sqrt{A_n}}{\left|\left|\mean{\bv{U}}_n\right|\right|_2},\label{eq:CFL}
\end{equation}%
\nomenclature[RA]{$A_n$}{average element area around gridnode $n$}%
\nomenclature[RC]{$C_\mathrm{CFL}$}{Courant-Friedrichs-Lewy constant (\autoref{sec:timestepping})}%
\nomenclature[RY]{$\bv{Y}$}{vector of particle properties / sample space variables ($V_1,V_2,V_3,\omega,\psi$)}%
\nomenclature[Gz]{$\Omega$}{turbulent frequency}%
where \(A_n\) is the average element area around gridnode \(n\). According to \Eqre{eq:CFL} we compute a characteristic timescale for each gridnode by dividing the characteristic edge length (defined by the square-root of the element area) by the length of the mean velocity vector at the given location. Then we choose the smallest characteristic timescale of all gridpoints for the next timestep multiplied by a CFL constant of $C_\mathrm{CFL}=0.7$. This ensures that during a single step no information will travel farther than the length of Eulerian elements.

\section[Solution of the Eulerian equations]{Solution of the Eulerian equations: mean pressure and elliptic relaxation}
\label{sec:Eulerian_equations}
In incompressible flows the pressure establishes itself immediately through the pressure-Poisson equation, which is a manifestation of the divergence constraint \Eqrs{eq:continuity} expressing mass conservation. The numerical difficulties arising from the straightforward discretization of this equation in finite difference, finite volume and finite element methods are reviewed by \citet{Lohner_01}. Several different methods have been devised to deal with these issues, which stem from the fact that the mass conservation equation decouples from the momentum equation and acts on it only as a constraint, which may result in the decoupling of every second gridpoint thereby numerically destabilizing the solution. Some of these methods are: the use of different functional spaces for the velocity and pressure discretization, artificial viscosities, consistent numerical fluxes, artificial compressibility and pressure projection schemes. For our purposes we adopt the pressure projection approach.

Additionally, in PDF methods due to the stochastic nature of the simulation, the Eulerian statistics and their derivatives are subject to considerable statistical noise. \citet{Fox_03} suggests three different ways of calculating the mean pressure in PDF methods. The first approach is to extract the mean pressure field from a simultaneous consistent Reynolds stress model solved using a standard CFD solver \citep{Correa_92}. This approach solves the noise problem although it leads to a redundancy in the velocity model. The second approach attacks the noise problem by computing the so-called `particle-pressure field' \citep{Delarue_97}. This results in a stand-alone transported PDF method and the authors succesfully apply it to compute a compressible turbulent flow. The third approach is the hybrid methodology mentioned in the \href{#chapter.1}{Introduction}, which uses an Eulerian CFD solver to solve for the mean velocity field and a particle-based code to solve for the fluctuating velocity \citep{Muradoglu_99,Givi_89}. These methods are made consistent by the careful selection of turbulence models in the Eulerian and Lagrangian frameworks and the use of consistency conditions.

A different approach is proposed here. We adopt a modified version of the pressure projection scheme originally proposed by \citet{Chorin_68} in the finite difference context, which has been widely used in laminar flows. The modification compared to the original projection scheme involves solving for the difference of the pressure between two consecutive timesteps, instead of the pressure field itself. This ensures that at steady state the residuals of the pressure correction vanish \citep{Lohner_01}. We adopt the scheme in the Lagrangian-Eulerian setting and combine the projection algorithm with the particle equations as follows.

The idea of pressure projection is to first predict the velocity using the current flow variables without taking the divergence constraint into consideration. Then in a second step, the divergence constraint is enforced by solving a pressure-Poisson equation. Finally the velocity is corrected using the new pressure field, resulting in a divergence-free velocity field. Thus, using full wall-resolution and explicit (forward Euler-Maruyama) time-integration of the particle velocity, one complete timestep (\(n\to n+1\)) is given by:
\begin{itemize}
\item \emph{Velocity prediction: \(\bv{\mathcal{U}}^n \to \bv{\mathcal{U}}^*\)}
\begin{equation}
\begin{split}
\mathcal{U}_i^*=\mathcal{U}_i^n-\frac{1}{\rho}\frac{\partial\mean{P}^n}{\partial x_i}\Delta t + 2\nu\frac{\partial^2\mean{U_i}^n}{\partial x_j\partial x_j}\Delta t + \left(2\nu\right)^{1/2}\frac{\partial\mean{U_i}^n}{\partial x_j}\Delta W_j\\
+G_{ij}\left(\mathcal{U}_j^n-\mean{U_j}^n\right)\Delta t + \left(C_0\varepsilon\right)^{1/2}\Delta W'_i;
\end{split}\label{eq:velocity-prediction}
\end{equation}
\item \emph{Pressure projection: \(\mean{P}^n \to \mean{P}^{n+1}\)}
\begin{eqnarray}
\nabla\cdot\mean{\bv{U}}^{n+1}&=&0,\label{eq:divergence}\\
\frac{\mean{\bv{U}}^{n+1}-\mean{\bv{U}}^*}{\Delta t}+\frac{1}{\rho}\nabla(\mean{P}^{n+1}-\mean{P}^n)&=&0,
\end{eqnarray}
which results in
\begin{equation}
\frac{1}{\rho}\nabla^2(\mean{P}^{n+1}-\mean{P}^n)=\frac{\nabla\cdot\mean{\bv{U}}^*}{\Delta t};\label{eq:pressure-projection}
\end{equation}
\item \emph{Mean velocity correction: \(\mean{\bv{U}}^* \to \mean{\bv{U}}^{n+1}\)}
\begin{equation}
\mean{\bv{U}}^{n+1} = \mean{\bv{U}}^* - \frac{1}{\rho}\Delta t\nabla(\mean{P}^{n+1}-\mean{P}^n).\label{eq:velocity-correction}
\end{equation}
\end{itemize}
Since the velocity field is fully represented by particles, the velocity prediction \Eqr{eq:velocity-prediction} and correction \Eqr{eq:velocity-correction} steps are applied to particles. The above procedure ensures that the Poisson equation for the mean pressure is satisfied at all times, thus the joint PDF representing an incompressible flow satisfies realizability, normalization and consistency conditions \citep{Pope_85} in every timestep. To stabilize the computation of the mean pressure a small artificial diffusion term is added to the divergence constraint in \Eqre{eq:divergence}
\begin{equation}
\nabla\cdot\mean{\bv{U}}^{n+1}=C_p\frac{1}{\rho}\nabla^2\mean{P}^n,\label{eq:divergence_ad}
\end{equation}%
\nomenclature[RC]{$C_p$}{constant for stabilizing the mean pressure, \Eqre{eq:pressure-projection-smooth}}%
where \(C_p\) is a small constant, e.g.\ \(C_p=10^{-3}\), which results in the stabilized version of the pressure projection step
\begin{equation}
\frac{1}{\rho}\nabla^2(\mean{P}^{n+1}-\mean{P}^n)=\frac{1}{\Delta t}\left(\nabla\cdot\mean{\bv{U}}^* - C_p\frac{1}{\rho}\nabla^2\mean{P}^n\right).\label{eq:pressure-projection-smooth}
\end{equation}

Both the elliptic relaxation \Eqr{eq:elliptic-relaxation-Lagrangian} and pressure projection \Eqr{eq:pressure-projection-smooth} equations are solved with the finite element method using linear shapefunctions on a grid consisting of triangles \citep{Lohner_01}. The grid is generated by the general purpose mesh generator, Gmsh, developed by \citet{Geuzaine_09}. The FEM coefficient matrices are stored in block compressed sparse row format \citep{Saad_03}. The resulting linear systems are solved by the method of conjugate gradients combined with a Jacobi preconditioner. While the elliptic equation \Eqrs{eq:elliptic-relaxation-Lagrangian} for the tensor \(\wp_{ij}\) may appear prohibitively memory-hungry and expensive for larger meshes, the equation is well-conditioned and the iterative solution converges in just a few iterations starting from an initial condition using the solution in the previous timestep.

\section{Estimation of Eulerian statistics}
\label{sec:Eulerian_statistics}
During the numerical solution of the governing equations, Eulerian statistics need to be estimated at different locations of the domain. Since the joint PDF contains information on all one-point statistics of the velocity, frequency and scalar concentration fields, these are readily available through appropriate averages of particle properties. For example, the mean velocity at a specific location in space and time is obtained as the integral over all sample space of the joint PDF \(f_{\bv{\scriptscriptstyle Y}}(\bv{Y};\bv{x},t)\)
\begin{equation}
\mean{U_i}\equiv\int V_i f_{\bv{\scriptscriptstyle Y}}(\bv{Y};\bv{x},t)\mathrm{d}\bv{Y},\label{eq:meanvelocity-definition}
\end{equation}%
\nomenclature[Rf]{$f_{\bv{\scriptscriptstyle Y}}(\bv{Y})$}{joint PDF of velocity, turbulent frequency and scalar ($f_{\bv{\scriptscriptstyle U}\scriptscriptstyle\Omega\phi}(\bv{V},\omega,\psi)$)}%
where \(\bv{Y}\) denotes the vector of all sample space variables \(\bv{Y}=(V_1,V_2,V_3,\omega,\psi)\). For brevity we omit (but assume) the space and time dependence of the statistics. In traditional particle-codes the calculation of statistics is usually performed by kernel estimation using weight-functions \citep{Pope_00}. In particle-in-cell methods \citep{Grigoryev_02} an Eulerian mesh covers the computational domain and means are computed in each element or gridpoint. The latter approach is followed here and \Eqre{eq:meanvelocity-definition} is computed by an ensemble average over all particle velocities in the vicinity of \(\bv{x}\)
\begin{equation}
\mean{U_i}\cong\frac{1}{N}\sum_{p=1}^{N}\mathcal{U}_i^p,
\end{equation}%
\nomenclature[Rx]{$\bv{x}$}{position}%
\nomenclature[RN]{$N$}{number of particles participating in local statistics (\autoref{sec:Eulerian_statistics})}%
\nomenclature[RU]{$\mathcal{U}_i^p$}{velocity vector of particle $p$}%
\begin{figure}[t]
\centering
\resizebox{7cm}{!}{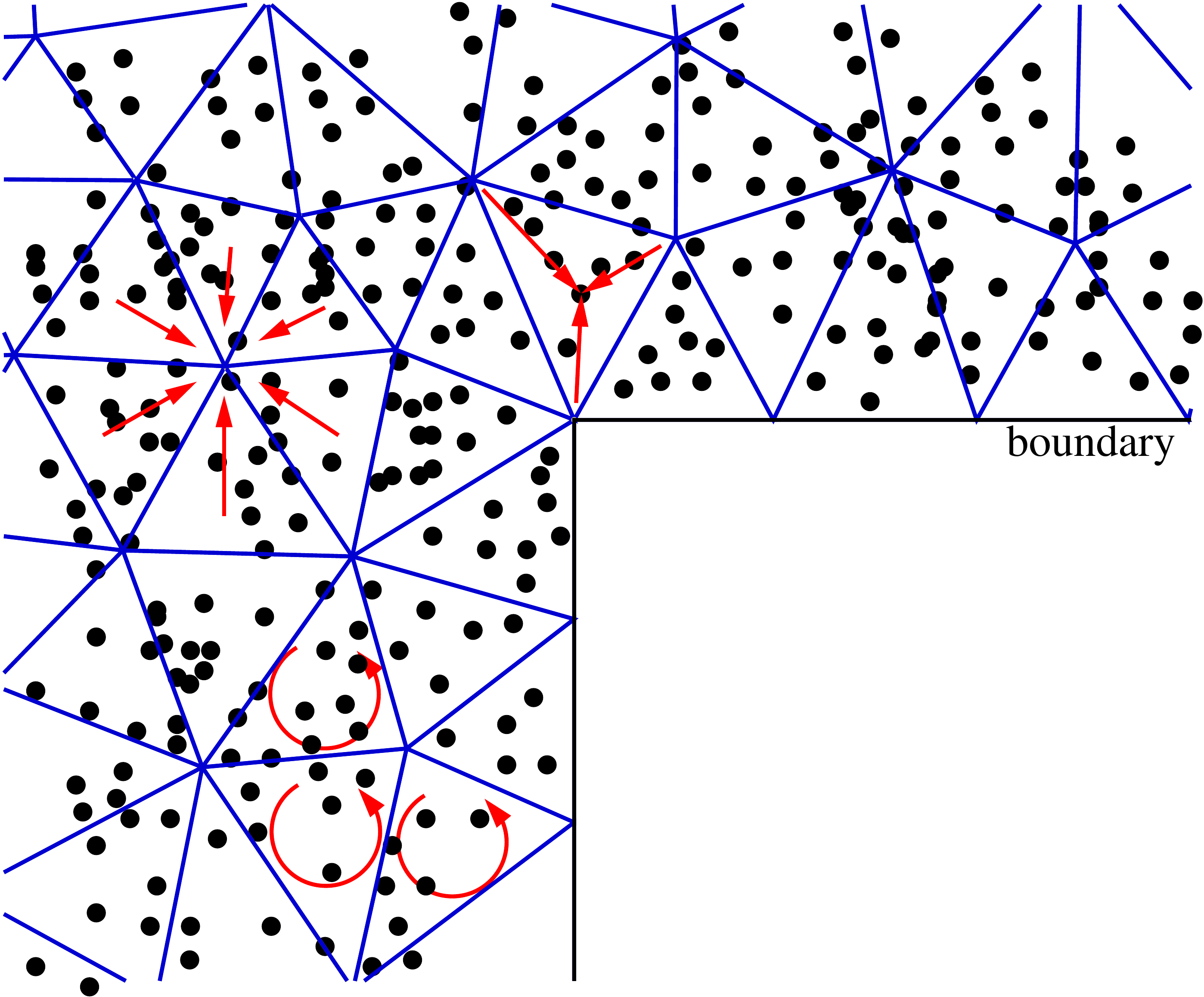}
\caption[Estimation of Eulerian statistics]{Estimation of Eulerian statistics on unstructured grids. In a first pass, element-based statistics are computed considering the particles residing in elements. In a second pass, element-based statistics are transferred to nodes by computing the averages of elements surrounding nodes. The nodal averages of each element are then used at particle locations in the Lagrangian governing equations.}
\label{fig:mesh-particles}
\figSpace
\end{figure}
where \(N\) is the number of particles participating in the local mean at \(\bv{x}\) and \(\mathcal{U}_i^p\) is the velocity vector of particle \(p\). In the first pass an element-based mean is computed considering the particles in a given element, \Fige{fig:mesh-particles}. In the second pass, these element-based means are transferred to nodes of the grid by calculating the average of the elements surrounding the nodes. Wherever Eulerian statistics are needed at particle locations, like in \Eqre{eq:Lagrangian-model}, the average of the nodal values are used for all particles residing in a given element. These node-based statistics are also used in the elliptic relaxation \Eqr{eq:elliptic-relaxation-Lagrangian} and pressure projection \Eqr{eq:pressure-projection-smooth} equations. An advantage of this two-pass procedure is that a natural smoothing is inherent in transferring statistics from elements to nodes. Using only nodal statistics to update particles also makes the method more robust, since it provides an efficient guard against the unwanted occurrence of empty elements, i.e.\ elements without any particles. The problem of high statistical error caused by an empty element is mitigated by the other elements surrounding the given node. Linked lists \citep{Lohner_01} provide an efficient access of unstructured-grid-based data from memory (e.g.\ elements surrounding points, points surrounding points, etc.). Once first-order statistics, like the mean velocity, are computed, higher order statistics are calculated by the same procedure. As an example, the Reynolds stress tensor is obtained by
\begin{equation}
\mean{u_iu_j}\equiv\int (V_i-\mean{U_i})(V_j-\mean{U_j})f_{\bv{\scriptscriptstyle Y}}(\bv{Y})\mathrm{d}\bv{Y}\cong\frac{1}{N}\sum_{p=1}^{N}\big(\mathcal{U}_i^p-\mean{U_i}\big)\big(\mathcal{U}_j^p-\mean{U_j}\big).
\end{equation}

\section{Derivatives of Eulerian statistics}
\label{sec:derivatives}
From finite element approximation theory, an unkown function \(q(\bv{x})\) given in nodes can be approximated over an element as
\begin{equation}
q(\bv{x})=\sum_{j=1}^n N^j(\bv{x})\hat{q}_j,\label{eq:approximation}
\end{equation}%
\nomenclature[Rq]{$q$}{an unknown function}%
\nomenclature[Rn]{$n$}{number of nodes of an Eulerian element}%
\nomenclature[Rq]{$\hat{q}_j$}{value of the function $q$ in node $j$}%
where \(n\) is the number of nodes of the element, \(\hat{q}_j\) is the value of the function \(q\) in node \(j\) and \(N^j\) are finite element shapefunctions. For speed and simplicity, we use only a single type of element (triangle) with linear shapefunctions, which are written in the local (\(\xi,\eta\)) coordinate system of the element as (see also \Fige{fig:element})
\begin{eqnarray}
N^\mathrm{A}&=&1-\xi-\eta,\nonumber\\
N^\mathrm{B}&=&\xi,\label{eq:shapefunctions}\\
N^\mathrm{C}&=&\eta.\nonumber
\end{eqnarray}%
\nomenclature[RN]{$N^j$}{finite element shapefunctions ($N^A,N^B,N^C$), \Eqre{eq:shapefunctions}}%
\nomenclature[Gn]{$\xi$}{local coordinate direction via nodes A and B of an Eulerian element, \Fige{fig:element}}%
\nomenclature[Gg]{$\eta$}{local coordinate direction via nodes A and C of an Eulerian element, \Fige{fig:element}}%
Employing the approximation in \Eqre{eq:approximation}, the spatial gradient of the expectation of any function \(Q(\bv{Y};\bv{x},t)\) can be computed over an element as
\begin{equation}
\frac{\partial Q}{\partial x_i} = \sum_{j=1}^n \frac{\partial N^j}{\partial x_i}\Hat{Q}_j,\label{eq:gradient}
\end{equation}%
\nomenclature[RQ]{$Q(\bv{Y})$}{random function with independent variables $\bv{Y}$}%
\nomenclature[RQ]{$\hat{Q}_j$}{nodal value of $Q$ at gridpoint $j$}%
where \(\hat{Q}_j\) denotes the nodal value of \(Q\) at gridpoint \(j\) of the element. The derivatives of the linear shapefunctions in \Eqre{eq:shapefunctions} in the global (\(x,y\)) coordinate system can be derived analytically \citep{Lohner_01}
\begin{equation}
\frac{\partial}{\partial x}
\begin{bmatrix}
N^\mathrm{A}\\
N^\mathrm{B}\\
N^\mathrm{C}\\
\end{bmatrix}
=
\frac{1}{2A_e}
\begin{bmatrix}
-y_{\mathrm{CA}} + y_{\mathrm{BA}}\\
y_{\mathrm{CA}}\\
-y_{\mathrm{BA}}\\
\end{bmatrix},
\quad\quad\quad\quad
\frac{\partial}{\partial y}
\begin{bmatrix}
N^\mathrm{A}\\
N^\mathrm{B}\\
N^\mathrm{C}\\
\end{bmatrix}
=
\frac{1}{2A_e}
\begin{bmatrix}
x_{\mathrm{CA}} - x_{\mathrm{BA}}\\
-x_{\mathrm{CA}}\\
x_{\mathrm{BA}}\\
\end{bmatrix},
\end{equation}%
\nomenclature[RA]{$A_e$}{area of element $e$, \Fige{fig:element}}%
\nomenclature[Rx]{$x_\mathrm{A}$}{$x$ coordinate of node A, \Fige{fig:element}}%
\nomenclature[Rx]{$x_\mathrm{CA}$}{$x_\mathrm{C}-x_\mathrm{A}$, \Fige{fig:element}}%
where \(A_e\) is the area of element \(e\). The derivatives are constant functions and are based only on the location of the gridpoints (see also \Fige{fig:element}), e.g.\ \(y_\mathrm{CA}=y_\mathrm{C}-y_\mathrm{A}\). If the grid does not change during computation, these derivatives can be pre-computed and stored in advance of timestepping.

Second derivatives are obtained using a two-pass procedure. In the first pass the first derivatives are computed using \Eqre{eq:gradient} and then transferred to nodes by computing the averages of the elements surrounding nodes. The same procedure is applied to the derivatives in gridpoints in the second pass to obtain second derivatives.

\section{Estimation of the velocity-conditioned scalar mean}
\label{sec:conditional_stats}
\Eqre{eq:IECM} requires the estimation of the scalar mean conditioned on the velocity field \(\mean{\phi|\bv{V}}\). In the current case, this is defined as
\begin{equation}
\mean{\phi|\bv{V}}\equiv\int \psi f_{\scriptscriptstyle\Omega\phi|\bv{\scriptscriptstyle U}}(\omega,\psi|\bv{V})\mathrm{d}\omega\mathrm{d}\psi,
\end{equation}%
\nomenclature[Rf]{$f_{\scriptscriptstyle\Omega\phi"|\bv{\scriptscriptstyle U}}(\omega,\psi"|\bv{V})$}{joint PDF of $\Omega$ and $\phi$ conditional on $\bv{U}=\bv{V}$, \Eqre{eq:defvelcondf}}%
\nomenclature[Rf]{$f_{\bv{\scriptscriptstyle U}}(\bv{V})$}{joint PDF of velocity}%
\nomenclature[Rf]{$f_{\bv{\scriptscriptstyle U}\scriptscriptstyle\Omega\phi}(\bv{V},\omega,\psi)$}{joint PDF of velocity, turbulent frequency and scalar ($f_{\bv{\scriptscriptstyle Y}}(\bv{Y})$)}%
where the conditional PDF \(f_{\scriptscriptstyle\Omega\phi|\bv{\scriptscriptstyle U}}\) is expressed through Bayes' rule using the full PDF \(f_{\bv{\scriptscriptstyle Y}}(\bv{Y})\) and the marginal PDF of the velocity \(f_{\bv{\scriptscriptstyle U}}(\bv{V})\) as
\begin{equation}
f_{\scriptscriptstyle\Omega\phi|\bv{\scriptscriptstyle U}}(\omega,\psi|\bv{V})\equiv\frac{f_{\bv{\scriptscriptstyle Y}}(\bv{Y})}{f_{\bv{\scriptscriptstyle U}}(\bv{V})}\equiv\frac{f_{\bv{\scriptscriptstyle U}\scriptscriptstyle\Omega\phi}(\bv{V},\omega,\psi)}{f_{\bv{\scriptscriptstyle U}}(\bv{V})}.\label{eq:defvelcondf}
\end{equation}
Mathematically, the conditional mean \(\mean{\phi|\bv{V}}\) defines a mean value for each combination of its conditional variables, i.e.\ in a three-dimensional flow, in every spatial and temporal location \(\mean{\phi|\bv{V}}\) is a function that associates a scalar value to a vector, \(\mean{\phi|\bv{V}}:\mathbb{R}^3\rightarrow\mathbb{R}\). In practice, this means that the velocity-sample space needs to be discretized (divided into bins) and different mean values have to be computed for each bin using the particles whose velocities fall into the bin. In order to keep the statistical error small this procedure would require a large number of particles in every element. To overcome this difficulty, \citet{Fox_96} proposed a method in which the three-dimensional velocity space is projected onto a one-dimensional subspace where the discretization is carried out. This substantially reduces the need for an extensive number of particles. This projection method is exact in homogeneous turbulent shear flows, where the joint velocity PDF is Gaussian. Nevertheless, in more complex situations it can still be incorporated as a modeling assumption.

A more general way of computing the conditional mean is to use three-dimensional binning of the veloctiy sample space \(\bv{V}\). In order to homogenize the statistical error over the sample space, the endpoints of the conditioning bins in each direction can be determined so that the distribution of the number of particles falling into the bins is as homogeneous as possible. For a Gaussian velocity PDF this can be accomplished by using statistical tables to define the endpoints \citep{Fox_96}. If the underlying velocity PDF is not known, however, another strategy is required. Note that there is absolutely no restriction on the distribution of the conditioning intervals. In other words they need not be equidistant, need not be the same (or even the same number) in every dimension and can also vary from element to element. Only some sort of clustering of the particles is needed, i.e.\ grouping them into subgroups of particles with similar velocities. A simple algorithm that accomplishes this task is as follows.

Without loss of generality, we assume that a sample-space binning of (\(2\times2\times2\)) is desired. In a first step all particles residing in the given element are sorted according to their \(\mathcal{U}\) velocity component. Then the first and the second halves of the group are separately sorted according to their \(\mathcal{V}\) component. After further dividing both halves into halves again, each quarter is sorted according to the \(\mathcal{W}\) component. Finally, halving the quarters once again we compute scalar means for each of these 8 subgroups. Naturally, the binning can be any other structure with higher (even unequal) number of bins if that is desirable, e.g.\ (\(5\times5\times5\)) or (\(4\times12\times5\)). This procedure defines the bins dynamically based on the criterion that the bin-distribution of the number of particles be as homogeneous as possible. By doing that, it homogenizes the statistical error over the sample space and also ensures that every bin will contain particles. This simple procedure is completely general, independent of the shape and extent of the velocity PDF and dynamically adjusts the bin-distribution to the underlying PDF in every element. It is also robust, since if the number of particles in an element happens to be very low compared to the desired binning, e.g.\ we only have 5 particles for the 125 bins of a (\(5\times5\times5\)) binning structure, the above sorting \& dividing procedure can be stopped at any stage and the subgroups defined up to that stage can already be used to estimate the conditioned means. In other words, if in the above example we require that at least 2 particles should remain in every subgroup we simply stop after the first sort and only use two groups. An algorithm that accomplishes the conditioning step after the particles have been sorted into subgroups is detailed in \autoref{app:velocity-conditioning}. The statistical error resulting from employing different number of conditioning bins is investigated in more detail in \autoref{chap:channel}.

\section{Particle tracking}
\label{sec:particle_tracking}
\begin{figure}
\centering
\input{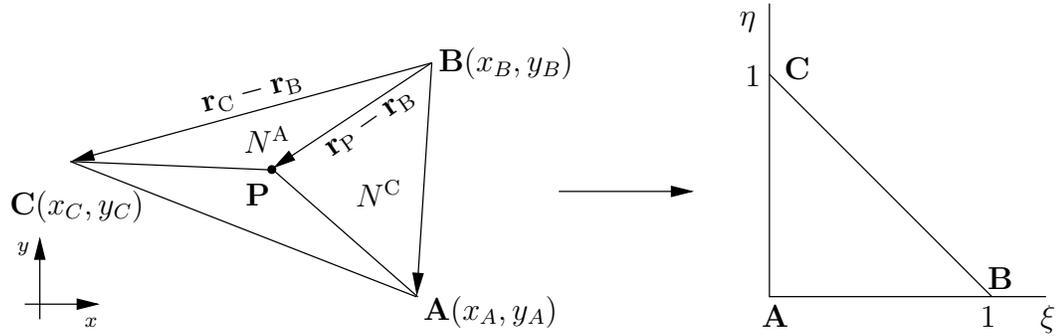}
\caption[Particle tracking with triangular elements; FEM mapping]{The decision whether a particle resides in a triangular element is made based on computing cross-products of element-edge vectors and vectors of vertex-particle coordinates. E.g.\ \(N^\mathrm{A}\) is half of the signed area of the parallelogram spanned by vectors \((\bv{r}_\mathrm{C}-\bv{r}_\mathrm{B})\) and \((\bv{r}_\mathrm{P}-\bv{r}_\mathrm{B})\). Also shown is the local coordinate system (\(\xi,\eta\)) of the triangle after a linear mapping with the finite element shapefunctions in \Eqres{eq:shapefunctions}.}
\label{fig:element}
\figSpace
\end{figure}

Particles have to be tracked continuously as they travel throughout the grid in order to identify which element they contribute to when local statistics are computed. A variety of algorithms with different characteristics have been developed to accomplish this task \citep{Grigoryev_02}. Since we use explicit timestepping, the particles will not jump over many elements in a timestep, thus the fastest way to track particles is some sort of known-vicinity algorithm \citep{Lohner_95}.  The two-dimensional particle tracking employed here is as follows. If a particle is not in its old element (where it was in the last timestep), it is searched in the \emph{next best element} of the surrounding elements. The knowledge of the next best element is a feature of the basic interpolation algorithm that is used to decide whether the particle resides in a given element. The interpolation algorithm is based on FEM shapefunctions, which are usually employed for approximating unknowns over elements (as it is used in \autoref{sec:Eulerian_equations} to discretize the Eulerian equations and in \autoref{sec:derivatives} to approximate functions and their derivatives) and correspond to a linear mapping between the global and local coordinates of the element, see also \Fige{fig:element}. We use these shapefunctions here for interpolation in two dimensions, but this procedure can also be used in a three-dimensional case with tetrahedra \citep{Lohner_95}. In the current two-dimensional case, evaluating two of them is sufficient to decide whether the particle is inside of the element. The decision is made by the following condition (see also \Fige{fig:element})
\begin{eqnarray}
&&\mathtt{if}\enskip\left\{\enskip\left(N^\mathrm{A} > 0\right)\quad\mathtt{and}\quad\left(N^\mathrm{C} > 0\right)\quad\mathtt{and}\quad\left(N^\mathrm{A}+N^\mathrm{C}\right)<A_e\enskip\right\}\label{eq:element-test}\\
&&\quad\mathrm{inside}\nonumber\\
&&\mathtt{else}\nonumber\\
&&\quad\mathrm{outside}\nonumber
\end{eqnarray}
where \(A_e\) is the total area of the element, while \(N^\mathrm{A}\) and \(N^\mathrm{C}\) are the signed half-lengths of the cross-products
\begin{eqnarray}
N^\mathrm{A}&=&\frac{1}{2}\big|(\bv{r}_\mathrm{C}-\bv{r}_\mathrm{B}) \times (\bv{r}_\mathrm{P}-\bv{r}_\mathrm{B})\big|,\\
N^\mathrm{C}&=&\frac{1}{2}\big|(\bv{r}_\mathrm{P}-\bv{r}_\mathrm{B}) \times (\bv{r}_\mathrm{A}-\bv{r}_\mathrm{B})\big|.
\end{eqnarray}%
\nomenclature[RP]{$\bv{\mathrm{P}}$}{particle location}%
\nomenclature[Rr]{$\bv{r}_\mathrm{A}$}{vector to node $\mathrm{A}$ of an Eulerian element, \Fige{fig:element}}%
\nomenclature[RA]{$\bv{\mathrm{A}}$}{label of an element vertex, \Fige{fig:element}}%
Note that these are also the area coordinates of the triangle corresponding to the nodes \(A\) and \(C\) and also the values of the finite element shapefunctions corresponding to the three nodes, \Eqres{eq:shapefunctions}, evaluated at the particle location \(\bv{\mathrm{P}}\). A convenient feature of this procedure is that once the values \(N^\mathrm{A},\) \(N^\mathrm{C}\) and \(N^\mathrm{B}=A_e-N^\mathrm{A}-N^\mathrm{C}\) are evaluated, in case the particle is not found in the element, they also give us a hint about the direction of the particle location that is outside of the element. If condition \Eqr{eq:element-test} is not satisfied, at least one of \(N^\mathrm{A}\), \(N^\mathrm{B}\) and \(N^\mathrm{C}\) is negative. The next best element is in the direction corresponding to the lowest of the three values. Combining this with a data structure (e.g.\ a linked list \citep{Lohner_01}) that stores the element indices surrounding elements, we can easily and efficiently identify which element is most likely to contain the particle or at least which direction to search next. Most of the time, the particles do not jump out of their host elements, but if they do, this procedure finds them in usually 2-3 steps.
\begin{figure}[t]
\centering
\resizebox{7cm}{!}{\input{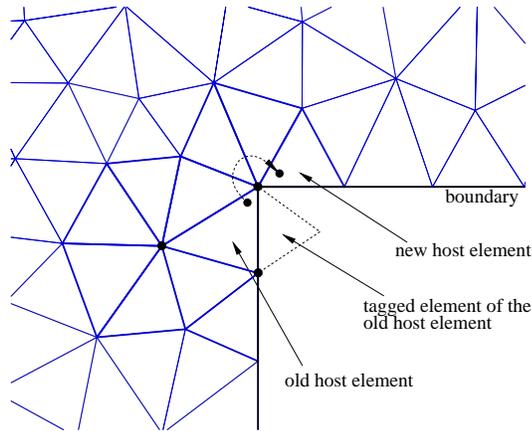}}
\caption[Particle tracking at concave boundaries]{A particle jumping over a concave corner on the boundary and the next best guess based on its old host element would be through the boundary, outside of the domain. A fall-back procedure finds the new host element of the particle by searching the elements surrounding the nodes (displayed with thicker edges) of its old host element.}
\label{fig:corner}
\figSpace
\end{figure}

The above neighbor-to-neighbor algorithm performs very well in the domain, but it may fail to jump over concave boundaries, resulting in a dead-lock \citep{Lohner_95}. In order to remedy this problem the following strategy is employed. An element on the boundary has two surrounding elements at most and the ones that would be outside of the domain are tagged in the data structure that stores the three element indices surrounding elements, see also \Fige{fig:corner}. If this tagged element is returned as the next best guess, the particle is on the other side of a concave section (or a corner) of the boundary. Since even in this case the particle must be close to its old host element, the particle is searched next in all elements surrounding the nodes of its old host element. (This is also stored in a linked list for fast access.) This fall-back procedure always finds the particle around a corner, thus a brute-force search is not necessary over all elements.

\section{Particle-number control}
\label{sec:particle-number-control}
In the setup phase an equal number of particles are uniformly generated into each element with the initial velocities \(\mathcal{U}_i\) sampled from a Gaussian distribution with zero mean and variance 2/3, i.e.\ the initial Reynolds stress tensor is isotropic with unit turbulent kinetic energy, \(\mean{u_iu_j}=\frac{2}{3}\delta_{ij}\). Initial particle frequencies \(\omega\) are sampled from a gamma distribution with unit mean and variance 1/4 and the scalar concentration \(\psi\) is set to 0.

During the timestepping procedure a sufficient number of particles have to be present in every element at all times to keep the deterministic error due to bias small \citep{Pope_95b}. However, the grid can be refined differently in different regions of the domain, as it is done at walls to resolve the boundary layer or around a concentrated source of a passive scalar to capture the high scalar gradients. Since the particles themselves model real fluid particles, at locations where the grid is refined more particles are necessary for an increased resolution. Therefore it is reasonable to keep the element-distribution of the number of particles as homogeneous as possible. Particle-number control is a delicate procedure in PDF methods, because external modification of the particle locations or properties may result in undesired changes of the local statistics and the joint PDF itself. Nevertheless, particle splitting and merging techniques are routinely applied to keep the particle distribution reasonable and to improve the efficiency and stability of the simulation \citep{Cassiani_07b}. \autoref{app:particle-redistribution} describes the algorithm that we developed to keep the number of particles per element above a certain treshold and to guard the simulation against the occurrence of empty elements (i.e. elements without particles).

In what follows, we describe a simple testcase that we use to investigate the error introduced by the particle redisitribution. Note that the traditional way of referring to this procedure is \emph{particle splitting and merging}. Since we do not change the total number of particles throughout the simulation (which is more memory efficient than splitting and merging) we refer to this as \emph{particle redistribution}. To investigate the error, we consider the simplified model equations
\begin{eqnarray}
\mathrm{d}\mathcal{X}_i &=& \mathcal{U}_i\mathrm{d}t,\label{eq:toy-position}\\
\mathrm{d}\mathcal{U}_i &=& -(\mathcal{U}_i-\alpha\mean{U_i})\mathrm{d}t + \sqrt{2}\mathrm{d}W_i,\label{eq:toy-velocity}
\end{eqnarray}%
\nomenclature[Ga]{$\alpha$}{scalar parameter in the simplified particle model, \Eqre{eq:toy-velocity}}%
where \(\alpha\) is a scalar parameter and the initial conditions for \(\mathcal{U}_i\) are taken to be independent, standardized, normally distributed random variables:
\begin{equation}
\mean{U_i}=0, \qquad \mean{u_iu_j} = \delta_{ij}.
\end{equation}
\Eqre{eq:toy-velocity} is characteristic of the Langevin equation \Eqr{eq:Lagrangian-model} without viscous effects, e.g.\ \Eqre{eq:Lagrangian-model-wf}, see also \citet{Xu_97}. The mean \(\mean{U_i}\) of the solution of the stochastic differential equation \Eqr{eq:toy-velocity} is the solution of the following linear deterministic differential equation \citep{Arnold_74}%
\begin{eqnarray}
\frac{\mathrm{d}\mean{U_i}}{\mathrm{d}t} &=& -(\mean{U_i}-\alpha\mean{U_i}),\label{eq:toy-velocity-mean}\\
\mean{U_i}(t=0) &=& 0.\label{eq:toy-ic-mean}
\end{eqnarray}%
\begin{figure}[t]
\centering
\resizebox{12cm}{!}{\input{ou-domain.pstex_t}}
\caption[Domain for a testcase to investigate the error of particle redistribution]{A rectangular domain with a stretched grid to test the error introduced by the particle redistribution algorithm using \Eqres{eq:toy-position} and \Eqr{eq:toy-velocity}.}
\label{fig:toy-domain}
\figSpace
\end{figure}%
It can be seen that the trivial solution \(\mean{U_i}=0\) satisfies the above deterministic initial value problem. For a nonzero initial condition the solution of \Eqre{eq:toy-velocity} is stable and reaches steady state if \(\alpha<1\) with \(\mean{U_i}=0\) and \(\mean{u_iu_j}=\delta_{ij}\). For \(\alpha>1\) the equation becomes unstable and the solution grows exponentially, while for \(\alpha=0\) the equation is neutrally stable. For our purposes we use \(\alpha=0.5\). \Eqres{eq:toy-position} and \Eqr{eq:toy-velocity} are advanced on a rectangular domain with two free-slip walls (from where particles are simply reflected) and a periodic inflow/outflow boundary-pair, see \Fige{fig:toy-domain}. The domain is highly stretched on purpose in the \(y\) direction. Initially, an equal number of particles are generated into every element, which in the current case results in a spatially inhomogeneous particle distribution. As the timestepping advances the particles naturally tend to evolve into a spatially homogeneous distribution, which may result in empty elements in the highly refined region if the number of particles is too small. This is circumvented by the particle redistribution algorithm. We will test the algorithm by calculating the time-evolutions of the spatial average of the diagonal components of \(\mean{u_iu_j}\), indicated by \(\overline{\mean{u_iu_j}}\), using different initial number of particles per element \(N_{p/e}\). In order to ensure that the particle redistribution algorithm intervenes on a same level in each case, the ratio%
\nomenclature[Ru]{$\overline{\mean{u_iu_j}}$}{spatial average of $\mean{u_iu_j}$ (\autoref{sec:particle-number-control})}%
\begin{equation}
\frac{N_{\scriptscriptstyle{p/e}}}{N^{\scriptscriptstyle{\mathrm{min}}}_{\scriptscriptstyle{p/e}}} \propto \textrm{number of particles moved}\label{eq:work}
\end{equation}%
\nomenclature[RN]{$N_{\scriptscriptstyle{\scriptscriptstyle{p/e}}}$}{number of particles per element initially}%
\nomenclature[RN]{$N^\mathrm{min}_{\scriptscriptstyle{\scriptscriptstyle{p/e}}}$}{required minimum number of particles per element (\autoref{sec:particle-number-control})}%
is kept constant. In other words, as the initial number of particles \(N_{\scriptscriptstyle{\scriptscriptstyle{p/e}}}\) is increased, we increase the required minimum number of particles per element \(N^\mathrm{min}_{\scriptscriptstyle{\scriptscriptstyle{p/e}}}\) as well, so that the number of particles that will have to be moved is approximately the same, hence the algorithm intervenes at the same level. To verify that this is the case, the number of times the redistribution algorithm is called (the number of particles moved in a timestep) is monitored and plotted in \Fige{fig:impcnt} for the different cases. \Fige{fig:solution} depicts \(\overline{\mean{u_iu_j}}\) for different values of \(N_{p/e}\). It can be seen in \Fige{fig:solution} (a) that the algorithm reproduces the analitical solution with a given numerical error. This error, which is always present in the numerical solution of stochastic differential equations, can be decomposed into three different parts: truncation error due to finite-size timesteps, deterministic error (or bias) due to the finite number of particles employed and random (or statistical) error \citep{Pope_95b}. The particle redistribution introduces an additional error which is directly visible by comparing \Figse{fig:solution} (a) and (d). It is also apparent that the bias decreases with increasing number of particles as it can be expected. However, \Figse{fig:solution} (b)-(f) also show that the additional error introduced by the particle redistribution also diminishes as the number of particles increase while the intervention of the redistribution, \Eqre{eq:work}, is kept at a constant level. This can be seen more directly in \Fige{fig:error}, which depicts the evolution of the total relative numerical error defined as
\begin{figure}
\centering
\resizebox{10cm}{!}{\input{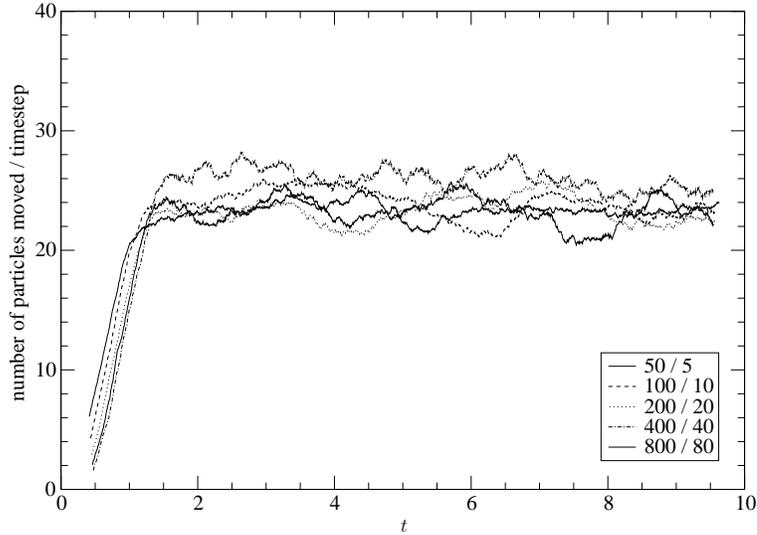}}
\caption[The number of particles moved by particle redistribution]{The number of particles moved in each timestep by the particle redistribution algorithm for different total number of particles. In the legend the constant \(N_{\scriptscriptstyle{p/e}}/N^\mathrm{min}_{\scriptscriptstyle{p/e}}\) ratio is displayed.}
\label{fig:impcnt}
\figSpace
\end{figure}%
\begin{figure}[t]
\centering
\resizebox{15cm}{!}{\input{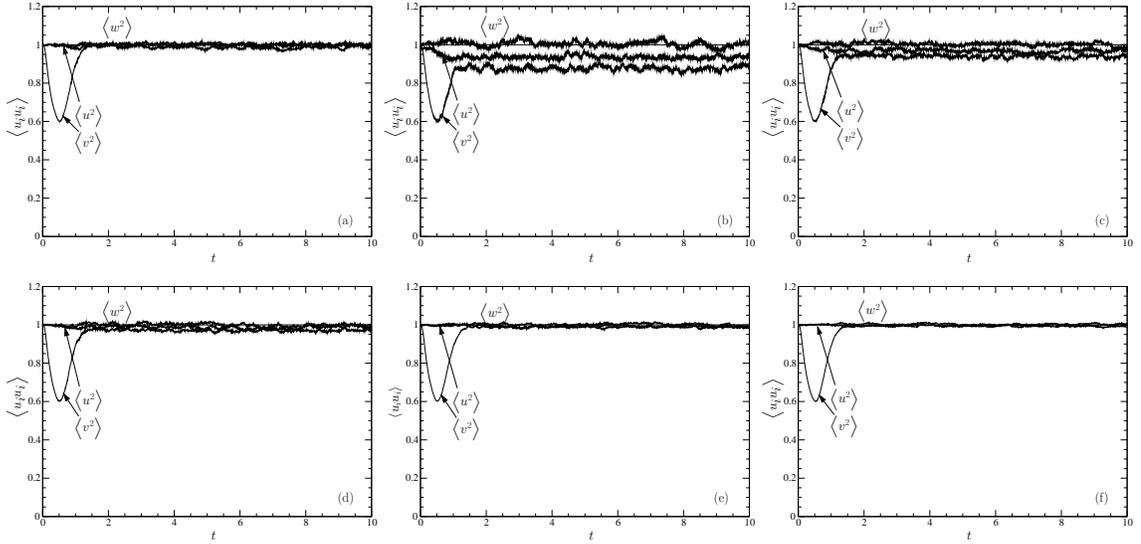}}
\caption[Results of a simplified testcase with particle redistribution]{Time-evolutions of the diagonal components of \(\overline{\mean{u_iu_j}}\) solving \Eqres{eq:toy-position} and \Eqr{eq:toy-velocity} employing different number of particles. (a) No redistribution with initial number of particles per element \(N_{p/e}\)=200; redistribution with (b) \(N_{p/e}\)=50, (c) \(N_{p/e}\)=100, (d) \(N_{p/e}\)=200, (e) \(N_{p/e}\)=400 and (f) \(N_{p/e}\)=800, respectively. The ratio \(N_{p/e}/N^\textrm{min}_{p/e}\)=10 is kept constant for cases (b) to (f). The horizontal line at the ordinate 1 depicts the analitical solution at steady state.}
\label{fig:solution}
\figSpace
\end{figure}%
\begin{figure}
\centering
\resizebox{10cm}{!}{\input{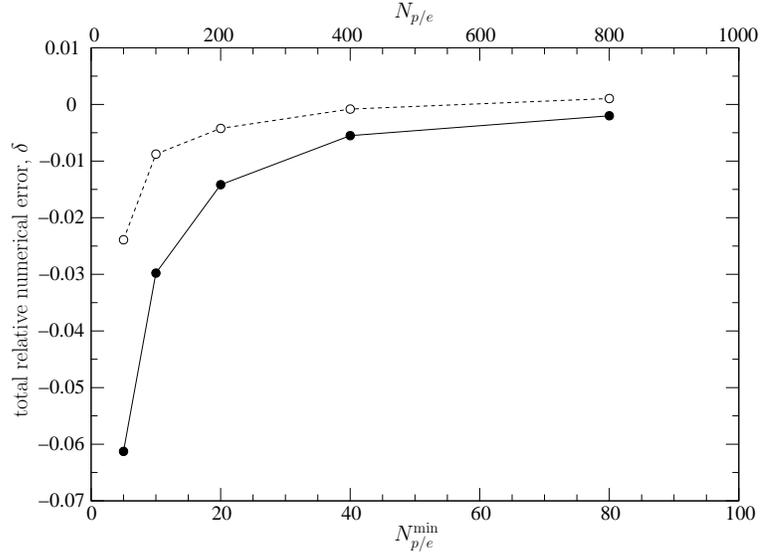}}
\caption[Evolution of total relative numerical error with particle redistribution]{Evolution of the total relative numerical error defined by \Eqre{eq:error} with increasing number of particles. Solid line -- with redistribution, dashed line -- without redistribution.}
\label{fig:error}
\figSpace
\end{figure}
\begin{equation}
\delta = \frac{k_c-k_a}{k_a},\label{eq:error}
\end{equation}%
\nomenclature[Gd]{$\delta$}{relative numerical error in kinetic energy, \Eqre{eq:error}}%
\nomenclature[Rk]{$k_a$}{analytical kinetic energy, \Eqre{eq:error}}%
\nomenclature[Rk]{$k_c$}{computed kinetic energy, \Eqre{eq:error}}%
where \(k_c\) and \(k_a\) denote the computed and analytical kinetic energy, respectively. This error incorporates both the usual numerical errors and the additional one due to the particle redistribution algorithm. For comparison, the evolution of the error without particle redistribution is also displayed. Since the total sum of the errors converges to zero, the error introduced by the redistribution algorithm also diminishes and the solution converges to the PDF without redistribution.

We have found that a particle redistribution algorithm of a similar sort (or particle splitting and merging) is essential to provide adequate numerical stability in modeling inhomogeneous flows especially in complex geometries. In addition, it also dramatically reduces the need for high number of particles per elements on stretched grids.

\section{No-slip wall-boundary conditions}
\label{sec:wall_conditions}
Over any given time-interval a particle undergoing reflected Brownian motion in the vicinity of a wall may strike the wall infinitely many times \citep{Dreeben_98}. This means that particles can follow three different trajectories when interacting with walls. The particle either (a) crosses the wall during the timestep and it is behind the wall at the end of the timestep or (b) crosses the wall during the timestep but it is not behind the wall at the end of the timestep or (c) does not cross the wall during the timestep. Therefore wall-conditions have to be enforced on particles that follow trajectory (a) and (b). The probability that the particle following trajectory (b) crossed the wall during timestep \(\Delta t\) can be calculated by \citep{Karatzas_91}
\begin{equation}
f_w=\exp\left(\frac{-d^nd^{n+1}}{\nu\Delta t}\right),\label{eq:fw}
\end{equation}%
\nomenclature[Rf]{$f_w$}{probability of a particle crossing the wall, \Eqre{eq:fw}}%
\nomenclature[Rd]{$d^n$}{distance of a particle from a wall at timestep $n$, \Eqre{eq:fw}}%
where \(d^n\) denotes the distance of the particle from the wall at timestep \(n\). Thus, particle wall-conditions are applied if
\begin{equation}
d^{n+1} < 0, \quad \textrm{trajectory (a),}
\end{equation}
or if
\begin{equation}
d^{n+1} \ge 0 \quad \mathrm{and} \quad \eta < f_w, \quad \textrm{trajectory (b),} \label{eq:exp-wall-condition}
\end{equation}%
\nomenclature[Gg]{$\eta$}{random variable with a standard uniform distribution, \Eqre{eq:exp-wall-condition}}%
where \(\eta\) is a random variable with a standard uniform distribution. The new particle location is calculated based on perfect reflection from the wall. The particle velocity is set according to the no-slip condition
\begin{equation}
\mathcal{U}_i = 0.
\end{equation}
A boundary condition on the characteristic turbulent frequency \(\omega\) has to ensure the correct balance of the turbulent kinetic energy at the wall \citep{Dreeben_98} and has to be consistent with the near-wall kinetic energy equation
\begin{equation}
\nu\frac{\partial^2k}{\partial n^2}+\varepsilon=0,
\end{equation}
where \(\bv{n}\) is the outward normal of the wall. Accordingly, the frequency for a particle striking the wall is sampled from a gamma distribution with mean and variance respectively
\begin{equation}
\mean{\omega} = \frac{1}{C_T}\frac{\mathrm{d}\sqrt{2k}}{\mathrm{d}y} \quad \mathrm{and} \quad \Big<\big(\omega-\mean{\omega}\big)^2\Big>=C_4\mean{\omega}^2.\label{eq:frequency-wallstats}
\end{equation}
For better performance the above particle conditions are only tested and enforced for particles that reside close to walls, i.e.\ in elements that have at least an edge or a node on a no-slip wall-boundary.

Following \citet{Dreeben_98}, the wall-boundary condition for the elliptic relaxation equation \Eqr{eq:elliptic-relaxation-Lagrangian} is set according to
\begin{equation}
\wp_{ij}=-4.5\varepsilon n_in_j.
\end{equation}
For the pressure-Poisson equation \eqref{eq:pressure-projection-smooth}, a Neumann-condition is obtained from the Eulerian mean-momentum equation
\begin{equation}
\frac{\partial\mean{U_i}}{\partial t} + \mean{U_j}\frac{\partial\mean{U_i}}{\partial x_j} + \frac{1}{\rho}\frac{\partial\mean{P}}{\partial x_i} =  \nu\nabla^2\mean{U_i} - \frac{\partial\mean{u_iu_j}}{\partial x_j} \label{eq:Reynolds},
\end{equation}
by taking the normal component at a stationary solid wall
\begin{equation}
\frac{1}{\rho}\frac{\partial\mean{P}}{\partial x_i}n_i = \nu\frac{\partial^2\mean{U_i}}{\partial x_j\partial x_j}n_i - \frac{\partial\mean{u_iu_j}}{\partial x_j}n_i.\label{eq:wall-pressure-condition}
\end{equation}
In the wall-functions case, when the boundary layers along no-lip walls are represented based on the ``law of the wall'', the advection term in \Eqre{eq:Reynolds} is non-zero at $y_p$, therefore the normal component of this term appears in the Neumann condition for the mean pressure
\begin{equation}
\frac{1}{\rho}\frac{\partial\mean{P}}{\partial x_i}n_i = \nu\frac{\partial^2\mean{U_i}}{\partial x_j\partial x_j}n_i - \frac{\partial\mean{u_iu_j}}{\partial x_j}n_i - \mean{U_j}\frac{\partial\mean{U_i}}{\partial x_j}n_i.
\end{equation}

\section{Parallel random number generation}
\label{sec:parallel_random_number_generation}
The solver has been parallelized and run on different shared memory architectures. Both the initialization and the timestepping require a large number of random numbers with different distributions and characteristics. Two components of the position \(\mathcal{X}_i\) and three components of the velocity \(\mathcal{U}_i\) are retained for a two-dimensional simulation, therefore the governing equations \Eqr{eq:Lagrangian-position-exact}, \Eqr{eq:Lagrangian-model} and \Eqr{eq:frequency-model} altogether require 6 independent Gaussian random numbers for each particle in each timestep. Since these 6 numbers per particle are always needed and are always Gaussian, they can be efficiently stored in a table, which is regenerated in each timestep. Different methods exist to efficiently sample pseudo-random numbers in parallel \citep{Mascagni_97}. In order to be able to reproduce the simulation results and to avoid surpassing cross-correlations between random number streams, we initialize a single stream and split it into \(k\) non-overlapping blocks, where \(k\) is the number of parallel threads. Then each of the threads generates from its own corresponding block, avoiding data races with other threads. This can be quite efficient, since a large amount of random numbers are generated at once and each thread accesses only its own portion of the stream. The same block-splitting technique is used to fill another table with uniform random numbers for the boundary condition \Eqre{eq:exp-wall-condition}. Using this sampling technique, an almost ideal speedup can be achieved when random numbers in tables are regenerated, see also Table \ref{tab:timestep_profile}. For those equations in which the number of random numbers is a priori unknown (e.g.\ sampling a gamma distribution for the wall-condition of \Eqre{eq:frequency-wallstats} for particles that struck the wall), a stream is split into \(k\) disjoint substreams and the leap-frog technique is used to sample from them in parallel \citep{Entacher_98}. The leapfrog technique could also be used for a priori known number of random numbers, but due to its higher cache-efficiency, block-splitting performs slightly better. (In block-splitting the sampling positions in the streams are much farther from each other and thus the processes are less likely to interfere with each other's caches.) These techniques have been found essential to achieve a good parallel performance for the loop advancing the particles, see also \autoref{sec:profile}.

\section{Solution procedure and execution profile}
\label{sec:profile}
\begin{table}[t!]
\caption[Profile and relative execution times for a timestep]{\label{tab:timestep_profile}Structure and profile of a timestep with relative execution times compared to the time spent on the full timestep and parallel performances of each step on a machine with two quad-core processors. The listing order corresponds to the order of execution. The performance data is characteristic of a case with 10M particles using a grid with 20K triangles, the simulation altogether requiring approximately 1.2GB memory. The processors are two quad-core CPUs (8 cores total), each pair sharing 4MB cache and the CPU-to-memory communication bandwidth.\\}
\ls{1}
\begin{tabular*}{1.0\textwidth}{p{5cm}@{\hspace{0.8cm}}r@{.}l@{\hspace{0.8cm}}r@{.}l@{\hspace{1.1cm}}r@{.}l@{\hspace{1.1cm}}r@{.}l@{\hspace{1.1cm}}r@{.}l}
\hline\hline
task & \multicolumn{2}{p{1.5cm}}{relative execution time} & \multicolumn{2}{p{1.0cm}}{speedup with 2 CPUs} & \multicolumn{2}{p{1.0cm}}{speedup with 4 CPUs} & \multicolumn{2}{p{1.0cm}}{speedup with 6 CPUs} & \multicolumn{2}{p{1.0cm}}{speedup with 8 CPUs} \\
\hline
\(\bullet\) compute the size of the next timestep, see \autoref{sec:timestepping} & 0&001 \% & \multicolumn{8}{c}{not parallelized} \\
\(\bullet\) solve elliptic relaxation equation \Eqr{eq:elliptic-relaxation-Lagrangian}, see \autoref{sec:Eulerian_equations} & 2&87 \% & 1&91 & 4&08 & 5&76 & 7&60 \\
\(\bullet\) advance particle properties according to \Eqres{eq:Lagrangian-position-exact}, \Eqr{eq:Lagrangian-model}, \Eqr{eq:frequency-model} and \Eqr{eq:IECM} & 73&2 \% & 2&02 & 4&12 & 6&16 & 8&20 \\
\(\bullet\) regenerate random number tables, see \autoref{sec:parallel_random_number_generation} & 19&01 \% & 2&01 & 3&99 & 5&79 & 7&50 \\
\(\bullet\) solve pressure-Poisson equation, see \autoref{sec:Eulerian_equations} & 2&0 \% & 1&86 & 3&49 & 4&55 & 5&02 \\
\(\bullet\) correct mean velocities, see \autoref{sec:Eulerian_equations} & 1&0 \% & 1&69 & 1&95 & 1&94 & 1&96 \\
\(\bullet\) compute Eulerian statistics, see Sections \ref{sec:Eulerian_statistics}--\ref{sec:conditional_stats} & 1&6 \% & 1&22 & 1&79 & 1&67 & 1&77 \\
\hline
one complete timestep & 99&68 \% & 1&98 & 3&95 & 5&55 & 7&20 \\
\hline\hline
\end{tabular*}
\tableSpace
\end{table}
The main stages of one complete timestep in their order of execution are displayed in Table \ref{tab:timestep_profile}. Also shown are the percentage of the execution times of each stage relative to a complete timestep and their speedups on a machine with two quad-core processors. The performance data were obtained by running a case that contained approximately 10 million particles and the Eulerian grid consisted of about 20 thousand triangles.

A significant portion of the execution time is spent on advancing the particle-governing equations. This is mostly a loop which can be constructed in two fundamental ways: in an element-based or in a particle-based fashion as displayed in Table \ref{tab:loops}.%
\begin{table}[t!]
\caption[Two designs of a loop to update particles]{\label{tab:loops}Two fundamental ways of constructing a loop to advance the particle-governing equations \Eqr{eq:Lagrangian-position-exact}, \Eqr{eq:Lagrangian-model}, \Eqr{eq:frequency-model} and \Eqr{eq:IECM}. (i) -- element-based loop, (ii) -- particle-based loop.\\}
\begin{tabular*}{1.0\textwidth}{l}
\hline\hline
\end{tabular*}
(i)
\begin{tabbing}
  \texttt{for} \= all \= Eulerian elements \texttt{e}\\
  \> gather Eulerian nodal statistics for element \texttt{e};\\
  \> compute element-average statistics;\\
  \\
  \> \texttt{for} all particles \texttt{p} in element \texttt{e} \textcolor{Gray}{// update particles in element \texttt{e}}\\
  \> \> advance particle \texttt{p};\\
  \> \texttt{end}\\
  \texttt{end}
\end{tabbing}
\begin{tabular*}{1.0\textwidth}{l}
\hline
\end{tabular*}
(ii)
\begin{tabbing}
  \texttt{for} \= all \= Eulerian elements \texttt{e} \textcolor{Gray}{// pre-compute element-average statistics}\\
  \> gather Eulerian nodal statistics for element \texttt{e};\\
  \> compute and store element-average statistics;\\
  \texttt{end}\\
  \\
  \texttt{for} \= all particles \texttt{p} \textcolor{Gray}{// update particles}\\
  \> obtain index \texttt{e} of host element for particle \texttt{p};\\
  \> get element-average Eulerian statistics for element \texttt{e};\\
  \> advance particle \texttt{p};\\
  \texttt{end}
\end{tabbing}
\begin{tabular*}{1.0\textwidth}{l}
\hline\hline
\end{tabular*}
\tableSpace
\end{table}%
\begin{figure}[t!]
\centering
\resizebox{11cm}{!}{\input{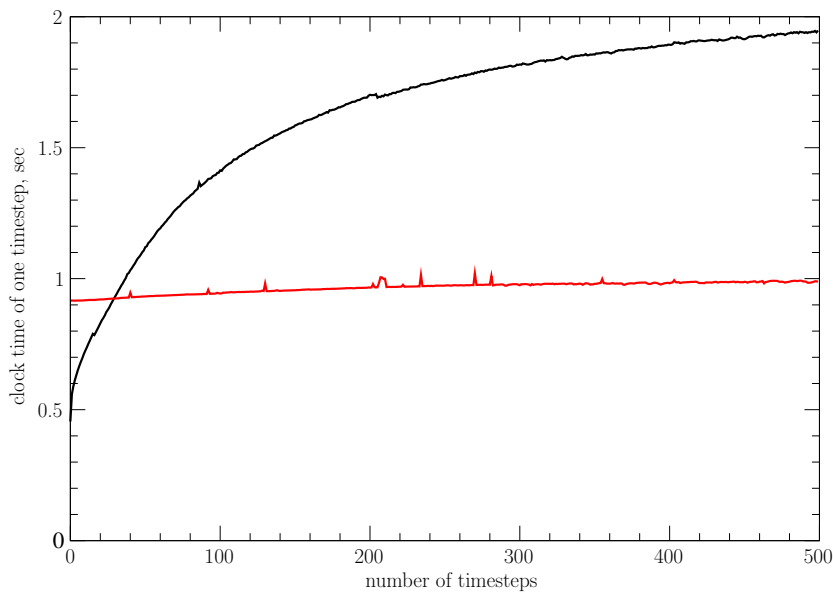}}
\caption[Comparison of cache misses for different loops]{Performance comparison of the two different loops (displayed in Table \ref{tab:loops}) to advance the particle governing equations \Eqr{eq:Lagrangian-position-exact}, \Eqr{eq:Lagrangian-model}, \Eqr{eq:frequency-model} and \Eqr{eq:IECM} for the first 500 timesteps using 8 CPUs. The almost horizontal (red) line represents the particle-based loop, while the curving (black) one is the element-based loop. The problem size is the same as in Table \ref{tab:timestep_profile}.}
\label{fig:cache-misses}
\figSpace
\end{figure}
The main advantage of the element-based loop is that once the Eulerian statistics are gathered for an element they can be used to update all particles in the element without recomputing them. However, it can be significantly off-balance in parallel, since it is not rare that the number of particles per element can differ by as much as two orders of magnitude at different regions of the domain. Another disadvantage of the element-based loop is that most of the time it accesses the arrays containing the particle properties, \(\mathcal{X}_i\), \(\mathcal{U}_i\), \(\omega\), \(\psi\), in an unordered fashion resulting in increasing cache misses as the timestepping progresses and the particles move throughout the domain, because they get scrambled in memory compared to their spatial locations. Conversely, the big advantages of the particle-based loop are its simplicity and excellent load-balance for parallel execution. The particle-based loop always accesses the arrays containing particle properties consecutively. The effect of the increasing cache misses and the different load-balance on the performance is displayed in \Fige{fig:cache-misses}, where the timings of the two loops are compared as the iteration progresses. The element-based loop slows down almost fourfold in just 500 timesteps, while the performance degradation of the particle-based loop is negligible. Also, this disparity increases as the number of threads increases, which is shown in Table \ref{tab:loop-performance}, where serial and parallel timings are displayed for both loops with different number of threads. While the element-based loop slightly outperforms the particle-based loop on a single CPU, the high scalability and cache-efficiency of the particle-based loop pays out very well in parallel. In fact its speedup is superlinear, which is due to the fact that as the number of processors increase, more and more data gathered from memory fit into the aggregate cache of the individual CPUs, resulting in faster processing than from central memory.

Cache misses may also be reduced by specifically optimizing for the architecture of shared caches on multi-core CPUs as it has been done in the current case. We have found that this guarantees a good performance on true shared memory machines as well, i.e.\ on machines whose CPUs do not share their caches and the communication bandwith between the CPU and memory. However, optimizing for non-shared caches and communication bandwidths does not necessarily guarantee optimal performance on multi-core CPUs. These findings clearly show the importance of efficient use of caches. This was also noted with Eulerian CFD codes computing a variety of flows by e.g.\ \citet{Camelli_07}.
\begin{table}
\caption[Comparison of serial and parallel loop performances]{\label{tab:loop-performance}A comparison of serial and parallel performances for a single timestep of the most time-consuming loop, implementing the governing equations to advance particles, \Eqres{eq:Lagrangian-position-exact}, \Eqr{eq:Lagrangian-model}, \Eqr{eq:frequency-model} and \Eqr{eq:IECM}, using the two different loop-structures displayed in Table \ref{tab:loops}. The data is obtained from the same test simulation as in Table \ref{tab:timestep_profile} using the same hardware. The timings are approximate values after the first 500 timesteps.}
\begin{tabular*}{1.0\textwidth}{l@{\hspace{1.0cm}}l@{\hspace{1.0cm}}r@{.}l@{\hspace{2.0cm}}l@{\hspace{1.0cm}}r@{.}l}
\hline\hline
 & \multicolumn{3}{l}{element-based loop} & \multicolumn{3}{l}{particle-based loop} \\
number of CPUs & time (ms) & \multicolumn{2}{l}{speedup} & time (ms) & \multicolumn{2}{l}{speedup} \\
\hline
1 & 6909 & 1&0 & 8068 & 1&00 \\
2 & 4122 & 1&68 & 3987 & 2&03 \\
4 & 2408 & 2&87 & 1943 & 4&12 \\
6 & 1979 & 3&49 & 1305 & 6&16 \\
8 & 1945 & 3&55 & 1000 & 8&20 \\
\hline\hline
\end{tabular*}
\tableSpace
\end{table}
\begin{figure}[t!]
\centering
\resizebox{9.5cm}{!}{\input{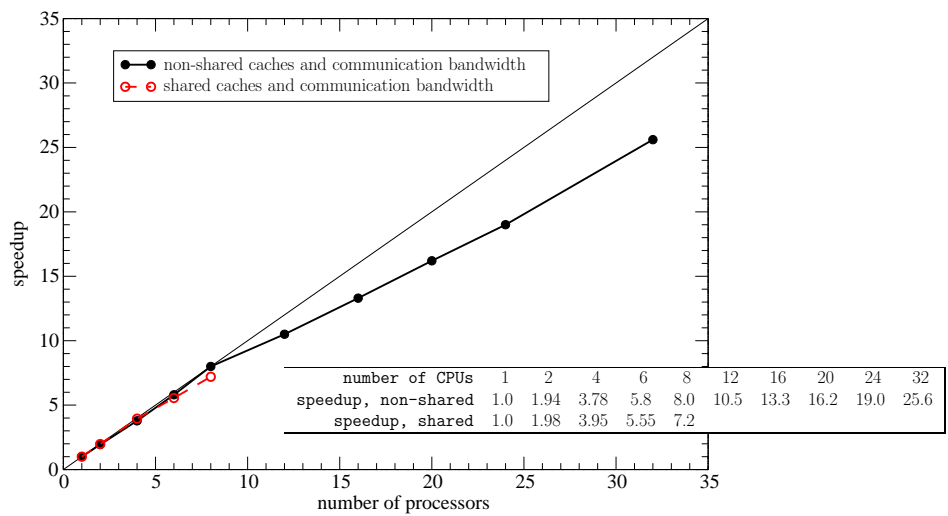}}
\caption[Overall speedups on different types of shared memory machines]{Overall parallel performance of 100 timesteps taken on two different types of shared memory machines. Solid line and symbols -- separate caches and memory-to-CPU bandwidths for each processor, dashed line and open symbols -- two quad-core CPUs (8 cores total) each pair sharing a cache and a memory-to-CPU bandwidth. The problem size is the same as in Table \ref{tab:timestep_profile}.}
\label{fig:profile.overall}
\figSpace
\end{figure}

The parallel performance on higher number of processors is plotted in \Fige{fig:profile.overall}. The size of the testproblem is the same as previously in Table \ref{tab:timestep_profile}, but the hardware is now a true shared memory machine with separate cache and memory-to-CPU bandwidth for each processor. The code performs reasonably well for this moderate-size problem and the parallel efficiency does not show a sign of leveling out up to the 32 CPUs tested. For comparison, the performance data in Table \ref{tab:timestep_profile} is also shown using mutlicore CPUs.

Table \ref{tab:timestep_profile} shows, that the second most time-consuming step in a timestep is the regeneration of the random number tables, which was discussed in \autoref{sec:parallel_random_number_generation}. Interestingly, the solution of the two Eulerian equations, namely the elliptic relaxation equation \Eqr{eq:elliptic-relaxation-Lagrangian} and the pressure-Poisson equation \Eqr{eq:pressure-projection-smooth}, only take up about 2-3\% of a timestep, respectively. It is worth noting, that the linear system for the elliptic relaxation is nine times larger than that of the pressure-Poisson equation. The former is very well conditioned, while the latter is usually the most time-consuming equation to solve in modeling laminar incompressible flows.


\chapter[Channel flow simulations: results and discussion]{\\Channel flow simulations: results and discussion}
\label{chap:channel}

\section{Introduction}
In this Chapter, the previously described PDF model is tested in a fully developed, turbulent, long-aspect-ratio channel flow, where a passive scalar is continuously released from concentrated sources. The joint PDF of velocity, characteristic turbulent frequency and concentration of a passive scalar is computed using stochastic equations. The flow is explicitly modeled down to the viscous sublayer by imposing only the no-slip and impermeability condition on particles without the use of damping, or wall-functions. The high-level inhomogeneity and anisotropy of the Reynolds stress tensor at the wall are captured by the elliptic relaxation method. A passive scalar is released from a concentrated source at the channel centerline and in the viscous wall-region. The effect of small-scale mixing on the scalar is mainly modeled by the IECM model. The performance and accuracy of the IECM model compared to the simpler, but more widely used IEM model are evaluated. Several one-point unconditional and conditional statistics are presented in both physical and composition spaces. An emphasis is placed on common approximations of those conditional statistics that require closure assumptions in concentration-only PDF methods, i.e.\ in methods that assume the underlying turbulent velocity field. The results are compared to the DNS data of \citet{Abe_04} and the experimental data of \citet{Lavertu_05}. The experiments were performed at two different Reynolds numbers (\(\textit{Re}_\tau\equiv u_\tau h/\nu=520\) and \(1080\) based on the friction velocity \(u_\tau\), the channel half width \(h\), and the kinematic viscosity \(\nu\)) in a high-aspect-ratio turbulent channel flow, measuring one point statistics of a scalar (temperature) emitted continuously at three different wall-normal source locations from concentrated line sources. Measurements were performed at six different downstream locations between \(4.0\leqslant x/h\leqslant22.0\).%
\nomenclature[Rh]{$h$}{channel half width (\autoref{chap:channel})}%
\nomenclature[RR]{$\textit{Re}_\tau$}{friction Reynolds number ($u_\tau h/\nu$)}%

The Chapter is organized as follows. A brief account of the underlying numerical methods with various implementation details specific to this flow are presented in \autoref{sec:channel-num_method}. In \autoref{sec:channel-results}, one-point velocity statistics are compared to direct numerical simulation data at \(\textit{Re}_\tau=1080\), and a comparative assessment of the two micromixing models with analytical and experimental data is also given. Detailed statistics of scalar concentration calculated with the IECM micromixing model are presented. \autoref{sec:channel-effect-of-numerics} presents a study of the effects of several numerical parameters on the computed results, including the effect of the Reynolds number, the type of velocity conditioning and the number of particles employed. An assessment of the computational cost of the current method is given compared to DNS in \autoref{sec:channel-computational-cost}. Finally, conclusions pertaining to the channel flow testcase and results are summarized in \autoref{sec:channel-conclusions}.

\section{Modeling specifics of channel flow}
\label{sec:channel-num_method}
The velocity field, in turbulent channel flow after an initial development time, becomes statisticially stationary and homogeneous in the streamwise direction, while it remains inhomogeneous in the wall-normal direction, i.e.\ the flow becomes statistically one-dimensional. The flow is assumed to be statistically symmetric about the channel centerline. A passive scalar released into this flow is inhomogeneous and three-dimensional. Assuming the channel cross section has a high aspect ratio, we confine our interest to the plane spanned by the wall-normal and streamwise directions, far from the spanwise walls. The computational scheme exploits these features by resolving only one spatial dimension for the velocity statistics and two dimensions for the passive scalar. Although this specialized implementation of the method includes flow-dependent features, it provides good indication of the total computational cost. The description is divided into sections that separately discuss the modeling of the fluid dynamics (\autoref{sec:fluid-dynamics-modeling}) and the transported scalar (\autoref{sec:scalar-modeling}). Both DNS and experimental data are used to validate the results.

\subsection{Modeling the fluid dynamics}
\label{sec:fluid-dynamics-modeling}
Since the transported scalar is inhomogeneous, both streamwise \(x\) and cross-stream \(y\) components of the particle positions are retained. A one-dimensional grid is used to compute Eulerian statistics of the velocity and turbulent frequency. An increasing level of refinement is achieved in the vicinity of the wall by obtaining the spacing of the gridpoints from the relation
\begin{equation}
y^+ = 1 - \cos\left(\frac{\pi}{2}a^{3/4}\right), \quad\quad 0\le a < 1,\label{eq:wall-refinement}
\end{equation}%
\nomenclature[Ry]{$y^+$}{distance from a wall in wall units ($u_\tau y/\nu$)}%
\nomenclature[Ra]{$a$}{loop variable in \Eqre{eq:wall-refinement}}%
where \(y^+=u_\tau y/\nu\) is the distance from the wall non-dimensionalized by the friction velocity \(u_\tau\) and the kinematic viscosity \(\nu\) and \(a\) is a loop-variable that equidistantly divides the interval between 0 and 1 (wall and centerline, respectively) into a desired number of gridpoints. The centerline symmetry of the flow is exploited, thus these statistics are only computed on half of the channel. Using this one-dimensional grid, Eulerian statistics are computed as described in \autoref{sec:Eulerian_statistics}. First and second derivatives of the mean velocity are calculated by first-order accurate finite difference formulas over each element and then transferred to nodes. A constant unit mean streamwise pressure gradient is imposed, which drives the flow and builds up the numerical solution. The cross-stream mean-pressure gradient is obtained by satisfying the cross-stream mean-momentum equation for turbulent channel flow
\begin{equation}
\frac{1}{\rho}\frac{\mathrm{d}\mean{P}}{\mathrm{d}y} = -\frac{\mathrm{d}\mean{v^{\scriptscriptstyle 2}}}{\mathrm{d}y},\label{eq:cross-stream-mean-momentum-channel}
\end{equation}
which implies that the pressure-projection is not necessary for this flow. Since the number of elements does not exceed 100, particle tracking in this one-dimensional case is simply a brute-force check on each element. This is a negligible fraction of the running time, thus there is no need for a more sophisiticated tracking algorithm.

Wall-boundary conditions for the particles are the same as described in \autoref{sec:wall_conditions}, only the situation is simpler here, since the wall is aligned with the coordinate line \(y=0\). The conditions for the centerline are symmetry conditions, i.e.\ particles trying to leave the domain through the centerline undergo perfect reflection and the sign of their wall-normal velocity is reversed. Consistently with these particle conditions, boundary conditions are imposed on the Eulerian statistics as well. At the wall, the mean velocity and the Reynolds stress tensor is forced to zero. The mean frequency \(\mean{\omega}\) is set according to \Eqre{eq:frequency-wallstats}. At the centerline, the shear Reynolds stress \(\mean{uv}\) is set to zero. At the wall in the elliptic-relaxation equation \Eqr{eq:elliptic-relaxation-Lagrangian}, \(\wp_{ij}\) is set according to \(\wp_{ij}=-4.5\varepsilon n_in_j\). In the current case the wall is aligned with \(y=0\) thus only the wall-normal component is non-zero: \(\wp_{22}=-4.5\varepsilon\). At the centerline, symmetry conditions are enforced on \(\wp_{ij}\), i.e.\ homogeneous Dirichlet-conditions are applied for the off-diagonal components and homogeneous Neumann-conditions for the diagonal components. The initial conditions for the particles are set according to \autoref{sec:particle-number-control}, however the current one-dimensional case enables the use of a sufficient number of particles so that there is no need for particle redistribution. The applied model constants for the joint PDF of velocity and frequency are displayed in Table \ref{tab:fd-constants}.
\begin{table}
\caption[Constants to model the joint PDF of velocity and frequency]{\label{tab:fd-constants}Constants for modeling the joint PDF of velocity and frequency.}
\begin{tabular*}{1.0\textwidth}{c@{\hspace{0.8cm}}c@{\hspace{0.8cm}}c@{\hspace{0.8cm}}c@{\hspace{0.8cm}}c@{\hspace{0.8cm}}c@{\hspace{0.8cm}}c@{\hspace{0.8cm}}c@{\hspace{0.8cm}}c@{\hspace{0.8cm}}c@{\hspace{0.8cm}}c}
\hline\hline
\(C_1\)&\(C_2\)&\(C_3\)&\(C_4\)&\(C_T\)&\(C_L\)&\(C_\eta\)&\(C_v\)&\(\gamma_5\)&\(C_{\omega1}\)&\(C_{\omega2}\)\\
1.85&0.63&5.0&0.25&6.0&0.134&72.0&1.4&0.1&0.5&0.73\\
\hline\hline
\end{tabular*}
\tableSpace
\end{table}

\subsection{Modeling the passive scalar}
\label{sec:scalar-modeling}
\begin{figure}[t!]
\centering
\input{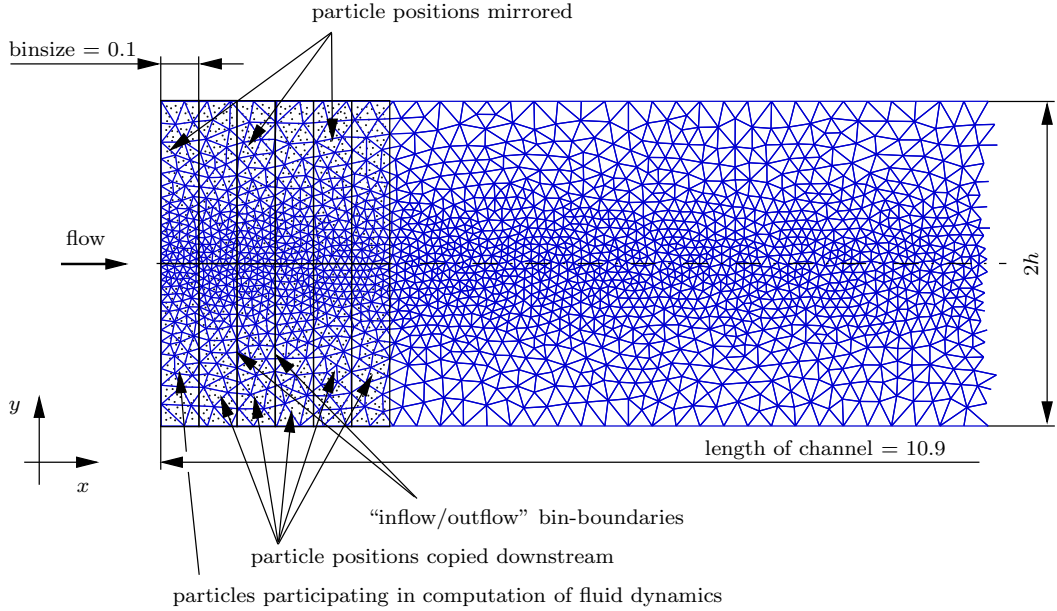}
\caption[Eulerian domain for fully developed turbulent channel flow]{The computational domain for the channel flow is subdivided into several bins to exploit the streamwise statistical homogeneity of the turbulent velocity and frequency fields. Particle positions are copied downstream and mirrored to the upper half. Particle scalar concentrations are exchanged through bin-boundaries and the centerline. Note, that the number of particles in the figure does not correspond to the actual number used in the simulation.}
\label{fig:domain}
\figSpace
\end{figure}

A passive, inert scalar is released from a concentrated source into the modeled fully developed turbulent channel flow, described above. Since the scalar field is inhomogeneous and, in general, not symmetric about the channel centerline, a second, two-dimensional grid is employed to calculate scalar statistics. Employing separate grids for the fluid dynamics and scalar fields enables the grid refinement to be concentrated on different parts of the domain, i.e.\ the scalar-grid can be refined around the source, while the fluid dynamics-grid is refined at the wall. The two-dimensional mesh is used to calculate Eulerian scalar statistics as described in \autoref{sec:Eulerian_statistics}. Since the scalar statistics are not homogeneous in the streamwise direction, the long rectangular domain is subdivided into several bins (thin vertical stripes, see \Fige{fig:domain}) and the following strategy is used to exploit these features. The velocity and turbulent frequency statistics are computed using the one-dimensional grid in which only particles in the first bin participate. The position of these particles are then copied to all downstream bins and (since the fluid dynamics is symmetric about the channel centerline) these particle positions are also mirrored to the upper half of the channel. This means that the particles (as far as positions are concerned) never leave the first bin physically. Since the scalar is passive, only one-way coupling between the two grids is necessary. This is accomplished by using the local velocity statistics computed in the 1d-elements for those 2d-elements that lie the closest to them in the wall-normal coordinate direction.%
At the wall and centerline boundaries the conditions on the particle properties have already been described in \autoref{sec:fluid-dynamics-modeling}. For particles trying to leave the bin through the ``inflow/outflow'' bin-boundaries a periodic boundary condition is applied, with leaving particles put back on the opposite side. This essentially means that the particle paths remain continuous (as they should), only the code accounts for them as different particles in the computer memory. In order to carry the scalar concentration through bin-boundaries, the particle-scalar \(\psi\) is copied downstream (upstream) when the particle tries to leave through the downstream (upstream) bin-boundary. If the particle hits the centerline, its concentration is exchanged with its mirrored pair on the upper half, facilitating a possible non-symmetric behaviour of the scalar. The computation of the velocity-conditioned scalar mean \(\mean{\phi|\bv{V}}\) required in the IECM model \Eqrs{eq:IECM} is carried out with the method described in \autoref{sec:conditional_stats}. The line-source, which in the current two-dimensional case is a point-source, is represented by a circular source with non-dimensional diameter $0.05\nu/u_\tau$. The scalar at the source has a constant distribution: particles passing through the source are assigned a constant normalized unit source strength, i.e.\ \(\psi=\phi_0=1\).%
\nomenclature[Gu]{$\phi_0$}{source strength}%

The micromixing timescale \(t_m\) required to model the viscous diffusion of the scalar is specified based on the following observations. In general, \(t_\mathrm{m}\) is assumed to be proportional to the timescale of the instantaneous plume \citep{Sawford_04}. Once the initial conditions are forgotten, theoretical results \citep{Franzese_07} show that the timescale of the instantaneous plume is linear in \(t\) in the inertial subrange and is proportional to the turbulence timescale in the far field, when the instantaneous plume grows at the same rate as the mean plume. Based on these considerations the micromixing timescale is computed according to
\begin{equation}
t_\mathrm{m} = \min\left[C_s\left(\frac{r_0^2}{\varepsilon}\right)^{1/3} + C_t\frac{x}{\mean{U}_c};\enskip\max\left(\frac{k}{\varepsilon}, C_T\sqrt{\frac{\nu}{\varepsilon}}\right)\right],\label{eq:micromixing-timescale-channel}
\end{equation}%
\nomenclature[Rr]{$r_0$}{radius of the source}%
\nomenclature[RU]{$\mean{U}_c$}{mean velocity at the centerline of the channel}%
\nomenclature[RC]{$C_s$}{micromixing model constant}%
\nomenclature[RC]{$C_t$}{micromixing model constant}%
where \(r_0\) denotes the radius of the source, \(\mean{U}_c\) is the mean velocity at the centerline of the channel, while \(C_s\) and \(C_t\) are micromixing model constants. This definition reflects the three stages of the spreading of the plume. In the first stage, the timescale of the plume is proportional to that of the source \citep{Batchelor_52}: accordingly, the first term in the \(min\) operator represents the effect of the source. In the second stage \(t_\mathrm{m}\) increases linearly as the scalar is dispersed downstream and the distance \(x\) from the source grows \citep{Franzese_07}. In the final stage, the timescale is capped with the characteristic timescale of the turbulence, which provides an upper limit in the third term of \Eqre{eq:micromixing-timescale-channel}. Following \citet{Durbin_91} this is defined as the maximum of the turbulent and Kolmogorov timescales: far from the boundaries it becomes \(k/\varepsilon\), whereas near a surface, where \(k \rightarrow 0\), the Kolmogorov timescale provides a lower bound as \(C_T(\nu/\varepsilon)^{1/2}\).

\section{Results}
\label{sec:channel-results}
The model has been run for the case of fully developed channel flow at \(Re_\tau=1080\) based on the friction velocity \(u_\tau\) and the channel half-width \(h\) with a passive scalar released from a concentrated source at the centerline \((y_s/h=1.0)\) and in the viscous wall region \((y_s/h=0.067)\). The results are divided into a discussion of the fluid dynamics statistics (\ref{sec:res_fd}), a comparison of the two micromixing models (\ref{sec:res_comp}) and a presentation of unconditional (\ref{sec:res_stat}) and conditional (\ref{sec:res_cond_stat}) scalar statistics.%
\nomenclature[Ry]{$y_s$}{cross-stream location of the source (\autoref{chap:channel})}%

\subsection{Fluid dynamics}
\label{sec:res_fd}
\begin{figure}[t!]
\centering
\resizebox{15cm}{!}{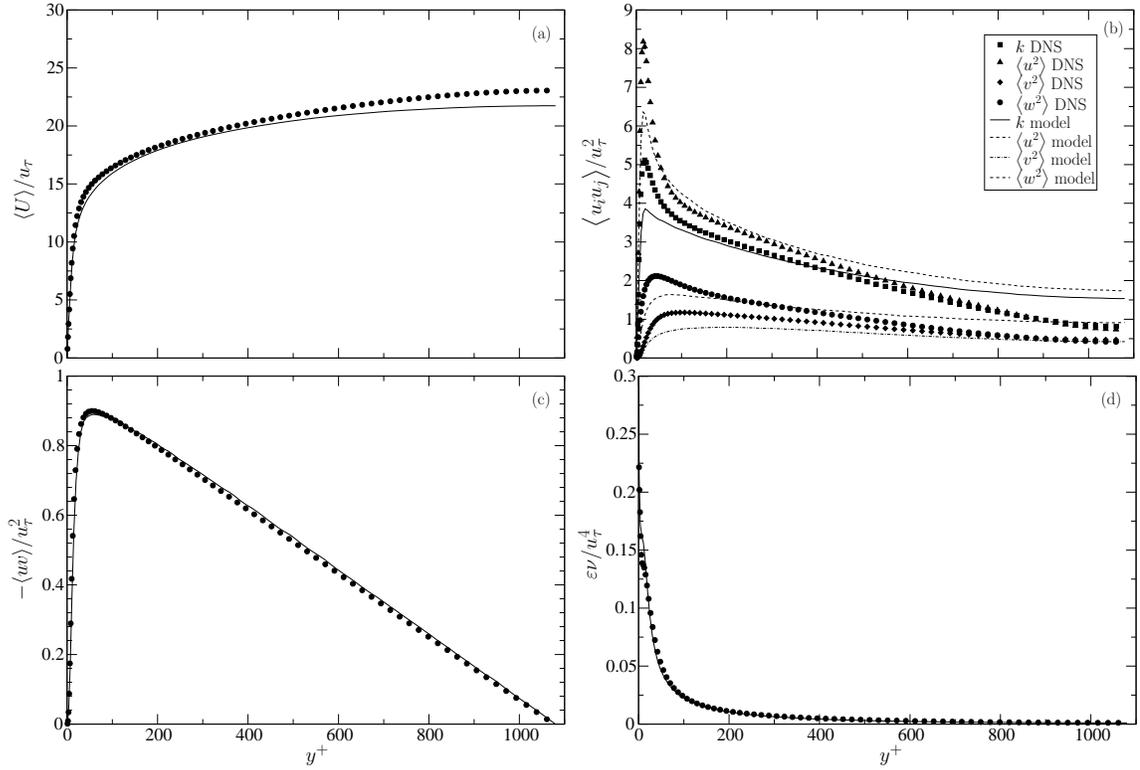}
\caption[Cross-stream profiles of velocity statistics in turbulent channel flow]{\label{fig:velocity}Cross-stream profiles of (a) the mean streamwise velocity, (b) the diagonal components of the Reynolds stress tensor, (c) the shear Reynolds stress and (d) the rate of dissipation of turbulent kinetic energy. Lines -- PDF calculation, symbols -- DNS data of \citet{Abe_04}. All quantities are normalized by the friction velocity and the channel half-width. The DNS data is scaled from \(\textit{Re}_\tau=1020\) to \(1080\).}
\figSpace
\end{figure}

The equations to model the velocity and turbulent frequency have been solved on a 100-cell one dimensional grid with 500 particles per cell. The applied model constants are displayed in Table \ref{tab:fd-constants}. The computed cross-stream profiles of mean streamwise velocity, the non-zero components of the Reynolds stress tensor and the rate of dissipation of turbulent kinetic energy are compared with the DNS data of \citet{Abe_04} at \(\textit{Re}_\tau=1020\) in \Fige{fig:velocity}. Previous PDF modeling studies employing the elliptic relaxation technique \citep{Dreeben_97,Dreeben_98,Waclawczyk_04} have been conducted up to \(\textit{Re}_\tau=590\). The high-level inhomogeneity and anisotropy in the viscous wall region are well represented by the technique at this higher Reynolds number as well. The purpose of including the parameter \(C_\xi\) in \Eqre{eq:L} of the wall-normal component of \(\wp_{ij}\) is to correct the overprediction of the wall-normal Reynolds stress component \(\mean{v^{\scriptscriptstyle 2}}\) at the centerline. This facilitates the correct behavior of the mean of the dispersed passive scalar in the center region of the channel (presented in \autoref{sec:res_comp}).

\subsection{Comparison of the IEM and IECM micromixing models}
\label{sec:res_comp}
An often raised criticism of the IEM model is that there is no physical basis for assuming the molecular mixing to be independent of the velocity field. This assumption gives rise to a spurious (and unphysical) source of scalar flux \citep{Pope_98}. This behavior of the IEM model has also been demonstrated for line sources in homogeneous grid turbulence \citep{Sawford_04}. The situation can be remedied by introducing the velocity-conditioned scalar mean \(\mean{\phi|\bv{V}}\), which leads to the IECM model. Often invoked as a desirable property of micromixing models is that the scalar PDF should tend to a Gaussian  for homogeneous turbulent mixing \citep{Pope_00,Fox_03} (i.e.\ statistically homogeneous scalar field in homogeneous isotropic turbulence).  While mathematically a Gaussian does not satisfy the boundedness property of the advection-diffusion scalar transport equation, it is generally assumed that the limiting form of the PDF can be reasonably approximated by a clipped Gaussian.  Also, \citet{Chatwin_02,Chatwin_04} argued that in most practical cases, where the flow is inhomogeneous, the scalar PDF is better approximated by non-Gaussian functions, which should ultimately converge to a Dirac delta function about the mean, \(\delta(\psi-\mean{\phi})\), where \(\mean{\phi}\) approaches a positive value in bounded domains and zero in unbounded domains. 

In fully developed turbulent channel flow the center region of the channel may be considered approximately homogeneous \citep{Brethouwer_01,Vrieling_03}. Thus for a centerline source, up to a certain downstream distance where the plume still lies completely in the center region, the mean scalar field can be described by Taylor's theory of absolute dispersion \citep{Taylor_21}. Likewise, numerical simulations are expected to reproduce experimental measurements of grid turbulence. According to the theory, the mean-square particle displacement \(\mean{\mathcal{Y}^2}\) is related to the autocorrelation function of the Lagrangian velocity \(R_L=\mean{v(t)v(t')}/\mean{v^{\scriptscriptstyle{2}}}\) as
\begin{equation}
\mean{\mathcal{Y}^2} = 2\mean{v^2}\int_0^t\int_0^{t'}R_L(\xi)\mathrm{d}\xi\mathrm{d}t',\label{eq:particle-displacement-Taylor}
\end{equation}%
\nomenclature[RY]{$\mathcal{Y}$}{particle position in the $y$ direction}%
\nomenclature[RR]{$R_L$}{autocorrelation function of the Lagrangian velocity, \Eqre{eq:expR}}%
where it is assumed that in stationary turbulence \(R_L\) depends only on the time difference \(\xi=t-t'\). Lagrangian statistics such as \(R_L(\xi)\) are difficult to determine experimentally. An analytical expression that is consistent with the theoretically predicted behavior of the Lagrangian spectrum in the inertial subrange is \citep{Arya_99}
\begin{equation}
R_L(\xi)=\exp\left(-\frac{|\xi|}{T_L}\right),\label{eq:expR}
\end{equation}%
\nomenclature[RT]{$T_L$}{Lagrangian integral timescale, \Eqre{eq:TL}}%
where \(T_L\) denotes the Lagrangian integral timescale. Substituting \Eqre{eq:expR} into \Eqre{eq:particle-displacement-Taylor} the following analytical expression can be obtained for the root-mean-square particle displacement
\begin{equation}
\sigma^2_y=\mean{\mathcal{Y}^2}=2\mean{v^2}T^2_L\left[\frac{t}{T_L}-1+\exp\left(-\frac{t}{T_L}\right)\right].\label{eq:ms-particle-displacement}
\end{equation}%
\nomenclature[Gr]{$\sigma^2_y$}{mean-square particle displacement in the $y$ direction, $\mean{\mathcal{Y}^2}$, \Eqre{eq:ms-particle-displacement}}%
This expression can be used to approximate the spread of the plume that is released at the centerline of the channel. As the Lagrangian timescale we take
\begin{equation}
T_L=\frac{2\mean{v^{\scriptscriptstyle 2}}}{C_0\varepsilon},\label{eq:TL}
\end{equation}
where \(C_0\) is usually taken as the Lagrangian velocity structure function inertial subrange constant \citep{Monin_Yaglom_75,Sawford_06}, which ensures consistency of the Langevin equation \Eqr{eq:Lagrangian-model} with the Kolmogorov hypothesis in stationary isotropic turbulence \citep{Pope_00}. In the current case the value of \(C_0\) is defined by \Eqre{eq:GandC} and is no longer a constant, but depends on the velocity statistics. For the purpose of the current analytical approximation, however, a constant value (0.8) has been estimated as the spatial average of \(C_0\) computed by \Eqre{eq:GandC}. For the cross-stream Reynolds stress \(\mean{v^{\scriptscriptstyle 2}}\) and the dissipation rate \(\varepsilon\) their respective centerline values are employed. In analogy with time \(t\) in homogeneous turbulence, we define \(t=x/\mean{U}_c\), where \(x\) is the downstream distance from the source and \(\mean{U}_c\) is the mean velocity at the centerline. Thus the cross-stream mean scalar profiles predicted by \Eqre{eq:particle-displacement-Taylor} are obtained from the Gaussian distribution
\begin{equation}
\mean{\phi(y)}=\frac{\phi_0}{\mean{U}_c\left(2\pi\sigma_y^2\right)^{1/2}}\exp\left[-\frac{\left(y-y_s\right)^2}{2\sigma_y^2}\right],\label{eq:Gaussians}
\end{equation}
where \(\phi_0\) is the source strength and \(y_s\) is the cross-stream location of the source.

After the velocity field converged to a statistically stationary state, a passive scalar is continuously released from a concentrated source. Two release cases have been investigated, where the scalar has been released at the centerline (\(y_s/h=1.0\)) and in the close vicinity of the wall (\(y_s/h=0.067\)). The viscous wall region experiences the most vigorous turbulent activity. The turbulent kinetic energy, its production and its dissipation and the level of anisotropy all experience their peak values in this region, see also \Fige{fig:velocity} (b). This suggests a significantly different level of turbulent mixing between the two release cases. Accordingly, the constants that determine the behavior of the micromixing timescales have been selected differently. Both the IEM and IECM models have been investigated with the micromixing timescale defined by \Eqre{eq:micromixing-timescale-channel} using the model constants displayed in Table \ref{tab:micromixing-constants}.

\begin{table}
\caption[Model constants for the micromixing timescale]{\label{tab:micromixing-constants}Model constants of the micromixing timescale \(t_\mathrm{m}\) defined by \Eqre{eq:micromixing-timescale-channel} for both the IEM and IECM models.}
\begin{tabular*}{1.0\textwidth}{r@{\hspace{1.8cm}}l@{\hspace{1.8cm}}l@{\hspace{1.8cm}}c@{\hspace{1.8cm}}c@{\hspace{1.8cm}}}
\hline\hline
\multicolumn{3}{c}{source location}&\(C_s\)&\(C_t\)\\
\hline
centerline&\(y_s/h=1.0\)&\(y^+=1080\)&0.02&0.7\\
wall&\(y_s/h=0.067\)&\(y^+=72\)&1.5&0.001\\
\hline\hline
\end{tabular*}
\tableSpace
\end{table}
\begin{figure}[t!]
\centering
\resizebox{15cm}{!}{\input{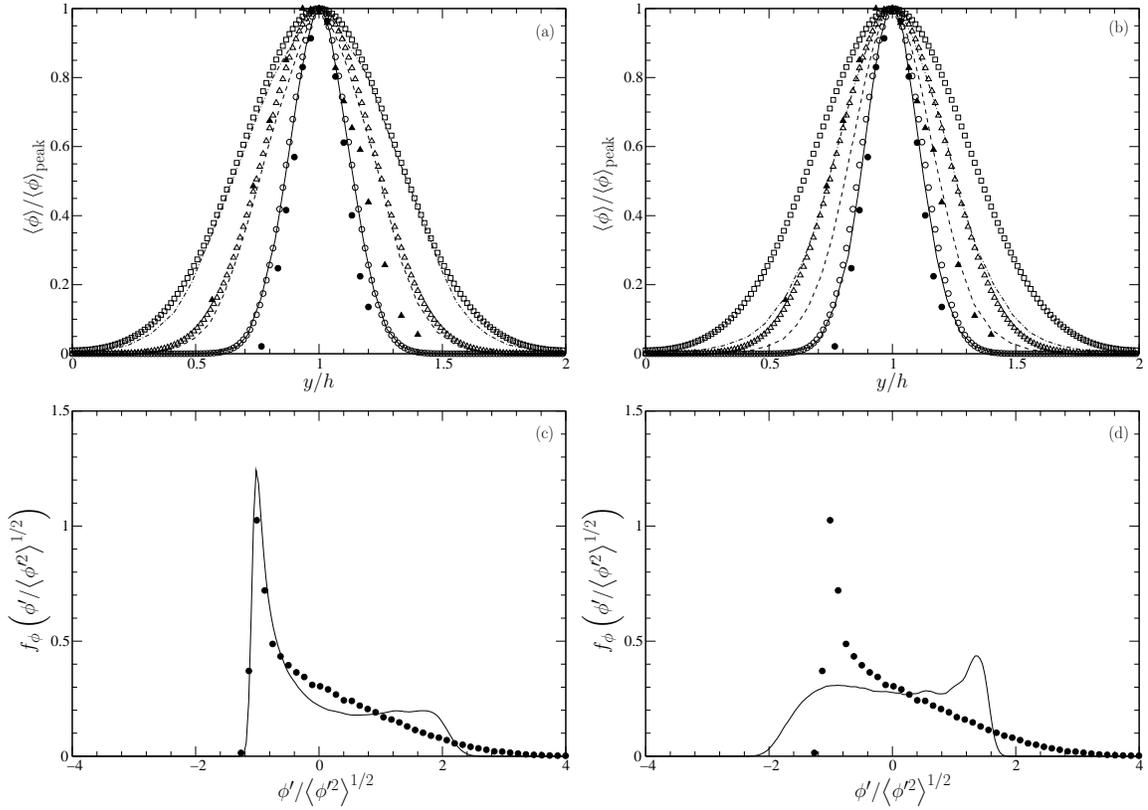}}
\caption[Comparison of the IEM and IECM models, scalar mean and PDFs]{\label{fig:comparison}Cross-stream mean concentration profiles normalized by their respective peak values at different downstream locations as computed by the (a) IECM and (b) IEM models for the centerline release. Lines -- PDF calculation at solid line, \(x/h=4.0\), dashed line, \(x/h=7.4\) and dot-dashed line, \(x/h=10.8\), hollow symbols -- analytical Gaussians using \Eqre{eq:Gaussians} at \(\circ\),~\(x/h=4.0\); \(\vartriangle\),~\(x/h=7.4\) and \(\Box\),~\(x/h=10.8\), filled symbols -- experimental data of \citet{Lavertu_05} at \(\bullet\),~\(x/h=4.0\) and \(\blacktriangle\),~\(x/h=7.4\).  Also shown, PDFs of scalar concentration fluctuations at (\(x/h=7.4\), \(y/h=1.0\)) for the (c) IECM and (d) IEM models. Lines -- computation, symbols -- experimental data.}
\figSpace
\end{figure}

The different behavior of the two models is demonstrated in \Fige{fig:comparison}, which shows mean concentration profiles for the centerline release computed by both the IEM and IECM models together with the analytical Gaussian solution (\ref{eq:Gaussians}) and the experimental data of \citet{Lavertu_05} for turbulent channel flow. Indeed, the downstream evolution of the cross-stream mean concentration profiles computed by the IECM model follows the Gaussians and is expected to deviate far downstream in the vicinity of the walls, where the effect of the walls is no longer negligible. It is also apparent in \Fige{fig:comparison} (b) that the IEM model changes the mean concentration, as expected. As discussed by \citet{Lavertu_05}, the measurements of the mean concentration experience the largest uncertainty due to inaccuracies in estimating the free-stream mean. Also, to improve the signal-to-noise ratio far downstream, a thicker wire had to be employed for measurements performed on the second half of the length considered, i.e.\ \(x/h>11.0\). These difficulties are probably the main source of the discrepancy between the experimental data and the agreeing analytical and numerical results for the case of the centerline release. Because of these inconsistencies only results for the first half of the measured channel length \((x/h<11.0)\) are considered in the current study.

The marginal PDF of scalar concentration can be obtained from the joint PDF \(f_{\bv{\scriptscriptstyle Y}}(\bv{Y})\) by integrating over the velocity and frequency spaces
\begin{equation}
f_{\scriptscriptstyle\phi}(\psi)=\int f_{\bv{\scriptscriptstyle U}\scriptscriptstyle\Omega\phi}(\bv{V},\omega,\psi)\mathrm{d}\bv{V}\mathrm{d}\omega.\label{eq:marginal-scalar-pdf}
\end{equation}%
\nomenclature[Rf]{$f_{\scriptscriptstyle\phi}(\psi)$}{scalar PDF}%
According to experimental data in grid turbulence \citep{Sawford_04} the skewness at the centerline is expected to be negative close to the source and to become positive only farther downstream. At \(x/h=7.4\), \(y/h=1.0\) the temperature PDF measured by \citet{Lavertu_05} suggests positive skewness in accordance with \citeauthor{Sawford_04}'s (\citeyear{Sawford_04}) data. In \Fige{fig:comparison} (c) and (d) the normalized PDFs of scalar concentration fluctuations at this location as computed by both models are depicted. As opposed to the IEM model prediction, both the location of the peak and the overall shape of the PDF are captured correctly by the IECM model.

The different behavior of the two micromixing models is apparent in all one point statistics considered, with the IECM model producing a closer agreement to experimental data. The price to pay for the higher accuracy is an additional 30-40\% in CPU time as compared to the IEM model. In the remaining section only the IECM model results are considered.

\subsection{Scalar statistics with the IECM model}
\label{sec:res_stat}
\begin{figure}[!]
\centering
\resizebox{15cm}{!}{\input{cross_stream_stats_condensed.pstex_t}}
\caption[Cross-stream profiles of the first four moments of scalar concentration]{\label{fig:cross-stream-stats}See next page for caption.}
\end{figure}%
\begin{figure}[t!]
Figure \thefigure: Cross-stream distributions of the first four moments of scalar concentration at different downstream locations for (a)--(d) the centerline release (\(y_s/h = 1.0\)) and (e)--(h) the wall release (\(y_s/h = 0.067\)). Lines -- calculations, symbols -- experimental data at solid line, \(\bullet\), \(x/h=4.0\); dashed line, \(\blacktriangle\), \(x/h=7.4\) and dot-dashed line, \(\blacksquare\),~\(x/h=10.8\). The horizontal dashed lines for the skewness and kurtosis profiles indicate the Gaussian values of 0 and 3, respectively. Note the logarithmic scale of the kurtosis profiles.
\figSpace
\end{figure}
Cross-stream distributions of the first four moments of the scalar concentration at different downstream locations are shown in \Fige{fig:cross-stream-stats} for both release scenarios. The results are compared to experimental data where available.

The mean and root-mean-square (r.m.s.) profiles are normalized by their respective peak values. The width of the mean concentration profiles is most affected by the wall-normal Reynolds stress component \(\mean{v^{\scriptscriptstyle 2}}\) which is responsible for cross-stream mixing. Due to the underprediction of this component by the velocity model throughout most of the inner layer \((y^+<800)\) and the uncertainties in the experimental data mentioned in \autoref{sec:res_comp}, the mean concentration profiles in \Fige{fig:cross-stream-stats} should be considered at most qualitative.

For the wall-release, the r.m.s.\ profiles display a clear drift of the peaks towards the centerline with increasing distance from the source \Fige{fig:cross-stream-stats} (f). This tendency has also been observed in turbulent boundary layers by \citet{Fackrell_82} and \citet{Raupach_83}. Since the scalar is statistically symmetric, in the case of the centerline release, no tranverse drift of the r.m.s.\ profiles is expected, \Fige{fig:cross-stream-stats} (b). Double peaking of the r.m.s.\ profiles has been observed in homogeneous turbulence by \citet{Warhaft_00} and \citet{Karnik_89}, noting that the profiles are initially double-peaked close to the source, then single-peaked for a short distance and then again double-peaked far downstream. \citet{Lavertu_05} found no double peaks in their measurements. Corresponding to the channel flow experiments, the PDF simulation exhibits no double-peaking in the r.m.s.\ profiles. Applying the projection method to compute \(\mean{\phi|\bv{V}}\) as mentioned in \autoref{sec:conditional_stats} and described in \autoref{sec:channel-effect-of-numerics} results in double peaking of the r.m.s.\ profiles, which is possibly due to a loss of statistical information due to its Gaussian assumption of the velocity PDF.

Skewness profiles are depicted in \Fige{fig:cross-stream-stats} (c) and (g). For both release cases, near the centers of the plumes the skewness is close to zero, indicating that the PDFs of the scalar concentration downstream of the sources are approximately symmetric. Towards the edges of the plumes however, the PDFs become very highly positively skewed, with a sudden drop to zero in the skewness outside of the plume. As observed by \citet{Lavertu_05}, the downstream evolutions of the skewness profiles indicate the eventual mixing of the plume, with the high peaks decreasing. In the current simulations the high skewness-peaks at the edge of the plumes start increasing first to even higher levels (up to about \(x/h=10.0\)) and only then start decreasing. In the case of the wall-release, the negative skewness in the viscous wall region (also apparent in the experimental data) becomes even more pronounced in the buffer layer and in the viscous sublayer, where experimental data is no longer available. The kurtosis values are close to the Gaussian value of 3 at the cross-stream location of the sources, but show significant departures towards the edges of the plume.

\begin{figure}[t!]
\centering
\resizebox{15cm}{!}{\input{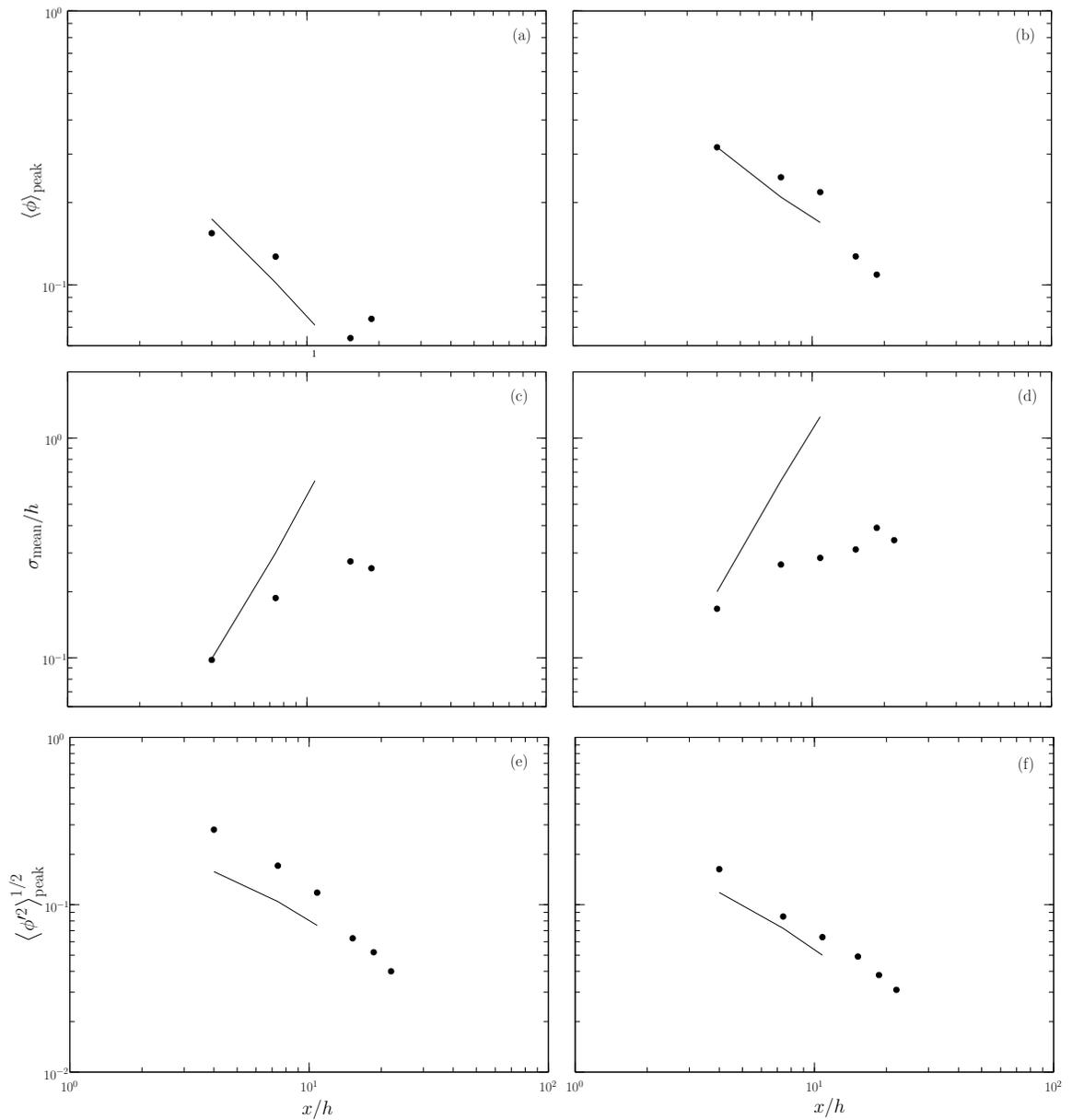}}
\caption[Downstream evolutions of scalar mean peak, mean width and r.m.s.\ width]{\label{fig:downstream-evolutions}Downstream evolutions of (a), (b) the peak mean scalar concentration, (c), (d) the width of the mean concentration and (e), (f) the peak of the r.m.s.\ profiles for the centerline and wall releases, respectively. Solid lines -- numerical results, symbols -- experimental data.}
\figSpace
\end{figure}
\begin{figure}[t!]
\centering
\resizebox{15cm}{!}{\input{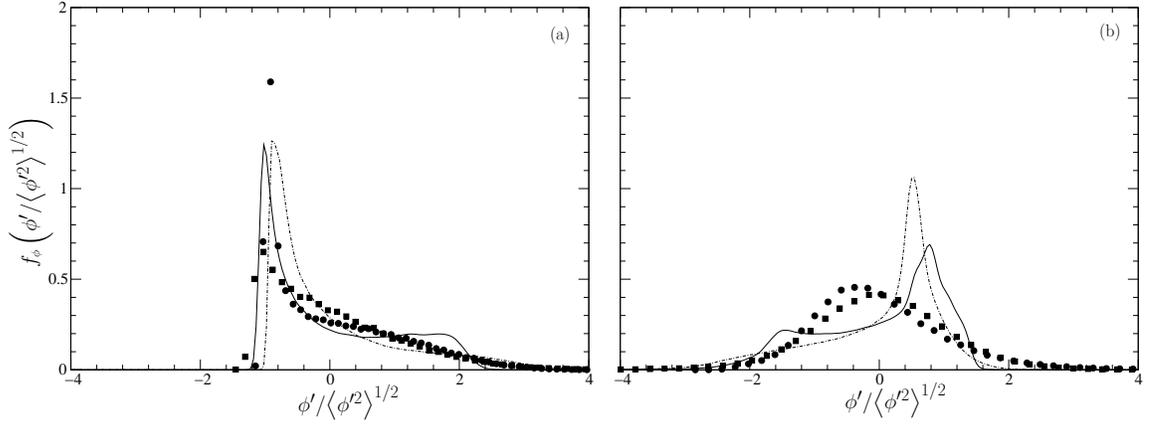}}
\caption[PDFs of scalar concentration fluctuations]{\label{fig:pdfs}Probability density functions of scalar concentration fluctuations at selected downstream locations for the (a) centerline and (b) wall-releases at the cross-stream location of their respective sources (i.e.\ \(y/h=1.0\) and \(y/h=0.067\), respectively). Lines -- calculation, symbols -- experimental data at solid line, \(\bullet\), \(x/h=4.0\) and dot-dashed line, \(\blacksquare\), \(x/h=10.8\).}
\figSpace
\end{figure}
Figure \ref{fig:downstream-evolutions} shows downstream evolutions of the peak of the mean and r.m.s.\ and the width of the mean concentration profiles. In homogeneous isotropic turbulence and homogeneous turbulent shear flow the decay rate of the peak of the mean concentration profiles is reasonably well described by a power law of the form \(\mean{\phi}_\mathrm{peak}\propto x^n\). In the present inhomogeneous flow \citet{Lavertu_05}, based on the experiments, suggest decay exponents of \(n\sim-0.7\) and \(-0.6\) for the wall and centerline sources, respectively. These evolutions are compared to experimental data in \Fige{fig:downstream-evolutions} (a) and (b). Downstream evolutions of the width of the mean concentration profiles \(\sigma_\mathrm{mean}\) are plotted in \Fige{fig:downstream-evolutions} (c) and (d) for the two releases. According to the experimental data, these do not exhibit power-law dependence, as is the case in homogeneous flows. Since the simulations are carried out only on the first half of the measured channel length, the three downstream locations are not sufficient to unambiguously decide whether the simulation data exhibits power-law behavior for the peaks and widths of the mean profiles.%
\nomenclature[Rn]{$n$}{decay exponent in channel flow (\autoref{chap:channel})}%
\nomenclature[Gr]{$\sigma_\mathrm{mean}$}{width of the mean concentration profile (\autoref{sec:res_stat})}%

The downstream decay of the peak values of the r.m.s.\ profiles can be well-approximated by a power-law of the form \(\mean{\phi'^{\scriptscriptstyle{2}}}^{\scriptscriptstyle 1/2}_\mathrm{\scriptscriptstyle{peak}}\propto x^n\), similarly to homogeneous shear flow and isotropic grid-generated turbulence, \Fige{fig:downstream-evolutions} (e) and (f). The experiments suggest \(n=-1\) for both releases.

Probability density functions of scalar concentration fluctuations are depicted in \Fige{fig:pdfs} for both release cases. The cross-stream location of these PDFs are chosen to coincide with that of their respective sources, i.e.\ \(y/h=1.0\) for the centerline release and \(y/h=0.067\) for the wall-release. Two downstream locations are plotted, at the first and at the third location from the sources measured by \citet{Lavertu_05}, at \(x/h=4.0\) and \(x/h=10.8\), respectively. While the PDFs for the centerline release are in reasonable agreement with the experiments, some discrepancies are apparent in the wall-release case. A possible reason behind this disparity is the ad-hoc specification of the mixing timescale in \Eqre{eq:micromixing-timescale-channel}, which is mostly based on theoretical considerations and experimental observations in homogeneous turbulence.

\subsection{Conditional statistics}
\label{sec:res_cond_stat}
The current model solves for the full joint PDF of the turbulent velocity, frequency and scalar concentration. Therefore we can also examine those quantities that require closure assumptions in composition-only PDF methods. These methods are often used in combustion engineering to model complex chemical reactions in a given turbulent flow or in dispersion modeling in the atmospheric boundary layer. In these cases the simplest approach is to assume the shape of the velocity PDF and numerically solve a set of coupled model equations that govern the evolution of the joint PDF of the individual species concentrations in composition space.

The conservation equation for a single reactive scalar is
\begin{equation}
\frac{\partial\phi}{\partial t} + \bv{U}\cdot\nabla\phi = \Gamma\nabla^2\phi + S(\phi(\bv{x},t)),
\end{equation}%
\nomenclature[RS]{$S(\phi)$}{chemical source term}%
where \(S(\phi)\) is the chemical source term. In high Reynolds number, constant property flow the PDF of a reactive scalar \(g(\psi;\bv{x},t)\) is governed by \citep{Dopazo_94,Pope_00}
\begin{equation}
\frac{\partial g}{\partial t} + \mean{U_i}\frac{\partial g}{\partial x_i} = \Gamma\nabla^2g - \frac{\partial}{\partial x_i}\left(g\mean{u_i|\psi}\right)-\frac{\partial^2}{\partial\psi^2}\left(g\mean{\Gamma\frac{\partial\phi}{\partial x_i}\frac{\partial\phi}{\partial x_i}\bigg|\psi}\right)-\frac{\partial}{\partial\psi}\left[g S(\psi)\right],
\end{equation}
or alternatively
\begin{equation}
\frac{\partial g}{\partial t} + \frac{\partial}{\partial x_i}\big[g\left(\mean{U_i}+\mean{u_i|\psi}\right)\big] = -\frac{\partial}{\partial\psi}\Big\{g\big[\mean{\Gamma\nabla^2\phi|\psi} + S(\psi)\big]\Big\}.
\end{equation}%
\nomenclature[Rg]{$g(\psi)$}{PDF of a reactive scalar}%
An attractive feature of these formulations is that the usually highly nonlinear chemical source term is in closed form. Closure assumptions, however, are necessary for the velocity fluctuations conditional on the scalar concentration \(\mean{u_i|\psi}\) and the conditional scalar dissipation \(\mean{2\Gamma\nabla\phi\cdot\nabla\phi|\psi}\) or the conditional scalar diffusion \(\mean{\Gamma\nabla^2\phi|\psi}\). Since for the current case \(S(\phi)=0\), the marginal scalar PDF \(f_{\scriptscriptstyle\phi}(\psi)\) defined in \Eqre{eq:marginal-scalar-pdf} is equal to \(g\), thus in the following we just use \(f_{\scriptscriptstyle\phi}\).

For the convective term \citet{Dopazo_75} applied the linear approximation
\begin{equation}
\mean{u_i|\psi} = \frac{\mean{u_i\phi'}}{\mean{\phi'^2}}\left(\psi-\mean{\phi}\right),\label{eq:linear-conditioned-velocity}
\end{equation}%
\nomenclature[Gu]{$\phi'$}{concentration fluctuation ($\psi-\mean{\phi}$)}%
to compute the centerline evolution of the temperature PDF in a turbulent axisymmetric heated jet. This linear approximation is exact for joint Gaussian velocity and scalar fluctuations. While many experiments \citep{Bezuglov_74,Golovanov_77,Shcherbina_82,Venkataramani_78,Sreenivasan_78} confirm the linearity of the conditional mean velocity around the local mean conserved scalar, \citet{Kuznetsov_90} observe that most of the experimental data show departure from this linear relationship when \(\left|\psi-\mean{\phi}\right|\) is large. Experimental data from \citet{Sreenivasan_78} and \citet{Bilger_91} also show that in inhomogeneous flows the joint PDF of velocity and scalar is not Gaussian, which makes the above linear approximation dubious in a general case. Nevertheless, this linear model is sometimes applied to inhomogeneous scalar fields because of its simplicity.

Another commonly employed approximation is to invoke the gradient diffusion hypothesis
\begin{equation}
-f_\phi\mean{u_i|\psi}=\Gamma_T\frac{\partial f_\phi}{\partial x_i},\label{eq:gradient-diffusion}
\end{equation}%
\nomenclature[Gc]{$\Gamma_T$}{turbulent diffusivity, \Eqre{eq:gradient-diffusion}}%
\nomenclature[Gm]{$\nu_T$}{turbulent viscosity, \Eqre{eq:GammaT}}%
\nomenclature[Gr]{$\sigma_T$}{turbulent Prandtl number, \Eqre{eq:GammaT}}%
where \(\Gamma_T(\bv{x},t)\) is the turbulent diffusivity. In the current case, we specify the turbulent viscosity \(\nu_T\) based on the traditional \(k-\varepsilon\) closure and relate it to \(\Gamma_T\) with the turbulent Prandtl number \(\sigma_T\) as
\begin{equation}
\Gamma_T=\frac{\nu_T}{\sigma_T}=\frac{C_\mu}{\sigma_T}\frac{k^2}{\varepsilon},\label{eq:GammaT}
\end{equation}
where \(C_\mu=0.09\) is the usual constant in the \(k-\varepsilon\) model and we choose \(\sigma_T=0.8\).

\begin{figure}[t!]
\centering
\resizebox{15cm}{!}{\input{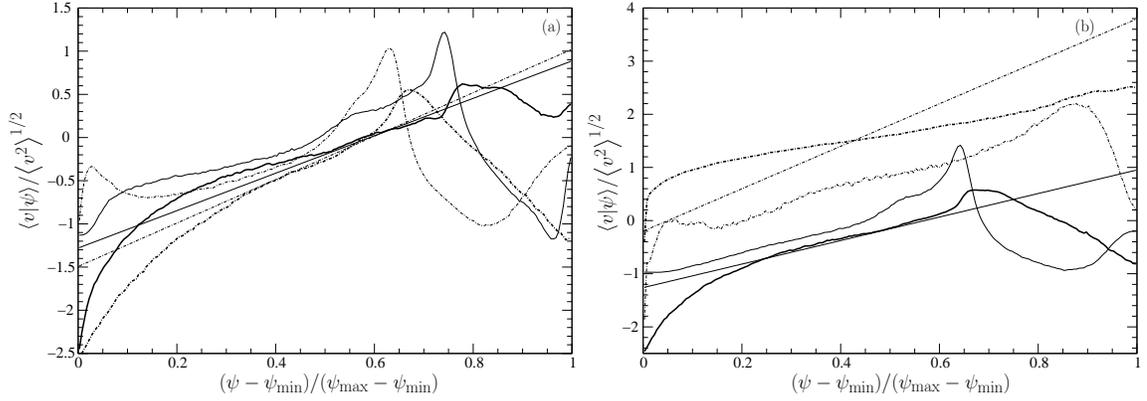}}
\caption[Velocity fluctuation conditioned on the scalar concentration]{\label{fig:conditioned-velocity}Cross-stream velocity fluctuation conditioned on the scalar concentration for the wall-release (\(y_s/h=0.067\)). Thick lines, IECM model; thin lines, gradient diffusion approximation of \Eqre{eq:gradient-diffusion}; straight sloping lines, linear approximation of \Eqre{eq:linear-conditioned-velocity}. (a) downstream evolution at the height of the source \(y/h=0.067\): solid lines, \(x/h=4.0\); dot-dashed lines, \(x/h=10.8\) and (b) cross-stream evolution at \(x/h=7.4\): solid lines, \(y/h=0.067\); dot-dashed lines, \(y/h=0.67\).}
\figSpace
\end{figure}

In \Fige{fig:conditioned-velocity} (a) the downstream evolution of the cross-stream velocity fluctuation conditioned on the scalar is depicted for the wall-release case. Both locations are at the height of the source, i.e.\ \(y/h=0.067\). The concentration axis for both locations is scaled between their respective local minimum and maximum concentration values, \(\psi_\mathrm{min}\) and \(\psi_\mathrm{max}\). Note that the model curves show higher negative velocity for low-concentration particles as the distance from the source increases. This is expected, since particles deep inside the plume can have very low concentrations only if they did not come from the source but traveled very fast from above, so that they did not have much time to exchange concentration with the source material. As the plume spreads, only particles with stronger negative velocity can maintain their low concentration values. Likewise, as the center of the plume moves towards the centerline of the channel, high-concentration particles also need to have stronger negative velocities to escape from exchange during their journey from the plume-center to our sensors, which is apparent on the right side of the figure. Obviously, the linear approximation (\ref{eq:linear-conditioned-velocity}) cannot be expected to capture the non-linearity of the model curves, but except for extremely low and high concentrations it performs reasonably well. On the other hand, the gradient diffusion approximation is capable of capturing most features of the IECM model behavior: it successfully reproduces the non-linearity, with some discrepancy at low and high concentrations. It is also apparent that the numerical computation of the derivatives of the PDFs in the gradient diffusion model (\ref{eq:gradient-diffusion}) is most sensitive to sampling errors at the concentration extremes due to lower number of particles falling into the concentration bins there.%
\nomenclature[Gw]{$\psi_\mathrm{min}$}{local minimum concentration}%
\nomenclature[Gw]{$\psi_\mathrm{max}$}{local maximum concentration}%

The cross-stream evolution of the conditioned velocity fluctuation is shown in \Fige{fig:conditioned-velocity} (b). Both sensors are now at the downstream location \(x/h=7.4\) with increasing distance from the wall at \(y/h=0.067\) and \(0.67\). As the sensor moves towards the channel centerline, the detected low-concentration particles need weaker negative velocity to maintain those low concentrations. The sensor locations relative to the plume centerline can be identified by examining the cross-stream velocity of the high concentration particles. The sensors at \(y/h=0.067\) and \(0.67\) are below and above the plume centerline, respectively, since high-concentration particles at these locations possess negative and positive cross-stream velocities. As is expected, the linear approximation reasonably represents the model behavior for mid-concentrations, while its performance degrades at locations with higher non-Gaussianity, i.e.\ towards the edge of the plume. The performance of the gradient diffusion model is reasonable, except at the concentrations extremes.

For the IECM micromixing model, the mean dissipation conditioned on the scalar concentration can be computed from \citep{Sawford_04b}
\begin{equation}
\mean{2\Gamma\frac{\partial\phi}{\partial x_i}\frac{\partial\phi}{\partial x_i}\bigg|\psi}f_{\scriptscriptstyle\phi} = -\frac{2}{t_\mathrm{m}}\int_0^\psi(\psi'-\Tilde{\phi})f_{\scriptscriptstyle\phi}(\psi')\mathrm{d}\psi',\label{eq:cond-diss}
\end{equation}
where
\begin{equation}
\Tilde{\phi}(\psi)=\int\mean{\phi|\bv{V}}f_{\bv{\scriptscriptstyle U}\scriptscriptstyle\Omega|\phi}(\bv{V},\omega|\psi)\mathrm{d}\bv{V}\mathrm{d}\omega,\label{eq:double-conditioned-scalar}
\end{equation}%
\nomenclature[Gu]{$\Tilde{\phi}(\psi)$}{mean scalar concentration conditional on $\bv{U}=\bv{V}$ and $\phi=\psi$, \Eqre{eq:double-conditioned-scalar}}%
\nomenclature[Rf]{$f_{\bv{\scriptscriptstyle U}\scriptscriptstyle\Omega"|\phi}(\bv{V},\omega"|\psi)$}{joint PDF of $\bv{U}$ and $\Omega$ conditional on $\phi=\psi$}%
in which the scalar-conditioned PDF is defined as \(f_{\bv{\scriptscriptstyle U}\scriptscriptstyle\Omega|\phi}(\bv{V},\omega|\psi)\equiv f_{\bv{\scriptscriptstyle U}\scriptscriptstyle\Omega\phi}(\bv{V},\omega,\psi)/f_{\scriptscriptstyle\phi}(\psi).\) The function \(\Tilde{\phi}(\psi)\) in \Eqre{eq:double-conditioned-scalar} can be obtained by taking the average of \(\mean{\phi|\bv{V}}\) over those particles that reside in the bin centered on \(\psi\). In other words, the concentration values are first conditioned on the velocity field, which is required to advance the particle concentrations according to the IECM model, then are conditioned again by dividing the concentration sample space into bins and computing separate means for each bin. We use a few bins for the velocity conditioning (\(N_c\)) and a significantly higher number of bins (200) for the scalar sample space in order to obtain a higher resolution. The integral in \Eqre{eq:cond-diss}, however, is more problematic. As \citet{Sawford_04b} notes, numerical integration errors that accumulate at extreme concentrations may be amplified when divided by the scalar PDF approaching zero at those locations. Since the integral over all concentrations vanishes, i.e.\ \(\mean{2\Gamma(\nabla\phi)^2|\psi_\mathrm{max}}f_\phi(\psi_\mathrm{max})=0\), for mid-concentrations it can be evaluated either from the left (\(\psi_\mathrm{min}\to\psi\)) or from the right (\(\psi_\mathrm{max}\to\psi\)). Thus the integration errors at the concentration extremes can be significantly decreased by dividing the domain into two parts, integrating the left side from the left and the right side from the right and merging the two results in the division-point. Due to statistical errors, however, the integral over all concentrations may not vanish. In that case, the nonzero value%
\nomenclature[RN]{$N_c$}{number of bins used for velocity conditioning}%
\begin{equation}
\int_{\psi_\mathrm{min}}^{\psi_\mathrm{max}}(\psi-\Tilde{\phi})f_\phi(\psi)\mathrm{d}\psi
\end{equation}
can be distributed over the sample space by correcting the integrand with the appropriate fraction of this error in each bin.

\begin{figure}[t!]
\centering
\resizebox{15cm}{!}{\input{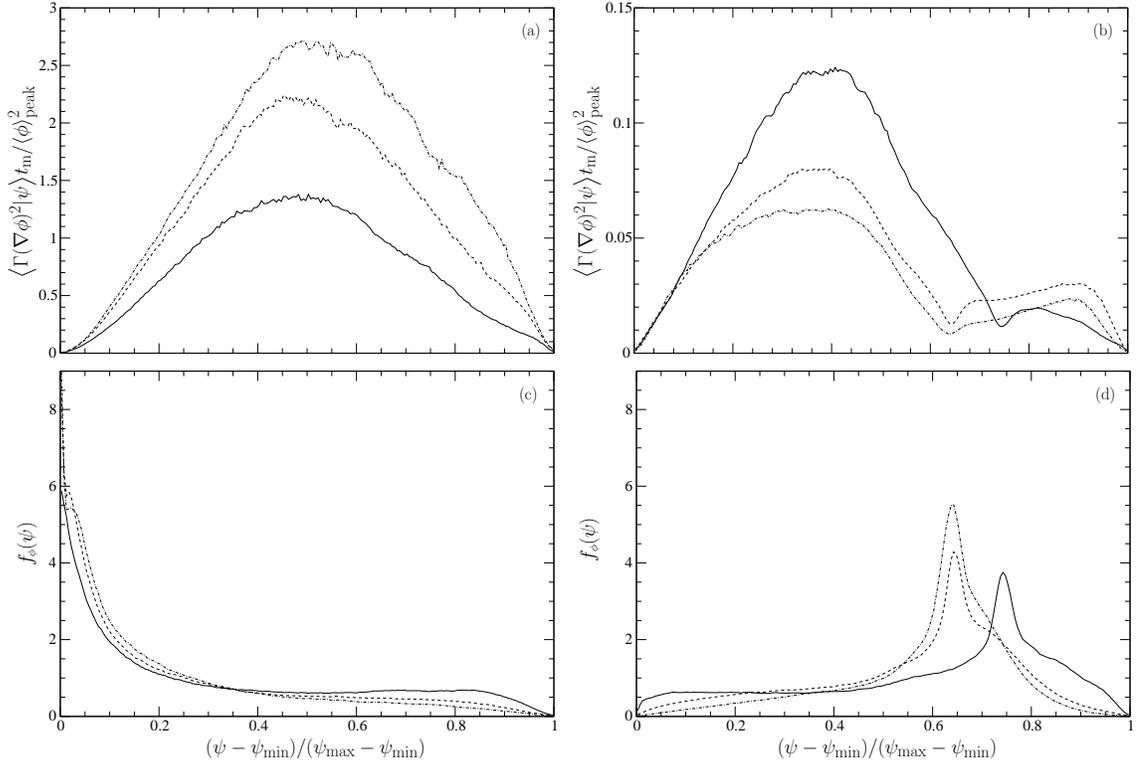}}
\caption[Mean scalar dissipation conditioned on the concentration]{\label{fig:conditional-dissipation}IECM model predictions for the mean scalar dissipation conditioned on the concentration for (a) the centerline release (\(y_s/h=1.0\)) and (b) the wall-release (\(y_s/h=0.067\)) at different downstream locations: solid line, \(x/h=4.0\); dashed line, \(x/h=7.4\) and dot-dashed line, \(x/h=10.8\). The cross-stream locations are the same as the respective source positions. Note the different scales for the dissipation curves between the different releases. Also shown are the scalar PDFs at the same locations for both releases in (c) and (d), respectively.}
\figSpace
\end{figure}
The conditional mean dissipation for three different downstream locations is depicted in \Fige{fig:conditional-dissipation} for both release cases. As for the conditional velocity, the abscissas here are also scaled between the local \(\psi_\mathrm{min}\) and \(\psi_\mathrm{max}\). The dissipation is normalized by the mixing timescale \(t_\mathrm{m}\) and the square of the mean scalar peak \(\mean{\phi}_\mathrm{peak}^2\) at the corresponding downstream locations. Note that in the case of the wall-release, the dissipation curves are an order of magnitude lower than in the centerline release case. This is mainly a result of the choice of the different micromixing model constants, especially \(C_t\).

In the case of the wall release, the curves exhibit bi-modal shapes at all three downstream locations. This tendency has also been observed by \citet{Kailasnath_93} in the wake of a cylinder and by \citet{Sawford_06} in a double-scalar mixing layer and, to a lesser extent, also in homogeneous turbulence \citep{Sawford_04b}. \citet{Sardi_98} suggest that in assumed-PDF methods of turbulent combustion a qualitative representation of the conditional dissipation can be obtained in terms of the inverse PDF. To examine this relationship, the corresponding scalar PDFs are also plotted in \Fige{fig:conditional-dissipation} with the same scaling on the concentration axis as the dissipation curves. It is apparent that these results support this reciprocal connection except at the extremes: high values of the PDF correspond to low dissipation (and vice versa). This can be observed for both releases, but it is most visible in the wall-release case, where the mid-concentration minimum between the two maxima of the bi-modal dissipation curves correspond to the peaks in the PDFs.

\begin{figure}[t!]
\centering
\resizebox{15cm}{!}{\input{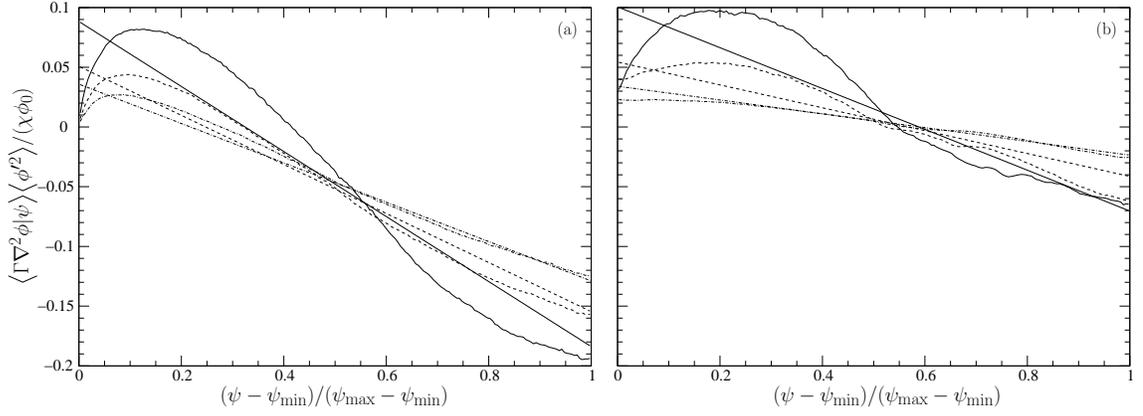}}
\caption[Mean scalar diffusion conditioned on the concentration]{\label{fig:conditional-diffusion}Mean scalar diffusion conditioned on the concentration as predicted by the IECM and IEM models for (a) the centerline release (\(y_s/h=1.0\)) and (b) the wall-release (\(y_s/h=0.067\)) at different downstream locations. The cross-stream locations are the same as the respective source positions. Solid line, \(x/h=4.0\); dashed line, \(x/h=7.4\) and dot-dashed line, \(x/h=10.8\). The straight lines are the linear predictions of the IEM model of \Eqre{eq:IEM-cdiff}.}
\figSpace
\end{figure}

The IECM model \Eqr{eq:IECM} implies a model for the mean diffusion conditioned on the scalar concentration as
\begin{equation}
\big<\Gamma\nabla^2\phi\big|\bv{V},\psi\big> = -\frac{1}{t_\mathrm{m}}\big(\psi-\mean{\phi|\bv{V}}\big).
\end{equation}
The downstream evolution of the conditional diffusion is depicted in \Fige{fig:conditional-diffusion} for both releases. The concentration axes are scaled as before and the curves are normalized by the scalar variance \(\mean{\phi'^2}\), the concentration at the source, \(\phi_0\) and the mean unconditioned dissipation \(\chi=\mean{2\Gamma(\nabla\phi)^2}\), which is computed by integrating \Eqre{eq:cond-diss} over the whole concentration  space. Also shown are the predictions according to the IEM model, which is given by the linear relationship \citep{Sawford_06}%
\nomenclature[Gv]{$\chi$}{mean scalar dissipation}%
\begin{equation}
\frac{\mean{\Gamma\nabla^2\phi|\psi}\mean{\phi'^2}}{\chi\phi_0} = \frac{1}{2}\left(\frac{\mean{\phi}}{\phi_0}-\psi\right).\label{eq:IEM-cdiff}
\end{equation}
Far downstream as the scalar gets better mixed, the predictions of the IEM and IECM models get closer. This behavior has been observed for other statistics, as well as for other flows such as the double-scalar mixing layer \citep{Sawford_06}. \citet{Kailasnath_93} report experimental data on similar shapes for the conditional diffusion in the turbulent wake of a cylinder.

\section{The effect of numerical parameters on the results}
\label{sec:channel-effect-of-numerics}
\begin{figure}
\centering
\resizebox{15cm}{!}{\input{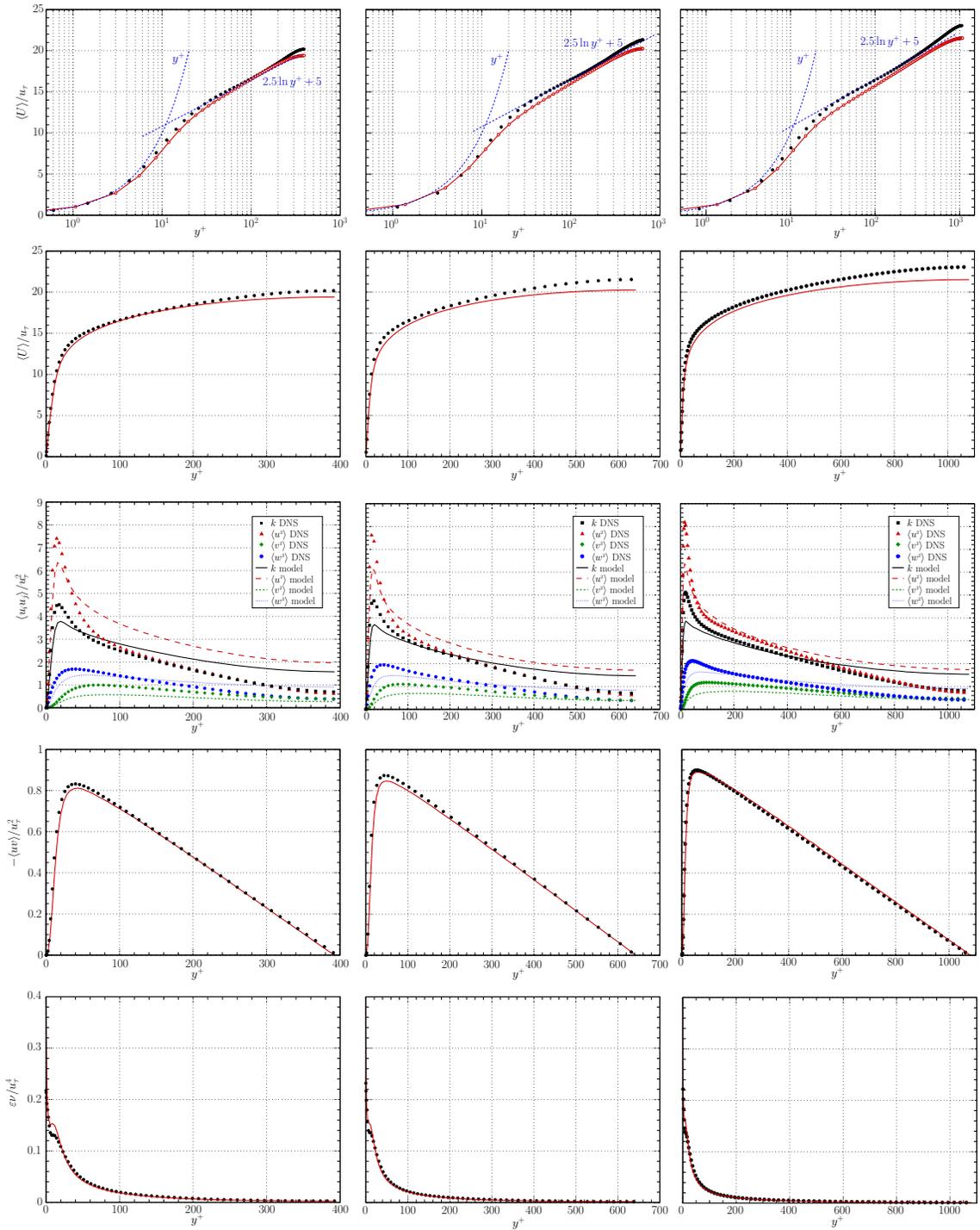}}
\caption[\textit{Re}-dependence of cross-stream velocity statistics in channel flow]{See next page for caption.}
\label{fig:RE-dependence}
\end{figure}
\begin{figure}[t!]
Figure \thefigure: Cross-stream velocity statistics for fully developed turbulent channel flow at (first column) \(\textit{Re}_\tau=392\), (middle column) \(\textit{Re}_\tau=642\) and (right column) \(\textit{Re}_\tau=1080\). Lines -- PDF calculation, symbols -- DNS data of \citet{Moser_99}, \citet{Iwamoto_02} and \citet{Abe_04} (scaled from \(\textit{Re}_\tau=1020\)), respectively. First two rows -- mean streamwise velocity, third row -- normal Reynolds stresses, fourth row -- shear Reynolds stress and fifth row -- rate of dissipation of turbulent kinetic energy. All quantities are normalized by the friction velocity \(u_\tau\) and the channel half-width \(h\).
\figSpace
\end{figure}
\begin{figure}[t!]
\centering
\resizebox{15cm}{!}{\input{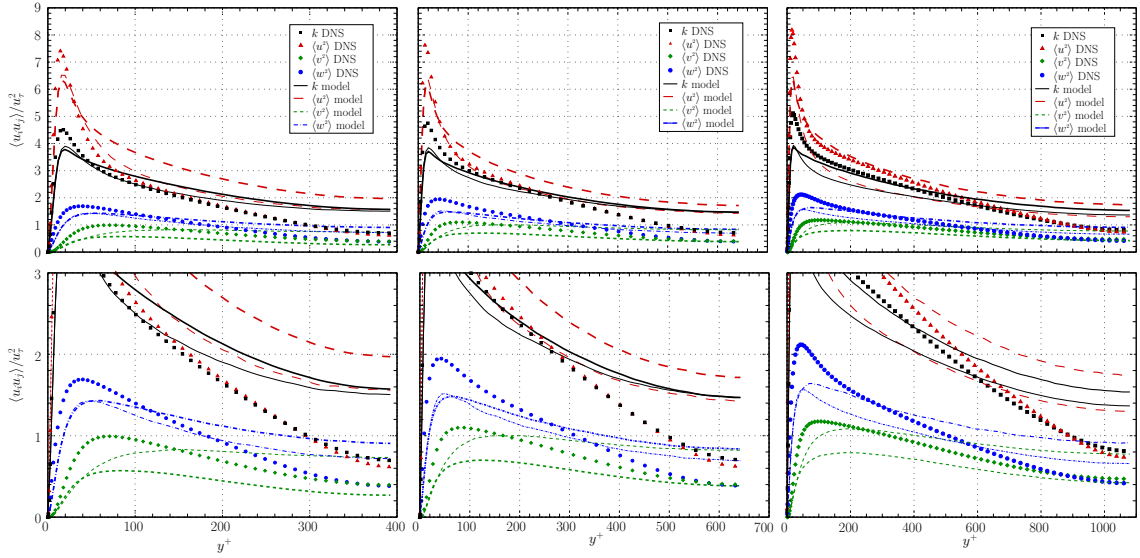}}
\caption[The effect of the anisotropic length-scale constant on the Reynolds stresses]{The effect of the modification of the characteristic lengthscale in \Eqre{eq:L} on the diagonal components of the Reynolds stress tensor by employing the additional model constant \(C_\xi\ne1\) at (first column) \(\textit{Re}_\tau=392\), (middle column) \(\textit{Re}_\tau=642\) and (right column) \(\textit{Re}_\tau=1080\). Thick lines, \(C_\xi=1.0+1.3n_in_i\); thin lines, \(C_\xi=1.0\); symbols, DNS data as in \Fige{fig:RE-dependence}.}
\label{fig:Cxi}
\figSpace
\end{figure}

Previous PDF modeling studies of channel flow in conjunction with elliptic relaxation have been reported at \(\textit{Re}_\tau=395\) \citep{Dreeben_98} and \(\textit{Re}_\tau=590\) \citep{Waclawczyk_04} based on the friction velocity \(u_\tau\) and the channel half-width \(h\). These works concentrate on model development and employ different methodologies with different model constants and numerical methods, which inevitably result in a different balance of model behavior and numerical errors. To assess the prediction at different Reynolds numbers the current model has been run at \(\textit{Re}_\tau=392\), 642 and 1080 using the model constants displayed in Table \ref{tab:fd-constants}. The velocity statistics for all three cases are depicted in \Fige{fig:RE-dependence}. The mean velocity is well represented in the viscous sublayer (\(y^+<5\)) for all three Reynolds numbers. In the buffer layer (\(5<y^+<30\)) there is a slight departure from the DNS data as the Reynolds number increases and from \(y^+>30\), where the log-law should hold, there exists approximate self-similarity, i.e.\ the universal slopes of the profiles are equally well-represented with a slight underprediction far from the wall at higher Reynolds numbers. The viscous wall region (\(y^+<50\)) contains the highest turbulent activity, where production, dissipation, turbulent kinetic energy and anisotropy reach their peak values. The location of the peaks of the Reynolds stress components are succesfully captured by the model at all three Reynolds numbers with their intensity slightly underpredicted. Previous studies using elliptic relaxation in the Reynolds stress framework (i.e.\ Eulerian RANS models) report excellent agreement for these second-order statistics \citep{Durbin_93,Whizman_96}. \citet{Waclawczyk_04} also achieve very good agreement with DNS data using a different version of a PDF model than the one applied here. A common characteristic of PDF models is the slight overprediction of the wall-normal Reynolds stress component \(\mean{v^{\scriptscriptstyle 2}}\) far from the wall. This component is responsible for the cross-stream mixing of a transported scalar released into a flow far from a wall. Therefore in applications where the mean concentration of scalars is important this quantity must be adequately captured. To improve on this situation we introduced a slight modification into the computation of the characteristic lengthscale \(L\) in the elliptic relaxation, \Eqre{eq:L} as
\begin{equation}
L=C_L\max\left[C_\xi\frac{k^{3/2}}{\varepsilon},C_\eta\left(\frac{\nu^3}{\varepsilon}\right)^{1/4}\right],
\end{equation}
with \(C_\xi=1.0+1.3n_in_i\), where \(n_i\) is the unit wall-normal of the closest wall-element pointing outward of the flow domain. This only affects the diagonal Reynolds stresses which can be seen in \Fige{fig:Cxi} for the different Reynolds numbers. Decreasing \(\mean{v^{\scriptscriptstyle 2}}\) at the centerline changes the relative fraction of energy distributed among the diagonal components of the Reynolds stress tensor, consequently the other two components, \(\mean{u^{\scriptscriptstyle 2}}\) and \(\mean{w^{\scriptscriptstyle 2}}\), are slightly increased. Obviously, these kind of flow-dependent modifications in the turbulence model are of limited value, since their effects in a general setting may not be easily predictable. The only nonzero shear stress component \(\mean{uv}\) in this flow and the turbulent kinetic energy dissipation rate \(\varepsilon\) are both in very good agreement with DNS data and even improve as the Reynolds number increases. It is apparent in both \Figse{fig:RE-dependence} and \ref{fig:Cxi} that the overall prediction of second order statistics improve as the Reynolds number increases. This tendency is expected to continue as the underlying high-Reynolds-number modeling assumptions become better fullfilled.

Into the fully developed flow, a passive scalar has been released from a concentrated source at the channel centerline. A general numerical procedure that can be used to compute the velocity-conditioned scalar mean \(\mean{\phi|\bv{V}}\) in the IECM model has been described in \autoref{sec:conditional_stats} and \autoref{app:velocity-conditioning}. Another method based on the projection of the three-dimensional velocity field onto a one-dimensional subspace, where the sample-spatial discretization can be carried out, has been developed and tested in homogeneous turbulence by \citet{Fox_96}. In that method, the projected velocity of a particle is found from
\begin{equation}
\mathcal{U}_\rho = \alpha_i \mathcal{U}_i,\label{eq:projection}
\end{equation}%
\nomenclature[RU]{$\mathcal{U}_\rho$}{projected particle velocity, \Eqre{eq:projection}}%
\nomenclature[Ga]{$\alpha_i$}{projection vector in the IECM model, \Eqre{eq:projection}}%
where the projection vector \(\alpha_i\) is obtained from the following linear relationship
\begin{equation}
\rho_i = \rho_{ij} \alpha_j\label{eq:linear-for-projection}
\end{equation}
between the normalized velocity-scalar vector and the velocity-correlation tensor (no summation on greek indices)
\begin{equation}
\rho_\alpha = \frac{\mean{u_\alpha\phi'}}{\langle u_\alpha^2\rangle^{1/2}\langle \phi'^2\rangle^{1/2}}, \quad\quad\quad
\rho_{\alpha\beta} = \frac{\mean{u_\alpha u_\beta}}{\langle u_\alpha^2 \rangle^{1/2}\langle u_\beta^2\rangle^{1/2}},\label{eq:correlation}
\end{equation}%
\nomenclature[Gq]{$\rho_\alpha$}{normalized velocity-scalar correlation vector, \Eqre{eq:correlation}}%
\nomenclature[Gq]{$\rho_{\alpha\beta}$}{normalized velocity-correlation tensor, \Eqre{eq:correlation}}%
where \(\phi'=\psi-\mean{\phi}\) denotes the scalar fluctuation. This projection method has been developed (and is exact for) Gaussian velocity PDFs, although it can still be used in inhomogeneous flows with the assumption that the local joint PDF of velocity is not too far from an approximate joint normal distribution. In order to assess the performances and the difference in the predictions, we implemented and compared both methods and tested them with different number of conditioning bins.

\begin{figure}[t!]
\centering
\resizebox{15cm}{!}{\input{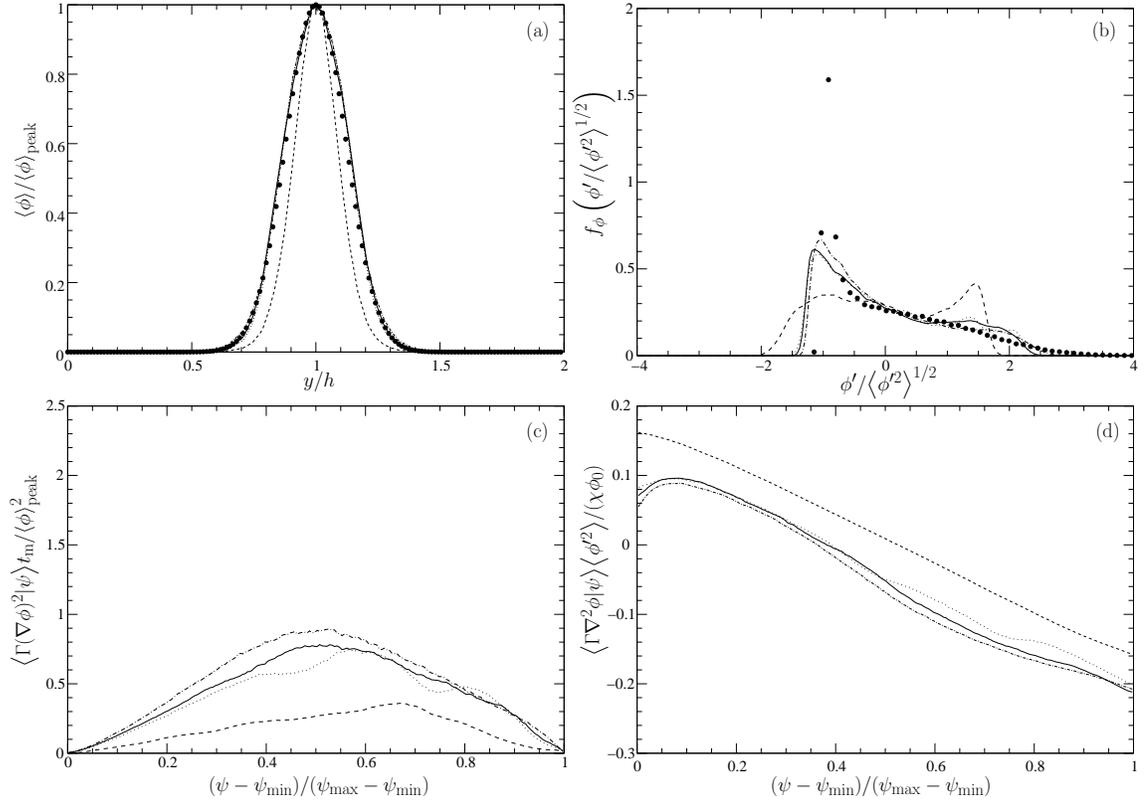}}
\caption[The effect of the number of conditioning bins using Fox's projection]{Scalar statistics affected by the number of conditioning intervals \(N_c\) with computing the velocity-conditioned mean \(\mean{\phi|\bv{V}}\) applying Fox's projection method using \Eqres{eq:projection}-\Eqr{eq:correlation}. (a) Cross-stream distribution of the scalar mean at \(x/h=4.0\), (b) PDF of scalar concentration fluctuations at (\(x/h=4.0,\) \(y/h=1.0\)), (c) mean scalar dissipation conditioned on the concentration at (\(x/h=4.0,\) \(y/h=1.0\)) and (d) mean scalar diffusion conditioned on the concentration at (\(x/h=4.0,\) \(y/h=1.0\)). Dashed line -- \(N_c\)=1 (IEM), dotted line -- \(N_c\)=3, solid line -- \(N_c\)=5, dot-dashed line -- \(N_c\)=20. Symbols on (a) analytical Gaussians according to \citet{Taylor_21} and on (b) experimental data of \citet{Lavertu_05}.}
\label{fig:CNBI}
\figSpace
\end{figure}
To investigate how the choice of the number of conditioning intervals \(N_c\) affects the solution with the projection method, several runs have been performed at the highest Reynolds number (\(\textit{Re}_\tau=1080\)) with different values for \(N_c\). Some of the unconditional and conditional statistics of the joint PDF are depicted in \Fige{fig:CNBI}. Note that employing \(N_c\)=1 corresponds to the special case of the IEM model, \Eqre{eq:IEM}. It is apparent that applying only a few intervals already makes a big difference compared to the IEM model in correcting the prediction of the mean concentration and the PDF of concentration fluctuations also moves towards the experimental data. Increasing \(N_c\) may be thought as an approach to increase the resolution of the conditioning (thus better exploiting the advantages of the IECM over the IEM model), however, as \citet{Fox_96} points out, this is of limited value, since the decreasing number of particles per interval increases the statistical error. The current test simulations have been carried out with an initial 500 particles per element and the total number of particles did not change during simulation. \Fige{fig:CNBI} shows that above \(N_c\)=5 there is no significant change in the statistics and even at \(N_c\)=20 the results do not deteriorate. Also displayed in \Fige{fig:CNBI} are the centerline normalized mean scalar dissipation and diffusion both conditional on the scalar concentration, \(\mean{\Gamma(\nabla\phi)^2|\psi}\) and \(\mean{\Gamma\nabla^2\phi|\psi}\), respectively. As before, the concentration axes in \Fige{fig:CNBI} (c) and (d) are scaled between the local minimum and maximum concentration values, \(\psi_\mathrm{min}\) and \(\psi_\mathrm{max}\), in order to zoom in on the interesting part of the concentration space. Using Fox's projection method, the choice of number of conditioning intervals on the velocity space (\(N_c\)) has a similar effect on the conditional dissipation and diffusion: they also support the earlier observation that the optimal number of conditioning intervals is at about \(N_c\)=3--5 to attain convergence.

\begin{figure}[t!]
\centering
\resizebox{15cm}{!}{\input{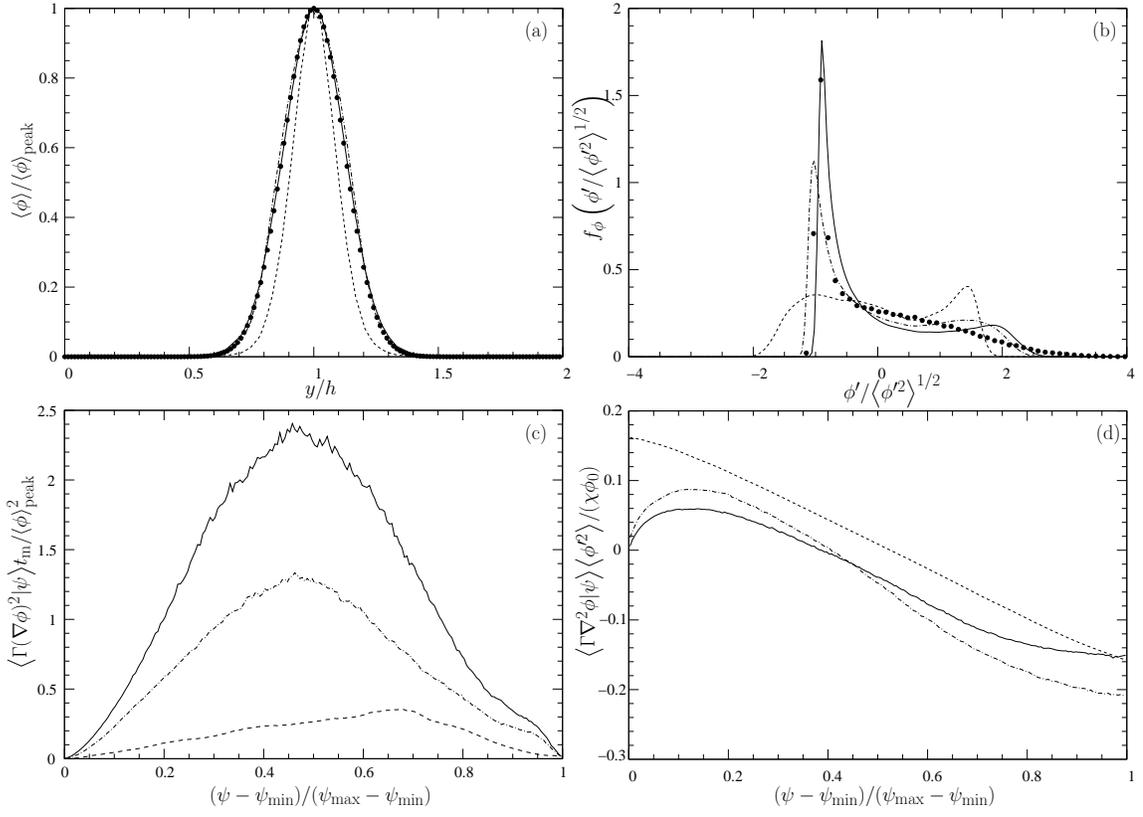}}
\caption[The effect of the number of conditioning bins with the proposed method]{Scalar statistics affected by the number of conditioning intervals when computing the velocity-conditioned mean \(\mean{\phi|\bv{V}}\) with the method described in \autoref{sec:conditional_stats} and \autoref{app:velocity-conditioning}. The quantities are the same as in \Fige{fig:CNBI}. Dashed line -- \(N_c\)=1 (IEM), dot-dashed line -- \(N_c\)=\((3\times3\times3)\), solid line -- \(N_c\)=\((5\times5\times5)\).}
\label{fig:BINd}
\figSpace
\end{figure}
A different picture reveals itself however, when \(\mean{\phi|\bv{V}}\) is computed with the current method instead of the projection that assumed Gaussianity of the underlying velocity field. The same statistics as shown in \Fige{fig:CNBI} are plotted in \Fige{fig:BINd} for different numbers of conditioning bins, but without employing the projection to compute \(\mean{\phi|\bv{V}}\). The mean profiles do not behave significantly differently, which underlines the earlier observation that employing only a few conditioning bins can already correct the prediction of the mean compared to the IEM model. The PDFs however show significantly higher spikes when compared to their counterparts with projection. The prediction of the conditional dissipation profiles are also different (overall they range about 150\% higher) as opposed to that with projection, while the conditional diffusion curves exhibit similar behavior both with and without projection. \Figse{fig:BINd} (b-d) also reveal that the currently employed finest conditional binning structure of (\(5\times5\times5\)) with an initial 500 particles per element is still not sufficient to achieve convergence for the PDF and these conditional statistics. It is also worth noting, that this is the case for a centerline release and that our sampling location is relatively close to the source and at the centerline, which lies in the ``approximately homogeneous'' region of the flow.

\begin{figure}
\centering
\resizebox{14.5cm}{!}{\input{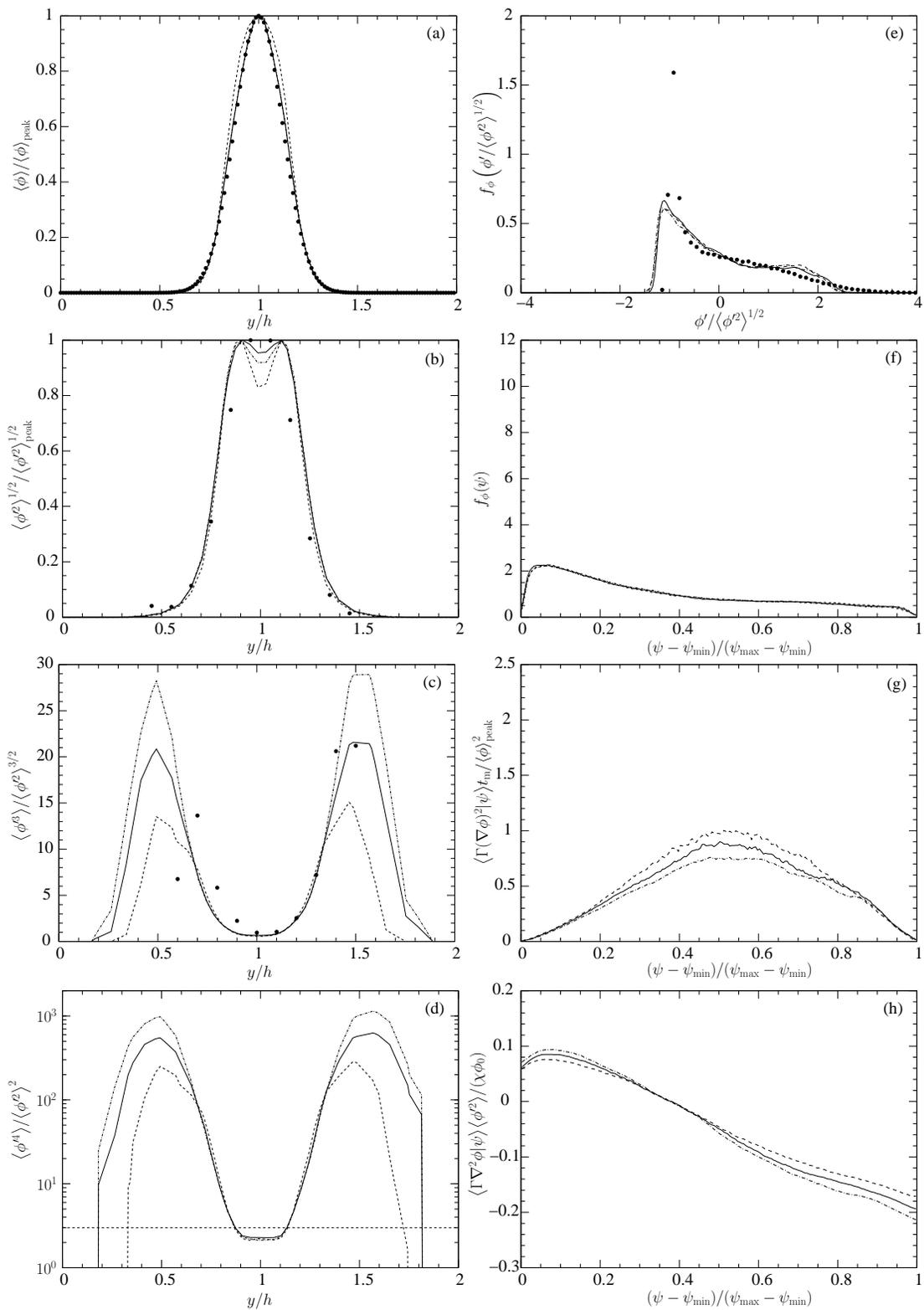}}
\caption[The effect of the number of particles using Fox's projection method]{See next page for caption.}
\label{fig:npar-proj}
\end{figure}

To examine the effect of the number of particles on the solution, several testruns have been performed with different number of particles employing both methods for computing \(\mean{\phi|\bv{V}}\). At the Reynolds numbers investigated, \(\textit{Re}_\tau=392\), 642 and 1080, we found the minimum number of particles per elements necessary for a numerically stable solution to be \(N_{p/e}\)=80, 100 and 150, respectively. Increasing \(N_{p/e}\) more than these minimum values would not be necessary to obtain a particle-number-independent velocity PDF, since running the simulation employing up to \(N_{p/e}\)=500 resulted in negligible change of the velocity statistics investigated. On the other hand, the scalar statistics exhibit significant differences when different number of particles are employed. \Fige{fig:npar-proj} shows unconditional and conditional statistics of the passive scalar field at \(\textit{Re}_\tau=1080\) using different numbers of particles employing the projection method with \(N_c\)=5. The cross-stream distribution of the first four moments show that the statistical error due to insufficient number of particles becomes higher towards the edge of the plume, where the joint PDF is most skewed. The discrepancy due to this error is more pronounced in the higher-order statistics. The PDFs of concentration fluctuations and the scalar at the centerline, where the flow can be considered approximately homogeneous, is nearly independent of the number of particles. The prediction of accurate conditional statistics usually requires a large number of particles. This is underlined by the mean conditional dissipation and diffusion in \Figse{fig:npar-proj} (g) and (h) in the center region, which show a slight dependence on \(N_{p/e}\). In summary, the velocity statistics are predicted independently of the number of particles. With the projection method to compute \(\mean{\phi|\bv{V}}\), the unconditional scalar statistics (including the PDFs) are predicted approximately independently of the number of particles in the homogeneous center region of the channel, however, the conditional statistics examined there still exhibit a slight particle-number-dependence even with \(N_{p/e}\)=500. We hypothesize that more complex inhomogeneous and highly skewed flows may require even larger number of particles than the currently employed maximum, 500.
\begin{figure}[t!]
Figure \thefigure: Unconditional and conditional statistics of the passive scalar field affected by the number of particles with \(\mean{\phi|\bv{V}}\) computed using the projection method of \Eqres{eq:projection}-\Eqr{eq:correlation} using \(N_c\)=5. (a)-(d) Cross-stream distribution of the first four moments at \(x/h=4.0\), (e) PDF of concentration fluctuations at (\(x/h=4.0,\) \(y/h=1.0\)), (f) PDF of concentration at (\(x/h=4.0,\) \(y/h=1.0\)), (g) mean scalar dissipation conditioned on the concentration at (\(x/h=4.0,\) \(y/h=1.0\)) and (h) mean scalar diffusion conditioned on the concentration at (\(x/h=4.0,\) \(y/h=1.0\)). Dashed line -- (initial number of particles per elements) \(N_{p/e}\)=150, solid line -- \(N_{p/e}\)=300 and dot-dashed line -- \(N_{p/e}\)=500. Symbols on (a) analytical Gaussians according to \citet{Taylor_21}, on (b), (c), (e) experimental data of \citet{Lavertu_05}. The horizontal dashed line on (d) indicates the Gaussian kurtosis value of 3.
\figSpace
\end{figure}
\begin{figure}
\centering
\resizebox{14.5cm}{!}{\input{effect_of_npar_noproj.pstex_t}}
\caption[The effect of the number of particles with the proposed method]{See next page for caption.}
\label{fig:npar-noproj}
\end{figure}
\begin{figure}[t!]
Figure \thefigure: Unconditional and conditional statistics of the passive scalar field affected by the number of particles with \(\mean{\phi|\bv{V}}\) computed with the method described in \autoref{sec:conditional_stats} and \autoref{app:velocity-conditioning} using a binning structure of (\(5\times5\times5\)). The legend is the same as in \Fige{fig:npar-proj}.
\figSpace
\end{figure}

In \Fige{fig:npar-noproj} the same scalar statistics as in \Fige{fig:npar-proj} are shown but with \(\mean{\phi|\bv{V}}\) computed with the current method instead of projection for different number of particles employing a binning structure of \((5\times5\times5)\). The technique described in \autoref{sec:conditional_stats} is robust enough to automatically use less conditioning intervals depending on the number of particles in a given element. Thus, when the simulations were run with \(N_{p/e}\)=150, 300 and 500, the average number of conditioning bins employed throughout the simulation has been automatically reduced to about 57, 100 and 124, respectively, as compared to the prescribed 125. The scalar mean is predicted equally well as with the projection method showing no sign of dependence on the number of particles, \Fige{fig:npar-noproj} (a). Interestingly, the r.m.s.\ curves do not double-peak if the projection is not used, \Fige{fig:npar-noproj} (b) and the width also agrees better with the experimental data. Thus the double-peaks on \Fige{fig:npar-proj} (b) may only be artifacts of the projection. Similarly to using projection, the skewness and kurtosis profiles are predicted with significant particle-number dependence at the edges of the plume. This shows that convergence has not yet been reached with \(N_{p/e}\)=500 for these higher-order statistics. Also, there is a pronounced flattening at the centerline in the skewness and kurtosis profiles using the projection technique, cf.\ \Figse{fig:npar-proj} and \ref{fig:npar-noproj} (c-d), which may also be a side-effect of the projection, since no flattening can be observed in the experimental data. The increasing peaks of the PDFs have already been observed before, when we compared the projection method to the general methodology using different values of \(N_c\). Both \Figse{fig:npar-noproj} (e) and (f) show that the PDFs have not converged yet, however, these figures may show the combined effect of increasing both \(N_{p/e}\) and \(N_c\), since the conditioning algorithm automatically reduces \(N_c\) in case of insufficient number of particles in an Eulerian element. Finally, the conditional dissipation and diffusion curves show a very light dependence on the number of particles applied.

We summarize the findings for the PDF algorithm related to a passive scalar released at the centerline of a fully developed turbulent channel flow as follows:
\begin{itemize}
\item the prediction of one-point velocity statistics becomes more accurate with increasing Reynolds number,
\item a stable numerical solution and a converged velocity field require about 80--150 particles per element depending on the Reynolds number,
\item the prediction of higher-order unconditional scalar statistics and concentration fluctuation PDFs are closer to experimental observations without employing the projection technique to compute \(\mean{\phi|\bv{V}}\),
\item conditioned statistics may exhibit a large difference (up to 150\%) depending on the application of the projection method, however the lack of experimental data currently prevents us to assess the true error in these quantities,
\item compared to the simpler IEM model, using the IECM model only with a few conditioning intervals already makes a big difference in correcting the prediction of the scalar mean, both with and without the projection method, for an increase in the overall computational cost of about 30--40\%,
\item the difference in computational costs of the projection and the current general method used to compute \(\mean{\phi|\bv{V}}\) is negligible,
\item \emph{with projection}, full convergence in the higher-order scalar statistics may require more particles than \(N_{p/e}\)=500, while \(N_c\)=3--5 was enough to reach convergence in all quantities investigated,
\item \emph{without projection}, full convergence in the higher-order scalar statistics and PDFs may require more particles than \(N_{p/e}\)=500, while the binning structure of (\(5\times5\times5\)) was enough to reach convergent unconditional statistics, but this was still not a sufficient conditioning-resolution to achieve convergent concentration PDFs and conditional statistics.
\end{itemize}
Table \ref{tab:particle-number-recommendations} lists the minimum number of particles per element necessary to accurately compute the one-point statistics investigated in this Chapter.
\begin{table}
\caption[Number of particles required for convergent statistics]{\label{tab:particle-number-recommendations}Minimum number of particles per element required to compute different statistics.}
\begin{tabular*}{1.0\textwidth}{l@{\hspace{0.5cm}}p{4.5cm}}
\hline\hline
quantity&particles per element\\
\hline
velocity statistics, \(\mean{U_i},\mean{u_iu_j},k,\varepsilon\) & 80--150, slightly increasing with the Reynolds number\\
first two scalar moments, \(\mean{\phi},\mean{\phi'^2}\) & 150\\
third, fourth and higher-order scalar moments, \(\mean{\phi'^3},\mean{\phi'^4}\) & 500+\\
scalar concentration PDFs, \(f_\phi(\phi'\mean{\phi'^{\scriptscriptstyle{2}}}^{\scriptscriptstyle{1/2}}),f_\phi(\psi)\) & 500+\\
mean conditional scalar dissipation, \(\mean{\Gamma(\nabla\phi)^2|\psi}\) & 300\\
mean conditional scalar diffusion, \(\mean{\Gamma\nabla^2\phi|\psi}\) & 150\\
\hline\hline
\end{tabular*}
\tableSpace
\end{table}

\section{Computational cost}
\label{sec:channel-computational-cost}
Quantitative assessments of the computational cost of PDF methods are sparse in the literature. There is no dedicated study to compare the different stand-alone and hybrid methods side by side or to compare PDF methods to other turbulence modeling techniques. However cost comparisons are useful even if they only provide limited information and are done between different methods at different levels of approximation.

\begin{figure}
\centering
\resizebox{15cm}{!}{\input{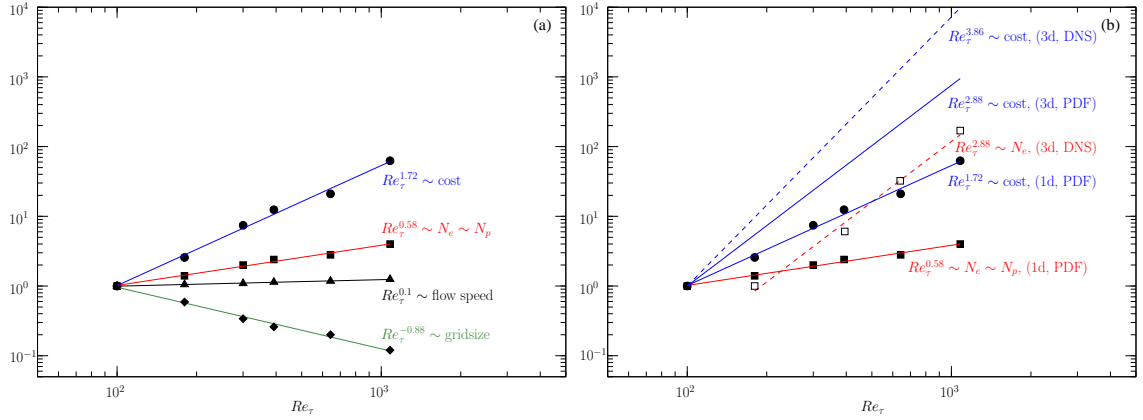}}
\caption[Comparison of computational cost vs.\ \textit{Re}]{Computational cost of (a) a measured one-dimensional and (b) an extrapolated three-dimensional PDF simulation. Filled symbols and solid lines -- PDF calculations, hollow symbols and dashed lines -- DNS of channel flow.}
\label{fig:cost}
\figSpace
\end{figure}

The computational cost of a simulation (the time required to reach convergence with a given accuracy) is largely determined by the resolution requirements, which in the case of a turbulent channel flow mostly amounts to adequately resolving the boundary layer. In an attempt to quantify the increase in cost of the current PDF methodology, several runs have been carried out at different Reynolds numbers between \(\textit{Re}_\tau=100\) and 1080. In all cases only the statistically one-dimensional velocity field has been computed reaching a statistically stationary state, without a scalar release and micromixing. As \(\textit{Re}_\tau\) is increased, the boundary layer becomes thinner and a finer Eulerian grid is needed to resolve the statistics, which inevitably results in the increase of the number of particles as well. Accordingly, keeping the Courant-number approximately constant, the size of the timestep has to be decreased to achieve the same level of accuracy and stability with increasing Reynolds numbers. This tendency can be examined in \Fige{fig:cost} (a), where the key factors affecting the computational cost vs.\ \(\textit{Re}_\tau\) are depicted. These are the smallest element (gridsize), the characteristic flow speed \(\mean{U}_c/u_\tau\), where \(\mean{U}_c\) is the mean velocity at the centerline, and the total number of elements \(N_e\) or equivalently, the total number of particles \(N_p\). All filled symbols on \Fige{fig:cost} represent the given quantity normalized by the quantity at \(\textit{Re}_\tau=100\). To an approximation, the number of floating-point operations, i.e.\ the computational cost, is proportional to the number of elements (and the number of particles) and the flow speed and inversely proportional to the gridsize (and the size of the timestep). Based on the slope of these three factors on a log-log scale, the approximate slope of the computational cost for the one-dimensional PDF simulation of channel flow can be estimated as%
\nomenclature[RN]{$N_e$}{number of elements}%
\nomenclature[RN]{$N_p$}{number of particles}%
\begin{equation}
\frac{\textit{Re}_\tau^{0.58} \times \textit{Re}_\tau^{0.1}}{\textit{Re}_\tau^{-0.88}} = \textit{Re}_\tau^{1.56}.\label{eq:costapp}
\end{equation}
This approximation based on the three key factors is in reasonable agreement with the measured slope, \(\textit{Re}_\tau^{1.72}\), which is based on actual timings. Employing the same arguments, the cost of a three-dimensional PDF simulation may be extrapolated as
\begin{equation}
\textit{Re}_\tau^{1.72} \times \textit{Re}_\tau^{2\times0.58} = \textit{Re}_\tau^{2.88},\label{eq:PDFcost}
\end{equation}
which is displayed in \Fige{fig:cost} (b). For comparison, the slope of the number of required elements for DNS simulations of turbulent channel flow is also displayed, based on the data reported by \citet{Abe_04}, normalized by \(N_e\) at \(\textit{Re}_\tau=180\). This gives the slope of \(\textit{Re}_\tau^{2.88}\) which reasonably agrees with \(\textit{Re}_\tau^{2.7}\), the prediction of \citet{Reynolds_90} for the total number of modes required for DNS of channel flow. (For comparison the cost of DNS in homogeneous turbulence grows as \(\textit{Re}_L^{2.25}\) \citep{Pope_00} based on the turbulence Reynolds number \(\textit{Re}_L=k^2/(\varepsilon\nu)\)). Based on the slope of \(N_e\) and \Eqre{eq:costapp} we approximate the increase in computational cost of DNS for the inhomogeneous channel flow as%
\nomenclature[RR]{$\textit{Re}_L$}{turbulence Reynolds number}%
\begin{equation}
\frac{\textit{Re}_\tau^{2.88} \times \textit{Re}_\tau^{0.1}}{\textit{Re}_\tau^{-0.88}} = \textit{Re}_\tau^{3.86}.
\end{equation}
Now we are in a position to quantitatively compare the computational requirements of a three-dimensional PDF to DNS simulations as it is displayed in \Fige{fig:cost} (b). A DNS simulation provides a great wealth of information on the turbulence for a steeply increasing cost at high Reynolds numbers by fully resolving all scales, including dissipation. A statistical technique, such as the current PDF method, approximates certain physical processes, thus it is expected to be less accurate. However, since it does not need to resolve the finest scales, it may be less computationally intensive. Based on \Fige{fig:cost} we observe that a three-dimensional PDF simulation will probably not be as expensive for higher Reynolds numbers as DNS. As depicted in \Fige{fig:cost} (b), the difference in computational cost between DNS and the three-dimensional PDF method is about a decade computing a fully resolved boundary layer at the Reynolds number $\textit{Re}_\tau=1080$. This means that at this Reynolds number DNS will produce the desired result in 10 times more computing hours than the PDF method. The figure also shows that extrapolating this result to more realistic Reynolds numbers will result in even larger differences in computational costs, DNS being increasingly more expensive than the current PDF method. As an example, resolving the boundary layer at $\textit{Re}_\tau=10^4$ will take 100 times more CPU time with DNS than with the PDF method.

It is worth noting that the cost in the current case largely amounts to adequately resolving the boundary layer. In general, any method that attempts to fully resolve the boundary layer will have to pay because of the required high resolution, and not necessarily because the method itself is inherently expensive. In a sense, the above assessment is even a bit unfair towards the PDF method since resolving walls is not its main advantage or purpose (although it still performs relatively well in comparison). Also, we estimate that our accounting for the increase in cost due to extrapolating from one to three spatial dimensions, \Eqre{eq:PDFcost}, is rather conservative, i.e.\ overpredictive -- while the Eulerian statistics are only extracted on a one-dimensional grid, all three components of the particle velocity are already retained, which consitutes as the majority of the computational cost, as it is shown in \autoref{sec:profile}.

We did not perform comparisons with other methods. A hybrid LES/FDF method for scalars may be expected to have a higher predictive power than the current stand-alone PDF method. However, we do not exclude that a stand-alone PDF method could be less expensive than a hybrid LES/FDF method, since resolution requirements may not have to be as stringent to achieve resolution-independent statistics. The Eulerian LES solution should be filter width and grid independent, which occurs only if a sufficient portion of the turbulent kinetic energy is resolved, i.e.\ in the case if only the dissipative scales are modeled and the majority of the inertial subrange and energy containing range is resolved. On the other hand, one-point statistical models, like the current stand-alone PDF method, do not need to resolve scales much below the integral scale \citep{Pope_00}.

Overall, the above assessment of the computational cost certainly cannot be taken in the most general sense as it is based on one simple flow, the fully developed turbulent channel flow, it extrapolates and compares to a method (DNS) that is quite different in both formulation and the results it obtains. Therefore reaching a final conclusion regarding the cost of the methodology is premature. Further assessments based on more flow topologies are needed to provide a better understanding of the computational cost of PDF methods compared to other methods, such as LES and other statistical approaches.

\section{Discussion}
\label{sec:channel-conclusions}
In this Chapter, the previously described stand-alone PDF method has been tested and validated against DNS, analytical and experimental data, computing the dispersion of passive scalars in fully developed turbulent channel flow. The complete PDF-IECM model computes the joint PDF of turbulent velocity, frequency and scalar concentration where the scalar is released from concentrated sources. The flow is represented by a large number of Lagrangian particles and the governing stochastic differential equations have been integrated in time in a Monte-Carlo fashion. The high anisotropy and inhomogeneity at the low-Reynolds-number wall-region have been captured through the elliptic relaxation technique, explicitly modeling the vicinity of the wall down to the viscous sublayer by imposing only the no-slip condition. \citet{Durbin_93} suggested the simple LRR-IP closure of \citet*{Launder_75}, originally developed in the Eulerian framework, as a local model used in the elliptic relaxation equation \Eqr{eq:elliptic-relaxation-Lagrangian}. Since then, several more sophisticated local Reynolds stress models have been investigated in conjunction with the elliptic relaxation technique \citep{Whizman_96}. In the PDF framework, the Lagrangian modified IP model of \citet{Pope_94} is based on the LRR-IP closure. We introduced an additional model constant \(C_\xi\) in the definition of the characteristic lengthscale \(L\) \Eqr{eq:L} whose curvature determines the behavior of the relaxation and, ultimately, the overall performance of the model in representing the Reynolds stress anisotropy. This resulted in a correction of the original model overprediction of the wall-normal component \(\mean{v^{\scriptscriptstyle 2}}\) far from the wall, which crucially influences the cross-stream mixing of the transported scalar. However, increasing the constant \(C_\xi\) adversely affects the level of anisotropy that can be represented by the technique. A more accurate treatment of the Reynolds stresses and scalar mixing should be achieved by a more elaborate second moment closure, such as the nonlinear C-L model of \citet{Craft_91} or the Lagrangian version of the SSG model of \citet*{Speziale_91} suggested by \citet{Pope_94}.

An unstructured triangular grid is used to compute Eulerian scalar statistics and to track particles throughout the domain. The main purpose of employing unstructured grids has been to prepare the methodology for more complex flow geometries. A similar particle-in-cell approach has been developed by \citet{Muradoglu_99,Muradoglu_01,Jenny_01,Zhang_04,Rembold_06} and by \citet{Ge_07} for the computation of turbulent reactive flows. These approaches combine the advantages of traditional Eulerian computational fluid dynamics (CFD) codes with PDF methods in a hybrid manner. Our aim here is to develop a method that is not a hybrid one, so the consistency between the computed fields can be naturally ensured. The emphasis is placed on generality, employing numerical techniques that assume as little as possible about the shape of the numerically computed joint PDF.

We compared the performance of the IEM and the IECM micromixing models in an inhomogeneous flow with strong viscous effects by modeling both the turbulent velocity field and the scalar mixing. The more sophisticated IECM model provides a closer agreement with experimental data in channel flow for the additional computational expense of 30-40\% compared to the IEM model.

Several conditional statistics that often require closure assumptions in PDF models where the velocity field is assumed were extracted and compared to some of their closures. In particular, our conclusions suggest that the scalar-conditioned velocity is well approximated by a linear assumption for mid-concentrations at locations where the velocity PDF is moderately skewed. The gradient diffusion approximation, however, captures most features including the nonlinearity and achieves a closer agreement with the IECM model in slightly more skewed regions of the flow as well. At local concentration extremes and in extremely skewed regions the gradient diffusion approximation markedly departs from the IECM model. The mean scalar dissipation conditioned on the scalar concentration may be well-approximated by the inverse relationship suggested by \citet{Sardi_98} in inhomogeneous flows with significant viscous effects as well, except at the concentration extremes. In computing the conditional scalar diffusion, both the IEM and the IECM models produce similar slopes due to the same scalar dissipation rate attained.

The effects of several numerical parameters on the computed results have also been investigated. We found that about a hundred particles per element are enough for a stable numerical solution. However, even 500 particles per element were not enough to obtain particle-number-independent higher-order scalar statistics. Moreover, to obtain accurate higher-order scalar statistics and concentration fluctuation PDFs in inhomogeneous flows, the use of the currently proposed method is advised to compute \(\mean{\phi|\bv{V}}\) as opposed to the projection method assuming Gaussianity.


\chapter[Street canyon simulations: results and discussion]{\\Street canyon simulations: results and discussion}
\label{chap:canyon}

\section{Introduction}
Regulatory bodies, architects and town planners increasingly use computer models in order to assess ventilation and occurrences of hazardous pollutant concentrations in cities. These models are mostly based on the Reynolds-averaged Navier-Stokes (RANS) equations or, more recently, large eddy simulation (LES) techniques. Both of these approaches require a series of modeling assumptions, including most commonly the eddy-viscosity and gradient-diffusion hypotheses. The inherent limitations of these approximations, even in the simplest engineering flows, are well known and detailed for example by \citet{Pope_00}. Therefore, there is a clear need to develop higher-order models to overcome these shortcomings. In pollutant dispersion modeling it is also desirable to predict extreme events like peak values or probabilities that concentrations will exceed a certain threshold. In other words, a fuller statistical description of the concentration is required \citep{Chatwin_93,Kristensen_94,Wilson_95,Pavageau_99}. These issues have been explored in the unobstructed atmosphere and models capable of predicting these higher-order statistics have also appeared \citep{Franzese_03,Cassiani_et_al_05,Cassiani_et_al_05b}, but more research is necessary to extend these capabilities to cases of built-up areas.%
\nomenclature[A]{RANS}{Reynolds-averaged Navier-Stokes}%

Probability density function (PDF) methods have been developed mainly within the combustion engineering community as an alternative to moment closure techniques to simulate chemically reactive turbulent flows \citep{Lundgren_69,Pope_85,Dopazo_94}. Because many-species chemistry is high-dimensional and highly nonlinear, the biggest challenge in reactive flows is to adequately model the chemical source term. In PDF methods, the closure problem is raised to a statistically higher level by solving for the full PDF of the turbulent flow variables instead of its moments. This has several benefits. Convection, the effect of mean pressure, viscous diffusion and chemical reactions appear in closed form in the PDF transport equation. Therefore these processes are treated mathematically exactly without closure assumptions eliminating the need for gradient-transfer approximations. The effect of fluctuating pressure, dissipation of turbulent kinetic energy and small-scale mixing of scalars still have to be modeled. The rationale is that since the most important physical processes are treated exactly, the errors introduced by modeling assumptions for less important processes amount to a smaller departure from reality. Moreover, the higher level description provides more information which can be used in the construction of closure models.

The PDF transport equation is a high-dimensional scalar equation. Therefore all techniques of solution rely on Monte Carlo methods with Lagrangian particles representing a finite ensemble of fluid particles, because the computational cost of Lagrangian Monte Carlo methods increases only linearly with increasing problem dimensionality, favourably comparing to the more traditional finite difference, finite volume or finite element methods. The numerical development in PDF methods has mainly centered around three distinctive approaches. A common numerical approach is the \emph{standalone Lagrangian} method, where the flow is represented by particles whereas the Eulerian statistics are obtained using kernel estimation \citep{Pope_00,Fox_03}. Another technique is the \emph{hybrid} methodology, which builds on existing Eulerian computational fluid dynamics (CFD) codes based on moment closures \citep{Muradoglu_99,Muradoglu_01,Jenny_01,Rembold_06}. Hybrid methods use particles to solve for certain quantities and provide closures for the Eulerian moment equations using the particle/PDF methodology. A more recent approach is the self-consistent \emph{non-hybrid} method \citep{Bakosi_07,Bakosi_08}, which also employs particles to represent the flow, and uses the Eulerian grid only to solve for inherently Eulerian quantities (like the mean pressure) and for efficient particle tracking. Since the latter two approaches extensively employ Eulerian grids, they are particle-in-cell methods \citep{Grigoryev_02}.

After an extensive testing of the methodology in a relatively simple setting, the fully developed turbulent channel flow (\autoref{chap:channel}), the current Chapter presents an application of the non-hybrid method to a simplified urban-scale case where pollution released from a concentrated line source between idealized buildings is simulated and results are compared to wind-tunnel experiments.

PDF methods in atmospheric modeling have mostly been focused on simulation of passive pollutants, wherein the velocity field (mean and turbulence) is assumed or obtained from experiments \citep{Sawford_04,Sawford_06,Cassiani_et_al_05,Cassiani_et_al_05b,Cassiani_07}. Instead, the current model directly computes the joint PDF of the turbulent velocity, characteristic turbulent frequency and scalar concentration, thus it extends the use of PDF methods in atmospheric modeling to represent more physics at a higher statistical level. Computing the full joint PDF also has the advantage of providing information on the uncertainty of the simulation on a physically sound basis.

In this Chapter the turbulent boundary layers developing along solid walls are treated in two different ways: either fully resolved or via the application of wall-functions (i.e.\ the logarithmic ``law of the wall''). The full resolution is obtained using Durbin's elliptic relaxation technique \citep{Durbin_93}, which was incorporated into the PDF methodology by \citet{Dreeben_97,Dreeben_98}. This technique allows for an adequate representation of the near-wall low-Reynolds-number effects, such as the high inhomogeneity and anisotropy of the Reynolds stress tensor and wall-blocking. Wall-conditions for particles based on the logarithmic ``law of the wall'' in the PDF framework have also been developed by \citet{Dreeben_97b}. These two types of wall-treatments are examined in terms of computational cost / performance trade-off, addressing the question of how important it is to adequately resolve the boundary layers along solid walls in order to obtain reasonable scalar statistics.

At the urban scale the simplest settings to study turbulent flow and dispersion patterns are street canyons. Due to increasing concerns for environmental issues and air quality standards in cities, a wide variety of canyon configurations and release scenarios have been studied both experimentally \citep{Hoydysh_74,Wedding_77,Rafailids_95,Meroney_96,Pavageau_99} and numerically \citep{Lee_94,Johnson_95,Baik_99,Huang_00,Liu_02}. Street canyons have a simple flow geometry, they can be studied in two dimensions and a wealth of experimental and modeling data are available for different street-width to building-height ratios. This makes them ideal candidates for testing a new urban pollution dispersion model. We validate the computed velocity and scalar statistics with the LES simulation results of \citet{Liu_02} and the wind tunnel measurements of \citet{Meroney_96}, \citet{Pavageau_96} and \citet{Pavageau_99}. The experiments have been performed in the atmospheric wind tunnel of the University of Hamburg, where the statistics of the pollutant concentration field have been measured in an unusually high number of locations in order to provide fine details inside the street canyon.

The Chapter is organized as follows. In \autoref{sec:canyon-bc} the specifics of the boundary conditions related to the street canyon are outlined. Several statistics computed using both full wall-resolution and wall-functions are compared to experimental data and large eddy simulation in \autoref{sec:canyon-results}. Finally, \autoref{sec:canyon-conclusions} draws some conclusions and elaborates on possible future directions.

\section{Modeling specifics of the street canyon}
\label{sec:canyon-bc}
The governing equations for both full wall-resolution and wall-functions cases together with boundary conditions have been described in \autoref{chap:governing_equations} and \autoref{sec:wall_conditions}, respectively. In \autoref{chap:numerics} we also elaborated on several aspects of the numerical techniques that are used to solve the equations. Thus here, only certain specific details that directly relate to the modeling of the street canyon case are described.

The flow geometry can be modeled as statistically two-dimensional if we suppose that the buildings are sufficiently long, like a long street. The particle copying-mirroring strategy used for the channel flow cannot be used here, so the general algorithm is applied. An additional complexity is the computation of the mean pressure in a general way, applying the pressure projection described in \autoref{sec:Eulerian_equations}. A non-homogeneous Neumann wall-boundary condition for the pressure projection \Eqr{eq:pressure-projection-smooth} has been described in \autoref{sec:wall_conditions} for both full wall-resolution and wall-functions representations of no-slip walls. The flow is expected to reach a statistically steady state and is driven by a mean-pressure difference between its inflow and outflow. This condition in the free stream (above the buildings) is imposed on the mean pressure as follows.

Assuming that the inflow and outflow are aligned with \(y\), as shown in \Fige{fig:canyon-geometry}, the two-dimensional steady state cross-stream mean-momentum equation holds
\begin{equation}
\frac{1}{\rho}\frac{\partial\mean{P}}{\partial y} = -\mean{U}\frac{\partial\mean{V}}{\partial x} - \mean{V}\frac{\partial\mean{V}}{\partial y} + \nu\left(\frac{\partial^2\mean{V}}{\partial x^2} + \frac{\partial^2\mean{V}}{\partial y^2}\right) - \frac{\mean{uv}}{\partial x} - \frac{\partial\mean{v^{\scriptscriptstyle{2}}}}{\partial y}.\label{eq:cross-stream-mean-momentum}
\end{equation}
If the inflow and outflow are far enough from the canyon, the flow can be assumed to be an undisturbed turbulent channel flow. Hence we can neglect all terms on the right hand side of \Eqre{eq:cross-stream-mean-momentum}, with the exception of the last term. Thus the inflow and outflow conditions for the mean pressure can be specified according to \Eqre{eq:cross-stream-mean-momentum-channel}. Flow-dependent non-homogeneous Dirichlet conditions have to be imposed in a way that the streamwise gradient \(\partial\mean{P}/\partial x\) is kept at a constant level. This can be achieved by specifying the values of \(\mean{P}\) at the inflow and outflow based on \(\mean{P}=-\rho\mean{v^{\scriptscriptstyle{2}}}\), which will equate their cross-stream derivatives as well. The streamwise gradient \(\partial\mean{P}/\partial x =\mathrm{const.}\) is applied by shifting up the values of \(\mean{P}\) at the inflow. Consistently with \Eqre{eq:pressure-projection-smooth} the above condition has to be imposed on the mean-pressure difference in time, \(\delta\mean{P}=\mean{P}^{n+1}-\mean{P}^n\). Thus we arrive at the inflow/outflow conditions
\begin{equation}
\delta\mean{P} = \left\{
\begin{aligned}
&-\Delta P \cdot L_x - \rho\mean{v^{\scriptscriptstyle{2}}} - \mean{P}^n, \quad & \textrm{for inflow points,}\\
&-\rho\mean{v^{\scriptscriptstyle{2}}} - \mean{P}^n, \quad & \textrm{for outflow points,}
\end{aligned}
\right.
\end{equation}%
\nomenclature[RL]{$L_x$}{length of domain in $x$ direction (\autoref{chap:canyon})}%
where \(\Delta P < 0\) denotes the imposed constant streamwise mean-pressure gradient over the streamwise length \(L_x\) of the domain. This inflow/outflow condition drives the flow and builds up a numerical solution that converges to a statistically stationary state. No conditions are imposed on particles leaving and entering the domain other than periodicity on their streamwise positions. This, in effect, will simulate the ``urban roughness'' case of \citet{Meroney_96}, which is a model for a series of street canyons in the streamwise direction. Wall-conditions are imposed on particles that hit wall-elements as described in \autoref{sec:wall_conditions}. On the top of the domain, free-slip conditions are imposed on particles, i.e.\ perfect reflection on their positions and a sign reversal of their normal velocity component. To model the small-scale mixing of the passive scalar the IECM model is applied with the \((5\times5\times5)\) binning structure without employing the projection method to compute \(\mean{\phi|\bv{V}}\). To define the micromixing timescale for a scalar released from a concentrated source in a geometrically complex flow domain bounded by no-slip walls, such as a street canyon, we follow \autoref{chap:channel} and \citet{Bakosi_07,Bakosi_08} and specify the inhomogeneous $t_\mathrm{m}$ as
\begin{equation}
t_\mathrm{m}(\bv{r}) = \min\left[C_s\left(\frac{r_0^2}{\varepsilon}\right)^{1/3} + C_t\frac{d_\bv{r}}{U_c(\bv{r})}; \enskip \max\left(\frac{k}{\varepsilon}; \enskip C_T\sqrt{\frac{\nu}{\varepsilon}}\right)\right],\label{eq:micromixing-timescale-canyon}
\end{equation}%
\nomenclature[Rr]{$\bv{r}$}{location}%
\nomenclature[RU]{$U_c$}{characteristic velocity, \Eqre{eq:micromixing-timescale-canyon}}%
\nomenclature[Rd]{$d_\bv{r}$}{distance of the point $\bv{r}$ from the source, \Eqre{eq:micromixing-timescale-canyon}}%
where $r_0$ denotes the radius of the source, $U_c$ is a characteristic velocity at $\bv{r}$ which we take as the absolute value of the mean velocity at the given location, $d_\bv{r}$ is the distance of the point $\bv{r}$ from the source, while $C_s$ and $C_t$ are model constants. The applied model constants for the micromixing timescale defined by \Eqre{eq:micromixing-timescale-canyon} are the same as for the centerline-release in channel flow, i.e.\ \(C_s=0.02\) and \(C_t=0.7\). 

The Reynolds number $\textit{Re}\approx12000$ based on the maximum free stream velocity $U_0$ and the building height $H$. This corresponds to $\textit{Re}_\tau\approx600$ based on the friction velocity and the free stream height, $h=H/2$, if the free stream above the buildings is considered as the lower part of an approximate fully developed turbulent channel flow. After the flow has reached a statistically stationary state, time-averaging is used to collect velocity statistics and a continuous scalar is released from a street level line source at the center of the canyon (corresponding to a point-source in two dimensions). The scalar field is also time-averaged after it has reached a stationary state.%
\nomenclature[RR]{$\textit{Re}$}{Reynolds number for the street canyon, based on $U_0$ and $H$ (\autoref{chap:canyon})}%
\nomenclature[RU]{$U_0$}{free stream velocity (\autoref{chap:canyon} and \autoref{chap:cylinder})}%
\nomenclature[RH]{$H$}{building height (\autoref{chap:canyon})}%
\nomenclature[Rh]{$h$}{free stream height (\autoref{chap:canyon})}%

\section{Results}
\label{sec:canyon-results}
\begin{table}
\caption[Model constants for the full joint PDF]{\label{tab:constants}Constants for modeling the joint PDF of velocity, characteristic turbulent frequency and transported passive scalar.}
\begin{tabular*}{1.0\textwidth}{c@{\hspace{0.57cm}}c@{\hspace{0.57cm}}c@{\hspace{0.57cm}}c@{\hspace{0.57cm}}c@{\hspace{0.57cm}}c@{\hspace{0.57cm}}c@{\hspace{0.57cm}}c@{\hspace{0.57cm}}c@{\hspace{0.57cm}}c@{\hspace{0.57cm}}c@{\hspace{0.57cm}}c@{\hspace{0.57cm}}c@{\hspace{0.57cm}}}
\hline\hline
\(C_1\)&\(C_2\)&\(C_3\)&\(C_4\)&\(C_T\)&\(C_L\)&\(C_\eta\)&\(C_v\)&\(\gamma_5\)&\(C_{\omega1}\)&\(C_{\omega2}\)&\(C_s\)&\(C_t\)\\
1.85&0.63&5.0&0.25&6.0&0.134&72.0&1.4&0.1&0.5&0.73&0.02&0.7\\
\hline\hline
\end{tabular*}
\tableSpace
\end{table}
\begin{figure}
\centering
\resizebox{13.7cm}{!}{\input{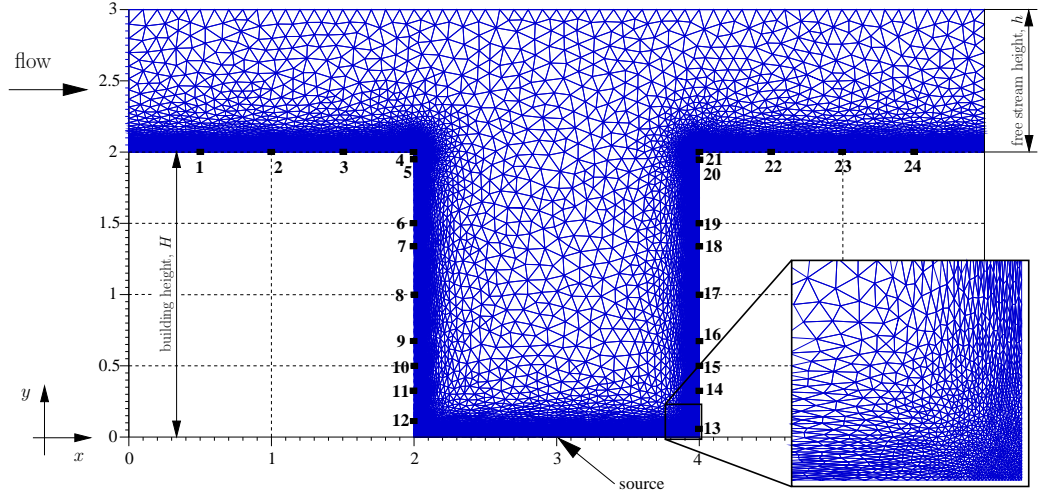}}
\caption[Eulerian grid to compute a street canyon with full resolution]{Geometry and Eulerian mesh for the computation of turbulent street canyon with full resolution of the wall-boundary layers using elliptic relaxation. The grid is generated by the general purpose mesh generator Gmsh \citep{Geuzaine_09}. The positions labeled by bold numbers indicate the sampling locations for the passive scalar, equivalent with the combined set of measurement tapping holes of \citet{Meroney_96}, \citet{Pavageau_96} and \citet{Pavageau_99}. In the zoomed area the refinement is depicted, which ensures an adequate resolution of the boundary layer and the vortices forming in the corner.}
\label{fig:canyon-geometry}
\figSpace
\end{figure}
\begin{table}
\caption[Concentration sampling locations along the street canyon walls]{\label{tab:measurement-locations}Concentration sampling locations at building walls and tops according to the experimental measurement holes of \citet{Meroney_96}, \citet{Pavageau_99} and \citet{Pavageau_96}. See also \Fige{fig:canyon-geometry}.}
\begin{tabular*}{1.0\textwidth}{r@{\hspace{0.27cm}}c@{\hspace{0.27cm}}c@{\hspace{0.27cm}}c@{\hspace{0.27cm}}c@{\hspace{0.27cm}}c@{\hspace{0.27cm}}c@{\hspace{0.27cm}}c@{\hspace{0.27cm}}c@{\hspace{0.27cm}}c@{\hspace{0.27cm}}c@{\hspace{0.27cm}}c@{\hspace{0.27cm}}c@{\hspace{0.27cm}}c@{\hspace{0.27cm}}c}
\hline\hline
\mbox{\textbf{\#} }&\textbf{1}&\textbf{2}&\textbf{3}&\textbf{4}&\textbf{5}&\textbf{6}&\textbf{7}&\textbf{8}&\textbf{9}&\textbf{10}&\textbf{11}&\textbf{12}&\textbf{13}&\textbf{14}\\
\mbox{\(x\) } & \mbox{0.5 } & \mbox{1.0 } & \mbox{1.5 } & \mbox{2.0 } & \mbox{2.0 } & \mbox{2.0 } & \mbox{2.0 } & \mbox{2.0 } & \mbox{2.0 } & \mbox{2.5 } & \mbox{2.0 } & \mbox{2.0 } & \mbox{4.0 } & \mbox{4.0 } \\
\mbox{\(y\) } & \mbox{2.0 } & \mbox{2.0 } & \mbox{2.0 } & \mbox{2.0 } & \mbox{1.93 } & \mbox{1.5 } & \mbox{1.33 } & \mbox{1.0 } & \mbox{0.67 } & \mbox{0.5 } & \mbox{0.33 } & \mbox{0.17 } & \mbox{0.17 } & \mbox{0.33 }\\
\hline
\mbox{\textbf{\#} }&\textbf{15}&\textbf{16}&\textbf{17}&\textbf{18}&\textbf{19}&\textbf{20}&\textbf{21}&\textbf{22}&\textbf{23}&\textbf{24}&&&\\
\mbox{\(x\) } & \mbox{4.0 } & \mbox{4.0 } & \mbox{4.0 } & \mbox{4.0 } & \mbox{4.0 } & \mbox{4.0 } & \mbox{4.0 } & \mbox{4.5 } & \mbox{5.0 } & \mbox{5.5 }&&&\\
\mbox{\(y\) } &\mbox{0.5 } & \mbox{0.67 } & \mbox{1.0 } & \mbox{1.33 } & \mbox{1.5 } & \mbox{1.93 } & \mbox{2.0 } & \mbox{ 2.0 } & \mbox{2.0 } & \mbox{ 2.0 }&&&\\
\hline\hline
\end{tabular*}
\tableSpace
\end{table}
\begin{figure}[t!]
\centering
\includegraphics[width=13.7cm]{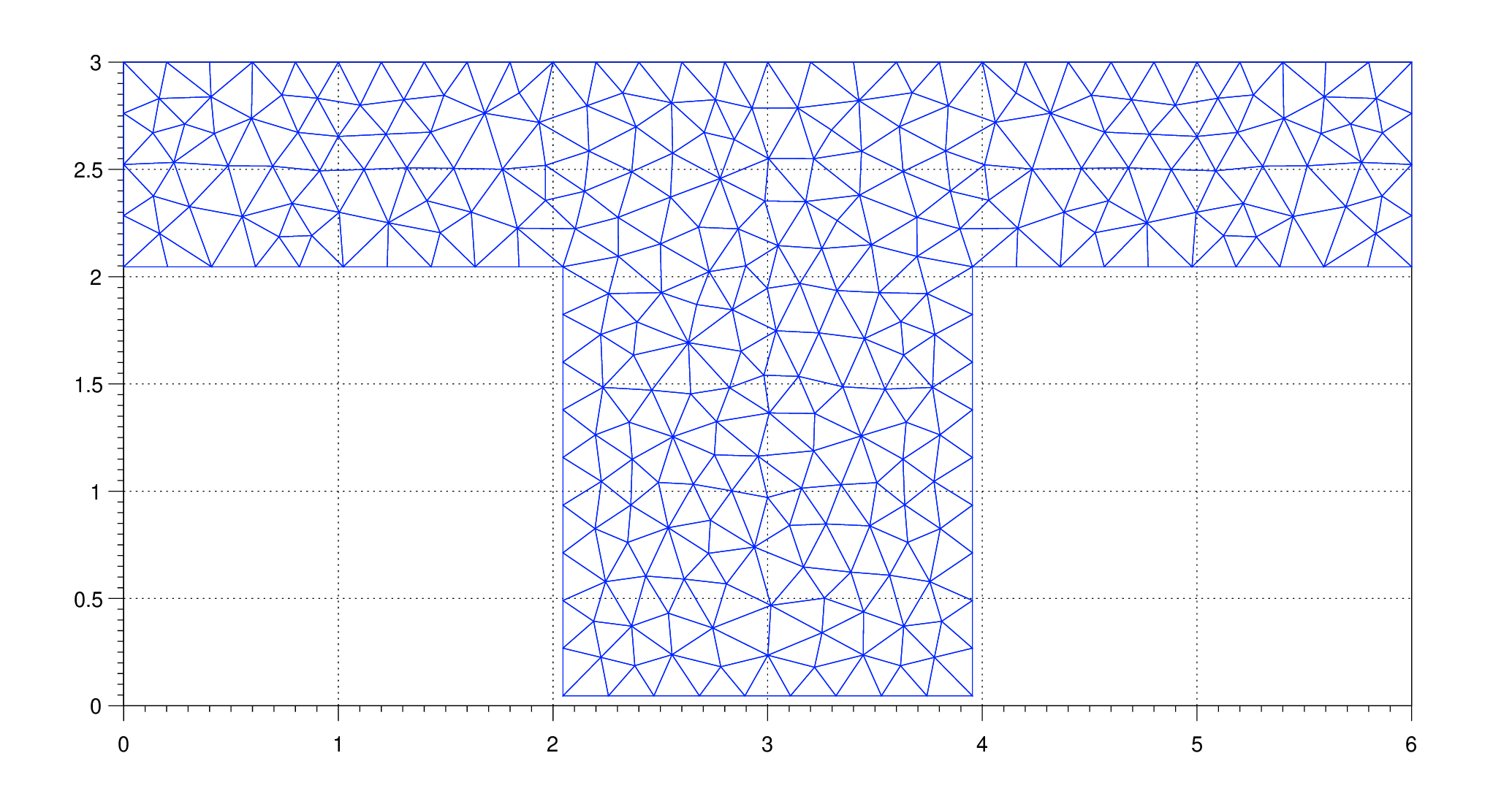}
\caption[Eulerian grid to compute a street canyon with wall-functions]{Geometry and Eulerian mesh for the computation of turbulent street canyon with wall-functions at $\textit{Re}\approx12000$. The domain is stripped at no-slip walls so that it does not include the close vicinity of the wall at $y^+<30$. The positions for sampling the scalar concentrations are the same as in \Fige{fig:canyon-geometry}.}
\label{fig:canyon-wf-geometry}
\figSpace
\end{figure}
The simulations with the full resolution model have been run with the constants given in Table \ref{tab:constants}, using 300 particles per element. The Eulerian mesh used for this simulation is displayed in \Fige{fig:canyon-geometry}, which shows the considerable refinement along the building walls and tops necessary to solve the boundary layers. In this case, the high anisotropy and inhomogeneity of the Reynolds stress tensor in the vicinity of walls are captured by the elliptic relaxation technique, using \Eqre{eq:elliptic-relaxation-Lagrangian}. 

The simulations using wall-functions were performed on the Eulerian mesh displayed in \Fige{fig:canyon-wf-geometry}, also using 300 particles per element. We implemented the particle-boundary conditions for arbitrary geometry described in \autoref{chap:governing_equations}. Note that the first gridpoint where the boundary conditions based on wall-functions are to be applied should not be closer to the wall than $y^+=u_\tau y/\nu=30$, where $y^+$ is the non-dimensional distance from the wall in wall-units, but sufficiently close to the wall to still be in the inertial sublayer \citep{Dreeben_97b}. Accordingly, the grid in \Fige{fig:canyon-wf-geometry} only contains the domain stripped from the wall-region at $y^+<30$.

Turbulence and scalar statistics are obtained entirely from the particles that represent both the flow itself and the scalar concentration field. The Eulerian meshes displayed in \Fige{fig:canyon-geometry} for the full resolution and in \Fige{fig:canyon-wf-geometry} for the wall-functions cases are used to extract the statistics, to track the particles throughout the domain and to solve the Eulerian equations: \Eqre{eq:elliptic-relaxation-Lagrangian} and the mean-pressure-Poisson equation \Eqr{eq:pressure-projection-smooth} in the fully resolved case and only the latter in the wall-functions case.
\begin{figure}
\centering
\psfragscanon
\includegraphics[width=7.5cm]{velocity.ps}
\includegraphics[width=7.5cm]{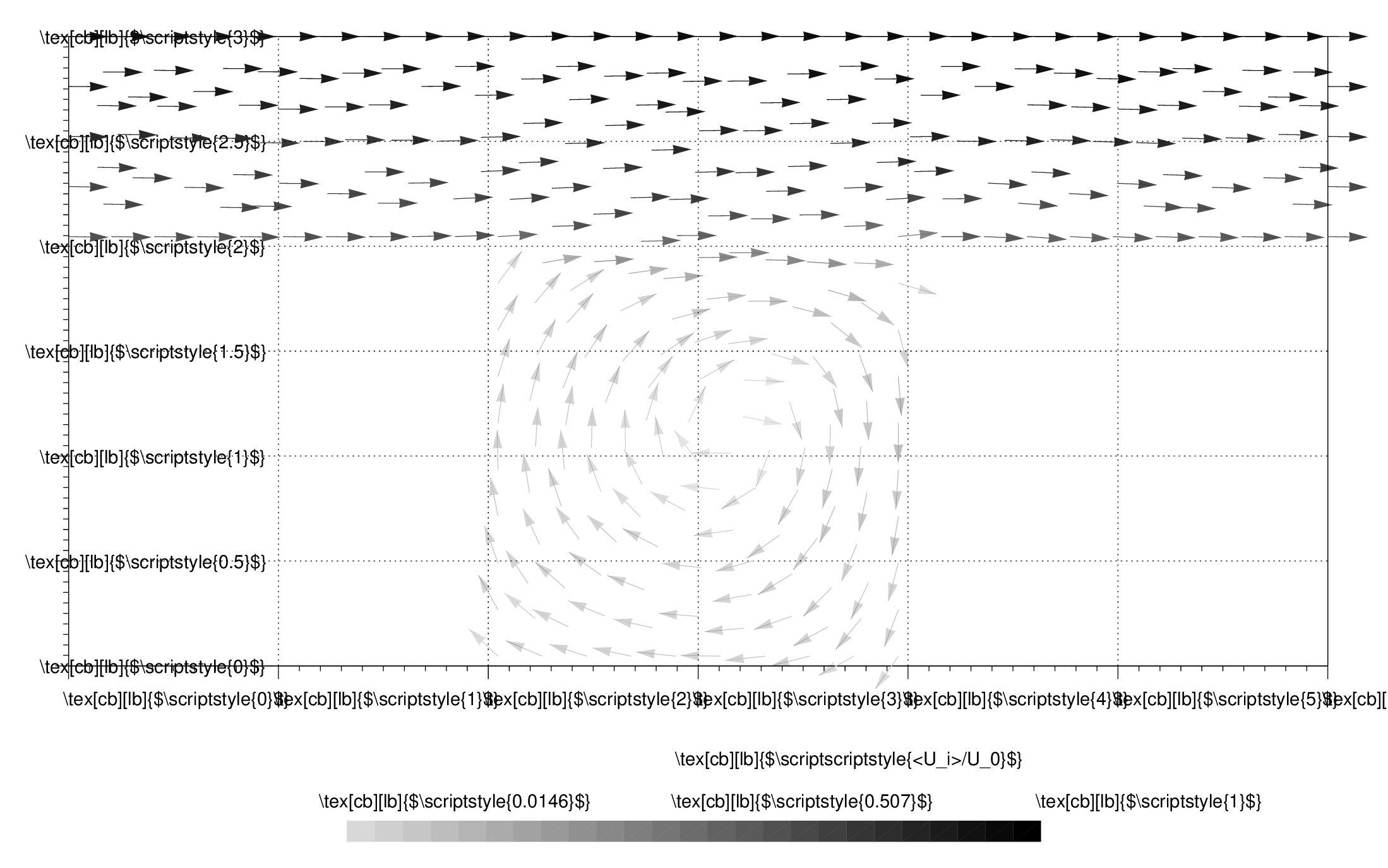}
\includegraphics[width=7.5cm]{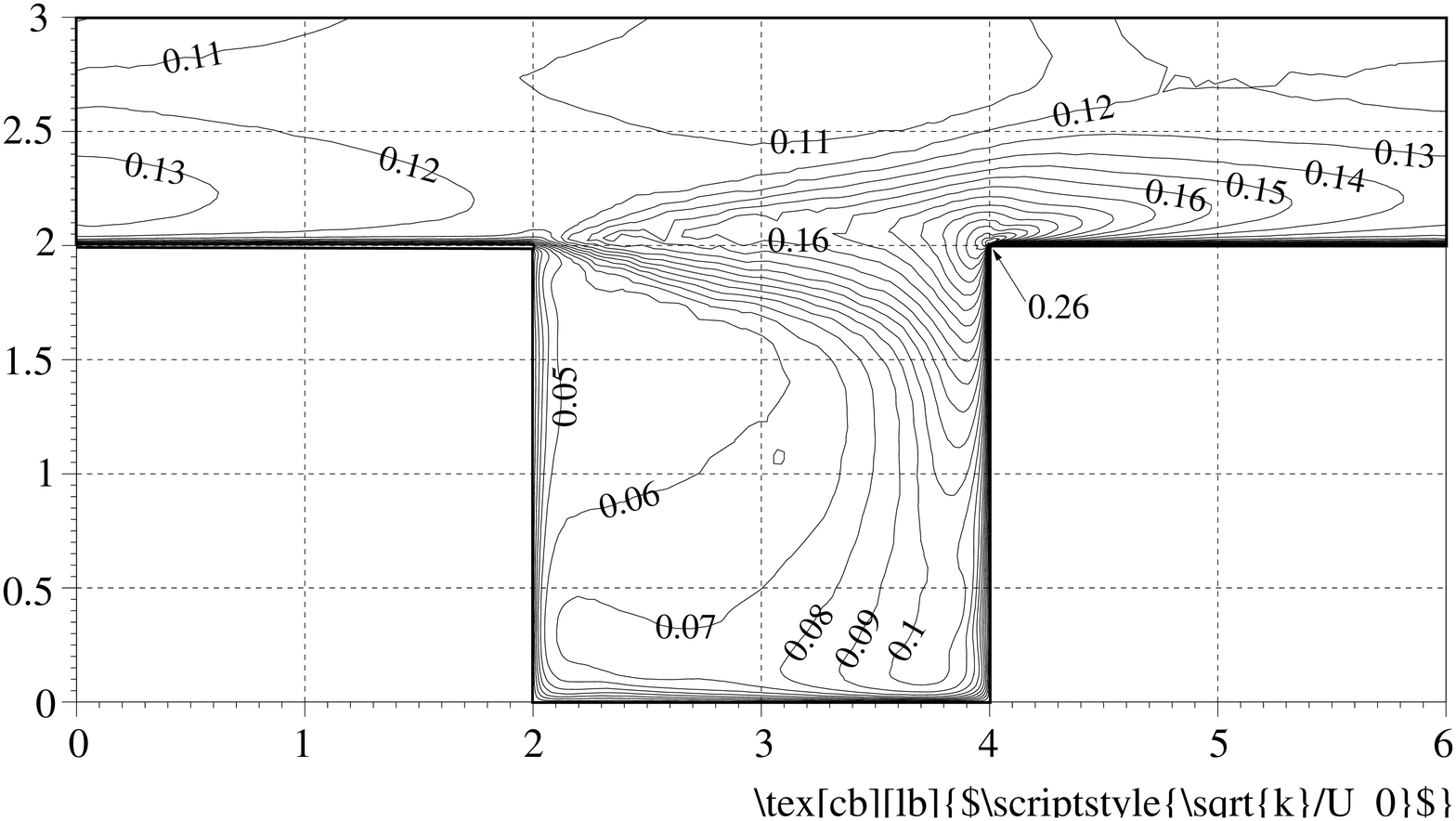}
\includegraphics[width=7.5cm]{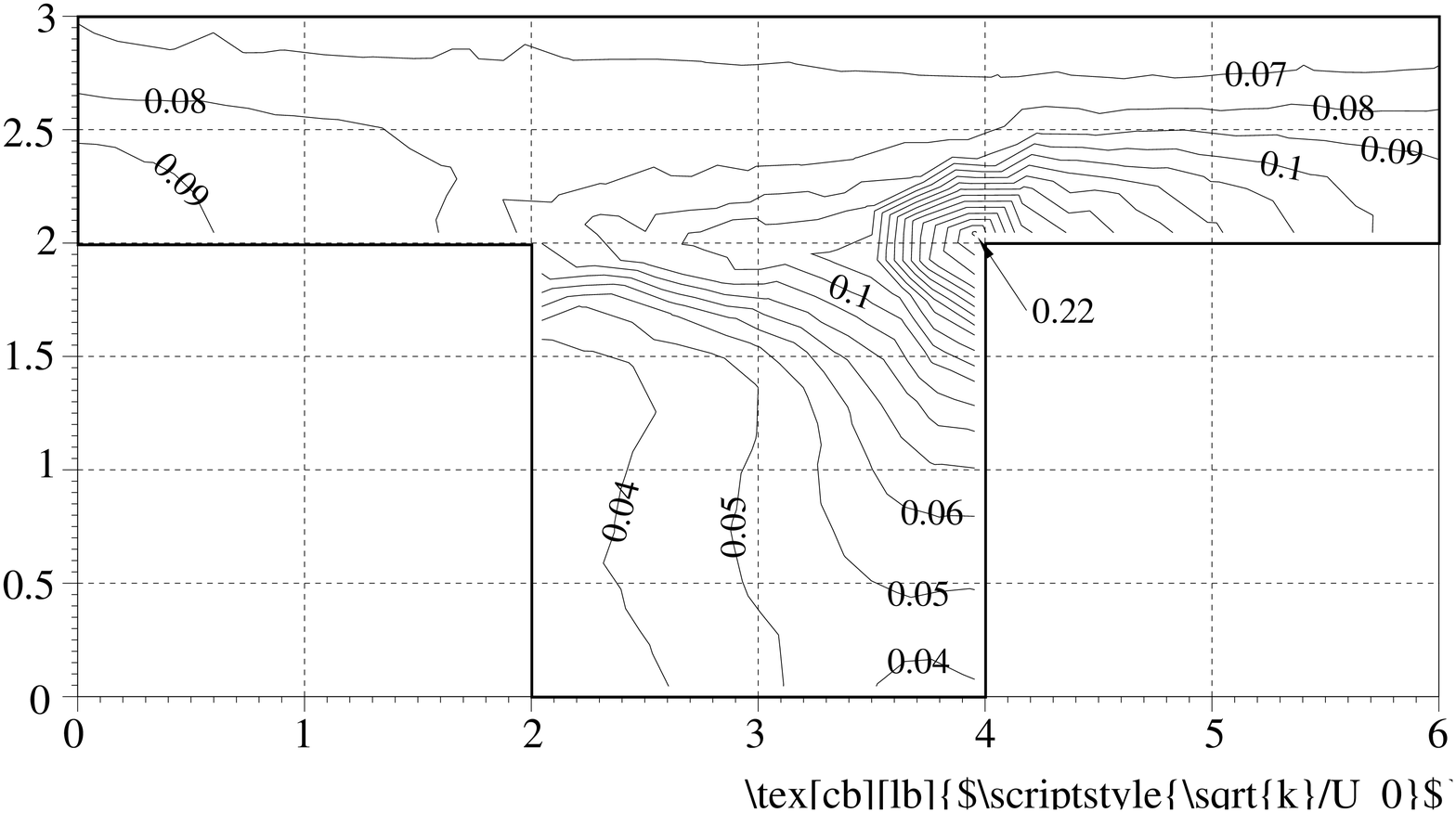}
\psfragscanoff
\caption[Velocty statistics in a street canyon]{Velocity vectors (first row) and iso-contours of turbulent kinetic energy (second row) of the fully developed turbulent street canyon at $\textit{Re}\approx12000$ based on the maximum free stream velocity $U_0$ and the building height $H$. Left -- full resolution with elliptic relaxation, right -- coarse simulation with wall-functions.}
\label{fig:canyon-velocity}
\figSpace
\end{figure}

In \Fige{fig:canyon-velocity}, the mean velocity vectorfield and the iso-contours of the turbulent kinetic energy are displayed for both fully resolved and wall-functions simulations.  It is apparent that the full resolution captures even the smaller counterrotating eddies at the internal corners of the canyon, while the coarse grid-resolution with wall-functions only captures the overall flow-pattern characteristic of the flow, such as the big steadily rotating eddy inside the canyon. The turbulent kinetic energy field is captured in a similar manner.  Both methods reproduce the highest turbulence activity at the building height above the canyon, with a maximum at the windward building corner.  The full resolution simulation shows a more detailed spatial distribution of energy, whereas the coarse resolution of the wall-functions simulation still allows to capture the overall pattern. 

\begin{figure}[t!]
\centering
\includegraphics[width=7.5cm]{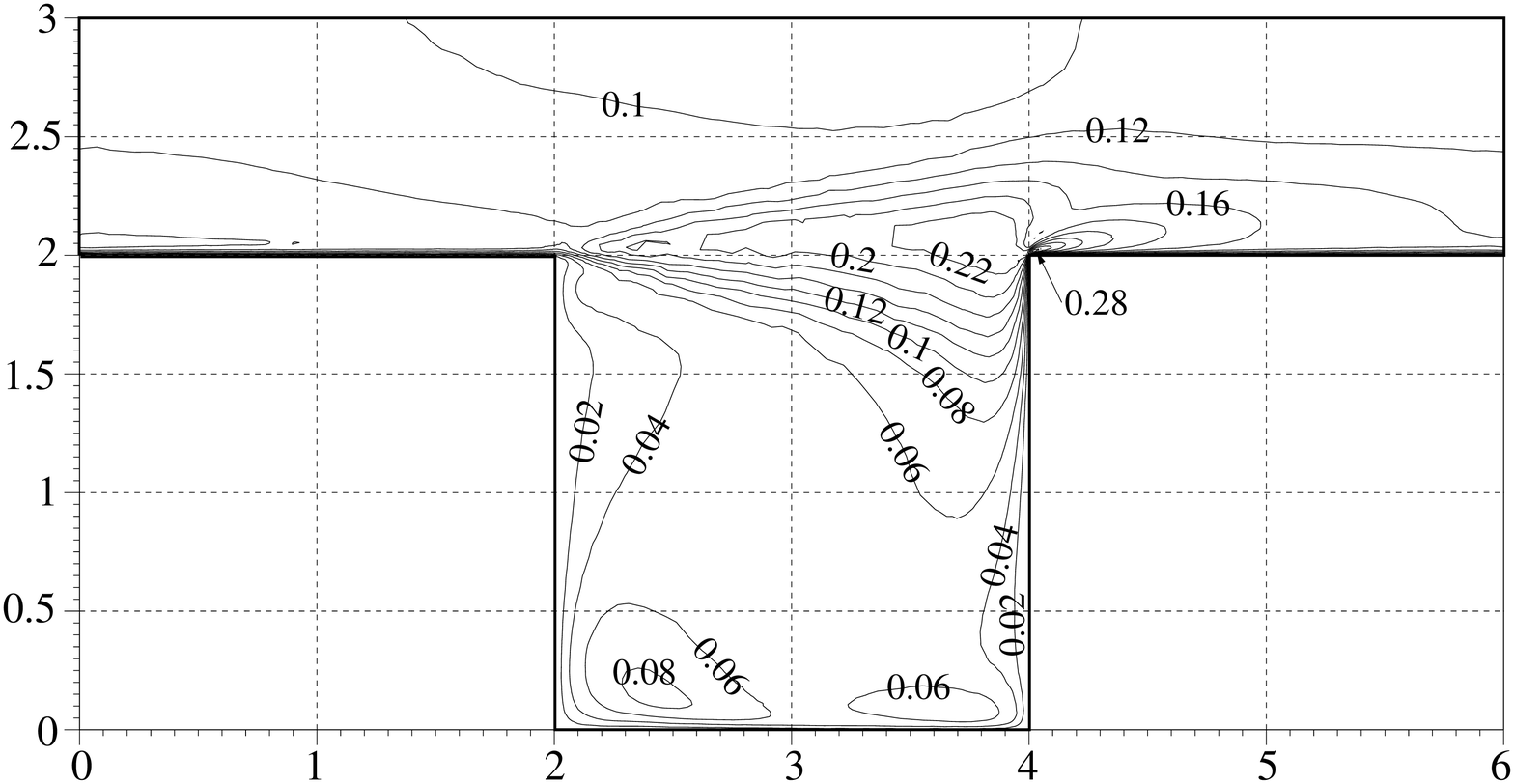}
\includegraphics[width=7.5cm]{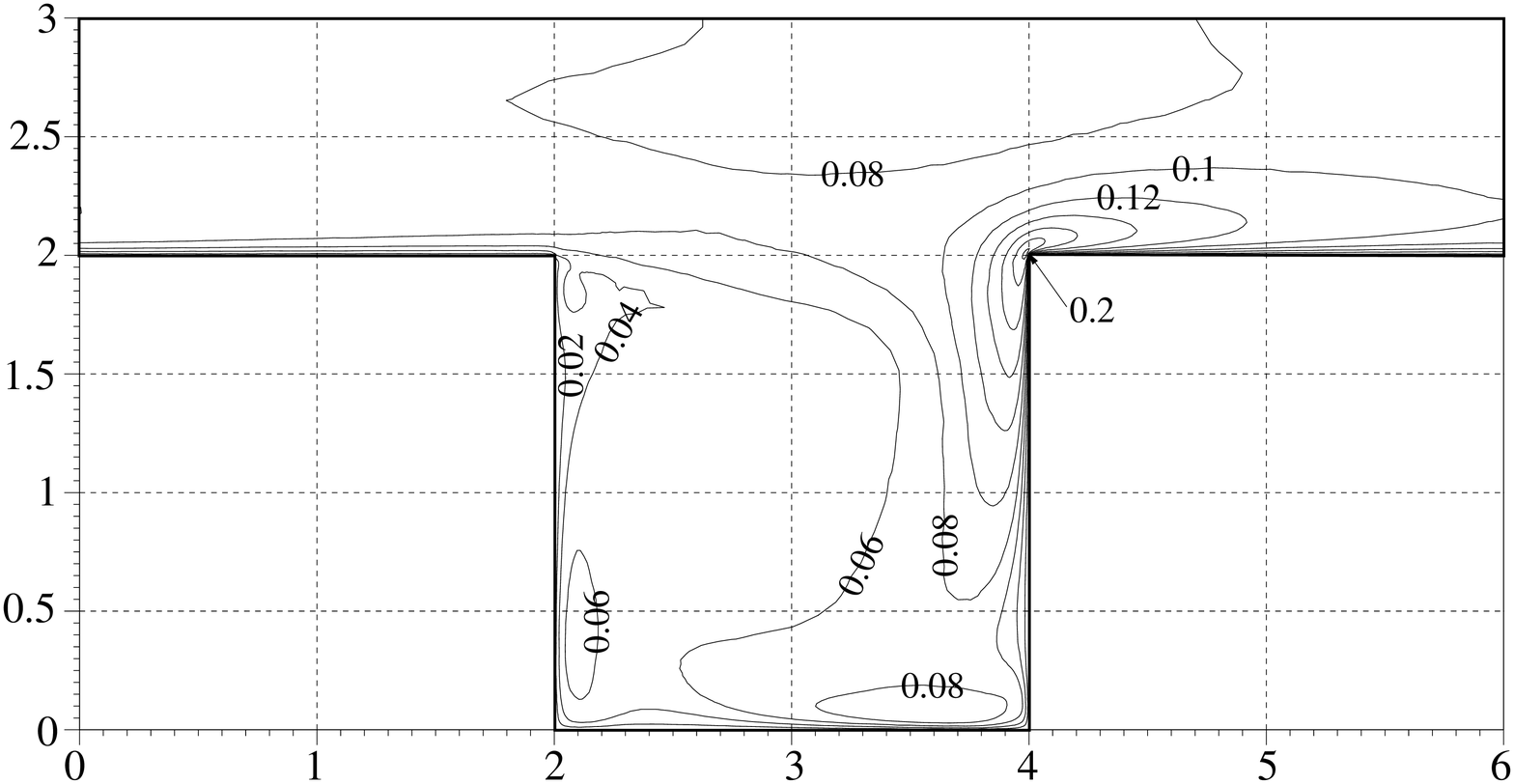}
\includegraphics[width=7.5cm]{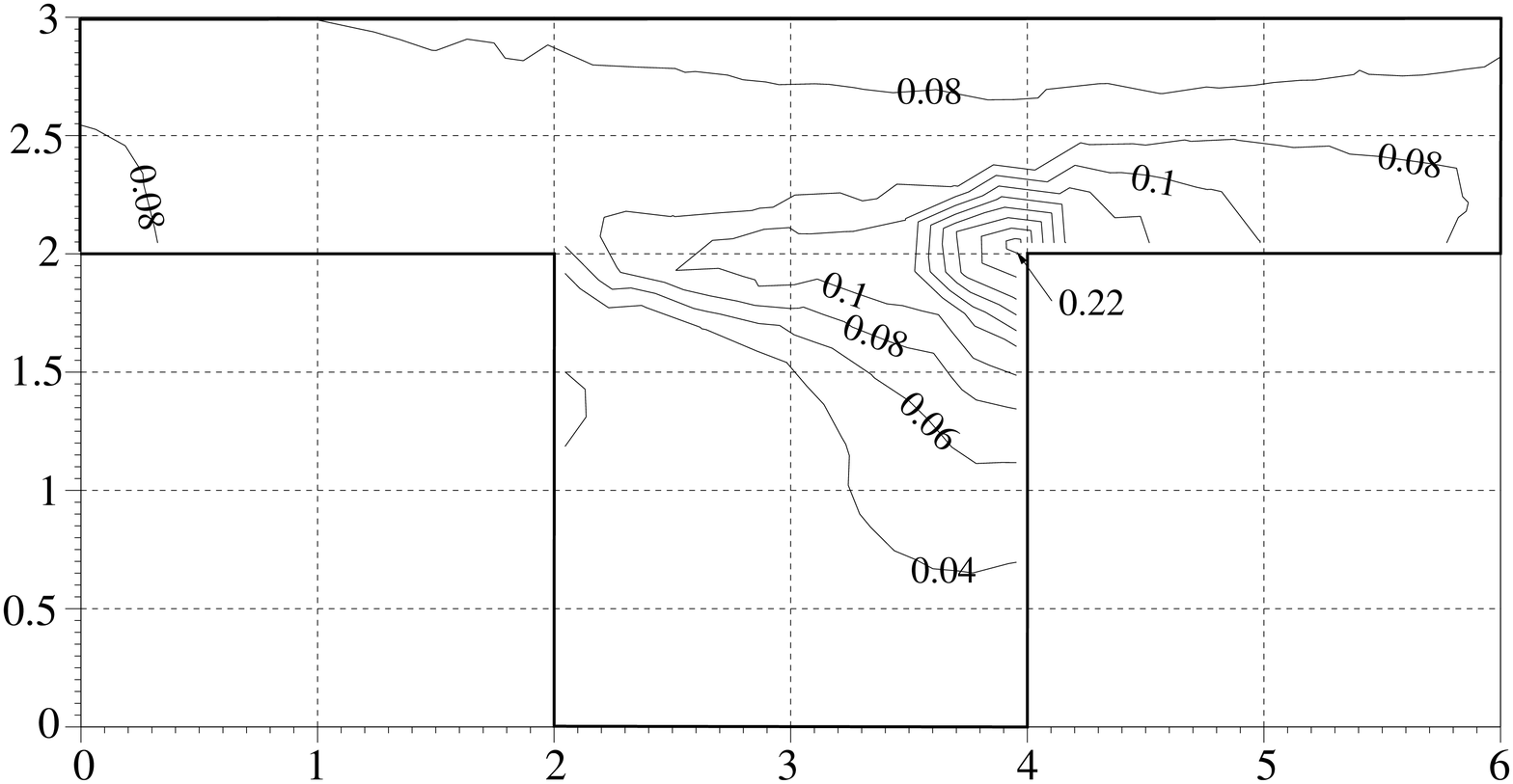}
\includegraphics[width=7.5cm]{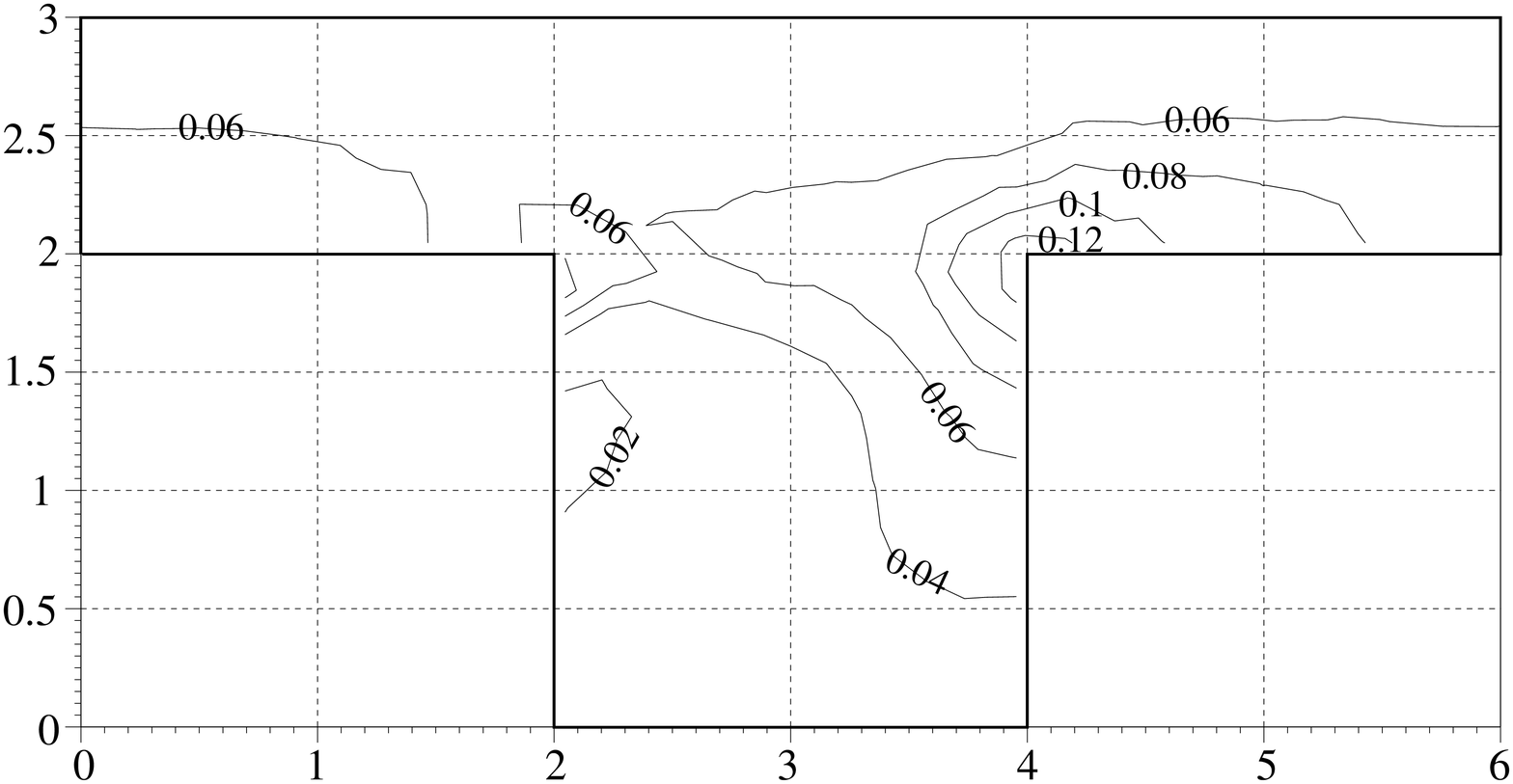}
\includegraphics[width=7.5cm]{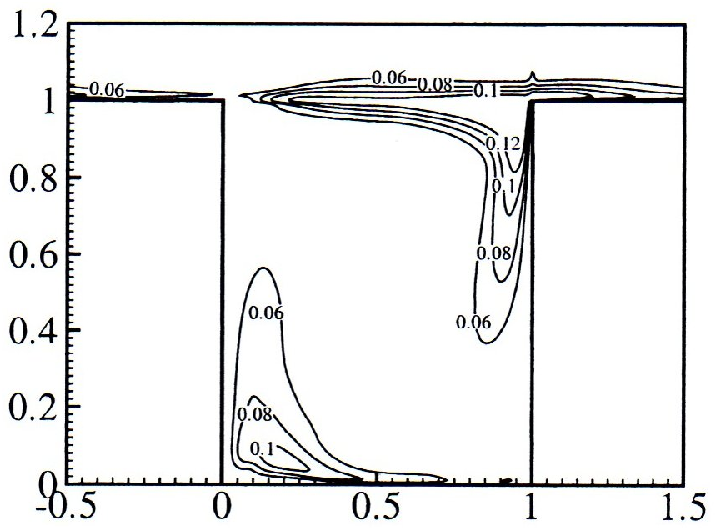}
\includegraphics[width=7.5cm]{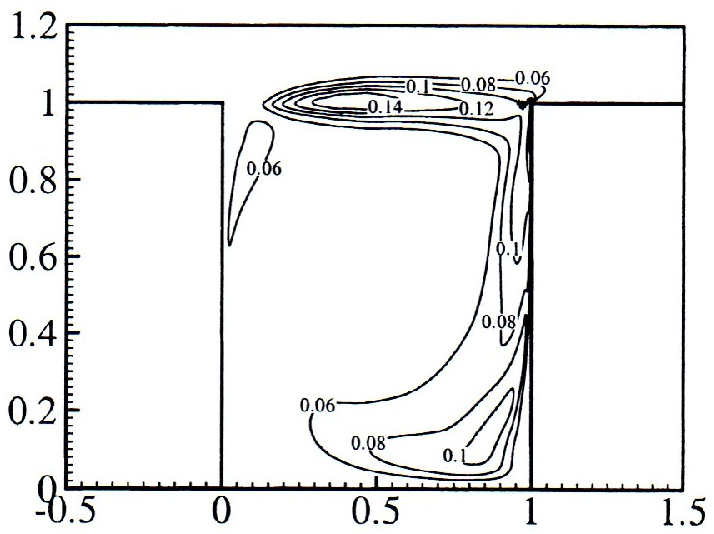}
\caption[Turbulent intensities in a street canyon]{Dimensionless turbulent intensities $\mean{u^2}^{\scriptscriptstyle 1/2}/U_0$ (first column) and $\mean{w^2}^{\scriptscriptstyle 1/2}/U_0$ (second column) computed using full wall-resolution (first row) and using wall-functions (second row) at $\textit{Re}\approx12000$ compared with the LES results (third row) of \citet{Liu_02}.}
\label{fig:canyon-Reynolds}
\figSpace
\end{figure}
In \Fige{fig:canyon-Reynolds}, two of the normalized turbulent intensities, $\mean{u^{\scriptscriptstyle 2}}^{\scriptscriptstyle 1/2}/U_0$ and $\mean{w^{\scriptscriptstyle 2}}^{\scriptscriptstyle 1/2}/U_0$, are displayed for both simulation cases and compared with the large eddy simulation results of \citet{Liu_02}. In the LES simulations the filtered momentum equations are solved by the Galerkin finite element method using brick three-dimensional elements, while the residual stresses are modeled by the Smagorinsky closure.

The full resolution simulation shows a very good agreement with the LES. The contour plots of $\mean{u^{\scriptscriptstyle 2}}^{\scriptscriptstyle 1/2}/U_0$ correctly display two local maxima, at the windward external and at the leeward internal corners. The contour plots of $\mean{w^{\scriptscriptstyle 2}}^{\scriptscriptstyle 1/2}/U_0$ show distributed high values at the building level above the canyon, along the windward internal corner and wall, and at the street level downstream of the source. By contrast, the wall-functions contour plots are in general less detailed, failing to reproduce the internal maximum of $\mean{u^{\scriptscriptstyle 2}}^{\scriptscriptstyle 1/2}/U_0$, and showing a more uniform representation of $\mean{w^{\scriptscriptstyle 2}}^{\scriptscriptstyle 1/2}/U_0$.

Several wind tunnel measurements have been carried out for this configuration, measuring concentration statistics above the buildings, at the walls and inside the canyon, for a scalar continuously released from a street level line source at the center of the canyon \citep{Meroney_96,Pavageau_99,Pavageau_96}. To examine the concentration values along the building walls and tops, we sampled the computed mean concentration field at the locations depicted in \Fige{fig:canyon-geometry} and listed in Table \ref{tab:measurement-locations}.

\begin{figure}[t!]
\centering
\psfragscanon
\rotatebox{270}{\includegraphics[width=10cm]{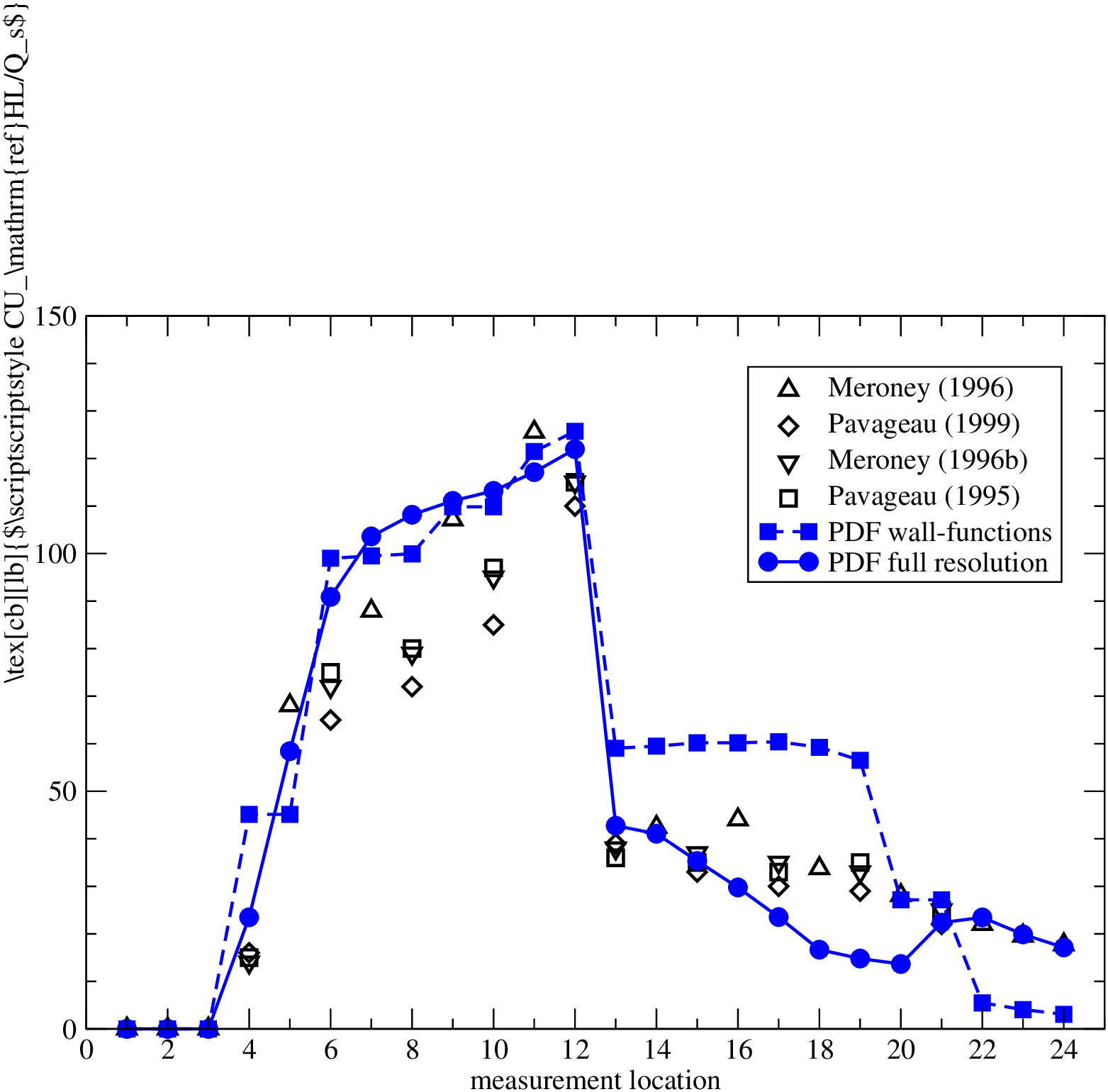}}
\psfragscanoff
\caption[Mean concentrations along the wall of a street canyon]{Distribution of mean concentrations at the boundary of the street canyon. The experimental data are in terms of the ratio $CU_\mathrm{ref}HL/Q_\mathrm{s}$, where $C$ is the actual measured mean concentration (ppm), $U_\mathrm{ref}$ is the free-stream mean velocity (m/s) taken at the reference height $y_\mathrm{ref}\approx11H$ and $Q_\mathrm{s}/L$ is the line source strength ($\mathrm{m}^2/\mathrm{s}$) in which $Q_\mathrm{s}$ denotes the scalar flow rate and $L$ is the source length. The calculation results are scaled to the concentration range of the experiments. References for experimental data: $\scriptscriptstyle{\triangle}$ \citet{Meroney_96}; $\diamond$, $\scriptscriptstyle\triangledown$, \citet{Pavageau_99}; $\scriptscriptstyle\square$ \citet{Pavageau_96}. See also \Fige{fig:canyon-geometry} and Table \ref{tab:measurement-locations} for the measurement locations.}
\label{fig:wall-concentrations}
\figSpace
\end{figure}

The excellent agreement of the results using both full resolution and wall-functions with a number of experiments is shown in \Fige{fig:wall-concentrations}. The concentration peak is precisely captured at the internal leeward corner and the model accurately reproduces the pattern of concentration along both walls including the higher values along the leeward wall.
\nomenclature[RC]{$C$}{measured concentration in the wind tunnel experiments (\autoref{chap:canyon})}%
\nomenclature[RU]{$U_\mathrm{ref}$}{measured free stream velocity at $y_\mathrm{ref}$ in the wind tunnel experiments (\autoref{chap:canyon})}%
\nomenclature[RL]{$L$}{length of line source in the wind tunnel experiments (\autoref{chap:canyon})}%
\nomenclature[RQ]{$Q_s$}{scalar flow rate in the wind tunnel experiments (\autoref{chap:canyon})}%
\begin{figure}
\centering
\includegraphics[width=7.5cm]{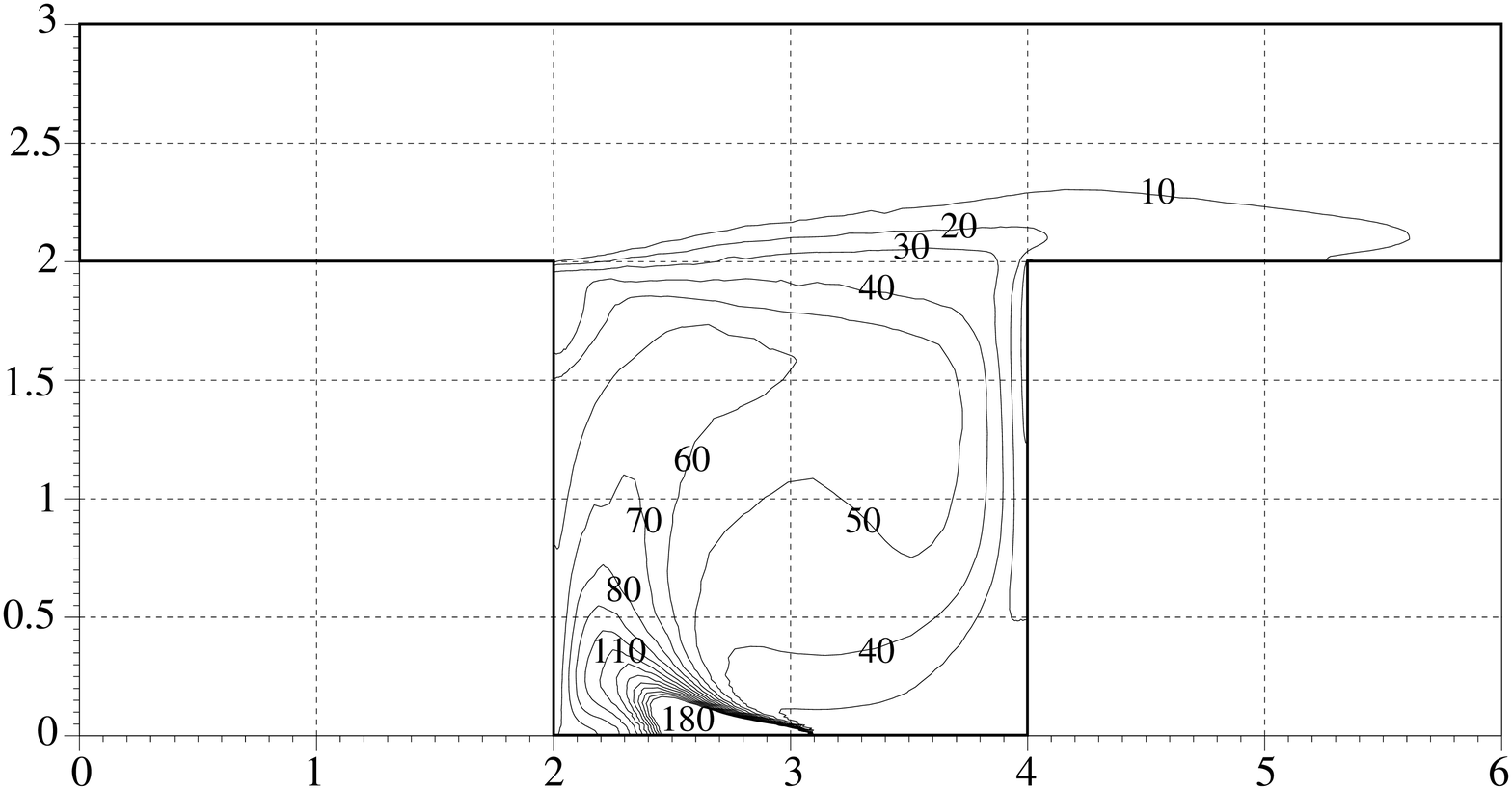}
\includegraphics[width=7.5cm]{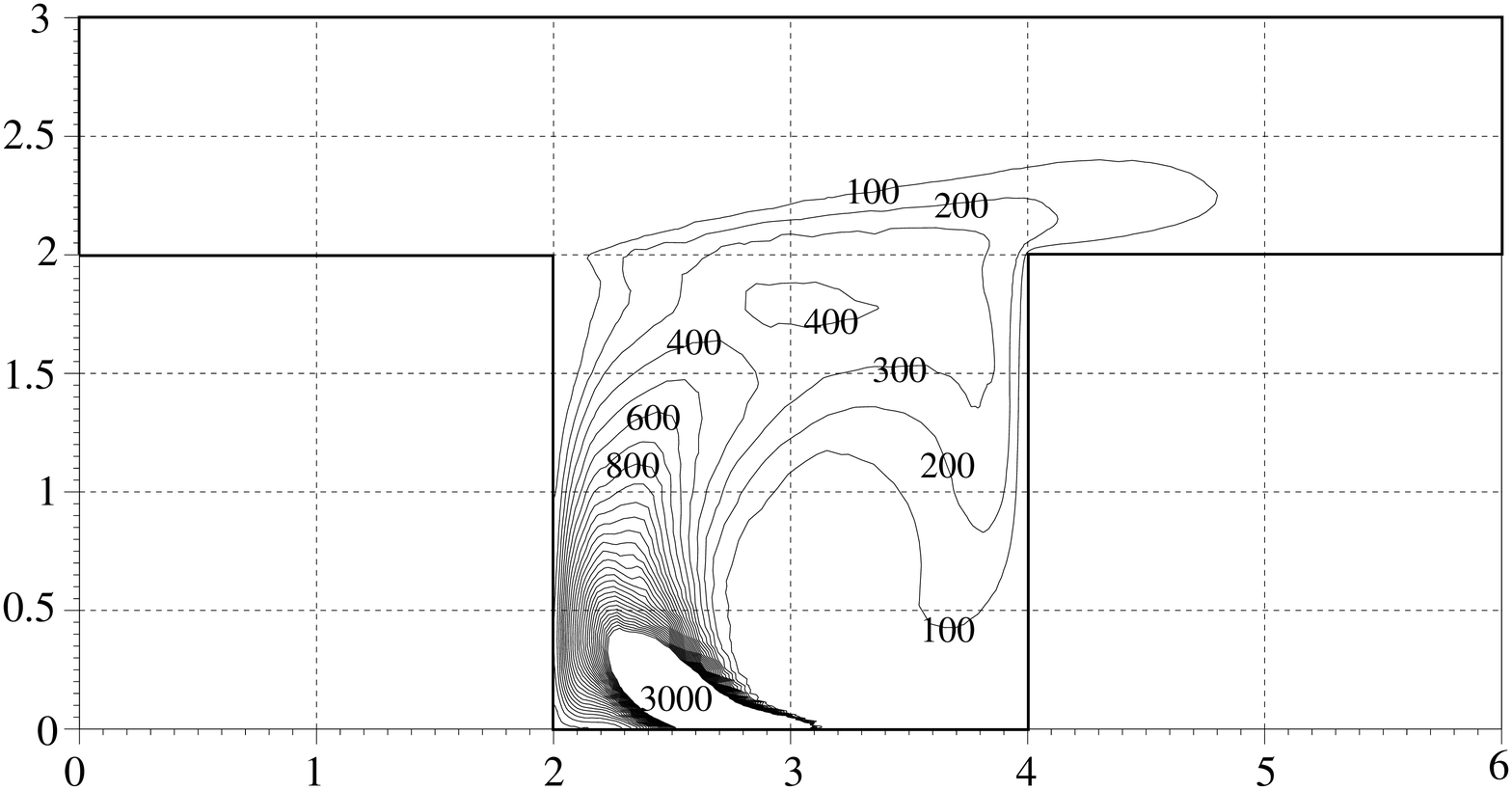}
\includegraphics[width=7.5cm]{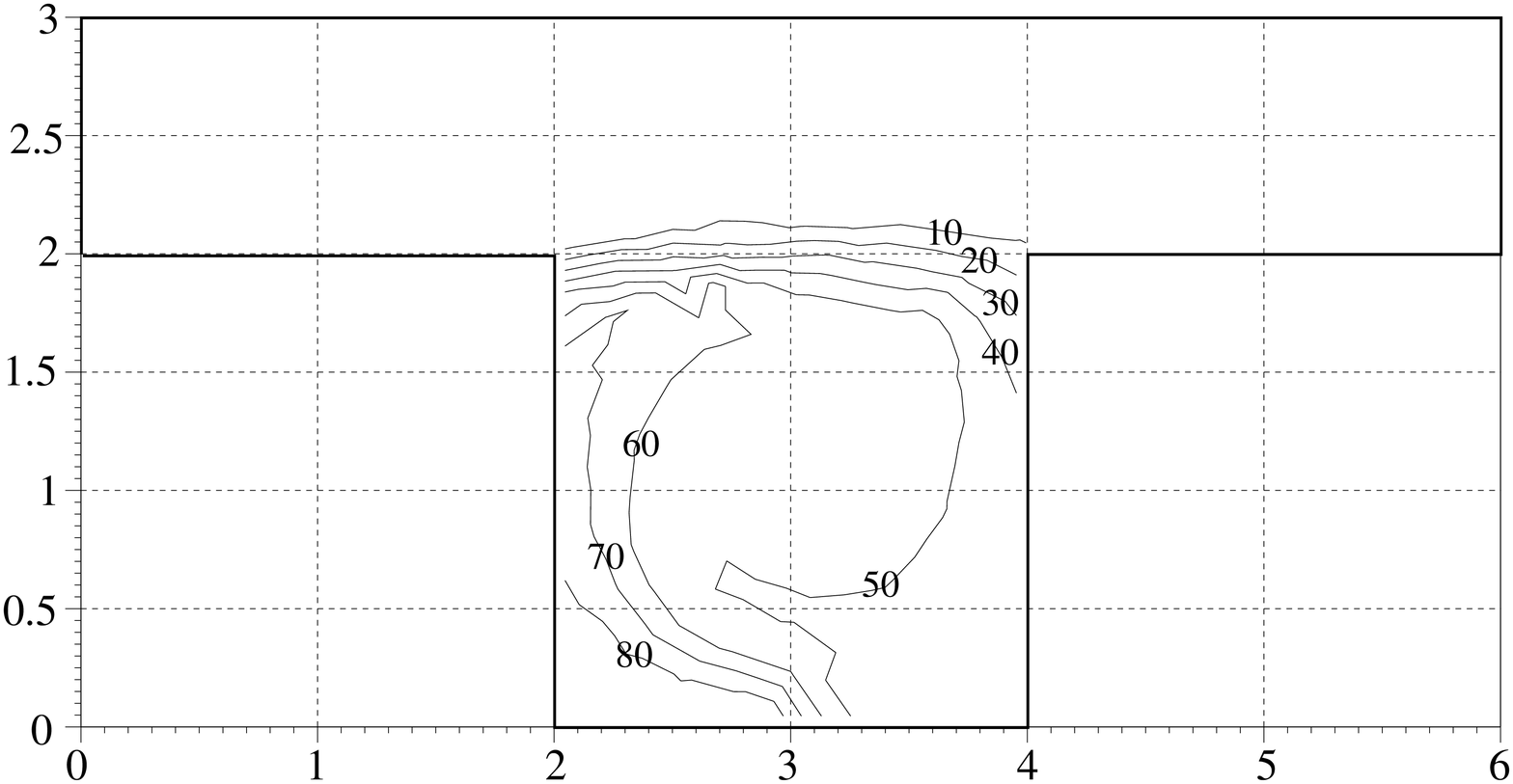}
\includegraphics[width=7.5cm]{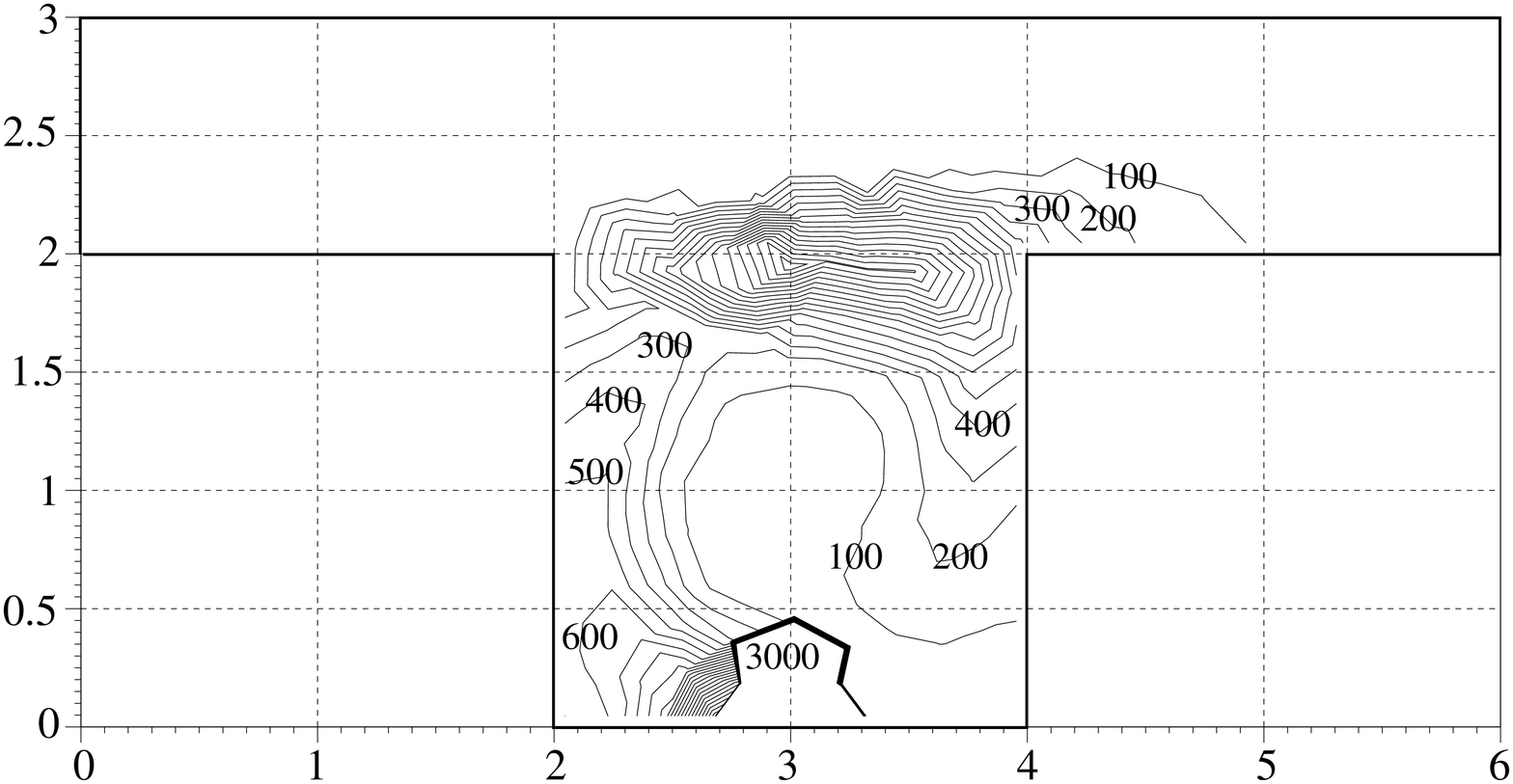}
\includegraphics[width=7.5cm]{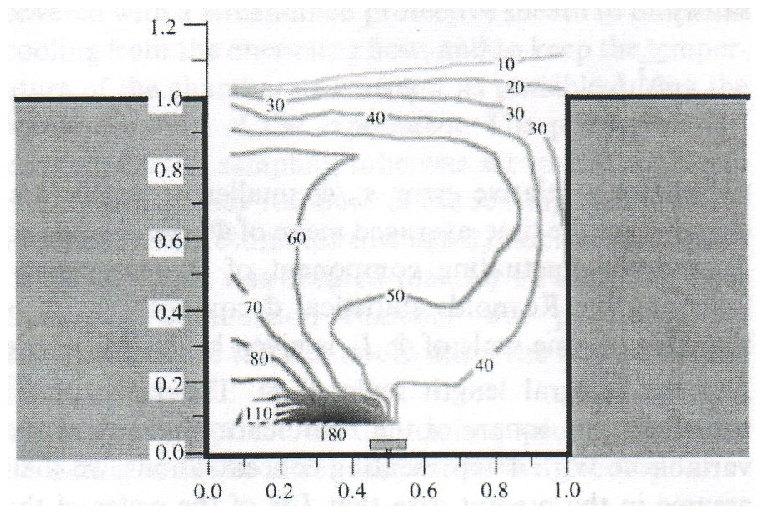}
\includegraphics[width=7.5cm]{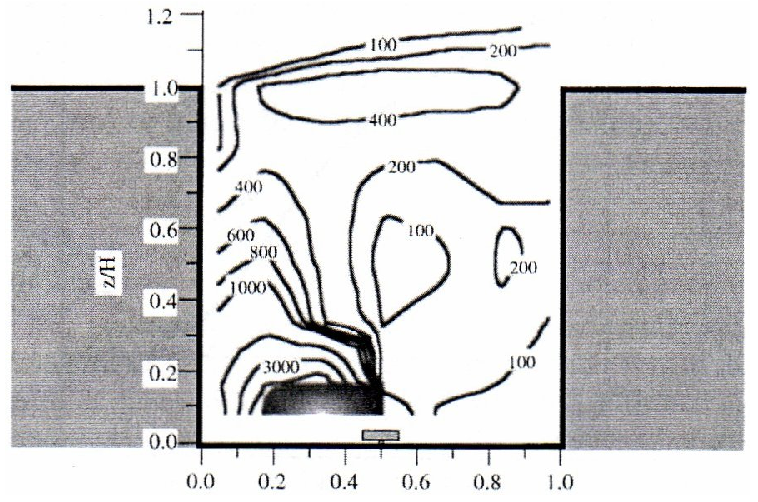}
\includegraphics[width=7.5cm]{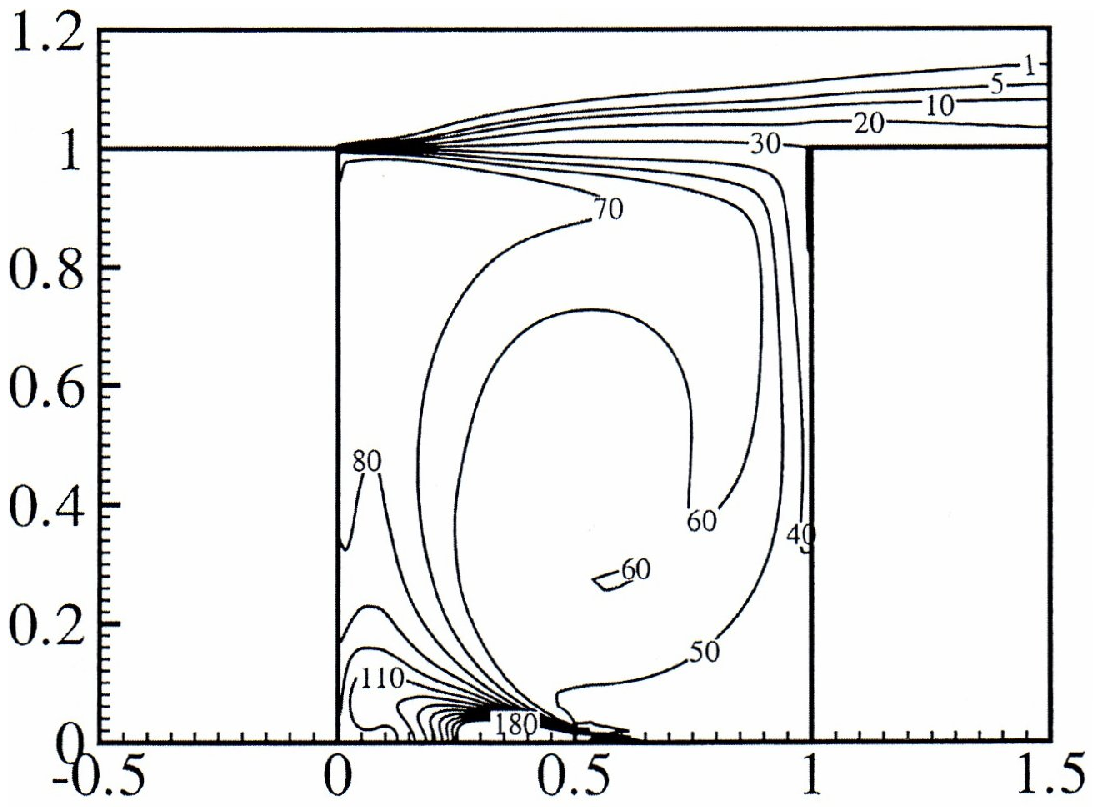}
\includegraphics[width=7.5cm]{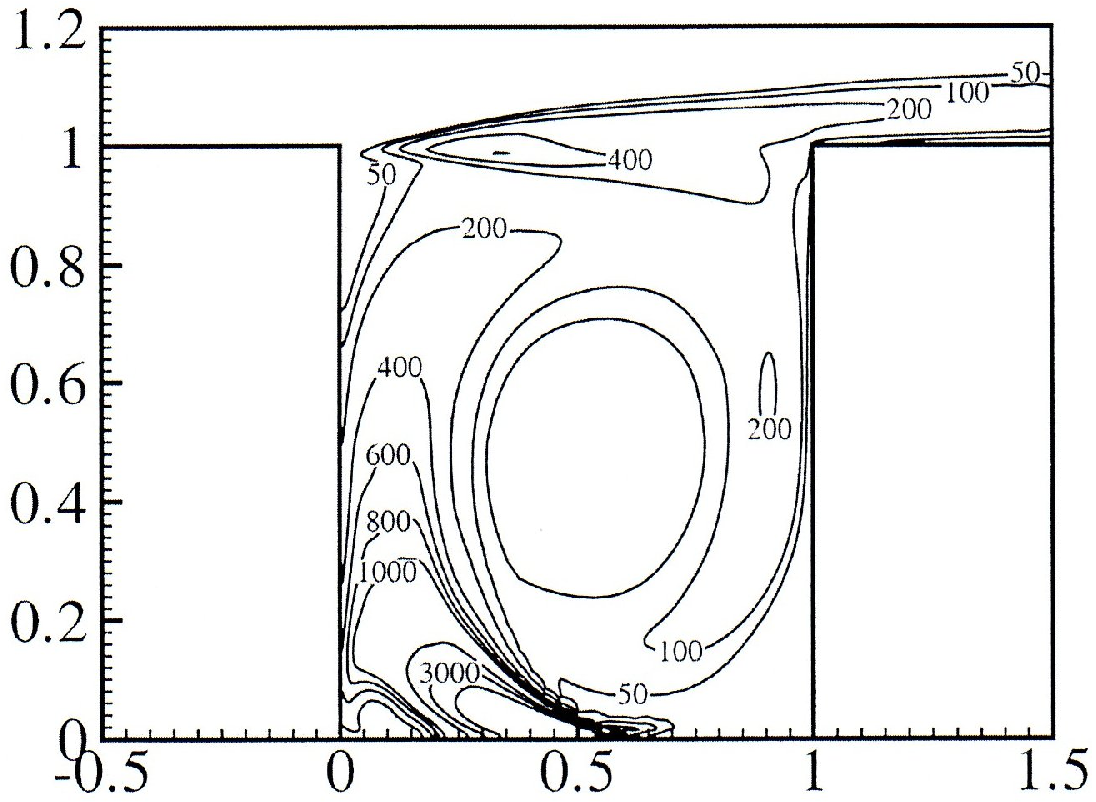}
\caption[Scalar concentration statistics in a street canyon]{See next page for caption.}
\label{fig:scalar-mean-variance}
\end{figure}
\begin{figure}[t!]
Figure \thefigure: Comparison of the spatial distribution of the normalized mean $CU_\mathrm{ref}HL/Q_\mathrm{s}$ (left column) and variance $\mean{c^2}(U_\mathrm{ref}HL/Q_\mathrm{s})^2$ (right column) of the scalar released at the center of the street level. The normalization and the scaling of the calculated results are the same as in \Fige{fig:wall-concentrations}. First row -- PDF calculations with full wall resolution, second row -- PDF calculations with wall-functions, third row -- experimental data of \citet{Pavageau_99} and fourth row -- LES calculations of \citet{Liu_02}.
\figSpace
\end{figure}

In \Fige{fig:scalar-mean-variance}, the first two statistical moments of the concentration inside the canyon are compared with experimental data and LES. The agreement with observations indicates that both the fluid dynamics and the micromixing components of the model provide a good representation of the real field. This is shown in the figures where one can observe the effects of the two driving mechanisms of transport of concentration by the large eddy inside the canyon as well as diffusion by the turbulent eddies.

\begin{figure}
\centering
\resizebox{15cm}{!}{\input{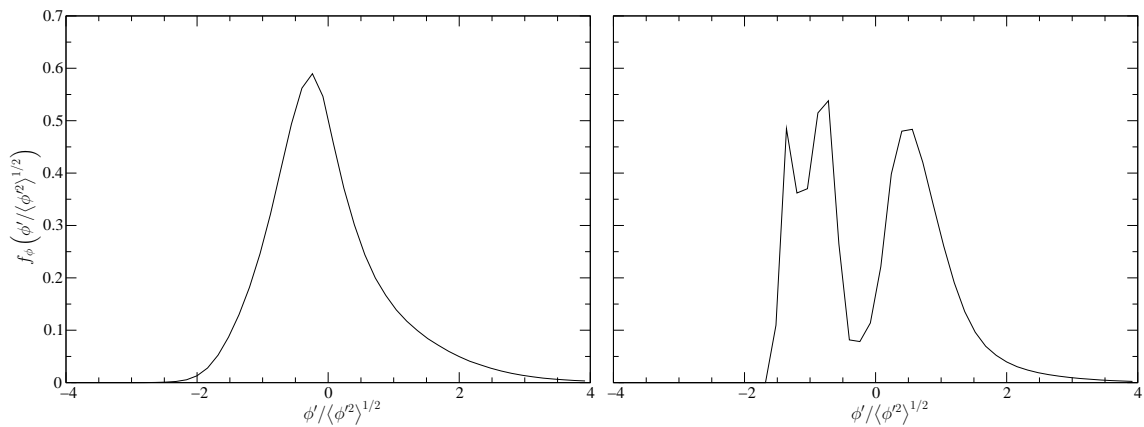}}
\caption[PDFs of concentration fluctuations in a street canyon]{Probability density functions of scalar concentration fluctuations (left) at $x=3,$ $y=0.2$ and (right) at $x=3,$ $y=2$ using full resolution at walls.}
\label{fig:pdfs-canyon}
\figSpace
\end{figure}

Because the one-point one-time joint PDF contains all higher statistics and correlations of the velocity and scalar fields resulting from a close, low-level interaction between the two fields, a great wealth of statistical information is available for atmospheric transport and dispersion calculations.  As an example, the time-averaged PDFs of scalar concentration fluctuations are depicted in \Fige{fig:pdfs-canyon} at selected locations of the domain for the full resolution case. While near the source (\Fig{fig:pdfs-canyon} left) the PDF is slightly skewed, but not far from a Gaussian, the distribution of fluctuations can become very complex especially due to intermittency effects, as shown by the multi-modal PDF in \Fig{fig:pdfs-canyon} right.

The performance gain obtained by applying wall-functions as opposed to full resolution was about two orders of magnitude already at this moderate Reynolds number.  The gain for higher Reynolds numbers is expected to increase more than linearly.  

\section{Discussion}
\label{sec:canyon-conclusions}
In this Chapter the PDF method described in the previous chapters was tested by computing the dispersion of a passive pollutant released from a point source. The Eulerian unstructured grid, consisting of triangular element type, is used to estimate Eulerian statistics, to track particles throughout the domain and to solve for inherently Eulerian quantities. The boundary layers developing close to solid walls are fully captured with an elliptic relaxation technique, but can also be represented by wall-functions, which use a coarser grid resolution and require significantly less particles, resulting in substantial savings in computational cost. We found that the one-point statistics of the joint PDF of velocity and scalar are well-captured by the wall-functions approximation.  In view of its affordable computational load and reasonable accuracy, this approximation appears to hold a realistic potential for application of the PDF method in atmospheric simulations, where the natural extension of the work is the implementation of the model in three spatial dimensions.

In hybrid PDF models developed for complex chemically reacting flows, numerical treatments for boundary conditions have been included for symmetric, inflow, outflow and free-slip walls employing the ghost-cell approach common in finite volume methods \citep{Rembold_06}. The representation of no-slip boundaries adds a significant challenge to the above cases. This is partly due to the increased computational expense because of the higher Eulerian grid resolution required if the boundary layers are to be fully resolved. In addition, there is an increased complexity in specifying the no-slip particle conditions for both fully resolved and wall-functions representations. We presented an implementation of both approaches to treat no-slip boundaries with unstructured grids in conjunction with the finite element method. This obviates further complications with ghost-cells.

In the case of full wall-resolution we employed the Lagrangian equivalent of a modified isotropization of production (IP) model as originally suggested by \citet{Dreeben_98}. The elliptic relaxation technique, however, allows for the application of any turbulence model developed for high-Reynolds-number turbulence \citep{Durbin_93,Whizman_96}. The standard test case for developing near-wall models is the fully developed turbulent channel flow. In this case, we explored the simpler \citet{Rotta_51} model, which is the Eulerian equivalent of the simplified Langevin model (SLM) in the Lagrangian framework \citep{Pope_94}. This is simply achieved by eliminating the term involving the fourth-order tensor $H_{ijkl}$ from the right hand side of \Eqre{eq:elliptic-relaxation-Lagrangian}. While the SLM makes no attempt to represent the effect of rapid pressure (in fact it is strictly correct only in decaying homogeneous turbulence), it is widely applied due to its is simplicity and robustness. Our experience showed a slight degradation of the computed velocity statistics (as compared to direct numerical simulation) using SLM for the case of channel flow. Since we experienced no significant increase in computational expense or decrease in numerical stability, we kept the original IP model.

Similarly, in the case of wall-functions, several choices are available regarding the employed turbulence model. The methodology developed by \citet{Dreeben_97b} uses the SLM, but it is general enough to include other more complex closures, such as the Haworth \& Pope models (HP1 and HP2) \citep{Haworth_86,Haworth_87}, the different variants of the IP models (IPMa, IPMb, LIPM) \citep{Pope_94} or the Lagrangian version of the SSG model of \citet*{Speziale_91}. All these closures can be collected under the umbrella of the generalized Langevin model, by specifying its constants as described by \citet{Pope_94}. These models have been all developed for high-Reynolds-number turbulence and need to be modified in the vicinity of no-slip walls. Including them in the wall-function formulation is possible by specifying the reflected particle frequency at the wall as $\omega_{\scriptscriptstyle R} = \omega_{\scriptscriptstyle I}\exp({-2\mathcal{V}_I\mean{\omega v}_{\scriptscriptstyle p}}/{\mean{\omega v^{\scriptscriptstyle 2}}_{\scriptscriptstyle p}})$ instead of \Eqre{eq:wall-omega}. This involves the additional computation of the statistics \(\mean{\omega v}\) and \(\mean{\omega v^{\scriptscriptstyle 2}}\) at \(y_p\), which does not increase the computational cost significantly, but may result in a numerically less stable condition since the originally constant parameter \(\beta\) which appears using the SLM has been changed to a variable that fluctuates during simulation. We implemented and tested all the above turbulence models using the wall-functions technique. Without any modification of the model constants we found the IPMa and SLM to be the most stable, providing very similar results. Thus we kept the original (and simplest) SLM along with \Eqre{eq:wall-omega}.%
\nomenclature[A]{HP1}{Haworth \& Pope model 1}%
\nomenclature[A]{HP2}{Haworth \& Pope model 2}%
\nomenclature[A]{IPMa}{isotropization of production model \emph{a}}%
\nomenclature[A]{IPMb}{isotropization of production model \emph{b}}%
\nomenclature[A]{LIPM}{Lagrangian isotropization of production model}%

The most widely employed closure to model the small scale mixing of the passive scalar in the Lagrangian framework is the interaction by exchange with the mean (IEM) model \citep{Villermaux_Devillon_72,Dopazo_OBrien_74}. This simple and efficient model, however, fails to comply with several physical constraints and desirable properties of an ideal mixing model \citep{Fox_03}.  The interaction by exchange with the conditional mean (IECM) model overcomes some of the difficulties inherent in the IEM model. In this Chapter we justify the sole use of the IECM model by its being more physical and more accurate, but we acknowledge that it markedly increases the computational cost.


\chapter[Cylinder flow simulations: results and discussion]{\\Cylinder flow simulations: results and discussion}
\label{chap:cylinder}

\section{Introduction}
\label{sec:introduction_cylinder}
As a third validation testcase we simulate the turbulent flow in the wake of a circular cylinder. This classical example has been widely studied both experimentally and numerically, therefore a large amount of data have been accumulated about its flow dynamics. Although the domain geometry is relatively simple, the flow exhibits a variety of vastly different behaviors depending on the Reynolds number, ranging from a steady laminar state through unsteady but periodic laminar vortex shedding to transitional and fully developed turbulence. We select the Reynolds number \(\textit{Re}_{\scriptscriptstyle{D}}=3900\) (based on the cylinder diameter and the free stream velocity), mainly because it corresponds to a transitional flow in the near wake behind the cylinder. Secondly, this Reynolds number has also been studied extensively with both LES and DNS, thus a quantitative comparison of several flow statistics computed by other methods is also possible. From the modeling viewpoint this Reynolds number is a challenging tasks to undertake. At this Reynolds number the separating boundary layers along the cylinder surface are fully laminar. Transition to turbulence occurs in the very near wake due to shear layer instabilities, which is followed by a region dominated by vortex shedding dynamics where the wake becomes fully turbulent and the coherent structures gradually give place to fully developed turbulence. Since these features require a solver to perform relatively well in all laminar, transitional and turbulent regions of the flow, this case appears to be a good candidate to identify the limitations of the current method. Another reason to compute this flow is to further evaluate the current PDF methodology using unstructured grids and no-slip walls with curvature in complex geometries.%
\nomenclature[RR]{$\textit{Re}_{\scriptscriptstyle{D}}$}{Reynolds number for cylinder, based on $U_0$ and $D$ (\autoref{chap:cylinder})}%
\nomenclature[RD]{$D$}{cylinder diameter}%

One of the key components simulating this flow is the adequate resolution of the separating boundary layers which decisively determines the flow behavior downstream and crucially influences the accuracy of the numerical solution. Accordingly, LES studies with sufficient wall resolution have been successful in predicting both cylinder surface and downstream wake statistics relatively accurately. On the other hand, RANS models, due to their inherent high-Reynolds-number assumption, have usually failed to predict both the wake and the mean integrated statistics along the cylinder surface, such as the drag (even with adequate wall resoution). Employing wall-functions at the cylinder surface may also be problematic, since wall-functions are built on the fundamental assumption that the boundary layer is turbulent and remains attached. Neither of these assumptions are correct along the cylinder surface at a sub-critical Reynolds number. The separating laminar boundary layers along a curved geometry provides a tough testcase for the elliptic relaxation technique as well. Although this type of wall-treatment can be tought of as a set of sophisticated blending functions for near-wall turbulence, its fundamental assumptions are less restrictive compared to wall-functions. It also represents all components of the Reynolds stress tensor at the wall instead of relying on the turbulent viscosity hypothesis. Although this technique has originally been developed for turbulent boundary layers, it seems compelling to investigate its performance modeling a separating laminar boundary layer transitioning to turbulence.

Another complication is that the flow is highly unsteady and the turbulence is mechanically generated in the domain by the obstacle. In such situations an adequately resolved LES/DNS may perform well both far and in the vicinity of walls since it solves both the small wall-generated vortices and the large eddies far from walls. In other situations, however, where a given level of turbulence is required to be present but the turbulence-generating obstacles are not required to be part of the domain, other means are necessary to provide the right level of fluctuations in LES which are not always obvious. A RANS model has no problems handling these latter situations since it represents the turbulent kinetic energy explicitly in its formulation. On the other hand, solving turbulence which is generated within the domain may be a difficult task for both RANS and URANS models, due to their above mentioned limitations close to walls.

Unsteady PDF methods have been developed based on the LES methodology defining the filtered density function (FDF) which is used to provide closure for the filtered equations. The development has resulted in the hybrid FV/particle methods. Following the same logical sequence that led to unsteady RANS based on RANS simulations, it seems relevant to investigate an unsteady PDF (UPDF) methodology based on steady PDF methods. Here we will apply the current model, developed and tested for steady flows, for a transient flow in the same way as RANS models are applied to obtain time-dependent statistics resulting in URANS.%
\nomenclature[A]{UPDF}{unsteady PDF}%

In \autoref{sec:review-cylinder-regimes} a short review of the circular cylinder flow regimes are given. This is followed by an overview of the literature regarding experimental and numerical studies investigating the cylinder near wake at sub-critical Reynolds numbers, \autoref{sec:overview-cylinder-studies}. In \autoref{sec:results-cylinder} several computed velocity statistics are examined and compared to LES, DNS and experimental data where available. Finally, \autoref{sec:discussion-cylinder} sums up the findings regarding this testcase.

\subsection{A short review of cylinder flow regimes}
\label{sec:review-cylinder-regimes}
Reviews on the physics of the cylinder flow have been compiled by \citet{Morkovin_64,Berger_72,Norberg_87} and more recently by \citet{Williamson_96}. Only a short overview is given in the following.

The single relevant parameter of the flow over a circular cylinder is the Reynolds number, defined here as \(\textit{Re}_{\scriptscriptstyle{D}}=U_0D/\nu\), where \(U_0\) is the free stream velocity, \(D\) is the cylinder diameter and \(\nu\) denotes the kinematic viscosity.

At \(\textit{Re}_{\scriptscriptstyle{D}}\) lower than approximately 40, the flow is laminar and steady. The boundary layer separates at \(\textit{Re}_{\scriptscriptstyle{D}}\approx3-5\) resulting in two symmetric counter-rotating vortices behind the cylinder. This recirculating region grows linearly with Reynolds number and the velocity profiles at the end of the recirculating region exhibit self-similarity.

At Reynolds numbers higher than 40 the vortices become unstable which initiates periodic vortex shedding resulting in a K\'arm\'an vortex street. The non-dimensionalized frequency of the separating vortices is the Strouhal number ($\textit{St}=nD/U_0$) which is used to characterize the unsteadyness of the flow related to the periodic vortex street. For up to about \(\textit{Re}_{\scriptscriptstyle{D}}=150\) the flow remains laminar and the Strouhal number increases with Reynolds number, then reaches a plateau of \(\sim0.21\). Transition to three-dimensionality starts at \(\textit{Re}_{\scriptscriptstyle{D}}=180-260\) due to the appearing streamwise vortices in the wake.

At the sub-critical Reynolds number range, between 300 and \(2\times10^5\), the separating boundary layers are still fully laminar along the cylinder surface and transition into turbulence occurs in the near wake due to shear layer instabilities. At the lowest Reynolds numbers in this range the flow becomes fully turbulent only about 40-50 diameters downstream, where the periodic vortices have been completely diffused. As the Reynolds number increases this transition moves closer to the cylinder. At the highest Reynolds numbers in this range the transition in the shear layers occurs very close to the separation points.

In the critical Reynolds number range, between \(2\times10^5\textrm{ and }3.5\times10^6\), two significant changes occur that crucially influence the drag on the cylinder. At \(\textit{Re}_{\scriptscriptstyle{D}}\approx3.6\times10^5\) the drag coefficient drops abruptly (from 1.2 to 0.3) due to a sudden increase in the base pressure behind the cylinder. The separating laminar boundary layer along the cylinder surface transitions to turbulence and reattaches then finally separates again. The separation point moves towards the downstream side of the cylinder and the width of the wake decreases to less than 1 cylinder diameter. In the range \(5\times10^5\textrm{ and }3.5\times10^6\) the base pressure decreases which increases the drag from 0.3 to 0.7 which remains at this value up to about \(\textit{Re}_{\scriptscriptstyle{D}}=10^7\). In the post-critical regime, above \(3\times10^6\), the boundary layer transitions to turbulence before separating and the Strouhal number stays approximately constant at 0.27.

\subsection{Past experimental and numerical studies}
\label{sec:overview-cylinder-studies}
\citet{Ma_00} divide the cylinder wake into three regions at sub-critical Reynolds numbers: the \emph{near wake} up to about 10 diameters downstream, the \emph{intermediate wake} up to fifty diameters and the \emph{far} or \emph{self-preserving wake} beyond that \citep{Matsumura_93}. There are relatively few experiments available in the near wake due to difficulties and special arrangements required in order to obtain accurate data, as in the experiments of \citet{Cantwell_83} who provided measurements up to \(x/D=8\) for the Reynolds number \(\textit{Re}_{\scriptscriptstyle{D}}=140\,000\). Employing particle image velocimetry (PIV) Lourenco \& Shih (1993, see \citeauthor{Beaudan_94} \citeyear{Beaudan_94}) have obtained data on the first two moments of the velocity field in the recirculation region at \(\textit{Re}_{\scriptscriptstyle{D}}=3900\). \citet{Ong_96} reported data on the first four moments of the velocity and its spectra based on hot-wire measurements conducted with an X-array probe between \(3 \le x/D \le 10\) at the same Reynolds number. Both the cylinder surface and near wake statistics are particularly sensitive to experimental disturbances, such as acoustic noise levels, cylinder vibrations, surface roughness and other geometric parameters in this Reynolds number range \citep{Norberg_87}. This is exemplified by the different lengths of the recirculation bubbles obtained by the experiments of \citet{Lourenco_93}, \citet{Ong_96} and Govardhan \& Williamson (2000, see \citeauthor{Ma_00} \citeyear{Ma_00}, Figure 1). The possible causes of the discrepancy among the experimental datasets are discussed in more detail by \cite{Noca_98}.%
\nomenclature[A]{PIV}{particle image velocimetry}%

A summary of the literature regarding numerical simulations of the cylinder flow up to the middle of the last decade at different Reynolds numbers is given by \citet{Beaudan_94}. Their overall conclusion is that two-dimensional Navier-Stokes simulations at transitional Reynolds numbers (between 150 and 300) are capable of predicting Strouhal numbers and drag coefficients, but become unreliable in the sub-critical regime. Although the flow geometry is nominally two-dimensional, three-dimensional effects at these higher Reynolds numbers become non-negligible. Steady RANS simulations employing the \(k-\varepsilon\) model predict inaccurate mean velocity and Reynolds stress distributions in the near wake and produce mixed results for the integrated statistics over the cylinder surface \citep{Beaudan_94}. This is perhaps little surprise, since in this flow the eddy-viscosity is anisotropic and negative in regions where history and transport effects dominate over production of Reynolds stresses, indicating the inadequacy of the turbulent-viscosity hypothesis for this flow \citep{Franke_89}. Underresolved DNS improve on RANS simulations by better capturing the drag coefficients up to the critical Reynolds number \(10^6\), which is attributed to better resolving the three-dimensionality of the flow \citep{Beaudan_94}. From this viewpoint it will be interesting to  see how the current PDF model performs: although the spanwise components of the particle positions are not retained, the velocity field is three-dimensional in the sense of fluctuations. In other words, while all three components of the particle velocities are retained to represent spanwise fluctuations, mean spanwise motions due to streamwise and cross-stream vorticity are not represented and it is assumed that $\mean{W}=0$.

A systematic LES study at \(\textit{Re}_{\scriptscriptstyle{D}}=3900\) has been undertaken by \citet{Beaudan_94} whose main objective was to evaluate the performance of the dynamic residual-stress model \citep{Germano_91} in a flow where RANS simulations have been known to have difficulties. They performed simulations without closure, with the fixed-coefficient Smagorinsky-model and with the dynamic model. Both two and three-dimensional cases have been computed to assess the importance of representing three-dimensional effects. Another goal was to evaluate the performance of higher order upwind schemes for the advection terms, including fifth and seventh order finite difference approximations. The work further demonstrates the necessity of three-dimensional calculations for this flow, documenting consistent improvements in all quantities examined when compared to two-dimensional simulations with no residual-stress model (other than the numerical diffusion inherent in upwind schemes). Regarding the spatial discretization, they conclude that even higher order upwind schemes are not suitable for LES due to their numerical diffusion which may be comparable to the subfilter-scale diffusion. Following this line of work \citet{Mittal_96} used central differencing in order to better control the numerical diffusion, while \citet{Kravchenko_00} employed a high order B-spline-based finite element method obtaining Reynolds stress distributions in closer agreement with their respective experimental profiles. The above series of LES studies show that the computed statistics in the near wake may be significantly influenced by the choice of the discretization scheme for the advection term. However, the choice of different models for the unresolved stress is clearly less important. Another concern in LES, just like in RANS, is that the use of eddy-viscosity-based models for the unresolved scales in non-equilibrium flows are questionable \citep{Liu_97}.

Preliminary results on DNS of the cylinder flow at sub-critical Reynolds numbers have been reported by \citet{Tomboulides_93} and \citet{Henderson_95}, but full resolution of the near wake has become possible only recently. \citet{Ma_97} have performed direct simulations based on hierarchical spectral methods employing unstructured grids. The study compared DNS and the LES results of \citet{Beaudan_94} and \citet{Mittal_96}. This work was followed by more detailed numerical studies by \citet{Ma_00} and more recently by \citet{Dong_06}, who combined experimental imaging (PIV) and DNS performing both experiments and numerical simulations at \(\textit{Re}_{\scriptscriptstyle{D}}=3900\) and \(10\,000\) in order to investigate the near wake focusing on the onset of shear-layer instabilities and Reynolds number effects.

We will compare results from the current PDF simulations with many of the experimental and numerical datasets mentioned above.

\section{Results}
\label{sec:results-cylinder}
\begin{figure}[!]
\centering
\resizebox{15cm}{!}{\input{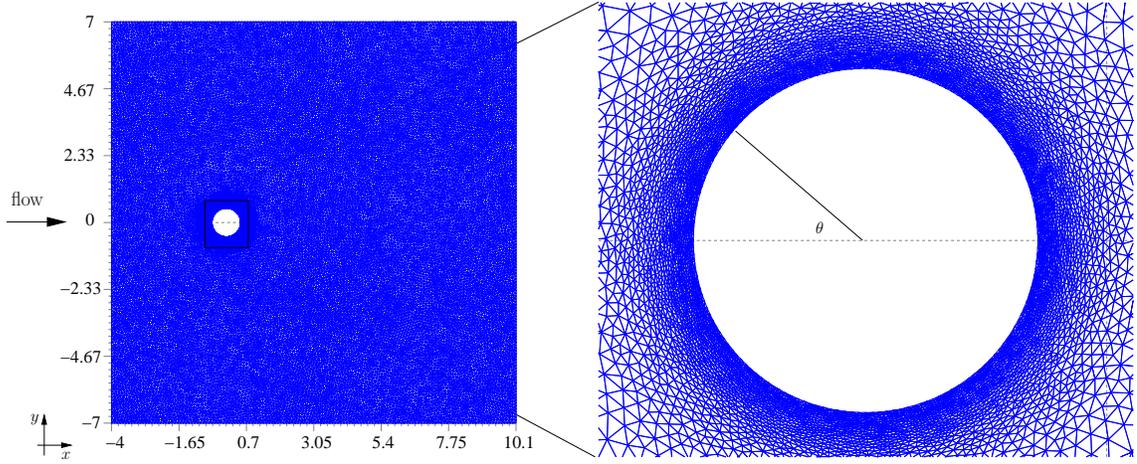}}
\caption[Eulerian mesh for flow around cylinder]{\label{fig:cylinder-domain}Eulerian mesh for computing the near wake of the cylinder flow at \(\textit{Re}_{\scriptscriptstyle{D}}=3900\) based on the cylinder diameter \(D\) and free stream velocity \(U_0\). The refinement along the wall amounts to 156 wall elements with an average size of \(4.5\times 10^{-3}D\) in the radial and \(0.02D\) in the circumferential direction.}
\figSpace
\end{figure}

Several PDF simulations have been carried out to model the unsteady flow around a circular cylinder at \(\textit{Re}_{\scriptscriptstyle{D}}=3900\) employing the grid displayed in \Fige{fig:cylinder-domain}. The refinement in  the vicinity of the cylinder amounts to 156 elements along the circumference with an average size of $4.5\times 10^{-3}D$ in the radial and \(0.02D\) in the circumferential direction. This is slightly coarser than the coarsest case in the LES simulations of \citet{Kravchenko_00} and corresponds to about half the resolution of the LES study of \citet{Mahesh_04}. The total number of elements is approximately 50K triangles. The number of particles per element initially is set to 50 and the CFL number is 0.8 (kept at this constant level) using forward Euler-Maruyama timestepping with the adaptive technique described in \autoref{sec:timestepping}. Once the boundary layers start separating from the cylinder surface, these parameters result in a particle redistribution of approximately 200--300 particles each timestep, requiring a minimum of 5 particles in each element. This extent of redistribution (only $\sim0.01\%$ of all the particles redistributed in each timestep) is sufficient enough to have a non-negligible negative effect on the overall performance of the code, so a more efficient particle redistribution procedure has been developed, which by itself is 200 times faster than the basic one described in \autoref{app:particle-redistribution} and results in an overall speedup of 15 times for the whole code. The details of this new algorithm are described in \autoref{app:particle-redistribution2}.

The initial conditions are as follows: the particle velocity is assigned a joint Gaussian distribution with a low-level turbulent kinetic energy, \(k/U_0^2=0.01\) homogeneously on the whole domain, while the particle frequencies are sampled from a gamma distribution with unit mean and variance \(1/4\). Free-slip conditions are imposed on the cross-stream boundaries, i.e.\ a particle hitting the wall is simply reflected with opposite cross-stream velocity $\mathcal{V}$. Particles leaving at the outflow are relocated at the inflow leaving their cross-stream position $\mathcal{Y}$ intact and setting their velocity to \(\mathcal{U}_i=(U_0,0,0)\), which corresponds to a joint delta distribution, i.e.\ incoming laminar flow with streamwise velocity \(U_0\). At the cylinder surface, no-slip conditions are imposed on particles as described in \autoref{sec:wall_conditions}. The no-slip wall-conditions are enforced on the extracted velocity statistics as well. A homogeneous Dirichlet condition is imposed on the mean pressure at the outflow and homogeneous Neumann conditions on every other outer boundary. At the cylinder the Neumann condition (\ref{eq:wall-pressure-condition}) is enforced for the mean pressure. The boundary conditions for the elliptic relaxation tensor are \(\wp_{ij}=-4.5\varepsilon n_in_j\) at the cylinder wall and homogeneous Neumann conditions along all other boundaries. The applied model constants are the same as before and displayed in Table \ref{tab:fd-constants}.

The current unsteady PDF simulations have been carried out in a similar fashion as an unsteady RANS simulation. In URANS the model equations developed for computing the time-averaged statistics for an inhomogeneous flow are solved in a time-accurate manner, sampling the solution at certain timesteps. This can be thought of as filtering in time with the filter width defined as the time between two consecutive timesteps. Similarly, in the current UPDF simulations we take the equations originally developed for steady flows and solve them with a time-accurate numerical algorithm and sample results at specified timesteps.

In the following, we examine flow statistics regarding the transient nature of the flow as well as integrated quantities along the cylinder surface (\autoref{sec:transient-surface-stats}) and time-averaged fields in the near wake (\autoref{sec:near-wake-stats}).

\subsection{Transient and cylinder surface statistics}
\label{sec:transient-surface-stats}
A common parameter used to examine the cylinder flow is the Strouhal number which is defined as the non-dimensional form of the vortex shedding frequency, $n$, as
\begin{equation}
\textit{St}=\frac{nD}{U_0}.
\end{equation}%
\nomenclature[RS]{$\textit{St}$}{Strouhal number}%
There are many quantities from which the Strouhal number can be extracted from a simulation, the time evolution of the cross-stream component of the force acting on the body, i.e.\ the lift, being the most common one. We evaluate the force $\bv{F}$ on the cylinder surface \(\mathcal{A}\) by
\begin{equation}
F_i=\int_\mathcal{A}\left(-\mean{P}\delta_{ij}+\rho\nu\frac{\partial\mean{U_i}}{\partial x_j}\right)n_j\mathrm{d}\mathcal{A},\label{eq:force-on-cylinder}
\end{equation}%
\nomenclature[RA]{$\mathcal{A}$}{surface area of cylinder}%
\nomenclature[RF]{$\bv{F}$}{force acting on the cylinder}%
where \(n_j\) is the wall-normal. The drag and lift can be obtained by taking the streamwise ($i=1$) and cross-stream ($i=2$) components of \(\bv{F}\). The drag and lift coefficients, \(C_D\) and \(C_L\), are the non-dimensional components of the force \(\bv{F}=F_x\bv{e}_x+F_y\bv{e}_y\) and can be decomposed into pressure and viscous parts:
\begin{eqnarray}
C_D&=&\frac{F_x}{(1/2)\rho U_0^2D^2} = C_{Dp} + C_{Dv},\\
C_L&=&\frac{F_y}{(1/2)\rho U_0^2D^2} = C_{Lp} + C_{Lv}.
\end{eqnarray}%
\nomenclature[Re]{$\bv{e}_x$}{unit basis vector in $x$ direction}%
\nomenclature[RD]{$C_D$}{total drag coefficient}%
\nomenclature[RC]{$C_L$}{total lift coefficient}%
\nomenclature[RC]{$C_{Dp}$}{pressure drag coefficient}%
\nomenclature[RC]{$C_{Dv}$}{viscous drag coefficient}%
\nomenclature[RC]{$C_{Lp}$}{pressure lift coefficient}%
\nomenclature[RC]{$C_{Lv}$}{viscous lift coefficient}%
\begin{figure}[!]
\centering
\resizebox{11cm}{!}{\input{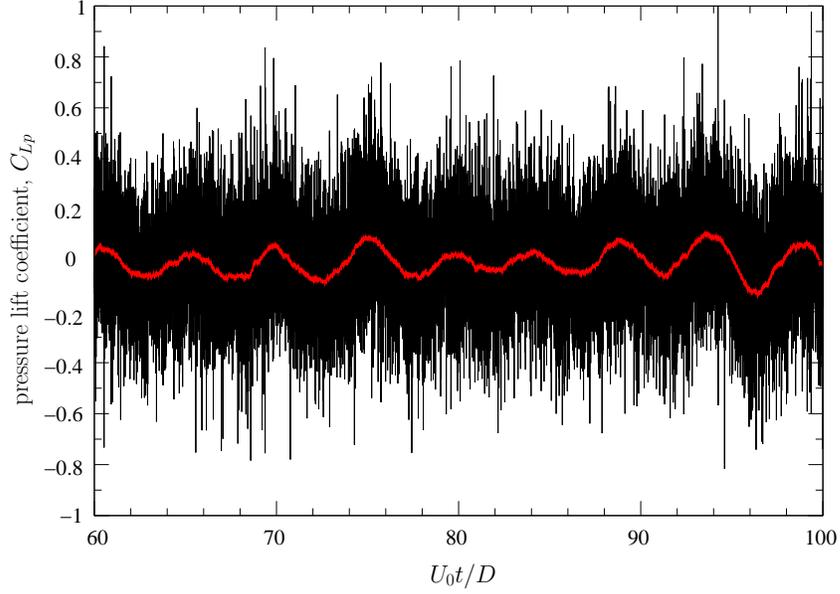}}
\caption[Time-evolution of the pressure lift coefficient]{\label{fig:CLp-computation}Time-evolution of the pressure lift coefficient. The black solid line denotes the computed instantaneous value in every timestep, while the red line is its 100-point running average.}
\figSpace
\end{figure}%
\begin{figure}[!]
\centering
\resizebox{15cm}{!}{\input{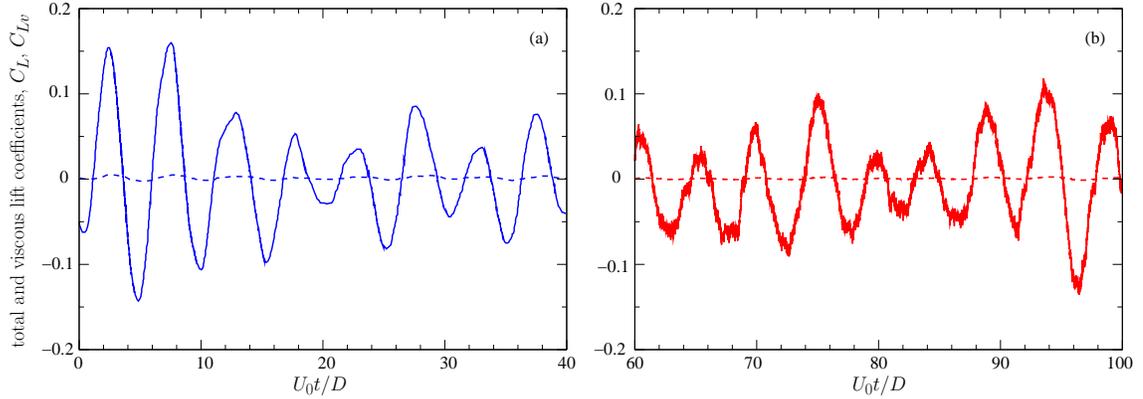}}
\caption[Time-evolution of the total and viscous lift coefficients]{\label{fig:CL}Time-evolution of the total and viscous lift coefficients at \(\textit{Re}_{\scriptscriptstyle{D}}=3900\). (a) three-dimensional LES of \citet{Beaudan_94} using a dynamic model for the unresolved scales, (b) current PDF simulations. The red lines on (b) indicate that they have been obtained using 100-point running averages from instantaneous data similar to the one in \Fige{fig:CLp-computation}. Solid lines -- total lift, dashed lines -- viscous lift.}
\figSpace
\end{figure}%
\begin{table}[t!]
\caption[Cylinder surface and recirculation bubble region statistics]{\label{tab:surface-summary}Cylinder surface and recirculation bubble region statistics. The large eddy simulation data employing upwind, central difference and B-spline schemes for the advection term are provided by \citet{Beaudan_94,Mittal_96} and \citet{Kravchenko_00}, respectively, all performed in three dimensions at \(\textit{Re}_{\scriptscriptstyle{D}}=3900\). The DNS data corresponds to the high resolution Case I of \citet{Ma_00} and \citet{Dong_06} also at \(\textit{Re}_{\scriptscriptstyle{D}}=3900\). References for the experimental data: $\textit{St}$ -- \citet{Ong_96}; $C_{Pb}$, $C_D$ -- \citet{Norberg_87} at \(\textit{Re}_{\scriptscriptstyle{D}}=4020\); $\theta_{\textrm{sep}}$ -- \citet{Son_69} at \(\textit{Re}_{\scriptscriptstyle{D}}=5000\); $L/D$, $\mean{U}_{\textrm{min}}/U_0$, $r_{\textrm{min}}/D$ -- PIV of \citet{Dong_06} at \(\textit{Re}_{\scriptscriptstyle{D}}=4000\). The last column labeled by ``PDF'' denotes the current PDF simulation.\\}
\ls{1}
\begin{tabular*}{1.0\textwidth}{p{5cm}@{\hspace{0.5cm}}c@{\hspace{0.5cm}}c@{\hspace{0.5cm}}c@{\hspace{0.5cm}}c@{\hspace{0.5cm}}c@{\hspace{0.5cm}}c}
\hline\hline
& Upwind & Central & B-spline & DNS & Expt. & PDF \\
& LES & LES & LES & & & \\
\hline
Strouhal number, $\textit{St}$ & 0.203 & 0.207 & 0.21 & 0.203 & 0.21 & 0.2 \\
Base pressure coefficient, $C_{Pb}$ & -0.95 & -0.93 & -0.94 & -0.96 & -0.99 & -0.79 \\
Total drag coefficient, $C_D$ & 1.0 & 1.0 & 1.04 & 0.981 & 0.98 & 1.04 \\
Separation angle, $\theta_{\textrm{sep}}$ & $85.8^\circ$ & $86.9^\circ$ & $88.0^\circ$ & $89.0^\circ$ & $86.0^\circ$ & $89.3^\circ$ \\
Length of recirculation bubble, $L/D$ & 1.36 & 1.4 & 1.35 & 1.12 & 1.47 & 1.53 \\
Minimum streamwise velocity in bubble, $\mean{U}_{\textrm{min}}/U_0$ & -0.32 & -0.35 & -0.37 & -0.291 & -0.252 &  -0.34 \\
Location of $\mean{U}_{\textrm{min}}$ in bubble, $r_{\textrm{min}}/D$ & 0.88 & & & 0.88 & 1.01 & 1.1 \\
\hline\hline
\end{tabular*}
\tableSpace
\end{table}%
\nomenclature[RC]{$C_{Pb}$}{base pressure coefficient}%
\nomenclature[Gh]{$\theta_{\textrm{sep}}$}{separation angle}%
\nomenclature[RL]{$L$}{length of recirculation bubble (\autoref{chap:cylinder})}%
\nomenclature[RU]{$U_{\textrm{min}}$}{minimum mean streamwise velocity in recirculation bubble (\autoref{chap:cylinder})}%
\nomenclature[Rr]{$r_{\textrm{min}}$}{radial location of the minimum mean streamwise velocity in the recirculation bubble (\autoref{chap:cylinder})}%
The time-evolution of the pressure lift coefficient \(C_{Lp}\) is plotted in \Fige{fig:CLp-computation}. Since Monte-Carlo PDF simulations are stochastic by nature, there is a considerable statistical noise in all quantities computed, especially in the ones based on the mean pressure. Nevertheless, applying a moving time-average with a window of 100 timesteps the Strouhal number of $\sim0.2$ can be easily extracted. Note that the value of $\textit{St}$ is not sensitive to the size of the window. Table \ref{tab:surface-summary} shows how this value compares to past experiments and numerical simulations. In \Fige{fig:CL} the evolution of the total and viscous lift coefficients are compared to the LES results of \citet{Beaudan_94} also performed at \(\textit{Re}_{\scriptscriptstyle{D}}=3900\). The PDF simulation successfully reproduces the irregularity of the vortex shedding at this Reynolds number, which is also apparent in the three-dimensional LES and has also been observed in experiments, such as the oil-flow visualizations of \citet{Schewe_86} at the critical Reynolds number $2.64\times10^5$ just before the drag crisis occurs. Also shown in Table \ref{tab:surface-summary} are the total drag $C_D$ and base pressure coefficients
\begin{equation}
C_{Pb}=\frac{\mean{P}_b-P_0}{(1/2)\rho U_0^2},
\end{equation}%
\nomenclature[RP]{$\mean{P}_b$}{pressure at the back stagnation point of the cylinder}%
\nomenclature[RP]{$\mean{P}_0$}{pressure at infinity}%
where $\mean{P}_b$ and $P_0$ are the pressures at the back stagnation point and at infinity, respectively, which are also quantities frequently examined in cylinder flow studies. In general, both of these quantities, calculated by the PDF method, are in good agreement with the LES, DNS and the experimental data at a slightly higher Reynolds number with the base pressure slightly overpredicted. \citet{Beaudan_94} report substantially lower base pressure ($-2.16$) and consequently higher drag (1.74) from a two-dimensional LES simulation with no subfilter-scale model. They found large discrepancies in other quantities as well, such as the amplitude and regularity of the lift, the skin-friction coefficient and the complete absence of an attached recirculation bubble. Since their three-dimensional simulations are in close agreement with experiments, they conclude that three-dimensional effects strongly influence the near-wake at this Reynolds number and that modeling the three-dimensionality of the flow is therefore essential. The current PDF simulations are two-dimensional in the sense of mean motions. While all three components of the particle velocities are retained, their positions are only allowed in the $x-y$ plane and the spanwise position is not represented. In other words, the turbulent fluctuations are modeled as three-dimensional, while the mean flow is two-dimensional. The generally close agreement of the current and subsequent PDF results with both three-dimensional simulations and experimental data suggests that retaining the three-dimensional fluctuations is essential.

Next we examine time-averaged statistics along the cylinder surface. These and subsequent time-averaged quantities have been collected after the quasi-periodic vortex shedding has been started, in the time-range of $60\le tU_0/D \le 200$, which amounts to approximately 28 vortex shedding cycles. The mean pressure coefficient
\begin{equation}
C_P = \frac{\mean{P}-P_0}{(1/2)\rho U_0^2}
\end{equation}%
\nomenclature[RC]{$C_P$}{mean pressure coefficient}%
is plotted in \Fige{fig:wallstats} (a) along with three-dimensional LES, DNS and experimental data. The overall agreement is very good, except for a slightly higher mean pressure at the front stagnation point, which is most likely due to the closeness of the inflow boundary to the cylinder -- only 3.5 diameters upstream, \Fige{fig:cylinder-domain}. The spanwise component of the mean vorticity, computed as
\begin{equation}
\Omega_z=\frac{1}{2}\left(\frac{\partial\mean{V}}{\partial x}-\frac{\partial\mean{U}}{\partial y}\right),
\end{equation}%
\nomenclature[Gz]{$\Omega_{z}$}{mean spanwise vorticity}%
is plotted in \Fige{fig:wallstats} (b). The PDF simulation accurately predicts the location of the boundary layer separation (where the vorticity becomes zero) indicated by the close agreement of the vorticity distribution with DNS data and by the correct separation angle of $\theta_{\mathrm{sep}}=89.3^\circ$, see also Table \ref{tab:surface-summary}.
\begin{figure}[t!]
\centering
\resizebox{15cm}{!}{\input{wall_stats.pstex_t}}
\caption[Time-averaged mean pressure and vorticiy along the cylinder]{\label{fig:wallstats}Time-averaged mean (a) pressure coefficient and (b) spanwise vorticity distributions along the cylinder wall. Red solid lines -- PDF simulation, dashed lines -- DNS of \citet{Ma_00}, blue dot-dashed lines -- LES of \citet{Kravchenko_00}, symbols -- experimental data of (a) \citet{Norberg_87} at \(\textit{Re}_{\scriptscriptstyle{D}}=3000\) and (b) \citet{Son_69} at \(\textit{Re}_{\scriptscriptstyle{D}}=5000\).}
\figSpace
\end{figure}

\subsection{Near wake statistics}
\label{sec:near-wake-stats}
Capturing the correct point of separation is crucial in predicting the correct statistics in the recirculation bubble as well, where transition to turbulence occurs as the thin shear layers become unstable. Predicting the transition has been a challenging task in both experiments and numerical simulations designed to obtain flow statistics at sub-critical Reynolds numbers. Measurements are found to be very sensitive to experimental disturbances as well as to other details, such as the cylinder aspect ratio (i.e.\ the spanwise length of the domain compared to the cylinder diameter). Some of the influencing factors in simulations are the type of the spatial and temporal discretization schemes which are directly related to the extent of numerical dissipation, the type and amount of subfilter-scale diffusion in LES and the spanwise size of the domain in both LES and DNS \citep{Beaudan_94,Ma_00}. Because of these difficulties, substantially different experimental datasets are available regarding the recirculation bubble. This is exemplified by \Fige{fig:vel-cen}, where the mean streamwise velocity from the current PDF simulation is plotted together with several experimental datasets and the LES of \citet{Kravchenko_00} employing a higher order B-spline-based finite element method and a dynamic subfilter model.
\begin{figure}[t]
\centering
\resizebox{11cm}{!}{\input{Ucen.pstex_t}}
\caption[Streamwise mean velocity behind the cylinder along the centerline]{\label{fig:vel-cen}Streamwise mean velocity behind the cylinder along the centerline at \(\textit{Re}_{\scriptscriptstyle{D}}~=~3900\). Red solid line -- PDF simulation, blue dot-dashed line -- LES of \citet{Kravchenko_00}, symbols -- experiments of $\Box$, Lourenco \& Shih (1993, see \citeauthor{Beaudan_94} \citeyear{Beaudan_94}), $\circ$, \citet{Ong_96} and $\times$, Govardhan \& Williamson (2000, see \citeauthor{Ma_00} \citeyear{Ma_00}).}
\figSpace
\end{figure}

Due to the aforementioned difficulties, major differences have also been found in the cross-stream shape of the mean streamwise velocity in the bubble. Moin and co-workers \citep{Beaudan_94,Mittal_96,Kravchenko_00} consistently obtained profiles closer to a \emph{U-shape} from their LES simulations, in disagreement with the experimental data of \citet{Lourenco_93} and the LES of \cite{Frohlich_98} generating \emph{V-shape} profiles. \citet{Kravchenko_00} discuss in length the possible sources of the differences. They point out that immediately behind the cylinder, at $x/D~=~0.58$, both the experiments and numerical simulations predict a U-shape profile, which evolves into a V-shape farther downstream. The shape of the mean velocity profiles is directly related to the level of fluctuations and therefore the transition in the shear layers. Smaller fluctuations result in U-shape, while larger fluctuations results in a more mixed and diffused V-shape profile for the mean velocity, see Figures 22 and 23 of \citet{Kravchenko_00}. Also, the length of the laminar shear layers is larger for U-shape and shorter for V-shape profiles, indicating that the onset of instability (the transition to turbulence) occurs farther and closer to the cylinder, respectively. In a direct simulation the precise point where the level of fluctuations becomes large enough to initiate the instability of the shear layers is influenced by many factors including the inherent numerical and the additional subfilter diffusion. Accordingly, simulations performed on coarser grids tend to produce V-shape profiles while finer grids result in U-shape profiles, see Figures 24 and 25 of \citet{Kravchenko_00}. Increasing the value of the Smagorinsky-constant also results in more diffusion, however its effect is more pronounced on the fluctuations, resulting in a U-shape profile for the mean velocity after the fluctuations have been attenuated. This major influence of the subfilter-scale diffusion in LES has been shown by the systematic DNS and LES study of \citet{Ma_00}, see Figures 7 and 8 therein, who also found the aspect ratio (i.e.\ the spanwise length of the domain) to be a decisive factor affecting the shape of the mean velocity profiles. From high resolution DNS simulations, they find two distinct converged states, arriving at either U or V-shapes, depending on the spanwise size of the domain employed, see their Figures 5 and 6. The narrower domain corresponds to the size used by Moin \emph{et al.}\ and converges to U-shape, while the twice as wider domain produces a V-shape profile in close agreement with the experiments of \citet{Lourenco_93}.

The series of studies mentioned above makes it clear that the ability of large eddy simulation to capture the precise point of instability heavily depends on the correct balance of phsyical, numerical and subfilter-scale diffusion. The same issue is present in PDF-type methods as well, with either subfilter diffusion (in LES/FDF) or modeled turbulent diffusion (in UPDF), therefore we can expect similar difficulties in these methods as well. Although modifying the model constants may improve certain predictions, it is always of limited value and we did not explore it. More importantly, grid-, and particle-number independence should be established.

Cross-stream distributions of the mean streamwise velocity $\mean{U}$ obtained from the current PDF simulation is plotted at different downstream locations in \Fige{fig:meanvel} (a) and (c) along with DNS and experimental data.  We see that the PDF simulation correctly predicts the V-shape of the streamwise velocity in the bubble with the minimum at the centerline slightly underpredicted towards the end of the bubble indicating a strong mean backflow there. Farther downstream, where the turbulence is dominated by vortex dynamics, the prediction is also very reasonable. \citet{Beaudan_94} also examine the errors in the experiments of Lourenco \& Shih based on the expected symmetries and anti-symmetries in the mean velocity and Reynolds stress components. They find that the errors in the mean streamwise velocities $\mean{U}$ are at 5\% of the maximum local velocity past 1 diameter downstream, while cross-stream velocities $\mean{V}$ exhibit errors comparable to their actual values, i.e.\ close to 100\%, in the first 3.5 diameters which increases to 200-300\% farther downstream. \citet{Ong_96} report experimental uncertainties of 2\% for their mean velocities.
\begin{figure}[t!]
\centering
\resizebox{15cm}{!}{\input{vel-near.pstex_t}}
\resizebox{15cm}{!}{\input{vel-far.pstex_t}}
\caption[Mean streamwise and cross-stream velocity behind the cylinder]{\label{fig:meanvel}Mean (a), (c) streamwise and (b), (d) cross-stream velocity at different downstream locations behind the cylinder at \(\textit{Re}_{\scriptscriptstyle{D}}=3900\). Red solid lines -- PDF simulation, black dashed lines -- DNS of \citet{Ma_00}, symbols -- experiments of $\Box$, \citet{Lourenco_93} and $\circ$, \citet{Ong_96}.}
\figSpace
\end{figure}

The time-averaged cross-stream velocities $\mean{V}$ produced by the PDF simulation are displayed in \Fige{fig:meanvel} (b) and (d). Up to the streamwise length examined, $x/D=10$, the anti-symmetric shape of the profiles are correctly captured with their magnitude gradually diminishing as the flow gets better mixed downstream. Immediately behind the cylinder, at $x/D=1.06$ and $1.54$, the profiles resemble both the DNS and the experimental data but with less pronounced extrema. The prediction of $\mean{V}$ improves farther downstream in the bubble, $x/D=2.02$, when compared to DNS data and stays close to the experiments of \citet{Ong_96} for $x/D\ge3$. It is worth noting that the two experimental datasets we use to compare the results do not match each other as shown in \Fige{fig:V3}. The maximum magnitude of the velocity and the spread of the wake are different with asymmetries evident in both datasets. Both the DNS data and the PDF simulation follow the experiments of \citet{Ong_96} more closely.
\begin{figure}[t!]
\centering
\resizebox{11cm}{!}{\input{V3.0.pstex_t}}
\caption[Cross-stream mean velocity behind the cylinder at $x/D=3.0$]{\label{fig:V3}Time-averaged cross-stream mean velocity behind the cylinder at $x/D=3.0$ at \(\textit{Re}_{\scriptscriptstyle{D}}=3900\). Red solid line -- PDF simulation, black dashed line -- DNS of \citet{Ma_00}, symbols -- experiments of $\Box$, \citet{Lourenco_93} and $\circ$, \citet{Ong_96}.}
\figSpace
\end{figure}

\begin{figure}[t!]
\centering
\resizebox{15cm}{!}{\input{rs-near.pstex_t}}
\resizebox{15cm}{!}{\input{rs-far.pstex_t}}
\caption[Streamwise and cross-stream Reynolds stress behind the cylinder]{\label{fig:rs}Streamwise (a), (c) and cross-stream (b), (d) Reynolds stress components at different downstream locations behind the cylinder at \(\textit{Re}_{\scriptscriptstyle{D}}=3900\). Red solid lines -- PDF simulation, black dashed lines -- DNS of \citet{Ma_00}, symbols -- experiments of $\Box$, \citet{Lourenco_93} and $\circ$, \citet{Ong_96}.}
\figSpace
\end{figure}
The streamwise and cross-stream components of the Reynolds stress tensor, $\mean{u^{\scriptscriptstyle{2}}}$ and $\mean{v^{\scriptscriptstyle{2}}}$, are displayed in \Fige{fig:rs} at several downstream locations. The error analysis of Beaudan \& Moin suggests 20\% error in the Reynolds stress components for the experimental data of \citet{Lourenco_93} between $1.0\le x/D\le2.5$ and slightly increasing farther downstream. \citet{Ong_96} provide the experimental error in these quantities as 2\% for all their measured length, $x/D\ge3$. The streamwise Reynolds stress component $\mean{u^{\scriptscriptstyle{2}}}$ in the recirculation bubble is in excellent agreement with the experiments and DNS data, \Fige{fig:rs} (a). The locations and the extent of the double peaks and their local minimum at the centerline are all predicted accurately, in close agreement with the experiments. Farther downstream, \Fige{fig:rs} (c), as the double peaks diminish in magnitude and gradually give place to single peaks indicating a more mixed state, the PDF predictions slightly diverge from the DNS data, underpredicting the level of fluctuations at $x/D\ge7$.

\begin{figure}[t]
\centering
\resizebox{11cm}{!}{\input{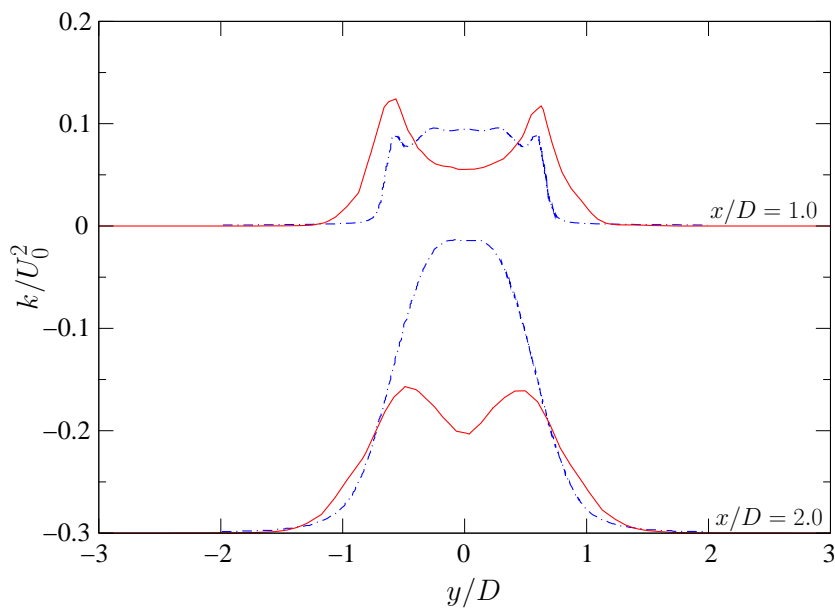}}
\caption[Turbulent kinetic energy in the recirculation bubble in the cylinder wake]{\label{fig:tke}Time-averaged cross-stream profiles of turbulent kinetic energy in the recirculation bubble in the wake of a cylinder at $x/D=1.0$ and $x/D=2.0$ at \(\textit{Re}_{\scriptscriptstyle{D}}=3900\). Red solid lines -- PDF simulation, blue dot-dashed lines -- LES of \citet{Beaudan_94}.}
\figSpace
\end{figure}
\Figse{fig:rs} (b) and (d) show that the cross-stream fluctuations $\mean{v^{\scriptscriptstyle{2}}}$ are severely underpredicted throughout the whole length of the wake examined. This may be due to the presence of an excessive level of turbulent and/or numerical diffusion originating from the turbulence model and a coarse grid (or overdiffusive spatial discretization), respectively. \citet{Ma_00} show that increasing the subfilter-scale diffusion in LES can have a disastrous effect on the second moments, especially in the recirculation bubble where the transition to turbulence occurs, where the turbulent kinetic energy reaches its highest levels. We can examine the turbulent kinetic energy, $k=(\mean{u^{\scriptscriptstyle{2}}}+\mean{v^{\scriptscriptstyle{2}}}+\mean{w^{\scriptscriptstyle{2}}})/2$, in the bubble in \Fige{fig:tke}. Close to the cylinder, at $x/D=1$, the level of energy is in reasonable agreement with the dynamic LES simulations of Beaudan \& Moin. In the LES simulations the energy grows almost threefold by the end of the bubble, $x/D=2$, while this growth in the PDF simulations is almost negligible or dissipated. We suspect that a high level of local diffusion (modeled and/or numerical) attenuates both cross-stream $\mean{v^{\scriptscriptstyle{2}}}$ and spanwise $\mean{w^{\scriptscriptstyle{2}}}$ fluctuations, which only dissipates further downstream. Another possible explanation for the discrepancy in the $\mean{v^{\scriptscriptstyle{2}}}$ and $\mean{w^{\scriptscriptstyle{2}}}$ components is the lack of representation of the mean motions in the third, spanwise dimension. Other than these factors, we also suspected another possible source of numerical dissipation, namely the way the local Eulerian statistics are computed, as described in \autoref{sec:Eulerian_statistics}. Computing ensemble averages in elements, then transferring them into gridpoints, and finally at the particle positions, employing nodal averages in elements, also have a smoothing/diffusive effect. Therefore another algorithm has been implemented in which the element-based ensemble averages are directly used in updating particles, without the intermediate step of transferring statistics to and from nodes. Originally, the main reason for the more complex two-step procedure was to mitigate the dire effects of elements without particles, however, this is not strictly necessary if one applies a particle redistribution procedure. A series of numerical tests with both algorithms, however, resulted in no significant change in the fields (i.e.\ it was not less diffusive). Because the simpler algorithm was not measurably more efficient than its current counterpart, we kept the two-step procedure. Further investigations are necessary to pinpoint the exact cause of the discrepancy between our simulations and the agreeing experimental and DNS data. These may include higher spatial refinement in the bubble, higher order (less diffusive) timestepping scheme and simulation in fully three-dimensional space.

\begin{figure}[t]
\centering
\resizebox{15cm}{!}{\input{uv.pstex_t}}
\caption[Shear Reynolds stress behind the cylinder]{\label{fig:uv}Time-averaged shear Reynolds stress at different downstream locations behind the cylinder at \(\textit{Re}_{\scriptscriptstyle{D}}=3900\). Red solid lines -- PDF simulation, black dashed lines -- DNS of \citet{Ma_00}, symbols -- experiments of $\Box$, \citet{Lourenco_93} and $\circ$, \citet{Ong_96}.}
\figSpace
\end{figure}
\Fige{fig:uv} displays the shear Reynolds stress component $\mean{uv}$ at different downstream locations in the wake. We see that the agreement with DNS and experimental data is quite good from the end of the recirculation bubble, $x/D=2.02$. Immediately behind the cylinder, $x/D=1.06$ and $x/D=1.54$, the predictions follow the DNS and experiments in shape, but the peaks of the profiles are diffused. Also, the anti-symmetric double peaks at $x/D=1.06$, apparent in both the DNS and the experiments, are only recognizable by the slightly flattening shear stress profile at the centerline. We suspect that a simulation with a more refined Eulerian grid would help in predicting the very near wake shear stress even more accurately.

We now turn our attention to higher order velocity statistics. In general, third and fourth order moments of the velocity field are rarely investigated in the literature regarding this flow. The most widely applied turbulence modeling techniques, $k-\varepsilon$ and Reynolds stress models, would not be economical and little is known about the reliance of the turbulent viscosity hypothesis at these higher level of closures. However, higher order moments of passive tracers (or the full concentration PDF) would be valuable in atmospheric pollution modeling, where there is a need to predict extreme events and probabilities in concentration fields. Lagrangian dispersion models are capable of providing this information and are routinely used to compute scalar fields, usually in conjunction with traditional CFD-type (URANS or LES) models that provide them the mean and fluctuating velocity. In these applications the micromixing of the scalar (which determines the scalar PDF) is commonly modeled without taking the velocity field into account, which is not justified theoretically \citep{Pope_98} and may result in erroneous predictions as we demonstrated in \autoref{chap:channel} for a very simple flow. PDF methods naturally account for the close interaction between the turbulent velocity and scalar fields, therefore potentially more accurate predictions can be achieved. In principle, direct simulations (LES and DNS) could also provide this higher level of velocity information for Lagrangian micromixing models, however, a more natural approach is to jointly model the PDF of velocity and concentrations, which requires an accurate representation of the higher order velocity statistics. Therefore we now examine the prediction of the skewness and flatness of the velocity field behind the cylinder.

\begin{figure}[t!]
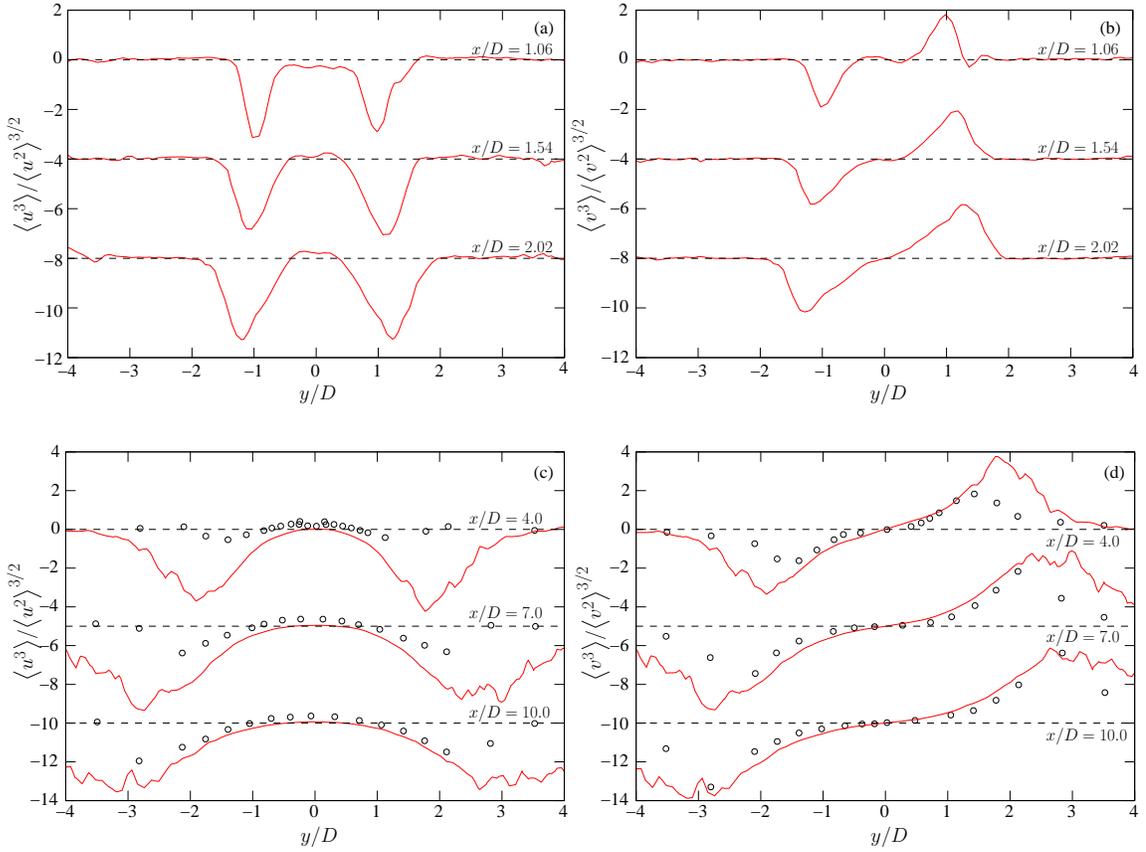

\centering
\resizebox{15cm}{!}{\input{skewness_near.pstex_t}}
\resizebox{15cm}{!}{\input{skewness_far.pstex_t}}
\caption[Skewness of the streamwise and cross-stream velocity behind the cylinder]{\label{fig:skewness}Skewness of the streamwise (a), (c) and cross-stream (b), (d) velocity components at different downstream locations behind the cylinder at \(\textit{Re}_{\scriptscriptstyle{D}}=3900\). Red solid lines -- PDF simulation, symbols -- experiments of \citet{Ong_96}. The horizontal dashed lines at each location indicate the Gaussian skewness value of 0.}
\figSpace
\end{figure}

\begin{figure}[t!]
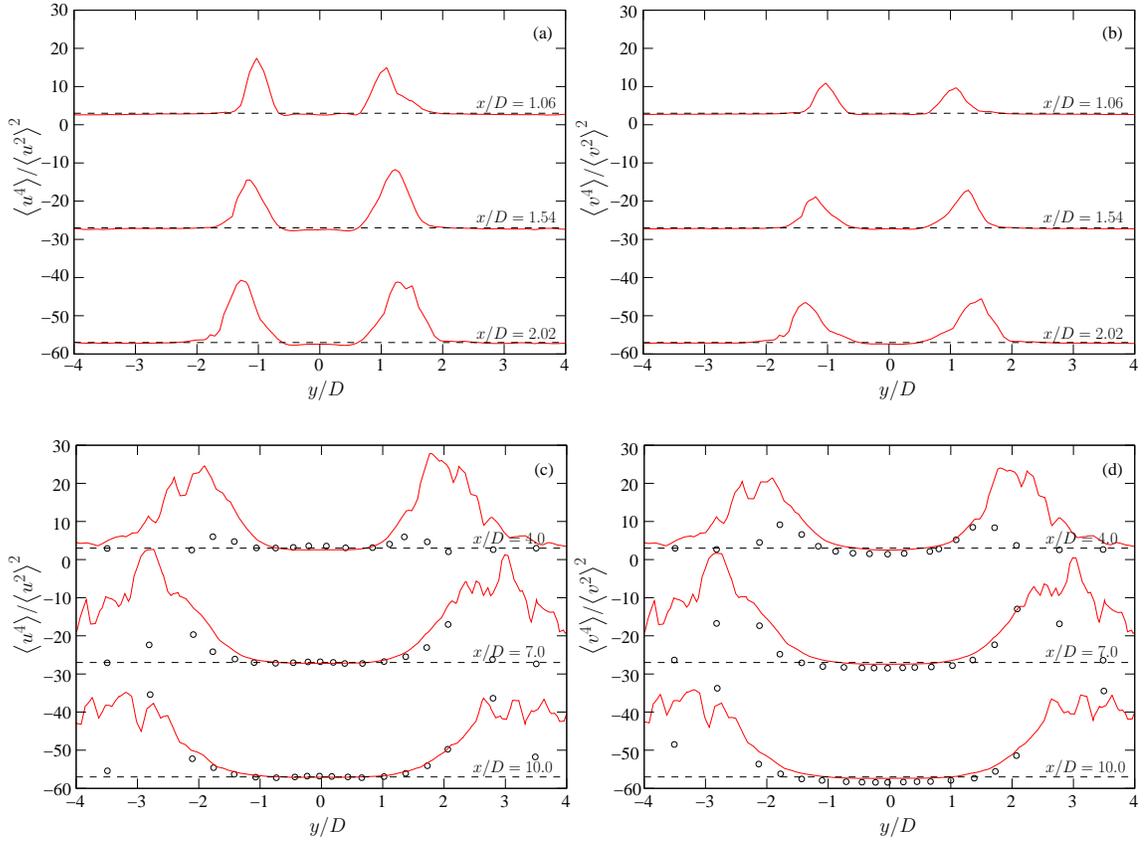

\centering
\resizebox{15cm}{!}{\input{flatness_near.pstex_t}}
\resizebox{15cm}{!}{\input{flatness_far.pstex_t}}
\caption[Flatness of the streamwise and cross-stream velocity behind the cylinder]{\label{fig:flatness}Flatness of the streamwise (a), (c) and cross-stream (b), (d) velocity components at different downstream locations behind the cylinder at \(\textit{Re}_{\scriptscriptstyle{D}}=3900\). Red solid lines -- PDF simulation, symbols -- experiments of \citet{Ong_96}. The horizontal dashed lines at each location indicate the Gaussian flatness value of 3.}
\figSpace
\end{figure}

Cross-stream profiles of the skewness of the streamwise and cross-stream velocity components are displayed in \Fige{fig:skewness} at several downstream locations. \citet{Ong_96} provide experimental data for the skewness and flatness for $x/D\ge3.0$. The profiles are plotted for the usual locations, both inside and outside the bubble. In general, all skewness predictions are in good agreement with the experimental data with the streamwise component sligthly underpredicted, especially at the edge of the wake close to the end of the recirculation bubble, $x/D=4.0$, \Fige{fig:skewness} (c). However, this discrepancy gradually diminshes farther downstream. The prediction of flatness is similary good as shown in \Fige{fig:flatness}. Although the flatnesses of both streamwise and cross-stream velocity are overpredicted immediately after the bubble in the highly skewed regions of the wake edges, the predictions greatly improve further downstream. It is worth noting here, that both skewness and flatness profiles are normalized by the velocity fluctuations which become numerically very small towards the edges of the wake, see \Fige{fig:rs}, thus the actual third and fourth moments, e.g.\ $\mean{u^{\scriptscriptstyle{3}}}/U_0^{3}$ and $\mean{u^{\scriptscriptstyle{4}}}/U_0^4$, would be better candidates for examining the accuracy of these higher order statistics. The good agreement of the skewness and flatness profiles indicates the model accurately captures the shape of the velocity PDF.

\section{Discussion}
\label{sec:discussion-cylinder}
The laminar-to-turbulent transitional flow in the near wake of a circular cylinder at the sub-critical Reynolds number of 3900 has been computed with a PDF method. The method has been applied for the first time to compute a flow with a complex geometry bounded by no-slip walls with significant curvature. The elliptic relaxation technique that is used to represent all components of the Reynolds stress tensor in the low-Reynolds-number wall-region has also been applied for the first time for highly curved boundaries with significant adverse pressure gradient resulting in boundary layer separation. Although mean spanwise motions are not represented in the current case thus only a two-dimensional Eulerian grid is employed to extract statistics, all three dimensions of the fluctuating velocity are retained.

Transient and time-averaged statistics of the joint PDF of the three velocity components have been compared to LES, DNS and experimental data. The predictions show significant improvement compared to past pure Eulerian RANS simulations. The quality and accuracy of the PDF results are comparable to three-dimensional LES and DNS predictions for all the quantities examined. One exception is the cross-stream Reynolds stress component $\mean{v^{\scriptscriptstyle{2}}}$, which is noticeably underpredicted. This can be due to an overdiffusive numerical scheme (the lack of resolution of the Eulerian grid and/or the low-order accuracy of the temporal discretization), the lack of representation of the spanwise mean motions or a too large modeled turbulent diffusion. The advantages of the method can be summarized as follows:
\begin{itemize}
\item higher level statistical description of the stochastic fields than traditional RANS-type closures,
\item a close interaction between the stochastic velocity and scalar fields,
\item mathematically exact representation of advection, viscous diffusion, the effect of mean pressure and complex chemical reactions; these physical processes are treated without closure assumptions,
\item the ability to follow highly distorted material surfaces accurately,
\item the possibility for a relatively straightforward inclusion of history-dependent constitutive relations,
\item excellent parallel performance.
\end{itemize}

A natural next step building on this work is to include the dispersion of passive scalars and develop a universal micromixing timescale that can be used in complex geometries in conjunction with the IEM/IECM models. Following this line a further step could be the inclusion of chemical reactions, in which the biggest advantage of the whole methodology lies, since it could be used to simulate chemically reactive turbulent flows surrounded by realistic no-slip walls in complex geometries, without the burden of the closure of the chemical source terms. Another obvious and straightforward way to expand on this work is to add the third dimension to the particle position and implement three-dimensional tetrahedra as the Eulerian grid, resulting in a fully three-dimensional code.

Further improvements to the code can be realized by employing edge-based data structures and solution techniques \citep{Barth_91,Mavriplis_91,Peraire_92,Luo_93} to build the finite element coefficient matrices and the right hand sides for the two Eulerian equations, the elliptic relaxation and mean pressure projection. Although currently the solution of these equations takes only a small fraction of the running time, see \autoref{sec:profile}, this will most likely change in three dimensions. In this case the solution will be more efficient with an edge-based solver because of the reduction of indirect addressing compared to the redundant element-based solution.

Currently, the simple Jacobi preconditioner is used to improve the convergence of the conjugate gradients solver for the mean pressure. More sophisticated preconditioners could also be explored to reduce the number of iterations, which will also significantly increase in three dimensions.

Porting the code to 3D will also result in excessive memory requirements if all nine components of the elliptic relaxation tensor are to be stored in every gridpoint as it is done throughout this study. More efficient elliptic relaxation could be achieved by employing different derivatives of the elliptic relaxation technique that store only a scalar variable instead of all 9 components \cite[e.g.][]{Waclawczyk_04} and/or storing and solving $\wp_{ij}$ only in the vicinity of walls, where the lengthscale $L$, \Eqre{eq:L}, has significant curvature.


\chapter[Summary and discussion]{\\Summary and discussion}
\label{chap:conclusions}

This work has presented a series of numerical methods to compute the one-point one-time joint PDF of turbulent velocity, characteristic frequency and scalar concentrations in high-Reynolds-number incompressible turbulent flows with complex geometries. Following the terminology coined by \citet{Muradoglu_99}, we call the current methodology \emph{non-hybrid} since an Eulerian CFD solver is not used in conjunction with the particle code to solve the PDF equations, i.e.\ the method is stand-alone. The method does belong to the familiy of particle-in-cell methods, where the Eulerian grid is used solely for: (i) estimating Eulerian statistics; (ii) tracking particles in the domain; and (iii) solving for quantities that are only represented in the Eulerian sense (i.e.\ mean pressure and elliptic relaxation). Compared to hybrid models, the current non-hybrid method assures that none of the fields are computed redundantly, therefore the simulation is kept consistent both numerically and at the level of turbulence closure without the need to enforce consistency conditions.

Adequate wall-treatment on the higher order statistics of the velocity field is achieved with an elliptic relaxation technique without damping or wall-functions, i.e.\ the boundary layers at solid (no-slip) walls are fully resolved. On the other hand, for application areas where full resolution of the turbulent boundary layers is not an option, we provide a treatment consistent with wall-functions that are commonly used in moment closures. The validation examples demonstrate the applicability of the algorithm in two-dimensional flows. A natural future direction along these lines is the extension to three spatial dimensions. This should be straightforward, since all the numerical methods are general enough and the extension only pertains to the Eulerian grid (e.g.\ tetrahedra instead of triangles) and an additional equation for the particle position, since all three components of particle velocities are already represented.

A significant challenge in stand-alone transported PDF methods is the accurate and stable computation of the mean pressure. This is mainly due to the following reasons: the mean velocity and Reynolds stresses have to be estimated from a noisy particle field and the pressure-Poisson equation requires their first and second derivatives, respectively, which are even noisier. We described a method to compute the mean pressure in conjunction with particle/PDF methods that only requires first derivatives of the mean velocity, which is based on a pressure-projection technique that has already been widely used and tested in laminar flows.

The two Eulerian equations needed by the algorithm are both solved on unstructured Eulerian grids with the finite element method. The last couple of decades have seen great strides in automatic unstructured grid generation, grid refinement and coarsening techniques and the development of highly sophisticated grid-based data structures that minimize cache misses. Using the algorithm presented in this work all this knowledge pertaining to unstructured meshes can be utilized in conjunction with the PDF equations and complex flow geometries. Employing finite elements together with particle/PDF methods also has the advantage of greatly simplifying boundary conditions for particles -- no ghost elements are required as in finite volume methods. Furthermore, finite element approximation functions are not only used for particle tracking but also provide an elegant way of estimating derivatives of statistics from particle fields.

We also described a general algorithm that can be used to calculate the velocity-conditioned scalar mean for the IECM micromixing model. The procedure homogenizes the statistical error over the sample space for arbitrary velocity PDFs by dynamically adjusting the number of bins and their distribution.

A particle-redistribution algorithm has also been described that provides stability by ensuring that no Eulerian elements remain without particles at any time during the simulation. This task has traditionally been accomplished via particle splitting and merging techniques. However, computationally it is more efficient not to introduce or eliminate particles during timestepping, so that the arrays storing physical properties can keep their original size and can remain consecutively accessible with minimal cache misses. Both particle splitting and merging algorithms and the current redistribution procedure do change the local PDF, and this is certainly an undesired effect. We are not aware of any algorithm in the literature which accomplishes particle-number control without altering the underlying numerically computed joint PDF. In particle splitting and merging algorithms mass may be conserved, but fulfillment of all mass, momentum and kinetic energy conservation is in general not possible in single events, only statistically \citep{Rembold_06}. Our method is no exception. We presented an error analyis employing a simplified set of particle equations on a homogeneous example. We believe, that further tests are certainly necessary to investigate the error introduced by the redistribution algorithm in inhomogeneous flows. Also it is worth pointing out that there is no clear or established benchmark to investigate the effects of redistribution algorithms, as the effects may be space-dependent, and their importance is relative to the application.

We also proposed a general form for the micromixing timescale that can be used in a flow-, and geometry-independent manner for modeling the effect of small-scale mixing on a transported passive scalar released from a concentrated source. Although the computed concentration results compare well with analytical and experimental data for the testcases, this is to be considered under heavy development regarding both its mathematical expression and modeling constants.

The solver has been parallelized with the OpenMP standard, which easily allows the exploitation of multi-core workstations mainly used for production codes. Our performance study has shown a good parallel speedup up to 32 CPUs tested on shared memory machines using single-, dual-, and quad-core CPUs. We also ported the code to Intel's Cluster OpenMP technology, which allows an OpenMP program to run on a beowulf-type cluster of networked workstations requiring a minor programing effort compared to an MPI-based implementation. However, we found that the algorithm with its current design is not suitable for Cluster OpenMP.

Three testcases of increasing complexity have been presented to demonstrate the applicability of the algorithm. The resulting fields show a good agreement when compared to DNS and experimental data where available. In the future, more tests with cases of different complexity will definitely need to be carried out. This is especially true for the micromixing timescale, which has to be tested in different flows and for different source scenarios to assess the validity of its form and its modeling constants.

The hybrid methods that combine existing Eulerian CFD solvers with the PDF methodology are based on RANS and LES methods. Both of these lines of development concentrate on the modeling of chemical reactions which appear in mathematically closed form in the PDF framework. The Eulerian governing equations consist of the fully compressible equations for conservation of mass, momentum and energy. This system is augmented by a set of stochastic equations for Lagrangian particles that represent species' concentrations and may also provide turbulence closure depending on how the fluctuating velocity field is represented. Furthermore, the mean pressure is obtained from an equation of state. In these hybrid methods, since the preferred way of representing flow variables is via Favre-averaging, the density must also appear explicitly at the Lagrangian particle level. Since the mean continuity equation is also required to solve the mean Eulerian governing equations, the density is represented redundantly. Consequently, consistency must be ensured explicitly. The currently proposed non-hybrid method is stand-alone and represents mass consistency without redundancy, only by ensuring that the mean velocity field is divergence-free at all times by solving a Poisson equation for the mean pressure. Thus particles (and particle number-density) do not represent mass (or real fluid density) and no additional mass-consistency condition is required. Therefore the problem of a high degree of variation in particle-number density between different regions of the flow field only amounts to different statistical errors, without the additional computational errors introduced by inconsistency. This is possible precisely, because we solve the fully incompressible equations, where the density is constant. This advantage, however, may certainly become a disadvantage in turbulent chemistry even in an otherwise incompressible flow if the stochastic density variations due to chemical reactions have to be represented accurately. Therefore the limitations of the current methodology regarding its applicability in conjunction with chemical reactions remain to be seen.

Regarding computational costs, \citet{Pope_00} places PDF methods somewhere between Reynolds stress closures and large eddy simulation. We showed that he most expensive parts of the current non-hybrid method are the advancement of particle properties and random number generation, which together account for more than 90\% of the computational cost. Both hybrid and non-hybrid methods need to advance and track particles, generate random numbers, estimate Eulerian statistics, solve for the mean pressure and ensure sufficient number of particles everywhere on the flow domain. Therefore we expect the computational costs of these components to be comparable for the two methodologies. In addition to the above, hybrid methods need to enforce consistency conditions and solve the Eulerian system of governing equations as well. This very approximate analysis suggests that there is no clear reason to think that the total cost of the two methods will be very different. However, it would be valuable to perform thorough side-by-side comparisons among the different stand-alone (fully Lagrangian), hybrid RANS, hybrid LES and non-hybrid methods in order to have a better picture on their relative computational costs.

Additionally to these approaches, it would be interesting to explore a method that solves the fully incompressible equations, just like the current non-hybrid method, but in the Eulerian framework, which represents only the scalar concentrations by Lagrangian particles. We expect the computational cost of such a method to be significantly less than the current method, since the velocity field would not be represented by particles. This would be beneficial in situations where higher order statistics of the velocity are not required and a close interaction between the stochastic velocity and scalar concentration fields is not important. The value of such a method may be limited in applications of turbulent chemistry, but the higher level statistical description of the scalar fields could be advantageous in atmospheric pollution modeling.

On the other hand, the close interaction between the stochastic velocity and scalar concentration fields (that both hybrid methods for velocity and scalars and the current non-hybrid method can provide) is important as a research tool to shed light on micromixing mechanisms; to provide information and data on higher order statistics; and in applications where accurate modeling of the micromixing of scalars is required.

\appendix
\appendixeqnumbering

\appchapter[Derivation of the Eulerian PDF transport equation]{\\Derivation of the Eulerian PDF transport equation}
\label{app:derivation}

In PDF methods of turbulent flows, the velocity and transported scalar fields are considered as time dependent, multivariate random fields \citep{Pope_00}. In other words, as opposed to a deterministic theory, the three components of the velocity and the scalar concentration are represented by a joint probability distribution function containing the full one-point, one-time statistics of the velocity and the scalar. It should be emphasized, however, that this one-point, one-time description does not contain information regarding other points in space and time, therefore -- as it will be shown -- does not provide a complete description of the random velocity and scalar fields. As a consequence, the one-point, one-time PDF contains no information about the length-scale or frequency of the fluctuations, thus appropriate models are necessary to supply this missing information in the form of models. In an incompressible turbulent flow containing scalar tracers, the state of the fluid at any location is described by the instantaneous Eulerian velocity \(\bv{U}(\bv{x},t)\), pressure \(P(\bv{x},t)\) and the species' mass concentrations \(\phi(\bv{x},t)\). In the following, the PDF transport equation (\ref{eq:EulerianPDF}) is derived starting from the system of Eulerian governing equations (\ref{eq:continuity}-\ref{eq:scalar}).

Let \(f(\bv{V},\psi;\bv{x},t)\) denote the one-point, one-time Eulerian joint PDF of the random velocity \(\bv{U}(\bv{x},t)\) and scalar \(\phi(\bv{x},t)\), where the three dimensional Euclidean space \((V_1,V_2,V_3)\) is the sample space of the random velocity vector \(\bv{U}=(U_1,U_2,U_3)\) and \(\psi\) is the sample space variable of the random scalar concentration \(\phi\). Table \ref{tab:PDFsamplespace} summarizes the random flow variables and their sample spaces in the joint PDF \(f(\bv{V},\psi;\bv{x},t)\). This can also be viewed as an eight-dimensional scalar-valued function having a unique value at each location of the eight-dimensional Euclidean state-space
\begin{equation}
f(V_1,V_2,V_3,\psi;x_1,x_2,x_3,t)\colon\mathbb{R}^8\rightarrow\mathbb{R},
\end{equation}
thus for every different set of eight independent variables \((V_1,V_2,V_3,\psi;x_1,x_2,x_3,t)\) the function \(f\) corresponds to a single scalar. An other way to look at this, is to have 3+1 scalar functions \((\mathbb{R}\rightarrow\mathbb{R})\) at each point in space and time. This increased dimensionality is characteristic of PDF methods, since every single random scalar variable is represented with its probability density distribution, so instead of a single scalar, a scalar function, its probability distribution, is taken into account.

The transport equation for the joint PDF \(f(\bv{V},\psi;\bv{x},t)\) can be derived from the conservation equations (\ref{eq:NavierStokes}-\ref{eq:scalar}), which are rewritten here in a more convenient form:
\begin{eqnarray}
\sd{U_i}&=&A_i,\mbox{\quad where\quad}A_i=\nu\nabla^2U_i-\frac{1}{\rho}\frac{\partial P}{\partial x_i},\label{eq:NavierStokes-sd}\\
\sd{\phi}&=&B,\mbox{\quad where\quad}B=\Gamma\nabla^2\phi,\label{eq:scalar-sd}
\end{eqnarray}%
\nomenclature[RA]{$A_i$}{fluid acceleration (\autoref{app:derivation})}%
\nomenclature[RB]{$B$}{scalar diffusion, \Eqre{eq:scalar-sd}}%
where the substantial derivative is denoted by
\begin{equation}
\sd{}\equiv\frac{\partial}{\partial t}+U_j\frac{\partial}{\partial x_j}.\label{eq:subder}
\end{equation}
There are several ways of deriving PDF transport equations. A useful method involving delta functions is described by \citet{Pope_00}, but here a different approach is followed, which has been used by \citet{Pope_85} and also by \citet{Fox_03}. The method is based on equating two independent expressions for \(\mean{\mathrm{D}Q/\mathrm{D}t}\), where \(Q(\bv{U},\phi)\) is ``almost'' any function\footnote{\(Q(\bv{U},\phi)\) is an arbitrary function, however, it has necessary properties so that its statistics, like \Eqres{eq:firstmoment} and \Eqrs{eq:secondmoment}, are not divergent.}. The first expression for \(\mean{\mathrm{D}Q/\mathrm{D}t}\) is obtained by employing the definition of the substantial derivative \Eqrs{eq:subder} and the mathematical expectation of a random function \(\mean{Q(\bv{U},\phi)}=\int Q(\bv{V},\psi)f(\bv{V},\psi)\mathrm{d}\bv{V}\mathrm{d}\psi\):%
\nomenclature[RQ]{$Q(\bv{U},\phi)$}{random function of $\bv{U}$ and $\phi$}%
\begin{equation}\begin{split}
\mean{\sd{Q}}&=\mean{\frac{\partial Q(\bv{U},\phi)}{\partial t}}+\mean{U_i\frac{\partial Q(\bv{U},\phi)}{\partial x_i}}\\ &=\frac{\partial}{\partial t}\int Q(\bv{V},\psi)f(\bv{V},\psi;\bv{x},t)\mathrm{d}\bv{V}\mathrm{d}\psi+\frac{\partial}{\partial x_i}\int V_iQ(\bv{V},\psi)f(\bv{V},\psi;\bv{x},t)\mathrm{d}\bv{V}\mathrm{d}\psi\\ &=\int Q(\bv{V},\psi)\left\{\frac{\partial f}{\partial t}+V_i\frac{\partial f}{\partial x_i}\right\}\mathrm{d}\bv{V}\mathrm{d}\psi.\label{eq:firstexpinc}
\end{split}\end{equation}
The second expression can be deduced by relating changes in \(Q\) to changes in \(\bv{U}\) and \(\phi\) as
\begin{table}[t]
\caption[Sample space of the Eulerian joint PDF]{Random flow variables and their corresponing sample spaces in the joint PDF \(f(\bv{V},\psi;\bv{x},t)\).}
\begin{tabular*}{1.0\textwidth}{l@{\hspace{2.0cm}}l@{\hspace{2.0cm}}l@{\hspace{2.0cm}}}
\hline\hline
Quantity&Random variable&Sample space\\
\hline
velocity&\(\bv{U}=(U_1,U_2,U_3)\)&\(\bv{V}=(V_1,V_2,V_3)\)\\
scalar&\(\phi\)&\(\psi\)\\
\hline\hline
\end{tabular*}
\label{tab:PDFsamplespace}
\tableSpace
\end{table}
\begin{equation}
\sd{Q}=\frac{\partial Q}{\partial U_i}\sd{U_i}+\frac{\partial Q}{\partial\phi}\sd{\phi},
\end{equation}
where the material derivatives can be replaced by \(A_i\) and \(B\) from \Eqres{eq:NavierStokes-sd} and \Eqrs{eq:scalar-sd} and we also take take the expectation as
\begin{equation}
\mean{\sd{Q}}=\mean{\frac{\partial Q}{\partial U_i}A_i}+\mean{\frac{\partial Q}{\partial\phi}B}.\label{eq:replacedinc}
\end{equation}
Note, that in general, \(A_i\) and \(B\) depend on multi-point information of the random fields \(\bv{U}\) and \(\phi\), for example, they depend on the velocity and scalar gradients and Laplacians. Since these quantities are not contained in the one-point, one-time PDF \(f(\bv{V},\psi;\bv{x},t)\), let these additional unknowns be collected into the vector \(\bv{Z}(\bv{x},t)\). Furthermore, let the joint PDF of \(\bv{U}\), \(\phi\) and \(\bv{Z}\) be \(f_{\bv{\scriptscriptstyle U}\scriptscriptstyle\phi\bv{\scriptscriptstyle Z}}(\bv{V},\psi,\bv{z};\bv{x},t)\). According to Bayes' rule \citep{vanKampen_04}, this can be written as the product of a conditional and a marginal PDF as%
\nomenclature[RZ]{$\bv{Z}$}{vector of multipoint variables in the joint PDF $f_{\bv{\scriptscriptstyle U}\scriptscriptstyle\phi\bv{\scriptscriptstyle Z}}(\bv{V},\psi,\bv{z})$}%
\nomenclature[Rz]{$\bv{z}$}{vector of sample space variables corresponding to $\bv{Z}$}%
\nomenclature[Rf]{$f_{\bv{\scriptscriptstyle U}\scriptscriptstyle\phi\bv{\scriptscriptstyle Z}}(\bv{V},\psi,\bv{z})$}{multi-point and multi-time joint PDF of $\bv{U}$, $\phi$ and $\bv{Z}$ (\autoref{app:derivation})}%
\nomenclature[Rf]{$f_{\bv{\scriptscriptstyle Z}"|\bv{\scriptscriptstyle U}\scriptscriptstyle\phi}(\bv{z}"|\bv{V},\psi)$}{multi-point and multi-time joint PDF of $\bv{Z}$ conditional on $\bv{U}=\bv{V}$ and $\phi=\psi$, \Eqre{eq:condz}}%
\begin{equation}
f_{\bv{\scriptscriptstyle U}\scriptscriptstyle\phi\bv{\scriptscriptstyle Z}}(\bv{V},\psi,\bv{z})=f_{\bv{\scriptscriptstyle Z}|\bv{\scriptscriptstyle U}\scriptscriptstyle\phi}(\bv{z}|\bv{V},\psi)f(\bv{V},\psi).\label{eq:condz}
\end{equation}
Thus the first term on the right hand side of \Eqre{eq:replacedinc} can be rewritten as
\begin{equation}\begin{split}
\mean{\frac{\partial Q}{\partial U_i}A_i}&=\int\frac{\partial Q(\bv{V},\psi)}{\partial V_i}A_i(\bv{V},\psi,\bv{z})f_{\bv{\scriptscriptstyle U}\scriptscriptstyle\phi\bv{\scriptscriptstyle Z}}(\bv{V},\psi,\bv{z})\mathrm{d}\bv{V}\mathrm{d}\psi\mathrm{d}\bv{z}\\ &=\int\frac{\partial Q(\bv{V},\psi)}{\partial V_i}\mean{A_i|\bv{V},\psi}f(\bv{V},\psi)\mathrm{d}\bv{V}\mathrm{d}\psi,
\end{split}\end{equation}
where the conditional expectation of the acceleration \(A_i\) is
\begin{equation}
\mean{A_i|\bv{V},\psi}=\int A_i(\bv{V},\psi,\bv{z})f_{\bv{\scriptscriptstyle Z}|\bv{\scriptscriptstyle U}\scriptscriptstyle\phi}(\bv{z}|\bv{V},\psi)\mathrm{d}\bv{z}.\label{eq:condaccel}
\end{equation}
Note, that \(\mean{A_i|\bv{V},\psi}\) is a function only of \(\bv{V}\) and \(\psi\) (additionally to the implicit dependence on \(\bv{x}\) and \(t\)) since all the unknowns \(\bv{Z}\) have been integrated out. Integration by parts yields
\begin{equation}\begin{split}
\mean{\frac{\partial Q}{\partial U_i}A_i}=\int\frac{\partial}{\partial V_i}\Big[Q(\bv{V},\psi)\mean{A_i|\bv{V},\psi}f(\bv{V},\psi)\Big]\mathrm{d}\bv{V}\mathrm{d}\psi\hspace{80pt}\\-\int Q(\bv{V},\psi)\frac{\partial}{\partial V_i}\Big[\mean{A_i|\bv{V},\psi}f(\bv{V},\psi)\Big]\mathrm{d}\bv{V}\mathrm{d}\psi,\label{eq:partterma}
\end{split}\end{equation}
where the first term on the right hand side vanishes provided that \(Q\) is monotonic as \(|\bv{V}|\) tends to infinity and the expectation \(\mean{A_iQ}\) exists.\footnote{These conditions are given in practically all flow conditions that one can encounter, see also \citep{Pope_85}.} A similar procedure can be followed for the term containing \(B\) in \Eqre{eq:replacedinc} to obtain
\begin{equation}
\mean{\frac{\partial Q}{\partial \phi}B}=-\int Q(\bv{V},\psi)\frac{\partial}{\partial \psi}\Big[\mean{B|\bv{V},\psi}f(\bv{V},\psi)\Big]\mathrm{d}\bv{V}\mathrm{d}\psi.\label{eq:parttermb}
\end{equation}
Substituting  \Eqres{eq:partterma} and \Eqrs{eq:parttermb} into \Eqre{eq:replacedinc} the second expression for \(\mean{\mathrm{D}Q/\mathrm{D}t}\) can be obtained:
\begin{equation}
\mean{\sd{Q}}=-\int Q(\bv{V},\psi)\left\{\frac{\partial}{\partial V_i}\Big[\mean{A_i|\bv{V},\psi}f(\bv{V},\psi)\Big]+\frac{\partial}{\partial\psi}\Big[\mean{B|\bv{V},\psi}f(\bv{V},\psi)\Big]\right\}\mathrm{d}\bv{V}\mathrm{d}\psi.\label{eq:secondexpinc}
\end{equation}
Substracting the second expression for \(\mean{\mathrm{D}Q/\mathrm{D}t}\) \Eqrs{eq:secondexpinc} from the first expression \Eqrs{eq:firstexpinc} we obtain
\begin{equation}
\int Q(\bv{V},\psi)\left\{\frac{\partial f}{\partial t}+V_i\frac{\partial f}{\partial x_i}+\frac{\partial}{\partial V_i}\Big[\mean{A_i|\bv{V},\psi}f\Big]+\frac{\partial}{\partial\psi}\Big[\mean{B|\bv{V},\psi}f\Big]\right\}\mathrm{d}\bv{V}\mathrm{d}\psi=0.
\end{equation}
Since this equation holds for ``almost'' any function \(Q(\bv{V},\psi)\), the term in the brackets must sum to zero, thus we obtain the transport equation for the velocity-scalar joint PDF \(f(\bv{V},\psi;\bv{x},t)\)
\begin{equation}
\frac{\partial f}{\partial t}+V_i\frac{\partial f}{\partial x_i}=-\frac{\partial}{\partial V_i}\Big[\mean{A_i|\bv{V},\psi}f\Big]-\frac{\partial}{\partial\psi}\Big[\mean{B|\bv{V},\psi}f\Big].\label{eq:pdftransportinc}
\end{equation}
As it can be seen, in physical space the joint PDF \(f\) evolves due to the velocity field \(\bv{V}\), in velocity space due to the conditional acceleration \(\mean{A_i|\bv{V},\psi}\) and in concentration space due to the conditional diffusion term \(\mean{B|\bv{V},\psi}\). Now, we substitute the right hand sides of the momentum and scalar conservation equations from \Eqrs{eq:NavierStokes-sd} and \Eqrs{eq:scalar-sd} and arrive at \Eqre{eq:EulerianPDF}
\begin{equation}
\frac{\partial f}{\partial t}+V_i\frac{\partial f}{\partial x_i}=-\frac{\partial}{\partial V_i}\bigg[\mean{\nu\nabla^2U_i-\frac{1}{\rho}\frac{\partial P}{\partial x_i}\bigg|\bv{V},\psi}f\bigg]-\frac{\partial}{\partial\psi}\Big[\mean{\Gamma\nabla^2\phi|\bv{V},\psi}f\Big].\label{eq:pdftransportphys}
\end{equation}
After decomposing the pressure \(P\) into its mean \(\mean{P}\) and fluctuating part \(p\), another useful form of this equation is derived by \cite{Pope_00}
\begin{equation}
\begin{split}
\frac{\partial f}{\partial t} + V_i\frac{\partial f}{\partial x_i} &= \nu\frac{\partial^2f}{\partial x_i\partial x_i} + \frac{1}{\rho}\frac{\partial\mean{P}}{\partial x_i}\frac{\partial f}{\partial V_i} - \frac{\partial^2}{\partial V_i\partial V_j}\left[f\mean{\nu\frac{\partial U_i}{\partial x_k}\frac{\partial U_j}{\partial x_k}\Bigg|\bv{V},\psi}\right]\\
&\quad+ \frac{\partial}{\partial V_i}\left[f\mean{\frac{1}{\rho}\frac{\partial p}{\partial x_i}\Bigg|\bv{V},\psi}\right] - \frac{\partial}{\partial\psi}\Big[f\big<\Gamma\nabla^2\phi\big|\bv{V},\psi\big>\Big],
\end{split}
\label{eq:EulerianPDFforviscous}
\end{equation}
where it is apparent that convection, the effect of mean pressure and viscous diffusion appear mathematically exactly. On the other hand, the physical processes of dissipation of turbulent kinetic energy, pressure redistribution and the small-scale mixing of the scalar, denoted by the three conditional expectations which require modelling assumptions, can also be recognized explicitly.


\appchapter[Computation of the velocity-conditioned scalar mean]{\\Computation of the velocity-conditioned scalar mean}
\label{app:velocity-conditioning}

In Section \ref{sec:conditional_stats} a numerical strategy to estimate the velocity-conditioned scalar mean \(\mean{\phi|\bv{V}}\) required in \Eqre{eq:IECM} is detailed. An algorithm that accomplishes the conditioning step after the particles have been sorted in element \texttt{e} into subgroups may be written as follows. Let \texttt{CNBI}(=\(N_c\)), \texttt{NELEM}(=\(N_e\)), \texttt{NPAR}(=\(N_p\)) and \texttt{MAXNPEL}(=\(N^\mathrm{max}_{\scriptscriptstyle{p/e}}\)) denote the number of conditioning bins, the total number of elements of the Eulerian grid, the total number of particles and the maximum number of particles per elements, respectively. Furthermore, let the arrays \texttt{np[CNBI]}, \texttt{vcce[NELEM*CNBI]}, \texttt{npel[NELEM]}, \texttt{parid[MAXNPEL]} and \texttt{parc[NPAR]} represent the number of particles in bins, the velocity-conditioned scalar concentration in bins of each element, the actual number of particles in each element, the indices of the particles residing in element \texttt{e} and the particle concentrations, respectively. (Note the use of C-style indexing, i.e.\ the array indices start from 0. Comments are initiated by ``\textrm{//}'' and typeset in \textcolor{Gray}{gray}.)
\begin{alltt}

  \textrm{sort} parid[0:MAXNPEL-1] \textrm{according to the sorting & dividing procedure};
  \textrm{initialize} np[0:CNBI-1] = vcce[0:CNBI-1] = n = 0;
  for \textrm{all particles in element} e
    \textcolor{Gray}{\textrm{// compute bin index}}
    i = CNBI*n/npel[e];
    \textcolor{Gray}{\textrm{// increase number of particles in bin i}}
    np[i] = np[i] + 1;
    \textcolor{Gray}{\textrm{// add particle concentration to bin i}}
    vcce[e*CNBI+i] = vcce[e*CBI+i] + parc[parid[n]];
    \textcolor{Gray}{\textrm{// store conditioning pointer for particle}}
    cp[parid[n]] = i;
    \textcolor{Gray}{\textrm{// increase number of particles considered}}
    n = n + 1;
  end
  \textcolor{Gray}{\textrm{// finish computing conditional mean in bin i}}
  for \textrm{all bin} i
    vcce[e*CNBI+i] = vcce[e*CNBI+i]/np[i];
  end

\end{alltt}
After this algorithm, the array \texttt{cp[NPAR]} will contain conditioning pointers for each particle relative to their host element, so that the velocity-conditioned scalar mean \(\mean{\phi|\bv{V}}\) for particle \texttt{p} in element \texttt{e} can be obtained as \texttt{vcce[e*CNBI+cp[p]]}.%
\nomenclature[AC]{\texttt{CNBI}}{number of conditioning bins, $N_c$}%
\nomenclature[AN]{\texttt{NELEM}}{number of elements, $N_e$, \texttt{nelem}}%
\nomenclature[AN]{\texttt{NPAR}}{total number of particles, $N_p$}%
\nomenclature[AM]{\texttt{MAXNPEL}}{maximum number of particles per element, $N^\mathrm{max}_{\scriptscriptstyle{p/e}}$}%
\nomenclature[An]{\texttt{np}}{number of particles in bins}%
\nomenclature[Av]{\texttt{vcce}}{velocity-conditioned scalar concentration in bins}%
\nomenclature[An]{\texttt{npel}}{number of particles per elements}%
\nomenclature[Ap]{\texttt{parid}}{indices of particles}%
\nomenclature[Ap]{\texttt{parc}}{particle concentration}%
\nomenclature[Ac]{\texttt{cp}}{conditioning pointer of a particle}%


\appchapter[Basic particle redistribution algorithm]{\\Basic particle redistribution algorithm}
\label{app:particle-redistribution}

In Section \ref{sec:particle-number-control} the need for a particle redistribution algorithm is emphasized. What follows is such an algorithm that we employ in order to keep the number of particles/element above a certain treshold.
\begin{alltt}

 do \{
     \textrm{find the elements (}mine\textrm{,} maxe\textrm{) containing the}
       \textrm{smallest and largest number of particles (}minnpel\textrm{,} maxnpel\textrm{)};
     if \{ (minnpel < MINNPEL) and (minnpel \(\ne\) maxnpel) \}
       \textrm{move a particle from element} maxe \textrm{to} mine;
     \textrm{regenerate array }npel\textrm{ and linked lists }psel1, psel2;
    \} while \{ (minnpel < MINNPEL) and (minnpel \(\ne\) maxnpel) \};

\end{alltt}
The loop stops if the required minimum number of particles/element \texttt{MINNPEL}(=\(N^\mathrm{min}_{\scriptscriptstyle{p/e}}\)) is reached or the element-distribution of particles becomes homogeneous over the elements. Any particle may be moved from element \texttt{maxe} to \texttt{mine} as long as the local statistics are not altered. In principle, this can be achieved if the properties \((\mathcal{U}_i,\omega,\psi)\) of the newly arriving particle in element \texttt{mine} are sampled from the local joint PDF. A quick way of doing this is to initialize the particle properties by copying a randomly chosen particle already residing in element \texttt{mine}. Since the joint PDF is represented by a finite number of particles, taking out a particle from element \texttt{maxe} and putting it into \texttt{mine} will alter the local statistics in both elements, even if the new properties are copied from a neighbor. Since \texttt{maxe} contained the largest number of particles on the whole domain, we are less concerned about the effect of a single leaving particle since the local PDF is well represented there. However, the effect of the newly introduced particle in element \texttt{mine}, where the joint PDF was already poorly represented, is of higher importance. Thus in Section \ref{sec:particle-number-control} we investigate the error introduced by the above particle redistribution using a simplified governing equation.

Array \texttt{npel} stores the number of particles in each element, while the linked lists \texttt{psel1} and \texttt{psel2} stores the particle indices in each element. These arrays are regenerated after each particle moved, since finding the elements with the smallest and largest number of particles requires \texttt{npel}, while moving a particle requires a particle index from the old and the new host element.%
\nomenclature[Am]{\texttt{mine}}{the element index with the least number of particles}%
\nomenclature[Am]{\texttt{maxe}}{the element index with the most number of particles}%
\nomenclature[Am]{\texttt{minnpel}}{number of particles in the element with the least number of particles}%
\nomenclature[Am]{\texttt{maxnpel}}{number of particles in the element with the most number of particles}%
\nomenclature[Ap]{\texttt{psel[12]}}{linked lists that store the particle indices of each element}%


\appchapter[A more efficient particle redistribution algorithm]{\\A more efficient particle redistribution algorithm}
\label{app:particle-redistribution2}

\autoref{app:particle-redistribution} introduces the basic idea of the algorithm that is used to ensure enough particles in every element at all times. That algorithm is simple and robust, however it is not very efficient because of the brute-force nature of finding the elements with the smallest and largest number of particles. It executes these searches before each particle is moved, which, in principle, may not be required. Also, after it moves each particle, it regenerates the linked lists that store the indices of particles in each element (\texttt{psel1} and \texttt{psel2}) and the array that stores the number of particles per elements (\texttt{npel}). Again, although this is not strictly required after moving each particle, the simple organization of that \texttt{while} loop requires it.

A much more efficient way of performing the above redistribution is as follows. First, in a temporary array we sort the indices of the elements into the order of increasing number of particles. Now we have all the elements that contain the smallest and the largest number of particles clustered in the bottom and top part of that array. Then the redistribution step consists of a nested loop over only the critical elements (that have less particles than \texttt{MINNPEL}) with an inner loop that iterates over the number of missing particles in the given element. The body of the loop is the same as before, i.e.\ removing a particle from an element that has many particles and adding it into one that does not have enough, copying the particle properties from a random particle that already resides there. In pseudo/C code this whole procedure is written as follows. (The array \texttt{elp[NPAR]} stores the element index of each particle. Array indexing starts from 0. The structure of the linked lists \texttt{psel1} and \texttt{psel2} follows \citet{Lohner_00}. Comments are initiated by ``\textrm{//}'' and typeset in \textcolor{Gray}{gray}.)

\begin{alltt}

  \textcolor{Gray}{\textrm{// make a copy of array }npel\textrm{ and its indices to} npels\textrm{ and }npeli}
  \textcolor{Gray}{\textrm{// also count the number of critical elements (the ones that have less particles than }MINNPEL\textrm{)}}
  nce = 0;
  for ( e = 0; e < nelem; e++ )            \textcolor{Gray}{\textrm{// loop over all elements}}
  \{
   if ( npel[e] < MINNPEL ) nce = nce + 1; \textcolor{Gray}{\textrm{// add up number of critical elements}}
    npels[e] = npel[e];                    \textcolor{Gray}{\textrm{// copy array element}}
    npeli[e] = e;                          \textcolor{Gray}{\textrm{// store index}}
  \}

  \textrm{sort temporary array }npels\textrm{ and drag along the indices }npeli;

  \textcolor{Gray}{\textrm{// now we have the elements with the least number of particles at the bottom of}}
  \textcolor{Gray}{\textrm{// array }npels\textrm{ and the elements with the most number of particles at the top}}

  \textcolor{Gray}{\textrm{// redistribute particles from the top to the bottom}}
  \textcolor{Gray}{\textrm{// loop over critical elements from the bottom up until we reach }MINNPEL}
  for ( e = 0; e < nce; e++ )
  \{
    \textcolor{Gray}{\textrm{// loop over the number of missing particles in each critical element}}
    for ( p = 0; p < MINNPEL-npels[e]; p++ )
    \{
      \textcolor{Gray}{\textrm{// get element index from the top where particle will be moved from}}
      i = npeli[nelem-e-1];
      \textcolor{Gray}{\textrm{// get a particle index from the top (this one will be moved to the bottom)}}
      pi = psel1[psel2[i]+1+p];  \textcolor{Gray}{\textrm{// starting from the first one, get next}}

      \textcolor{Gray}{\textrm{// get element index in the bottom where particle will be moved to}}
      j = npeli[e];
      \textcolor{Gray}{\textrm{// get a particle index in the bottom (whose properties will be used to}}
      \textcolor{Gray}{\textrm{// initialize the newly arriwing particle)}}
      if (npels[e] > 0)          \textcolor{Gray}{\textrm{// if there is at least one particle}}
        \textcolor{Gray}{\textrm{// starting from the first get the next one, restart if exhausted}}
        pj = psel1[psel2[j]+1+(p\%npels[e])];
      else                       \textcolor{Gray}{\textrm{// if there are no particles at all}}
        pj = pi;                 \textcolor{Gray}{\textrm{// keep the properties of the newly arriving particle}}
     
      \textrm{copy particle properties from particle }pj\textrm{ to }pi;

      elp[pi] = j;               \textcolor{Gray}{\textrm{// store particle's new element number}}
    \}

    npel[i] = npel[i] - p;       \textcolor{Gray}{\textrm{// take out p particles from top element}}
    npel[j] = npel[j] + p;       \textcolor{Gray}{\textrm{// put p particles into bottom element}}
 \}

 \textrm{regenerate array }npel\textrm{ and linked lists }psel1, psel2;

\end{alltt}

The above procedure removes particles from elements at the top of array \texttt{npels} and adds them into elements at the bottom, initializing the newly arriving particle properties with one of those already in the critical element. In essence, this is the same as in \autoref{app:particle-redistribution}, but now we only operate on the elements that contain the smallest and largest number of particles. The big advantages are that now the brute-force searches are completely eliminated, we only access data which we have to modify (further reducing a large number of cache misses) and the array \texttt{npel} and linked lists \texttt{psel1} and \texttt{psel2} have to be regenerated only once and not for all particles moved. Additionally, the parallelization of the new algorithm is simpler. The brute-force searches required at least one synchronization point (when a new minimum or maximum was found and had to be updated), while parallelization of the new algorithm is trivial and requires no synchronization at all.

\begin{table}[t!]
\caption[Particle redistribution benchmarks]{\label{tab:improvepar-speedup}Timings for the two particle redistribution algorithms, described in the current chapter and in \autoref{app:particle-redistribution}. The case is the computation of the cylinder flow detailed in \autoref{chap:cylinder}, employing an Eulerian mesh of approximately 50K triangles, 2.5 million particles with requiring a minimum of 5 particles per elements at all times (\texttt{MINNPEL}=5), which results in a redistribution of 200-300 particles moved in every timestep. The data are relevant to a single timestep using 8 processor cores.\\}
\ls{1}
\begin{tabular*}{1.0\textwidth}{p{1.8cm}@{\hspace{0.75cm}}p{2.9cm}@{\hspace{0.75cm}}p{3.25cm}@{\hspace{0.75cm}}p{3.2cm}@{\hspace{0.75cm}}p{3cm}}
\hline\hline
algorithm & number of particles redistributed & clock time for redistribution only, ms & total time of a whole timestep, ms & \% \\
\hline
basic & 261 & 16\,881.6 & 18\,371 & 91.8 \\
improved & 270 & 85.8 & 1\,251 & 6.8 \\
\hline
speedup & & 196.8 & 14.7 & \\
\hline\hline
\end{tabular*}
\tableSpace
\end{table}
Simple tests indicate that this algorithm in itself is about 200 times faster than the one described in \autoref{app:particle-redistribution} using only about 2.5 million particles with 50K Eulerian elements. Table \ref{tab:improvepar-speedup} shows some timings comparing the two different algorithms computing the cylinder case, where continuous redistribution of about 200-300 particles per timestep is required after the vortex shedding has been initiated. We see that the old algorithm accounts for more than 90\% of the total running time, while the new one a mere 6.8\%, resulting in an overall speedup of the code of almost 15 times. The improvement is expected to be even more significant with larger cases, more complex flows and more processors.%
\nomenclature[Ae]{\texttt{elp}}{element indices of particles, ``element of particle''}%
\nomenclature[An]{\texttt{npels}}{number of particles per elements, temporary array}%
\nomenclature[An]{\texttt{npeli}}{temporary index array for \texttt{npels}}%
\nomenclature[An]{\texttt{nce}}{number of critical elements}%

\bibliographystyle{elsart-harv.bst}
\bibliography{jbakosi}

\cvpage

\noindent J\'ozsef Bakosi was born on September 24, 1975, in Ny\'iregyh\'aza, Hungary and completed his Masters in Mechanical Engineering at the University of Miskolc in 1999. After working for a couple of years in Hungary as a research and teaching assistant, then later as a development engineer, he did a full-time internship in precision manufacturing at the Center for Manufacturing Innovation of Fraunhofer USA Inc.\ in Boston, MA, from 2001 to 2003. He joined the Comprehensive Atmospheric Modeling Program of the School of Computational Sciences at GMU in September, 2003.

\end{document}